\documentclass{scrbook}
\KOMAoption{fontsize}{12pt}
\usepackage{tikz}
\usepackage{float} 
\usepackage{algorithmic}
\usepackage{algorithm}
\usepackage[ansinew]{inputenc} 
\usepackage{amsthm}

\usepackage{etoolbox}
\usepackage{textcomp}

\usepackage{url}

\usepackage{amsmath,amsfonts,amssymb,mathrsfs,amsthm,amsbsy,graphicx,color,bm,balance,mathtools,xfrac,nccmath,siunitx}
\usepackage{multirow}

\usepackage[titletoc]{appendix}
\usepackage{tabulary}

\newcommand{\q}[1]{{\mathbf{#1}}}
\newcommand{\mt}[1]{\mathtt{#1}}

\newcommand{\E}[1]{{\mathbb{E} \left\{ #1 \right\}}}

\newcommand{\tr}{\text{Tr}}

\renewcommand{\Re}{\text{Re}}
\newcommand{\diag}{\text{diag}}
\newcommand{\bdiag}{\text{blgdiag}}

\newcommand\blfootnote[1]{%
	\begingroup
	\renewcommand\thefootnote{}\footnote{#1}%
	\addtocounter{footnote}{-1}%
	\endgroup
}

\begin{document}
	\begin{titlepage}
		\title{Cell-Free Integrated Sensing and Communication}
		\author{Diluka  Galappaththige and Chintha Tellambura}
		

	\end{titlepage}
	\maketitle

	\blfootnote{~Diluka Galappaththige and Chintha Tellambura (2025), ``Cell-Free Integrated Sensing and Communication'', Foundations and Trends in Communications and Information Theory: Vol. xx, No. xx, pp 1-18. DOI: 10.1561/XXXXXXXXX.\\
		\textcopyright2025 D. Galappaththige and C. Tellambura}

	\tableofcontents

	\chapter*{Preface}\label{chp_preface}

Cell-free (CF) integrated sensing and communication (ISAC) merges the CF architecture with ISAC functionalities. CF-ISAC  leverages distributed access points, removes cell boundaries, and enhances coverage, spectral efficiency, and reliability. It also improves energy efficiency, enabling robust multi-user communication, distributed multi-static sensing, and seamless resource optimization.

A comprehensive survey on CF-ISAC has been lacking. This monograph addresses that gap by covering the foundational principles, cooperative transmission, radar cross-section, target parameter estimation, ISAC integration levels, sensing metrics, and key applications. It also explores the advantages of multi-static sensing.

Performance analysis, resource allocation, security, and user/target-centric designs are discussed. Finally,  synchronization, multi-target detection, interference management, and fronthaul limitations are discussed. Advanced antenna technologies, network-assisted systems, near-field CF-ISAC, cross-technology integration, and machine learning approaches are presented. 

\newpage 
\section*{Acknowledgment}
The preparation of this book has been both a challenging and rewarding endeavor, requiring sustained effort over several months. The authors are deeply indebted to all those who have provided guidance, encouragement, and support throughout this process.

The ideas and perspectives developed here have also been enriched through collaborations and scientific discussions with Dr. Shayan Zargari, Dr. Mohammadali Mohammadi, and Prof. Gayan Aruma Baduge. Their contributions to our joint research, as well as their constructive insights, have significantly influenced the development of the concepts and results presented in this volume.

We are grateful to the University of Alberta, and in particular the Department of Electrical and Computer Engineering, for providing an excellent academic environment and the necessary resources. 

Finally, the first author would like to extend his most profound appreciation to his mother (Swarna Solomans), his wife (Dilini Meddegoda), and Mr. Kelum Saumyasiri. Their unwavering support, patience, and encouragement have been a constant source of strength throughout his academic journey. 

The second author would like to dedicate this work to his spouse, Kyoko, and our children, Kenta and Hannah, for their constant love. \\

\rightline{Diluka Galappaththige}
\rightline{Chintha Tellambura} 
	\chapter{Introduction}\label{chp_intro}
Integrated Sensing and Communication (ISAC) is an emerging paradigm that unifies wireless communication and radar sensing within a single system. By sharing the same waveform, spectrum, and hardware, ISAC enables simultaneous data transmission and environmental sensing. This integration enhances spectral efficiency, reduces hardware redundancy, and positions ISAC as a key enabler for next-generation intelligent wireless systems.

ISAC is expected to be a key component of future wireless standards \cite{Liu2022ISAC, Wang2022ISAC, Zhang2022, Azar2024, liu2023integratedbook}. The integration allows the base station (BS) to sense and analyze its surroundings and transmit information. Sensing refers to the detection, localization, and tracking of objects or environmental features within the BS coverage area by utilizing transmitted signals to gather information about targets, such as range, velocity, angle of arrival (AoA), shape, material composition, or orientation \cite{Liu2022ISAC, Wang2022ISAC, Zhang2022, Azar2024, liu2023integratedbook}. 

\subsection{Key Features}
The ISAC architecture enhances spectrum efficiency (SE) and energy efficiency (EE), reduces hardware redundancy, and tightens the integration of wireless communication and situational awareness \cite{Liu2022ISAC, Wang2022ISAC, Zhang2022, Azar2024, liu2023integratedbook}. This unification is no longer just a theoretical possibility; recent breakthroughs in signal processing, machine learning (ML), antenna technologies, and real-time computing have made it a reality \cite{Liu2022ISAC, Wang2022ISAC, Zhang2022, Azar2024, liu2023integratedbook}. As wireless networks evolve to accommodate massive connectivity and low-latency applications in dynamic environments, the integration of sensing capabilities becomes increasingly essential \cite{Liu2022ISAC, Wang2022ISAC, Zhang2022, Azar2024, liu2023integratedbook}. Sensing provides the network with real-time knowledge of user locations, mobility patterns, and radio propagation conditions. This awareness enables dynamic resource allocation, proactive interference management, and rapid beam alignment, which are critical for supporting large numbers of devices simultaneously while minimizing delay \cite{Liu2022ISAC, Wang2022ISAC, Zhang2022, Azar2024, liu2023integratedbook}. Consequently, sensing not only improves reliability in dense deployments but also shortens response times in mission-critical scenarios \cite{Liu2022ISAC, Wang2022ISAC, Zhang2022, Azar2024, liu2023integratedbook}.

The key features of ISAC are as follows: 
\begin{itemize}
    \item \textbf{Shared Waveform:} ISAC systems use the same signal to transmit data and to sense the environment.
    \item \textbf{Hardware Efficiency:} By reusing antennas, radio frequency (RF) chains, and processing units, ISAC reduces system cost, complexity, and energy consumption.
    \item \textbf{Spectrum Sharing:} Communication and sensing coexist in the same frequency band, improving spectral efficiency.
\end{itemize}
The core motivation for ISAC lies in its ability to enhance spectrum utilization and system-level efficiency \cite{Liu2022ISAC, Wang2022ISAC, Zhang2022, Azar2024, liu2023integratedbook}. Traditionally, communication and sensing require separate spectral bands, hardware, and signal processing chains. ISAC eliminates this separation by enabling the co-design of these functions, allowing shared use of antennas, transceivers, spectrum, and power resources. For example, a multi-antenna BS can simultaneously transmit data and sense the environment using the same waveform. The BS can also estimate channels from reflected signals to optimize beamforming. Sensing outputs, such as target location or motion, can further enhance communication functions like user tracking and link adaptation \cite{Kumari2018, Lu2024}. For instance, IEEE 802.11ad-based vehicular systems use radar-like sensing to improve beam alignment and reduce latency in vehicle-to-everything (V2X) communications \cite{Kumari2018}. Moreover, human activity detection can assist in adaptive beamforming and dynamic link scheduling, demonstrating that sensing data can directly improve communication performance \cite{Lu2024}.
By jointly designing waveforms, beamforming, and resource allocation, ISAC mitigates co-channel interference and unlocks new degrees of freedom (DoF), improving overall system performance.

\subsection{Applications}
ISAC is a core component of future wireless systems, including:
\begin{itemize}
    \item \textbf{Autonomous Vehicles:} Simultaneous communication with other vehicles and object detection via sensing.
    \item \textbf{Smart Cities:} Traffic monitoring, public safety, and environmental sensing integrated with wireless connectivity.
    \item \textbf{6G Networks:} ISAC is a foundational technology envisioned for native support in 6G.
    \item \textbf{Military Systems:} Joint data transmission and target detection using shared infrastructure.
\end{itemize}This approach supports a wide variety of emerging use cases, including the Internet of Things (IoT), V2X communication, human activity recognition, industrial automation, autonomous navigation, smart infrastructure, immersive extended reality (XR) applications, and more \cite{Liu2022ISAC, Wang2022ISAC, Zhang2022, Azar2024, liu2023integratedbook}.

\subsection{Technical Challenges}
ISAC systems must balance conflicting performance goals:
\begin{itemize}
    \item \textbf{High Data Rate:} Essential for communication services.
    \item \textbf{High Sensing Accuracy:} Required for precise environmental awareness.
\end{itemize}

\begin{figure}[!t]\vspace{-0mm}
    \centering
    \includegraphics[width=0.7\textwidth]{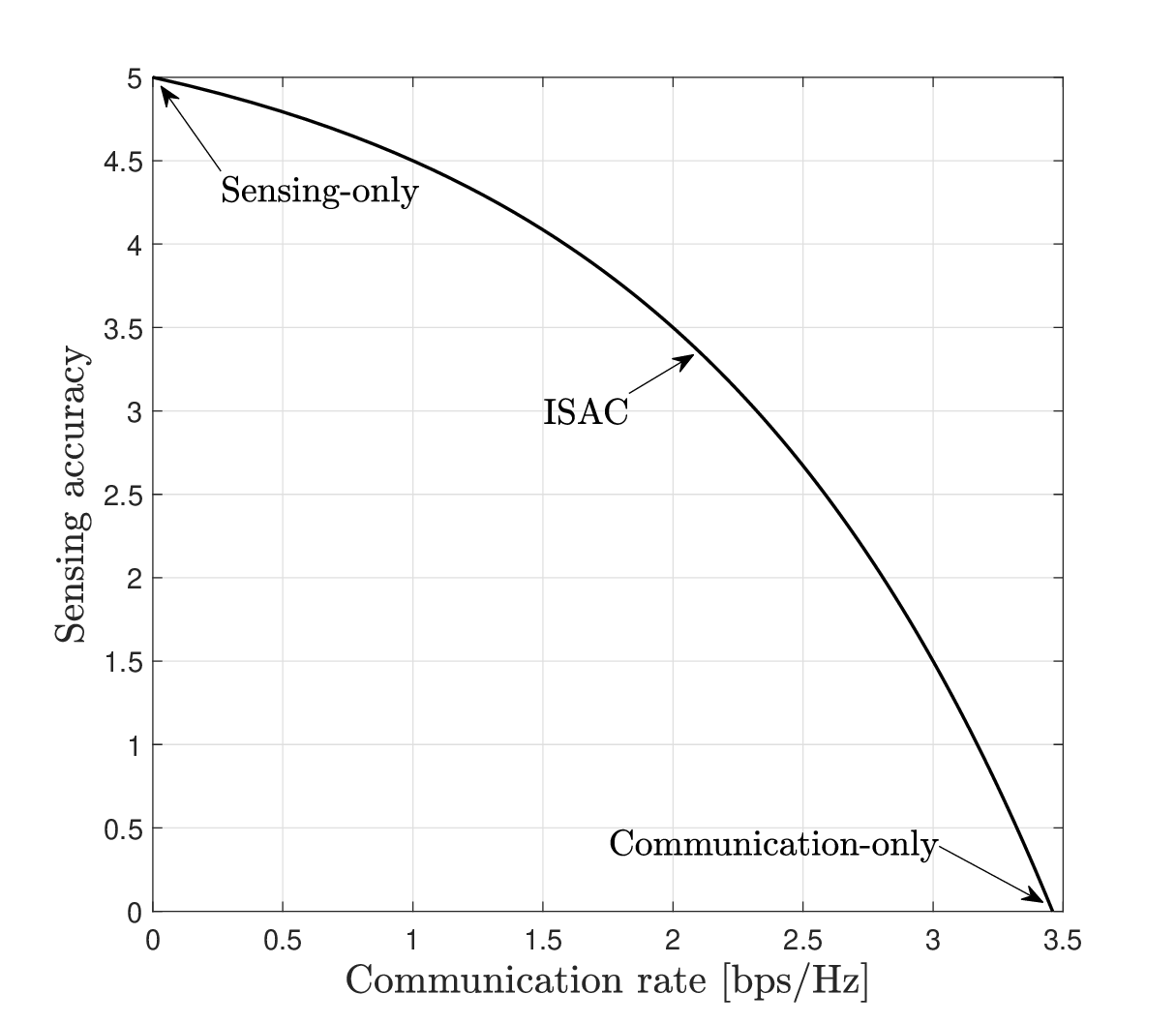}
    \vspace{-0mm}
    \caption{Pareto boundary between the communication and sensing performance in an ISAC system under a fixed BS transmit power constraint.}
    \label{fig_ParetoBoundary} \vspace{-0mm}
\end{figure}
This dual functionality introduces a fundamental trade-off in resource allocation, particularly under constraints such as limited transmit power, bandwidth, and time. A practical ISAC system must jointly optimize its waveform, beamforming, and power distribution to serve both communication and sensing objectives.

This trade-off can be characterized using a Pareto boundary analysis. Consider a simplified model with a single user, single target, and a BS, where the total BS transmit power ($p_t$) is fixed, and a portion of it is allocated to communication ($p_c$) and the remainder to sensing ($p_s = p_t-p_c$).  Let the communication performance be defined as the data rate, i.e., $\mathcal{R} = \log_2(1 + p_c\alpha)$, and the sensing performance be defined as a linear approximation for detection accuracy, i.e., $\mathcal{S} = p_s \beta$, where $\alpha$ accounts for the normalized communication channel gain and $\beta$ models the normalized sensing channel gain and radar cross section (RCS) of the target relative to noise. The sensing metrics used in ISAC systems will be discussed in Chapter~\ref{chp_isac}. The resulting multi-objective optimization problem can be given as
\begin{subequations}\label{prob_P1}
\begin{eqnarray}
(\mathcal{P}):~&& \max_{p_c} \quad [\log_2(1 + p_c\alpha),  (p_t-p_c)\beta],  \\
\text{s.t.} \,\,\, &&   0\leq p_c \leq p_t.
\end{eqnarray}
\end{subequations} \par \vspace{-0mm}
\noindent This problem is convex for the power allocation variable and can be efficiently solved using convex optimization tools such as CVX in MATLAB \cite{grant2014cvx}.

Figure~\ref{fig_ParetoBoundary} illustrates the Pareto boundary for an ISAC system, highlighting the fundamental trade-off between communication rate and sensing (detection) performance under a fixed BS transmit power constraint. As more transmit power is allocated to enhance communication rate, sensing accuracy degrades, and vice versa. This intrinsic conflict highlights a critical technological difficulty in ISAC system design, i.e., striking an ideal balance in which neither functionality is significantly compromised. The figure illustrates that improving one function comes at the expense of the other, highlighting the need for joint optimization strategies tailored to application-specific requirements.

ISAC research has mainly focused on co-located multiple-input multiple-output (MIMO) or massive MIMO (mMIMO) cellular systems, where one or two BSs perform mono-static or bi-static sensing in addition to communication \cite{Liu2022ISAC, Wang2022ISAC, Zhang2022, Azar2024, liu2023integratedbook}. These systems with mono-static and bi-static sensing, while conceptually straightforward, often suffer from limited spatial coverage and sensing diversity. On the other hand, in dense urban environments, indoor settings, or regions with heavy scattering and obstructions, a single BS may not have a clear line of sight (LoS) to either the communication users or the sensing targets \cite{Mao2023, Demirhan2023, Huang2022Coordinated, Cao2023Design, Wang2023, Sakhnini2022Uplink, Silva2023, Behdad2022, Behdad2024Interplay}. Additionally, in practice, multiple BSs will operate in the same geographical region, sharing the frequency and timing resources. Consequently, performance degradation is often observed, especially at the cell-edge users/targets, where weak signal strength, path loss, frequent cell switching, and interference impair both communication reliability and sensing accuracy \cite{Mao2023, Demirhan2023, Huang2022Coordinated, Cao2023Design, Wang2023, Sakhnini2022Uplink, Silva2023, Behdad2022, Behdad2024Interplay}.

For a summary of the technical details of ISAC, please see Chapter 3.  

\section{Cell-free (CF) integrated sensing and communication (CF-ISAC)}

These limitations have motivated cooperative and distributed architectures, culminating in cell-free (CF) integrated sensing and communication (CF-ISAC)  \cite{Mao2023, Demirhan2023, Huang2022Coordinated, Cao2023Design, Wang2023, Sakhnini2022Uplink, Silva2023, Behdad2022, Behdad2024Interplay}. CF-ISAC extends ISAC from a centralized paradigm to a distributed one, where multiple spatially separated access points (APs) jointly serve all users and detect environmental targets within the network area \cite{Mao2023, Demirhan2023, Huang2022Coordinated, Cao2023Design, Wang2023, Sakhnini2022Uplink, Silva2023, Behdad2022, Behdad2024Interplay}. In contrast to conventional cellular topologies, CF-ISAC removes the notion of fixed cells and allows user/target-centric associations with all surrounding APs. This offers seamless connectivity, reduces inter-cell interference, and enables high-resolution sensing even in highly dynamic or obstructed environments \cite{Mao2023, Demirhan2023, Huang2022Coordinated, Cao2023Design, Wang2023, Sakhnini2022Uplink, Silva2023, Behdad2022, Behdad2024Interplay}. On the other hand, recent advances in multi-BS/AP communication cooperation, such as coordinated multi-point (CoMP) transmission/reception, cloud-radio access networks, distributed MIMO radar sensing, and networked ISAC, lay the groundwork for addressing the aforementioned challenges in conventional ISAC systems \cite{Gesbert2010, Dahrouj2010, Jun2015, Ngo2017, Fishler2004, Haimovich2008}. 
 
A defining feature of CF-ISAC is its ability to perform multi-static sensing, where transmitters and receivers are spatially distributed across the network. Unlike mono-static or bi-static systems that rely on limited geometric diversity, multi-static configurations leverage multiple, spatially diverse propagation paths to detect, track, and localize targets with enhanced robustness and precision \cite{richards2005fundamentals}. The diversity gain from uncorrelated sensing channels leads to better target detection under multipath fading. Moreover, distributed beamforming and joint processing across APs can provide enhanced signal gains and greater coverage consistency, especially for weak or distant targets \cite{Mao2023, Demirhan2023, Huang2022Coordinated, Cao2023Design, Wang2023, Sakhnini2022Uplink, Silva2023, Behdad2022, Behdad2024Interplay}. However, CF-ISAC also faces challenges such as synchronization, complex signal processing, and deployment costs.

Nevertheless, integrating CF and ISAC is a logical and timely evolution, aligning with the broader vision of distributed intelligence, cooperative networks, and the fusion of communication and environmental awareness \cite{Mao2023, Demirhan2023, Huang2022Coordinated, Cao2023Design, Wang2023, Sakhnini2022Uplink, Silva2023, Behdad2022, Behdad2024Interplay}. Hence, CF-ISAC offers a scalable and resilient roadmap for future wireless systems. \cite{Mao2023, Demirhan2023, Huang2022Coordinated, Cao2023Design, Wang2023, Sakhnini2022Uplink, Silva2023, Behdad2022, Behdad2024Interplay}. Further details of CF-ISAC will be provided in Chapter 5. 

\begin{figure}[!t]\vspace{-0mm}
    \centering
    \includegraphics[width=1.0\linewidth,trim=2 2 2 2,clip]{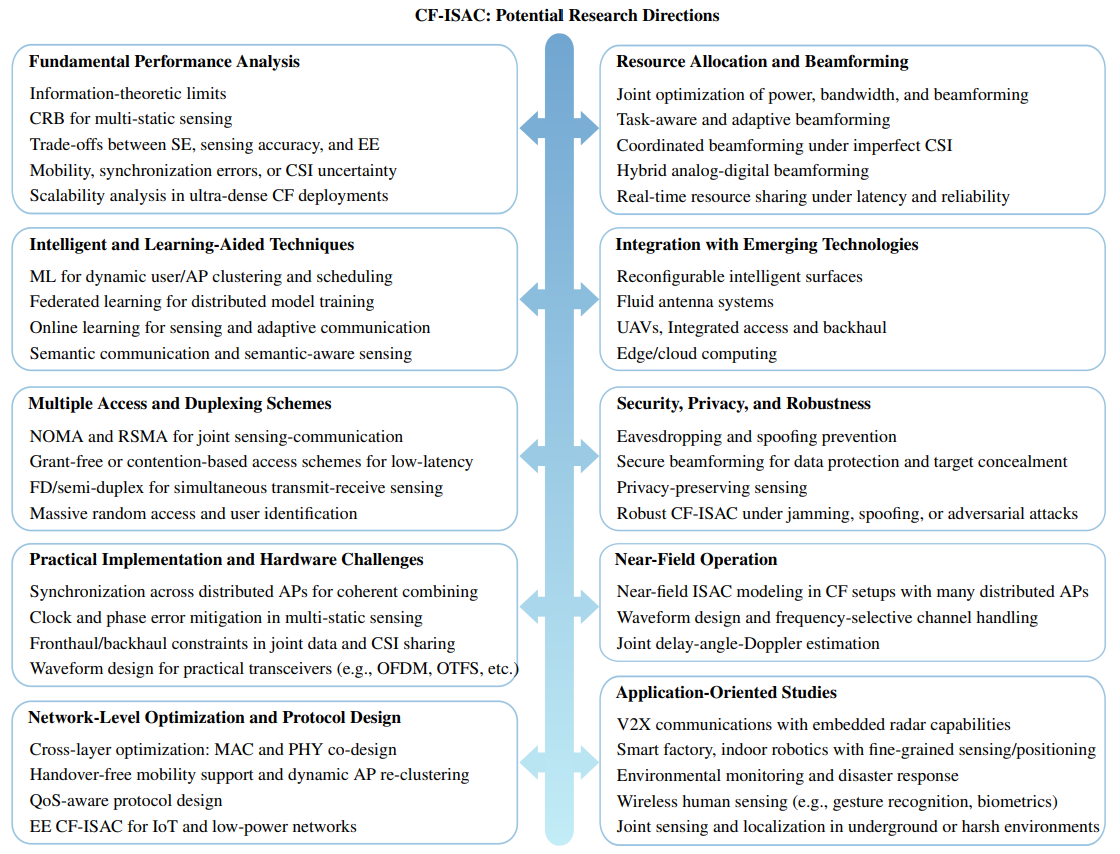}
    \vspace{-0mm}
    \caption{Potential research directions of CF-ISAC for supporting future wireless networks.}
    \label{fig_OutlineV4} \vspace{-0mm}
\end{figure}

Figure~\ref{fig_OutlineV4} illustrates a wide range of potential and largely unexplored research directions in CF-ISAC networks. These directions span from fundamental performance analysis and joint resource allocation strategies to advanced topics such as intelligent and learning-aided optimization, integration with emerging technologies (e.g., RIS, fluid antennas, semantic communication), and innovative multiple access and duplexing schemes (e.g., non-orthogonal multiple access (NOMA), rate-splitting multiple access (RSMA), frequency Division Multiple Access (FDMA)). Other critical areas include network-level design and protocol development, secure and privacy-preserving operations, near-field and wideband challenges, and application-driven system architectures for environments such as vehicular networks and industrial automation. While this monograph primarily focuses on performance characterization, beamforming, resource and interference management, as well as security and network-assisted scalability, the figure highlights the scope of opportunities that remain open for future exploration in the CF-ISAC paradigm.

\section{Roadmap of this Monograph}
Chapter \ref{chp_CF} introduces the fundamentals of CF mMIMO (CFMM) systems. Chapter \ref{chp_radar} explores key concepts in radar sensing, including signal transmission, system configurations, and target parameter estimation techniques. Chapter \ref{chp_isac} reviews the principles of conventional ISAC systems along with recent advancements. Chapter \ref{chp_CF_isac} provides an in-depth overview of CF-ISAC systems, emphasizing their core concepts, distinctive features, and benefits. In Chapter \ref{chp_state}, the current technical contributions are presented. Chapters \ref{chp_CF_isac_perfor} to \ref{chp_NA_CF_isac} offer case studies on CF-ISAC systems, focusing on performance analysis, beamforming design, secure system design, and network-assisted (NA) strategies, respectively. Chapter \ref{chp_key_challenges} highlights the main challenges in CF-ISAC implementation. Finally, Chapter \ref{chp_future} outlines potential future research directions and opportunities. 
	\chapter{Fundamentals of Cell-Free Massive MIMO}\label{chp_CF} 
Before delving into CF-ISAC, this chapter introduces the fundamentals of CFMM.  It highlights the motivations for its adoption and key features such as the elimination of cell boundaries, cooperative transmission and reception, synchronization, fronthaul infrastructure, and the advantages of multiple antenna effects. These foundations illustrate how CFMM enables more uniform service quality, improved spectral and energy efficiency, and greater robustness against interference, positioning it as a strong candidate for future wireless communication and sensing systems.

\section{Distributed Access Points} 
\begin{figure}[!t]\vspace{0mm}
    \centering 
    \def\svgwidth{220pt} 
    \fontsize{8}{8}\selectfont 
    \graphicspath{{Figures/}}
   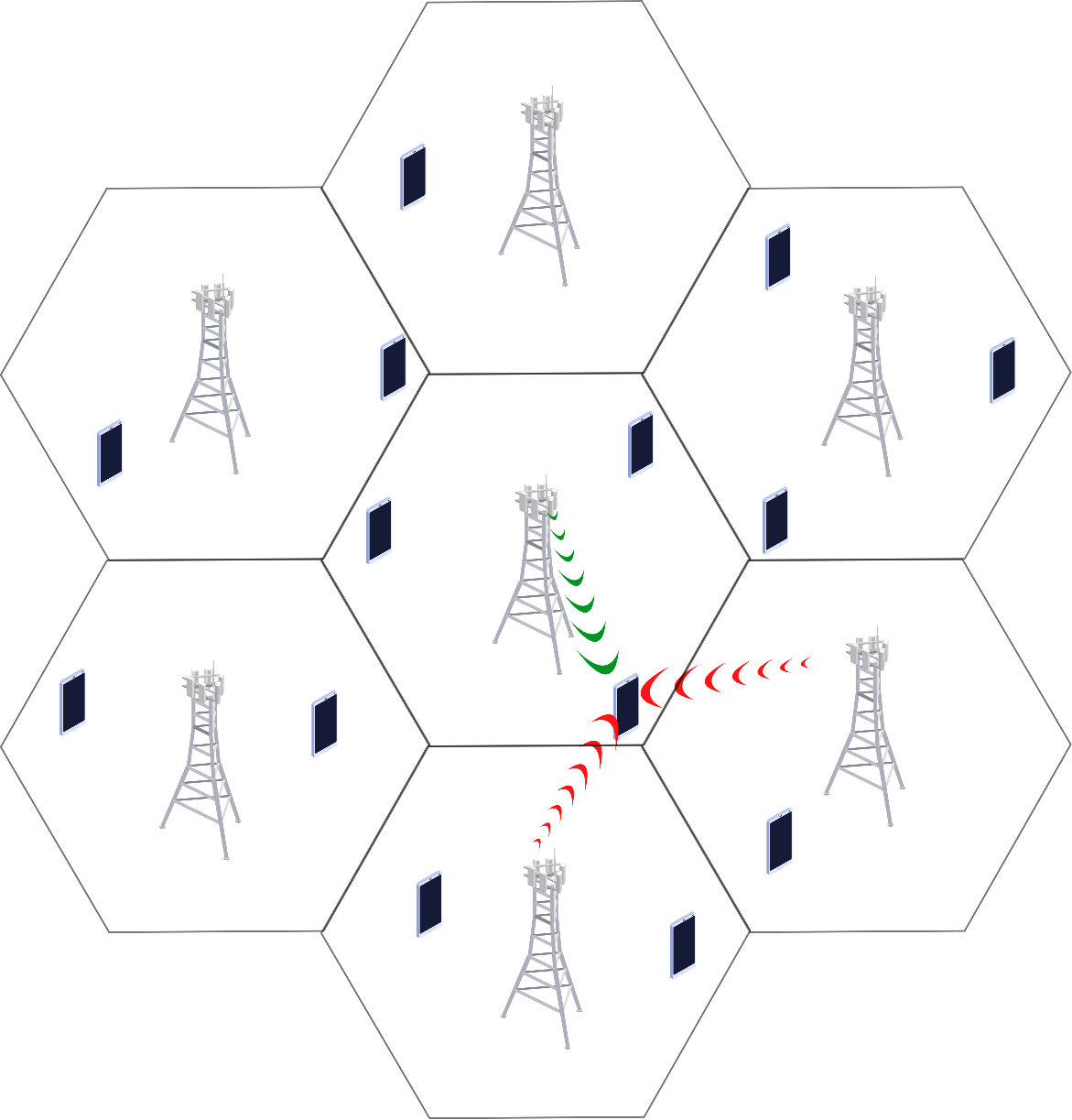 \vspace{0mm}
    \caption{A cell-based cellular system.}\vspace{0mm} \label{fig_Colocated_network}
\end{figure}

\begin{figure}[!t]\vspace{0mm}
    \centering 
    \def\svgwidth{420pt} 
    \fontsize{8}{8}\selectfont 
    \graphicspath{{Figures/}}
\begingroup%
  \makeatletter%
  \providecommand\color[2][]{%
    \errmessage{(Inkscape) Color is used for the text in Inkscape, but the package 'color.sty' is not loaded}%
    \renewcommand\color[2][]{}%
  }%
  \providecommand\transparent[1]{%
    \errmessage{(Inkscape) Transparency is used (non-zero) for the text in Inkscape, but the package 'transparent.sty' is not loaded}%
    \renewcommand\transparent[1]{}%
  }%
  \providecommand\rotatebox[2]{#2}%
  \newcommand*\fsize{\dimexpr\f@size pt\relax}%
  \newcommand*\lineheight[1]{\fontsize{\fsize}{#1\fsize}\selectfont}%
  \ifx\svgwidth\undefined%
    \setlength{\unitlength}{1097.15466309bp}%
    \ifx\svgscale\undefined%
      \relax%
    \else%
      \setlength{\unitlength}{\unitlength * \real{\svgscale}}%
    \fi%
  \else%
    \setlength{\unitlength}{\svgwidth}%
  \fi%
  \global\let\svgwidth\undefined%
  \global\let\svgscale\undefined%
  \makeatother%
  \begin{picture}(1,0.31455419)%
    \lineheight{1}%
    \setlength\tabcolsep{0pt}%
    \put(0,0){\includegraphics[width=\unitlength]{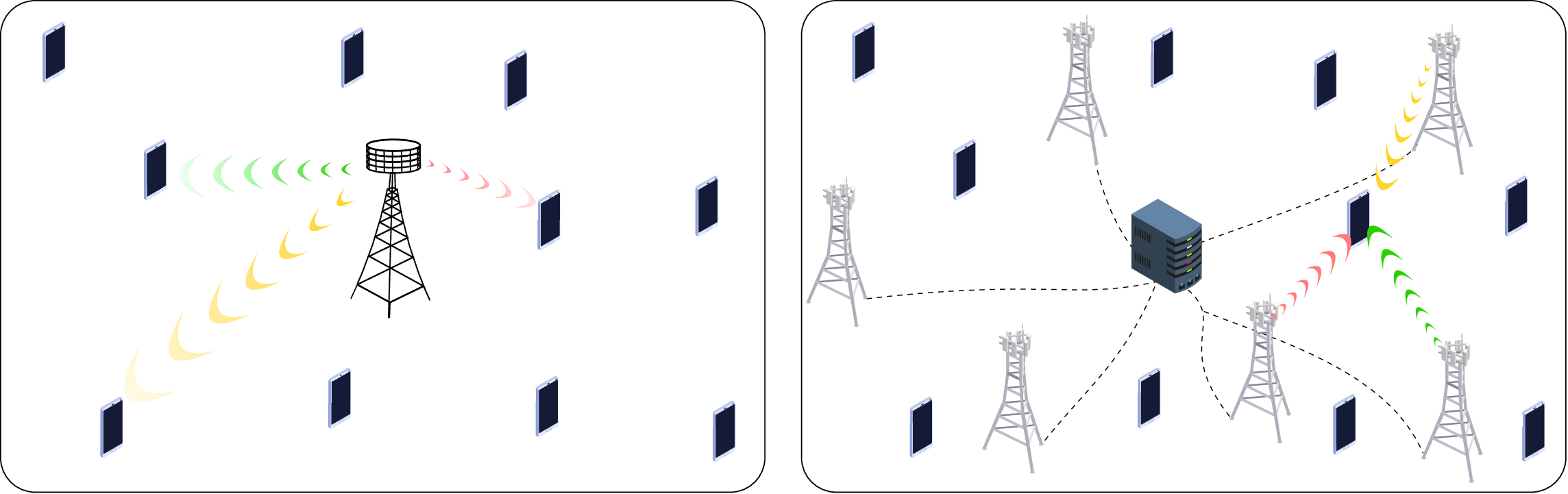}}%
    \put(0.24139406,0.09740737){\color[rgb]{0,0,0}\makebox(0,0)[lt]{\lineheight{1.25}\smash{\begin{tabular}[t]{l}BS\end{tabular}}}}%
    \put(0.57598611,0.1345756){\color[rgb]{0,0,0}\makebox(0,0)[lt]{\lineheight{1.25}\smash{\begin{tabular}[t]{l}Front-haul link\end{tabular}}}}%
    \put(0.52919092,0.09068363){\color[rgb]{0,0,0}\makebox(0,0)[lt]{\lineheight{1.25}\smash{\begin{tabular}[t]{l}AP\end{tabular}}}}%
    \put(0.72782665,0.19220303){\color[rgb]{0,0,0}\makebox(0,0)[lt]{\lineheight{1.25}\smash{\begin{tabular}[t]{l}CPU\end{tabular}}}}%
    \put(0.71635715,0.0255973){\color[rgb]{0,0,0}\makebox(0,0)[lt]{\lineheight{1.25}\smash{\begin{tabular}[t]{l}User\end{tabular}}}}%
    \put(0.16965427,-0.01688315){\color[rgb]{0,0,0}\makebox(0,0)[lt]{\lineheight{1.25}\smash{\begin{tabular}[t]{l}(a) Co-located mMIMO\end{tabular}}}}%
    \put(0.70421957,-0.01688315){\color[rgb]{0,0,0}\makebox(0,0)[lt]{\lineheight{1.25}\smash{\begin{tabular}[t]{l}(b) CFMM\end{tabular}}}}%
  \end{picture}%
\endgroup%
 \vspace{0mm}
    \caption{An architectural comparison between co-located mMIMO and CFMM systems.}\vspace{0mm} \label{fig_CF_network}
\end{figure}
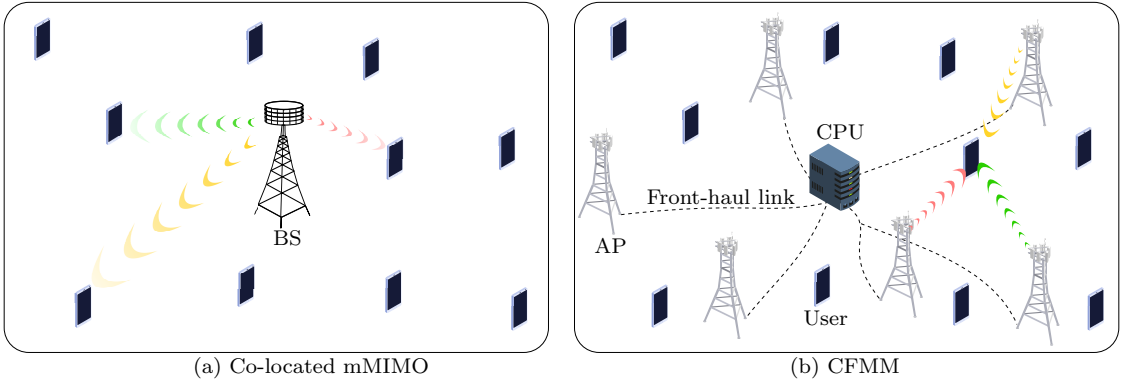

Traditional cellular mobile networks divide the coverage area into multiple geographically defined cells, each served by a BS equipped with a large antenna array (see Fig~\ref{fig_Colocated_network}). These BSs centrally manage communication with users located within their respective cell boundaries. However, such cell-based architectures face several limitations, including inter-cell interference, uneven load distribution, and frequent handovers as users move across cell boundaries. To address these challenges, mMIMO systems have been introduced, where each BS is equipped with a large number of antennas (usually \num{64} or more) to serve many users simultaneously through spatial multiplexing (see Figure~\ref{fig_CF_network}a). This enhances spectral efficiency (SE) and reliability within each cell.

However, CFMM replaces this traditional cellular architecture, introducing a user-centric, collaborative approach to connectivity \cite{Demir2021book, Ngo2017, Zhang2019cellfree, Diluka2019, Galappaththige2021, Diluka2020, Galappaththige2024, Galappaththige2021Cellfree, Diluka2021}. In contrast to traditional co-located mMIMO systems, where each BS centrally serves users within a defined cell boundary, CFMM relies on many spatially distributed APs. These APs are connected to a CPU and jointly serve all users in the network without cell boundaries (see Figure~\ref{fig_CF_network}b).

This distributed and CF architecture delivers several key advantages over traditional cellular architectures, including improved SE, energy efficiency (EE), uniform quality of service across the coverage area, and increased resilience to interference, path loss, and shadowing~\cite{Ngo:TGCN:2018}. By decoupling the association between users and fixed BSs, CFMM improves user fairness and performance at the network edge, making it an especially appealing solution for ISAC with multi-static sensing applications that require reliable service and spatial diversity.

\section{Lack of  Cell Boundaries}

Since there are no cell borders, multiple APs collaboratively cover a large geographic area without dividing it into discrete cells \cite{Demir2021book, Ngo2017, Zhang2019cellfree, Diluka2019, Galappaththige2021, Diluka2020, Galappaththige2024, Galappaththige2021Cellfree, Diluka2021}. The network operates as a cooperative infrastructure, where all APs jointly serve all users, regardless of location. This enables uniform service quality and system performance independent of a user's proximity to any specific AP.

The absence of cell boundaries eliminates the need for handovers, which can introduce latency and service interruptions in traditional systems. Instead, users experience seamless connectivity through coordinated transmission and reception by distributed APs \cite{Demir2021book, Ngo2017, Zhang2019cellfree, Diluka2024CFBiBC}. This results in lower latency, reduced signaling overhead, and enhanced mobility support. Additionally, CFMM exploits spatial diversity by enabling multiple nearby APs to serve each user simultaneously, reducing average path loss and improving reliability, especially in challenging environments with shadowing, blockages, or severe propagation conditions \cite{Demir2021book, Ngo2017, Zhang2019cellfree, Diluka2024CFBiBC}.

Nevertheless, these advantages come with unique challenges. First, CFMM introduces substantial fronthaul signaling requirements for channel state information (CSI) acquisition and user data exchange among APs, leading to higher communication overhead. Second, the processing complexity increases as coordination among many APs and users becomes computationally intensive. For instance, fronthaul bandwidth demands and processing load scale linearly (or faster) with the number of APs and served users \cite{Emil:TCOM:2020, Parida2023}.

Despite these challenges, CFMM significantly outperforms conventional alternatives such as distributed antenna systems (DAS) and coordinated multipoint (CoMP). DAS typically involves static, loosely coordinated clusters, while CoMP suffers from inter-cluster interference and limited dynamic cooperation \cite{Elhoshy2016, Irmer2011, Venkatesan2007, Simeone2008}. In contrast, CFMM enables full-network coordination with dynamic AP-user associations, achieving substantial performance gains, especially in dense, heterogeneous, or highly mobile user environments \cite{Demir2021book, Ngo2017, Zhang2019cellfree}.

\section{Coherent Joint Transmission and Synchronization} 
This involves the joint operation of spatially distributed APs, coordinated by a central processing unit (CPU), to serve all users simultaneously \cite{Demir2021book, Ngo2017, Zhang2019cellfree}. In cooperative operation, the APs share CSI and user data with the CPU via fronthaul links. The CPU then performs centralized signal processing, such as precoding and power allocation, and instructs the APs to transmit phase-aligned signals. This ensures coherent joint transmission, effectively eliminating inter-cell interference and enhancing SE, EE, and service fairness, particularly in dynamic and densely populated environments \cite{Demir2021book, Ngo2017, Zhang2019cellfree}.

Precise synchronization between distributed APs is the lynchpin of cooperative operation \cite{Demir2021book, Ngo2017, Zhang2019cellfree}. Specifically, time, frequency, and phase synchronization are essential \cite{Demir2021book, Ngo2017, Zhang2019cellfree}.
\begin{itemize}
    \item \textit{Time synchronization:} This ensures that all APs transmit and receive signals on a common time grid \cite{Demir2021book, Ngo2017, Zhang2019cellfree}. In downlink (DL) transmission, synchronized timing guarantees that signals from different APs arrive at the users simultaneously, enabling constructive combining and avoiding inter-symbol interference. In TDD systems, tight time alignment is also critical to preserve UL-DL channel reciprocity, which is foundational for accurate channel estimation and efficient precoding in cell-free massive MIMO \cite{Demir2021book, Ngo2017, Zhang2019cellfree,8957510}.
    
    \item \textit{Frequency synchronization:} This prevents carrier frequency offsets between APs that could cause inter-carrier interference, especially in orthogonal frequency-division multiplexing (OFDM)-based CFMM systems \cite{Demir2021book, Ngo2017, Zhang2019cellfree}. Frequency offsets must be corrected using techniques like centralized reference signals or distributed phase-locked loops (PLLs).

    \item \textit{Phase synchronization:}
    The phases of multiple AP transmit signals must be aligned, which is critical for coherent beamforming in the DL. Without phase coherence, these signals may add destructively at the user, drastically reducing array gain. Thus, signal phases must be accurately matched to maximize constructive interference at the users, thereby enhancing signal strength and system performance \cite{Demir2021book, Ngo2017, Zhang2019cellfree}. Phase alignment is typically achieved through GPS-disciplined oscillators or fronthaul-based synchronization protocols.
\end{itemize}

On the other hand, real-world impairments, such as oscillator drift, phase noise, and propagation delay variations, introduce residual synchronization errors, which must be mitigated via periodic calibration and phase tracking algorithms \cite{Demir2021book, Ngo2017}.

\section{CPU Coordination and CSI-Based Operation}
The CPU serves as the centralized intelligence hub of the network, managing coordination among APs and ensuring coherent multi-user transmission and reception \cite{Demir2021book,  Ngo2017, Zhang2019cellfree}. It collects globally or locally estimated CSI from each AP to perform centralized beamforming, resource allocation, and other tasks. Unlike conventional cellular systems, where inter-cell interference limits performance, this coordinated framework minimizes interference while maximizing SE and EE and optimizing the network performance. Key functions of the CPU include:
\begin{itemize}
    \item \textit{CSI acquisition:} 
    Each AP locally estimates CSI for its served users, typically using UL pilots under time-division duplexing (TDD), and forwards it to the CPU. The CPU aggregates this CSI to construct DL precoding, UL signal detection, and user-specific scheduling. This ensures that APs collaboratively transmit coherent signals, reducing interference and improving SE \cite{Demir2021book,  Ngo2017, Zhang2019cellfree}. 
    
    \item \textit{Centralized processing:} In the DL, the CPU computes precoding vectors such as zero-forcing (ZF) or minimum mean square error (MMSE) beamforming vectors that exploit the joint CSI across APs. These precoders ensure that signals from multiple APs constructively combine at the intended users while nullifying interference at others \cite{Demir2021book,  Ngo2017, Zhang2019cellfree}. In the UL, the CPU performs joint decoding, applying combining strategies like maximum ratio combining (MRC) or MMSE to enhance SINR using macro-diversity gain \cite{Demir2021book,  Ngo2017, Zhang2019cellfree}.

    \item \textit{Dynamic resource allocation:} 
    The CPU executes dynamic user scheduling and allocates power, bandwidth, and time resources based on traffic demand, CSI quality, and user priority \cite{Demir2021book,  Ngo2017, Zhang2019cellfree}. Power control algorithms (e.g., max-min fairness or weighted sum-rate optimization) help balance network-wide SE and EE objectives while enforcing quality-of-service (QoS) constraints for all users, including those at cell edges \cite{Demir2021book,  Ngo2017, Zhang2019cellfree,9024101,9978919}.
\end{itemize}

However, this centralized coordination requires precise synchronization and a fronthaul infrastructure.

\section{Fronthaul Networks} 
Fronthaul networks interconnect APs with the CPU, enabling data and control signal exchange necessary for centralized processing, i.e., cooperative operation requires a robust fronthaul network \cite{Demir2021book, Ngo2017, Zhang2019cellfree}. In dense networks, frequent CSI, user data, and control exchanges generate significant fronthaul traffic. Thus, the fronthaul network must support low latency, high reliability, and significant throughput to handle (i) forwarded CSI and pilot signals from APs to the CPU, (ii) user data for DL precoding and UL decoding, and (iii) control information for synchronization and scheduling \cite{Demir2021book, Ngo2017, Zhang2019cellfree}.

Fronthaul architectures can be broadly categorized into centralized and distributed systems \cite{intel2021fronthaul}. In centralized architectures (e.g., Cloud-RAN), all baseband processing occurs at the CPU. This simplifies coordination and global optimization but places stringent requirements on fronthaul bandwidth and latency, especially in mMIMO settings \cite{Peng2015}. In contrast, distributed architectures (e.g., Fog-RAN or Edge-RAN) offload part of the processing to the APs. This reduces fronthaul load but requires distributed CSI processing and synchronization, increasing edge complexity and coordination effort \cite{Mao2017}.

To address fronthaul constraints, several optimization strategies are employed:
\begin{enumerate}
    \item \textit{Compression techniques:} 
    Efficient quantization and compression of CSI and user data can reduce bandwidth usage. Examples include vector quantization, Wyner-Ziv coding, or sparsity-based compression aligned with compressed sensing theory \cite{intel2021fronthaul}.
    
    \item \textit{Functional splitting:} 
    The processing burden is adaptively split between the CPU and APs (i.e., functional splits) based on current traffic load, latency budgets, and CSI dynamics. For instance, precoding may be centralized, while decoding is performed at the AP level for latency-sensitive UL traffic.
    
    \item \textit{Resource allocation:} 
    Efficient scheduling of fronthaul resources ensures that high-priority users (e.g., with low-latency requirements) receive timely updates. Techniques include traffic shaping, prioritization of CSI updates, and queue-aware scheduling.
    
    \item \textit{Hybrid fronthaul solutions:} 
    A combination of fiber (for high capacity) and wireless (for flexibility) fronthaul/backhaul technologies, such as mmWave or Free-Space Optics (FSO), can offer a cost-effective, resilient infrastructure. These can be adaptively used depending on AP location, user density, and environmental constraints.
\end{enumerate}
Additionally, synchronization protocols across the fronthaul network, such as the Institute of Electrical and Electronics Engineers (IEEE) 1588 Precision Time Protocol (PTP), are critical to ensure the timely and coherent exchange of signals, especially for advanced applications like CoMP and CFMM.

\begin{figure}[!t]\vspace{-0mm}
    \centering
    \includegraphics[width=0.7\textwidth]{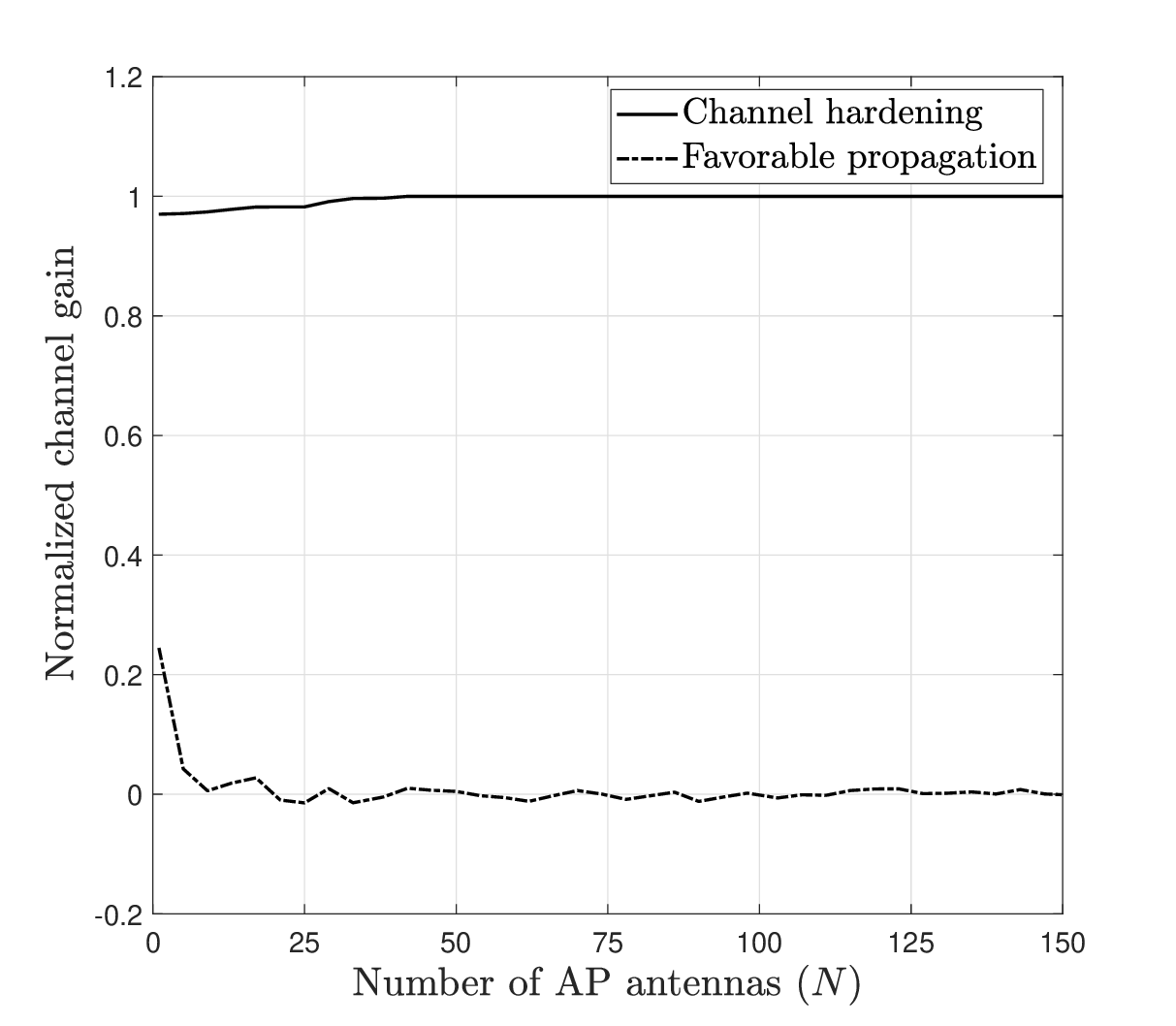}
    \vspace{-0mm}
    \caption{Channel hardening and favorable propagation versus the number of AP antennas, $N$, assuming i.i.d. Rayleigh fading.}
    \label{fig_CH_FP_numAntenna} \vspace{-0mm}
\end{figure}

\section{Channel Hardening and Favorable Propagation}
When APs use multiple antennas to communicate with distributed user terminals, two critical phenomena arise: Channel hardening and Favorable propagation \cite{Demir2021book}. 

\subsection{Channel Hardening}
The effect of small-scale fading diminishes as the number of AP antennas increases.  Thus,  the channel tends to a deterministic gain with reduced variability, resulting in negligible fluctuations around the mean of the channel \cite{Demir2021book, Chen2018ChannelHardening, Zhang2019cellfree, Polegre2020}. To illustrate this process, consider the channel between the $m$-th AP with $N$ antennas and the $k$-th single-antenna user denoted  $\q{h}_{mk} = [h_{mk,1},\dots, h_{mk, N}]^{\mathrm{T}} \in \mathbb{C}^{N \times 1}$, where $h_{mk,n}$ is the channel gain between the $n$-th antenna element and the user. For simplicity, $h_{mk,n}$'s are assumed independent and identically distributed (i.i.d.) Rayleigh fading coefficients, i.e., $h_{mk,n}\sim \mathcal{CN}(0, \beta_{mk})$, where $\beta_{mk}$ accounts for the large-scale path-loss and shadowing. Then, the coefficient of variation of the instantaneous channel gain, i.e., $\Vert \q{h}_{mk} \Vert^2$, which is a measure of the variability relative to the mean, is given as \cite{Demir2021book}
\begin{eqnarray}
    {\mathrm{CV}} = \frac{\Vert \q{h}_{mk} \Vert^2}{\E{\Vert \q{h}_{mk} \Vert^2} } \rightarrow 1 \quad \text{as} \quad N \rightarrow \infty.
\end{eqnarray}\par \vspace{-0mm}
\noindent As $N \rightarrow \infty$, the coefficient of variation approaches one,  indicating that the channel gain $\Vert\q{h}_{mk} \Vert^2$ becomes a deterministic constant (Figure~\ref{fig_CH_FP_numAntenna}) \cite{Demir2021book}. In particular, since $\Vert\q{h}_{mk} \Vert^2$ is the sum of $N$ independent and i.i.d. random variables $\vert h_{mk,n} \vert^2$, each with finite means and variances, according to the law of large numbers, the sum converges to its expected value as $N$ increases \cite{papoulis2002probability}. Thus, the fluctuations due to the small-scale fading diminish because the randomness of individual channel gains averages out over a large number of antennas \cite{Demir2021book}. This averaging effect causes the channel gain to stabilize around its mean, i.e.,  channel hardening.

This phenomenon can also be observed in co-located mMIMO and CFMM. Fortunately,  channel hardening enhances communication and sensing performance in CF-ISAC systems. The reason is that it leads to more predictable and stable wireless channels as the number of distributed AP antennas increases, reducing the impact of small-scale fading \cite{Demir2021book}. This results in deterministic channel gains, simplifying channel estimation and reducing the need for frequent CSI updates \cite{Demir2021book}. This is particularly beneficial in CF-ISAC systems, where accurate and real-time CSI is essential for reliable data transmission and precise sensing tasks such as localization and target detection. The deterministic channels also improve the effectiveness of linear processing techniques like maximum ratio transmission (MRT) and maximum ratio combining (MRC), enabling efficient beamforming for communication and sensing without complex algorithms \cite{Demir2021book, liu2023integratedbook}. Additionally, channel hardening also allows more efficient power control and resource allocation based on large-scale fading, optimizing the trade-off between sensing accuracy and communication throughput \cite{Demir2021book, liu2023integratedbook}. Overall, channel hardening in CF-ISAC systems leads to simplified system design, improved SE, and enhanced performance for joint sensing and communication tasks \cite{Mao2024}.

\subsection{Favorable Propagation} 
This refers to the phenomenon where different user-AP channels become virtually orthogonal as the number of AP antennas increases \cite{Demir2021book, Chen2018ChannelHardening, Zhang2019cellfree, Polegre2020}. Orthogonality refers to the fact that two vectors have a dot product of zero.  This property reduces inter-user interference, enhancing system capacity and performance in multi-user communication. Mathematically, the inner product of the two-channel vectors between the $m$-th AP and the $k$-th and $l$-th users (i.e., $\q{h}_{mk}$ and $\q{h}_{ml}$) is $\q{h}_{mk}^{\mathrm{H}} \q{h}_{ml} = \sum_{j=1}^{N} h_{mk, j}^* h_{ml, j}$, which is a sum of $N$ i.i.d. random variables with zero mean and variance $\beta_{mk} \beta_{ml}$ \cite{Demir2021book}. As $N\rightarrow \infty$, according to the law of large numbers, $\q{h}_{mk}^{\mathrm{H}} \q{h}_{ml}$ converges to 0, implying that the two-channel vectors become orthogonal (Figure~\ref{fig_CH_FP_numAntenna}), i.e.,
\begin{eqnarray}
    \frac{\q{h}_{mk}^{\mathrm{H}} \q{h}_{ml}}{\sqrt{\E{\Vert \q{h}_{mk} \Vert^2} \E{\Vert \q{h}_{ml} \Vert^2} }} \rightarrow 0 \quad \text{as} \quad N \rightarrow \infty.
\end{eqnarray}\par \vspace{-0mm}
\noindent Favorable propagation aligns with small-scale fading models like Rayleigh and Rician, especially when fading is uncorrelated across AP antennas and users \cite{Demir2021book, Chen2018ChannelHardening, Zhang2019cellfree, Polegre2020}. However, it weakens in highly correlated fading, severe shadowing, or densely deployed APs, leading to high spatial correlations. With a large number of AP antennas (\numrange{50}{100}), it can make user-AP channels nearly orthogonal, reducing inter-user interference by up to \qty{100}{\percent} \cite{Demir2021book, Chen2018ChannelHardening, Zhang2019cellfree, Polegre2020}.

By minimizing interference, favorable propagation enables APs to serve more users, improving capacity in CFMM \cite{Demir2021book}. While it does not directly lower AP transmit power, it enhances EE by reducing the power needed to combat interference \cite{Demir2021book, Chen2018ChannelHardening, Zhang2019cellfree, Polegre2020}. It also facilitates effective user grouping with nearly orthogonal channels. Overall, it enhances resource allocation, user fairness, SE, EE, and user selection/grouping while minimizing interference \cite{Demir2021book, Chen2018ChannelHardening, Zhang2019cellfree, Polegre2020}.

While the benefits of channel hardening and favorable propagation have been extensively analyzed in the context of communication systems, their specific impact on sensing performance in ISAC networks has not been thoroughly investigated, presenting an open research area.

\begin{figure}[!t]\vspace{-0mm}
    \centering
    \includegraphics[width=0.7\textwidth]{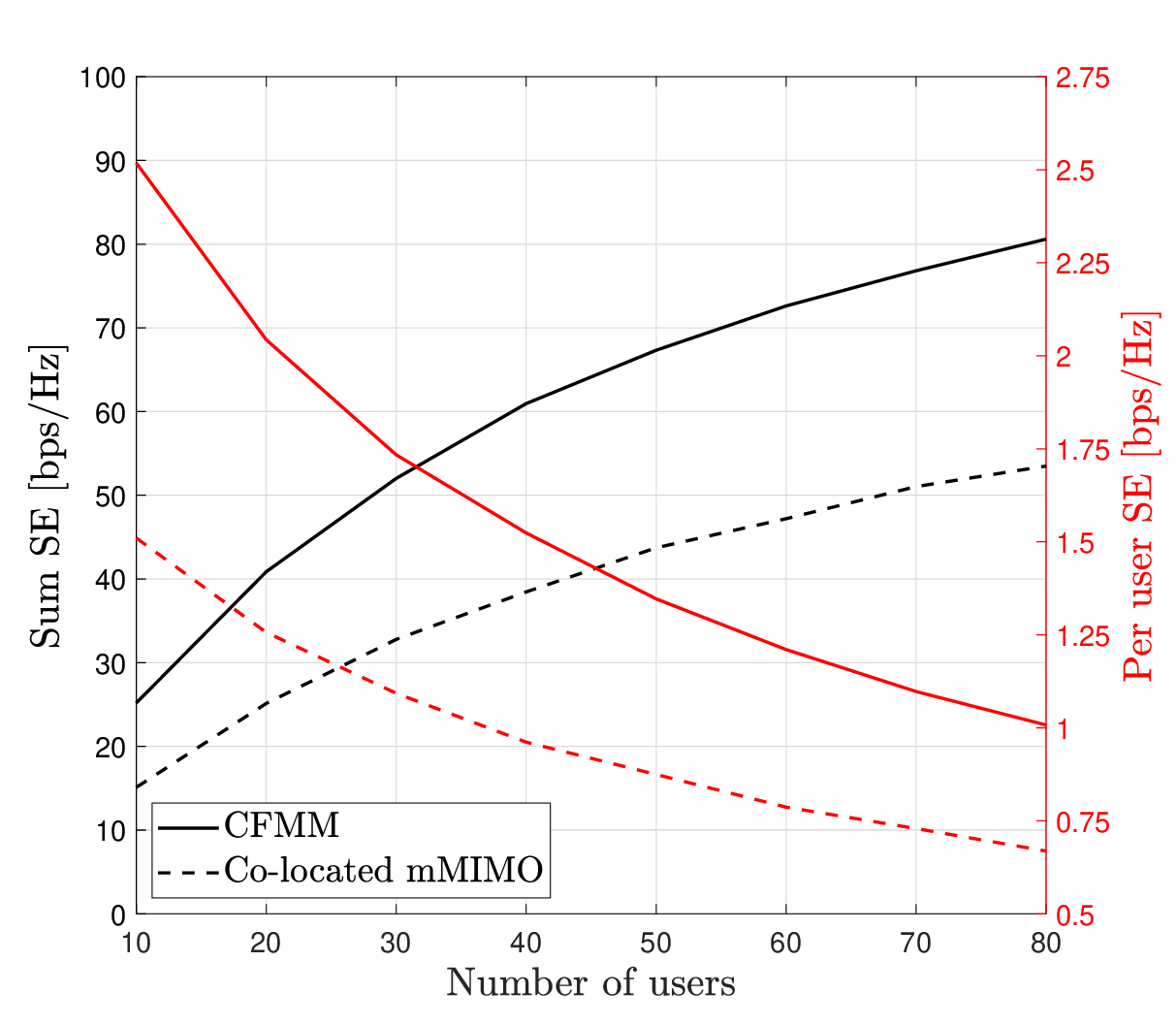}
    \vspace{-0mm}
    \caption{Sum SE (left $y$-axis) and per-user SE (right $y$-axis) comparison between CFMM and co-located mMIMO systems with \num{100} antennas in a coverage area of \qty{1}{\km^2}. In the CF system, the \num{100} single-antenna APs are uniformly distributed, whereas in the co-located system, a \num{100}-antenna BS is placed in the cell center. The DL SEs are achieved by assuming conjugate beamforming and statistical CSI knowledge at the users.}
    \label{fig_SumAndPerUsereSE_Comp} \vspace{-0mm}
\end{figure}

\section{High Robustness}
CFMM is inherently robust to fading, shadowing, and interference due to spatial diversity and cooperative processing. Therefore, it offers significant performance advantages over conventional cellular and co-located mMIMO systems. These advantages include: 
\begin{itemize}
    \item \textit{Enhanced coverage:} 
    Spatially distributed APs provide ubiquitous coverage across the network, ensuring that users, regardless of their location, are served by nearby APs. This mitigates coverage holes and shadowing effects in challenging environments, such as rural, dense urban, and indoor environments. For example, CFMM can reduce outage probability by up to \qty{90}{\percent} compared to traditional cellular systems, particularly in poor line-of-sight (LoS) or obstructed environments \cite{Ngo2017EE, Papazafeiropoulos2020}.
       
    \item \textit{Improved SE:}
    CFMM significantly boosts per-user and system-wide SE through coordinated spatial multiplexing and macro-diversity. In particular, CFMM can achieve \numrange{3}{5} times greater SE than traditional mMIMO systems, as each user is serviced by numerous APs with favorable channel conditions. For example, under ideal conditions, such as perfect CSI and uniform AP distribution, SEs exceeding \qty{100}{bps/\Hz} \cite{Elhoushy2022}. As illustrated in Figure~\ref{fig_SumAndPerUsereSE_Comp}, CFMM outperforms co-located MIMO in both per-user and sum SE, even in dense user scenarios.

    \item \textit{Reduced interference:} 
    In conventional systems, inter-cell interference, particularly at cell boundaries, is a primary limitation. CFMM mitigates this by removing cell boundaries and allowing coordinated joint transmission across the network. CFMM can spatially suppress interference via beamforming and cooperation, resulting in over \qty{50}{\percent} reduction in interference power, especially benefiting users at the cell edge \cite{Yang2018}.
    
    \item \textit{Improved EE:} 
    The proximity of APs to users enables CFMM to achieve communication requirements at significantly lower transmit powers, reducing overall energy consumption and effective path loss. Additionally, adaptive power control and user-specific beamforming further improve EE. For example, energy savings of \qtyrange{30}{50}{\percent} are achieved compared to co-located mMIMO \cite{Ngo2017EE, Yang2018}. Another option to improve EE is the use of energy detection methods \cite{Wang2020,5429879,5208031,6987540}. 
\end{itemize}

As the fundamentals of CFMM are well known, interested readers are referred to \cite{Demir2021book} and the references therein for further information.

	\chapter{Fundamentals of Radar Sensing}\label{chp_radar}

This chapter presents an overview of key radar concepts, including signal transmission, system configurations, and target parameter estimation techniques. It begins with the basics of electromagnetic (EM) signal propagation and distinguishes between continuous wave and pulsed radar. Various radar architectures, mono-static, bi-static, and multi-static, are then discussed, highlighting their design implications. The chapter concludes with methods for estimating range, velocity, and angle of arrival using signal processing. The goal is to provide a clear understanding of radar system fundamentals as a foundation for CF-ISAC.

\section{Introduction}
\begin{figure}[!h]\centering \vspace{0mm}
    \def\svgwidth{370pt} 
    \fontsize{8}{8}\selectfont 
    \graphicspath{{Figures/}}
\begingroup%
  \makeatletter%
  \providecommand\color[2][]{%
    \errmessage{(Inkscape) Color is used for the text in Inkscape, but the package 'color.sty' is not loaded}%
    \renewcommand\color[2][]{}%
  }%
  \providecommand\transparent[1]{%
    \errmessage{(Inkscape) Transparency is used (non-zero) for the text in Inkscape, but the package 'transparent.sty' is not loaded}%
    \renewcommand\transparent[1]{}%
  }%
  \providecommand\rotatebox[2]{#2}%
  \newcommand*\fsize{\dimexpr\f@size pt\relax}%
  \newcommand*\lineheight[1]{\fontsize{\fsize}{#1\fsize}\selectfont}%
  \ifx\svgwidth\undefined%
    \setlength{\unitlength}{862.2008667bp}%
    \ifx\svgscale\undefined%
      \relax%
    \else%
      \setlength{\unitlength}{\unitlength * \real{\svgscale}}%
    \fi%
  \else%
    \setlength{\unitlength}{\svgwidth}%
  \fi%
  \global\let\svgwidth\undefined%
  \global\let\svgscale\undefined%
  \makeatother%
  \begin{picture}(1,0.24877914)%
    \lineheight{1}%
    \setlength\tabcolsep{0pt}%
    \put(0,0){\includegraphics[width=\unitlength]{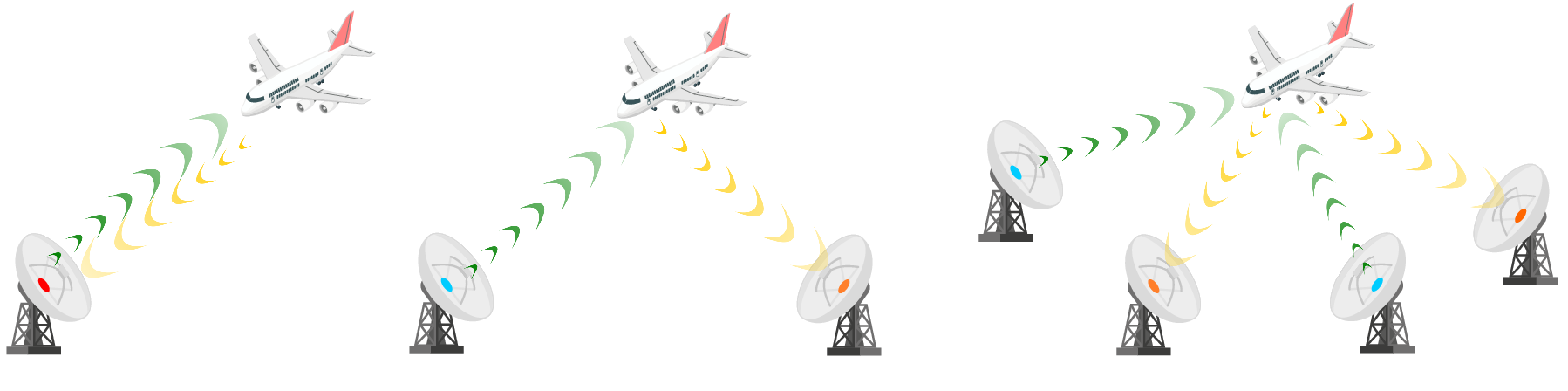}}%
    \put(0.05802213,0.00460573){\makebox(0,0)[lt]{\lineheight{1.25}\smash{\begin{tabular}[t]{l}(a) Mono-static\end{tabular}}}}%
    \put(0.3659554,0.00460573){\makebox(0,0)[lt]{\lineheight{1.25}\smash{\begin{tabular}[t]{l}(b) Bi-static\end{tabular}}}}%
    \put(0.77131394,0.00460573){\makebox(0,0)[lt]{\lineheight{1.25}\smash{\begin{tabular}[t]{l}(c) Multi-static\end{tabular}}}}%
  \end{picture}%
\endgroup%
 \vspace{0mm}
    \caption{Types of radar/sensing.}\vspace{0mm} \label{fig_RadarConfigurations}
\end{figure}
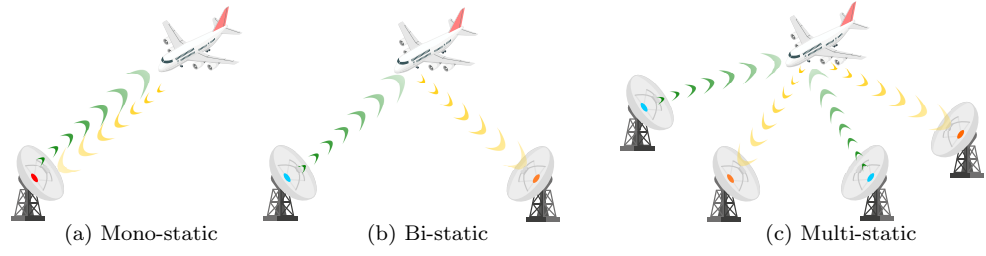

Radar (short for radio detection and ranging) uses EM waves to detect and estimate object properties. It transmits EM signals toward targets and analyzes their echoes \cite{Mark2010RadarBook, richards2005fundamentals}. A radar transmitter emits a narrow beam scanning the expected target area, including aircraft, ships, spacecraft, vehicles, astronomical bodies, birds, insects, and rain. When the beam strikes a target, some energy scatters or reflects back to the radar receiver, which can be co-located with the transmitter (mono-static) or at a separate site (bi-static/multi-static) (Figure~\ref{fig_RadarConfigurations}). Some systems time-share a single antenna for transmission and reception \cite{Mark2010RadarBook, richards2005fundamentals}.

Reflections from targets form the signal of interest, while those from other sources, like the ground or rain, act as interference, degrading detection performance \cite{Mark2010RadarBook, richards2005fundamentals}. The radar receiver processes echo signals to estimate a target's presence, location, velocity, range, direction, size, and shape. By tracking its position over time, the target's trajectory and path can be predicted. Radar is widely used in defense, automotive, and weather forecasting due to its ability to operate in diverse conditions and measure critical parameters like distance, velocity, and angle.

\section{Radar Cross-Section}
The RCS ($\sigma$) is a critical parameter that quantifies how much EM energy an object reflects back toward a radar receiver, effectively measuring a target's detectability and its reflecting and scattering characteristics \cite{Mark2010RadarBook}. Let's consider the following radar equation:
\begin{eqnarray}\label{eqn_radar_eqn}
    P_r = \frac{P_t G_t G_r \lambda^2 \sigma}{(4\pi)^3 r^4},
\end{eqnarray} \par \vspace{-0mm}
\noindent where $P_r$ is the power received back from the target by the radar (watt), $P_t$ is the transmitter's input power (watt), $G_t$ is the gain of the radar transmit antenna (dimensionless), $G_r$ is the gain of the radar receiver antenna (dimensionless), and $r$ is the distance between radar and target. Moreover, $\sigma$, i.e., RCS,  represents the magnitude of the echo signal returned to the radar by the target.

Mathematically, RCS ($\sigma$) is defined as the ratio of scattered power to incident power at a given distance, i.e., 
\begin{eqnarray}
    \sigma = \lim_{r\rightarrow \infty} 4 \pi r^2 \frac{P_{\mathrm{scatter}}}{P_{\mathrm{incident}}},
\end{eqnarray} \par \vspace{-0mm}
\noindent where $P_{\mathrm{scatter}}$ is the power scattered by the target back toward the radar and $P_{\mathrm{incident}}$ the power incident on the target \cite{Mark2010RadarBook}. Alternatively, in terms of electric fields, the RCS can also be given as
\begin{eqnarray}
    \sigma = \lim_{r\rightarrow \infty} 4 \pi r^2 \frac{E_{\mathrm{scatter}}}{E_{\mathrm{incident}}},
\end{eqnarray} \par \vspace{-0mm}
\noindent where $E_{\mathrm{scatter}}$ and $E_{\mathrm{incident}}$ are the magnitudes of the scattered and incident electric fields, respectively \cite{Mark2010RadarBook}.

While RCS is often thought of as the physical area of an object visible to the radar, it is actually an effective area that depends not only on size but also on shape, material composition, radar frequency, incidence angle, and polarization \cite{Mark2010RadarBook}. 

\noindent \textbf{Example:} Let us consider a radar system operating at a frequency corresponding to a wavelength of $\lambda = \qty{0.03}{\m}$ (i.e., \qty{10}{\GHz}). The radar transmits power $P_t=\qty{1000}{\watt}$, and both the transmit and receive antennas have a gain of $G_t=G_r = \num{1000}$ (equivalent to \qty{30}{\dB}). A target is located at a distance $r=\qty{e4}{\m}$ from the radar, and the received echo power is measured as $P_r = \qty{4.54 e-14}{\watt}$. What is the RCS value of the target?

\noindent \textit{Solution:}
Using the radar range equation in \eqref{eqn_radar_eqn}, we can solve for the radar cross-section $\sigma$ as
\begin{eqnarray} 
\sigma = \frac{P_r (4\pi)^3 r^4}{P_t G_t G_r \lambda^2}. 
\end{eqnarray}
By substituting the known values
\begin{eqnarray*} 
\sigma &=& \frac{4.53 \times 10^{-14} \times (4\pi)^3 \times (10^4)^4}{1000 \times 1000 \times 1000 \times (0.03)^2} \\
&=& \frac{4.53 \times 10^{-14} \times 1984.4 \times 10^{16}}{10^9 \times 9 \times 10^{-4}} \\ 
&=& \frac{8.9863 \times 10^5}{9 \times 10^5} \approx 0.998 \approx 1 \text{m}^2. \end{eqnarray*}
The calculated RCS of the target is approximately $\sigma=\qty{1}{\m^2}$, which corresponds to a moderately reflective object, such as a small vehicle or metallic sphere at the given frequency.

\subsection{Size and Shape}
Larger objects have a higher RCS, but their geometric features determine how much energy is reflected. For example, a flat metallic surface perpendicular to the incoming radar wave produces a strong reflection, making it highly detectable \cite{Mark2010RadarBook}. Its RCS under ideal conditions is approximated by $\sigma = 4\pi A^2/\lambda^2$, where $A$ is the physical area of the plate and $\lambda$ is the wavelength of the radar signal. 

However, if the surface is tilted, much of the reflected energy will be deflected away from the radar, reducing its apparent RCS. Stealth technology exploits this principle, where aircraft and military vehicles are designed with faceted surfaces that scatter radar waves in multiple directions instead of back toward the transmitter \cite{Mark2010RadarBook}. Conversely, objects with concave structures or right-angled corners, such as buildings or aircraft landing gear, act as corner reflectors, concentrating and reflecting energy to the source, leading to a much higher RCS.

Table~\ref{tab_rcs_eqn} provides approximate RCS expressions for some common shapes/objects.

\begin{table}[t!]
\centering
\caption{RCS expressions for some common shapes.}\label{tab_rcs_eqn}
\begin{tabular}{|p{4cm}|p{4cm}|p{5cm}|}
\hline
\textbf{Object} & \textbf{RCS Expression} & \textbf{Description } \\ \hline \hline
Sphere     &  $\sigma=\pi r^2$  &  $r$ - radius; for $r \gg \lambda$ \\ \hline
Flat square plate   &  $\sigma = 4\pi A^2/\lambda^2$   &  $A$ - physical area; for normal incidence  \\ \hline
Circular flat plate     &  $\sigma = (\pi D^2)^2/(4\lambda^2)$     &  $D$ - diameter; for normal incidence   \\ \hline
Cylinder     &   $\sigma = 2\pi L^2/\lambda$  & $L$ - cylinder length; for side incidence and $L\gg \lambda$    \\ \hline
Cone     &   $\sigma = \pi a^2 (\sin(\theta))^2/\lambda^2$    &  $a$ - base radius, $\theta$ - cone half-angle; dependent on incidence angle  \\ \hline
Trihedral corner reflector     &  $\sigma = 12 \pi a^4 /\lambda^2$     & for ideal trihedral with edge length $a$   \\ \hline
\end{tabular}
\end{table}

\subsection{Material properties}
Metallic objects with high electrical conductivity strongly reflect radar waves and generally have a large RCS. On the other hand, dielectric materials, such as fiberglass or certain composites, tend to scatter and absorb radar energy, leading to a lower RCS \cite{Mark2010RadarBook}. In advanced radar-evading technologies, specialized radar-absorbing materials are used to coat military aircraft and ships, converting the incident radar energy into heat or dissipating it within the material itself \cite{Mark2010RadarBook}. Some stealth systems use plasma cloaking, where ionized gases envelop the object, altering the EM properties of the surrounding space to reduce radar reflections \cite{Mark2010RadarBook}.

\subsection{Frequency}
The interaction of an object with radar waves depends on its size relative to the wavelength of the incident wave. When the wavelength is much larger than the object's dimensions, the object reflects very little energy, making it difficult to detect \cite{Mark2010RadarBook}. Conversely, when size is comparable to the wavelength, complex interactions occur, leading to fluctuations in RCS depending on the frequency \cite{Mark2010RadarBook}. Moreover, when the wavelength is much smaller than the object, reflections become more predictable, similar to how visible light reflects off everyday surfaces \cite{Mark2010RadarBook}. For instance, the RCS of a metallic sphere of radius $r$ can be approximated in different regimes as
\begin{eqnarray}
    \sigma = \begin{cases}
        9\pi r^6/\lambda^4, & \text{for} \quad r \ll \lambda,\\
        \text{fluctuates}, & \text{for} \quad r \approx \lambda,  \\   
         \pi r^2, & \text{for} \quad r \gg \lambda.
    \end{cases}
\end{eqnarray}

Due to this frequency dependence, an object might be nearly invisible to low-frequency radars but highly detectable at higher frequencies, a fact that is carefully considered in stealth technology and radar countermeasure design.

\subsection{Incidence Angle and Polarization}
The RCS of an object varies dramatically based on the angle at which the radar signal strikes it. A target that appears highly reflective from one angle may be nearly undetectable from another due to changes in surface orientation relative to the incoming wave \cite{Mark2010RadarBook}. In addition, the polarization of the radar signal, whether it is linearly, circularly, or elliptically polarized, influences the amount of reflected energy received by the radar \cite{Mark2010RadarBook}. Some stealth designs exploit polarization effects by using materials and structures that preferentially absorb or redirect specific polarization states, further reducing detectability \cite{Mark2010RadarBook}.

Table~\ref{tab_rcs} presents approximate RCS values for various objects, illustrating how different materials and structures influence radar detectability \cite{knott2004radar, Rezende2002RCS, skolnik2001introduction}. For example, a human standing upright has an RCS of approximately \qty{1}{\m^2}, while a commercial airliner like a Boeing 747 can have an RCS ranging from \qty{30}{\m^2} to \qty{100}{\m^2} depending on its orientation and altitude. Large ships and cargo vessels can exhibit RCS values exceeding \qty{e6}{\m^2}, making them highly visible to radar even at long ranges. On the other hand, stealth fighter jets are engineered to have an RCS as low as \qty{0.001}{\m^2}, making them appear no larger than a small bird on enemy radar screens.

\begin{table}[t!]
\centering
\caption{Approximated RCS values of common objects (\num{8}-\qty{12}{\GHz}).}\label{tab_rcs}
\begin{tabular}{|l|c|}
\hline
\textbf{Object} & \textbf{Approx. RCS (\qty{}{\m^2})} \\ \hline \hline
Insect &  $\num{e-6}-\num{e-5}$    \\ \hline
Bird (e.g., pigeon) &  \num{0.01}       \\ \hline
Human (standing, broadside)  &  \num{1} \\ \hline
Car (sedan, broadside)     &   $\num{10}-\num{100}$  \\ \hline
Large truck  &  $\num{100}-\num{200}$  \\ \hline
Commercial aircraft (e.g., Boeing 747)  &   $\num{30}-\num{1000}$  \\ \hline
Cargo aircraft   & Up to \num{100}   \\  \hline
Small combat aircraft   & $\num{2}-\num{3}$   \\  \hline
Large combat aircraft   & $\num{5}-\num{6}$   \\  \hline
Large ship (e.g., cargo ship, tanker)   & $\num{e5}-\num{e6}$   \\  \hline
\end{tabular}
\end{table}

\section{Clutter}
In radar systems, clutter refers to unwanted echoes or reflections originating from objects in the environment that are not of interest to the radar operator \cite{Mark2010RadarBook}. These include reflections from the ground (referred to as ground clutter), buildings, vegetation, terrain, and atmospheric phenomena such as rain, snow, or even birds and insects. While clutter is not inherently random in the same sense as thermal noise, its impact on radar detection can be equally disruptive. Clutter can obscure or mask the echoes from intended targets, particularly if the target has a low RCS or is located within or near clutter-dense environments. This is particularly problematic for radar systems attempting to detect small aircraft, drones, or missiles flying at low altitudes over complex terrain \cite{Mark2010RadarBook}.

The received signal at the radar receiver, i.e., $r(t)$, can be expressed as the sum of three components, i.e., the echo from the target, the clutter return, and additive noise:
\begin{eqnarray}
    r(t) = s_t(t) + c(t) + n(t),
\end{eqnarray} \par \vspace{-0mm}
\noindent where $s_t(t)$ is the reflected signal from the target, $c(t)$ is the clutter component, and $n(t)$ is the noise. In the range-Doppler domain, assuming narrow-band pulse-Doppler processing, the signal model can be expressed as:
\begin{eqnarray}
    R(f, \tau) = S_t(f, \tau) + C(f, \tau) + N(f, \tau),
\end{eqnarray} \par \vspace{-0mm}
\noindent where $f$ is the Doppler frequency and $\tau$ is the time delay (range bin) \cite{Mark2010RadarBook}.

Although the effect of clutter on detection can be comparable to that of noise, clutter is influenced by the environment, frequency, geometry, and waveform characteristics, which may result in a distinct profile from noise \cite{Mark2010RadarBook}. Unlike noise, which is typically uniform and statistically well-behaved, clutter has spatial and temporal structure \cite{Mark2010RadarBook}. For instance, sea clutter is strongly influenced by wave dynamics and wind conditions, while ground clutter may exhibit angular variation depending on terrain features and urban density \cite{Mark2010RadarBook}. Consequently, accurate modeling of clutter is essential for optimizing detection and tracking performance.

\subsection{Statistical Models for Clutter}
Several statistical models have been developed to model the amplitude or power distribution of clutter. The Weibull, log-Weibull, log-normal, and $K$-distributions models are among the most widely used \cite{Sayama2001, Haykin2002Book}. Let $x$ denote the clutter amplitude or intensity (power). These models can be mathematically given as
\begin{itemize}
    \item \textit{Weibull distribution:} 
    \begin{eqnarray}
        f(x) = \frac{k}{\mu} \left( \frac{x}{\mu}\right)^{k-1} e^{-(x/\mu)^k}, \quad x \geq 0,
    \end{eqnarray} \par \vspace{-0mm}
    \noindent where $k > 0$ is the shape parameter and $\mu > 0$ is the scale parameter.

    \item \textit{Log-normal distribution:} 
    \begin{eqnarray}
        f(x) = \frac{1}{x \sigma_c \sqrt{2\pi}}  e^{-(\ln x-\mu_c)^2/2\sigma_c^2}, \quad x > 0,
    \end{eqnarray} \par \vspace{-0mm}
    \noindent where $\mu_c$ and $\sigma_c$ are the mean and standard deviation of logarithmic clutter power, i.e., $\ln x$.

    \item \textit{$K$-distribution:} 
    \begin{eqnarray}
        f(x) = \frac{2b^{(a+1)/2} x^{(a-1)/2}}{\Gamma(a)} K_{a-1}\left(2\sqrt{bx} \right), \quad x \geq 0,
    \end{eqnarray} \par \vspace{-0mm}
    \noindent where $a$ is the shape parameter, $b$ is the scale, $\Gamma(\cdot)$ is the gamma function, and $K_n(\cdot)$ is the modified Bessel function of the second kind.
\end{itemize}
These models help radar systems in adapting detection thresholds and distinguishing targets from background clutter \cite{Sayama2001}. The Weibull distribution, for example, is effective for modeling land clutter and sea clutter under certain conditions, while the $K$-distribution is suited for describing spiky or heterogeneous clutter often encountered in synthetic aperture radar (SAR) or low-grazing angle maritime scenarios.

\subsection{Clutter Mitigation Techniques}
To mitigate the adverse effects of clutter, modern radar systems employ several clutter rejection techniques \cite{Mark2010RadarBook}. These include moving target indication (MTI), Doppler filtering, constant false alarm rate (CFAR), clutter maps, polarization diversity, and SAR \cite{Mark2010RadarBook}. For example, the MTI technique exploits the stationary nature of clutters, allowing moving targets to be separated based on the Doppler shift. Doppler filtering, implemented using Fast Fourier Transform (FFT) techniques, enables the separation of echoes based on velocity \cite{richards2005fundamentals}. However, both MTI and Doppler filters can suffer from blind speeds, where targets moving at specific velocities cannot be effectively distinguished from clutter. Moreover, CFAR dynamically adjusts the detection threshold based on the local noise and clutter statistics, maintaining a consistent false alarm probability even in the presence of variable clutter. Clutter maps are another strategy, where the radar system builds a spatial and temporal model of expected clutter returns and uses it to suppress future returns from known clutter regions.

More advanced methods, such as polarization diversity, take advantage of the different scattering behaviors of clutter and targets under varying polarizations. SAR systems, through high-resolution imaging, can also discriminate targets from clutter using spatial context. In weather radars, clutter suppression is aided by dual-polarization techniques that allow the radar to differentiate between meteorological and non-meteorological scatterers \cite{richards2005fundamentals}.

Effective clutter suppression is fundamental in many applications, including military and civilian radar. In air traffic control, it ensures the accurate detection of aircraft in complex environments. In defense systems, it enhances the ability to detect low-flying threats in the presence of terrain masking or urban cover. With the emergence of cognitive radar and ML-based processing, future clutter rejection strategies will become more adaptive and intelligent, learning environmental patterns in real time to further reduce false alarms and improve target discernibility.

\section{Signal Transmission}
Radar systems emit EM signals in the radio or microwave frequency bands \cite{Mark2010RadarBook, richards2005fundamentals, levanon2004radar, Blunt2016}. These signals are transmitted through antennas, which convert electrical energy into propagating EM waves. The transmitted signals travel through the environment, interacting with objects/targets, and a portion of the energy is reflected or scattered back toward the radar receiver \cite{Mark2010RadarBook}. 

The frequency of the transmitted signal can vary depending on the application, but typical radar systems use frequencies ranging from several megahertz (MHz) to gigahertz (GHz). The signal frequency has a direct influence on both range resolution and propagation characteristics. For instance, higher frequencies generally provide finer spatial resolution due to their shorter wavelengths, but they also experience greater attenuation and reduced penetration through obstructions or adverse weather \cite{Mark2010RadarBook, richards2005fundamentals, levanon2004radar, Blunt2016}.

Two common types of radar exist based on the transmitted signals, i.e., continuous wave radar and pulsed radar \cite{levanon2004radar, Blunt2016, skolnik2008radar}.

\subsection{Continuous Wave Radar}
A continuous signal is transmitted, and the reflected signal is continuously measured. This is often used for measuring velocity (Doppler radar).

These systems transmit a constant/continuous signal over time and continuously measure the frequency shifts in the received signal to extract motion-related information. These systems are particularly effective for measuring relative velocity via the Doppler effect, making them ideal for speed detection and short-range tracking applications \cite{levanon2004radar, Blunt2016}.

The Doppler frequency shift $f_d$ caused by a target moving with radial velocity $v_t$ can be given by
\begin{eqnarray}
    f_d = \frac{2v_t}{\lambda} = \frac{2v_tf}{c},
\end{eqnarray}\par \vspace{-0mm}
\noindent where $c$ is the speed of light in free space and $f$ is the frequency of the radar signal. This shift is used to infer the target's velocity. However, continuous wave radars cannot directly measure range unless frequency modulation is used, as in frequency-modulated continuous wave (FMCW) systems \cite{levanon2004radar, Blunt2016}.

In FMCW radar, the transmitted signal varies in frequency (e.g., linearly chirped). The round-trip time delay $\tau$ introduces a frequency shift $\Delta f$ that is is proportional to the range $R$, i.e., 
\begin{eqnarray}
    \Delta f = \gamma \tau = \frac{2R\gamma}{c},
\end{eqnarray}\par \vspace{-0mm}
\noindent where $\gamma$ is the frequency sweep rate (Hz/s). The range can then be calculated as $R = c \Delta f/2\gamma$ \cite{levanon2004radar, Blunt2016}.

\subsection{Pulsed Radar}
These systems transmit discrete bursts or pulses of EM energy, followed by intervals of silence during which the receiver listens for echoes. This allows the system to measure the time delay between transmission and reception, which can be used to calculate/estimate distance or range \cite{levanon2004radar, Blunt2016}.

The basic range equation in pulsed radar can be given as
\begin{eqnarray}
    R = \frac{c \tau}{2}
\end{eqnarray}\par \vspace{-0mm}
\noindent where $R$ is the target range and $\tau$ is the round-trip time delay.

The duration of the transmitted pulse ($T_p$) affects the radar's range resolution ($\Delta R$), with shorter pulses enabling finer discrimination between closely spaced targets, i.e., $\Delta R = c T_p/2$ \cite{levanon2004radar, Blunt2016}.

To maintain high resolution without sacrificing signal energy, pulse compression techniques are used. For example, using a chirp waveform with the time-bandwidth product $BT\gg 1$, the effective range resolution is given by
\begin{eqnarray}
    \Delta R = \frac{c}{2B}
\end{eqnarray}\par \vspace{-0mm}
\noindent where $B$ is the signal bandwidth. This allows long pulses (high energy) with short effective resolution.

The choice between continuous and pulsed transmission depends on multiple factors, including system complexity, resolution requirements, power constraints, and the desired range-velocity ambiguity trade-offs \cite{levanon2004radar, skolnik2008radar}.

\section{Types of Sensing}\label{sec_radartypes}
Radar sensing/systems can be classified into three groups based on the spatial relationship between the transmitter and receiver, i.e., mono-static, bi-static, and multi-static sensing/configurations (Figure~\ref{fig_RadarConfigurations}) \cite{Mark2010RadarBook, richards2005fundamentals, li2009mimoradar}. Each configuration presents unique advantages, limitations, and challenges in radar signal processing, hardware design, and system performance.

\subsection{Mono-static radar} 
This is characterized by a co-located transmitter and receiver pair, often sharing the same antenna (array) for transmission and reception (Figure~\ref{fig_RadarConfigurations}a). It thus requires full-duplex (FD) operation \cite{Mark2010RadarBook}. However, the simultaneous transmission and reception can introduce strong self-interference (SI), necessitating SI cancellation techniques \cite{Mohammadi2023, Diluka2024CFFD}. Alternatively, the receiver should be isolated from the transmitter to protect it from the high SI. However, some modern systems use sophisticated duplexers to allow simultaneous transmission and reception at different frequencies or polarization states \cite{Mark2010RadarBook}. This configuration is widely used in most radar applications due to its simplicity in design and ease of signal processing \cite{Mark2010RadarBook}. Due to the co-located or shared antennas, this has less hardware complexity compared to other configurations. Mono-static radars benefit from well-developed signal processing algorithms, including pulse compression, Doppler processing, and clutter suppression \cite{Mark2010RadarBook}. The transmitter and receiver co-location allows for precise range measurements, with accuracy dependent on the pulse width or bandwidth \cite{Mark2010RadarBook}. Mono-static radar often fails to detect objects having a low RCS, which does not reflect much radar energy to the source \cite{Mark2010RadarBook}. Mono-static sensing is widely used in conventional ISAC systems with co-located BSs \cite{Liu2022ISAC, Wang2022ISAC, Zhang2022, Azar2024, Diluka2024NF, Zhenyao2023}.  

\subsection{Bi-static radar} 
The transmitter and receiver are separated, i.e., placed at different locations (Figure~\ref{fig_RadarConfigurations}b) \cite{Mark2010RadarBook}. The separation can vary from short distances to several kilometers, making it suitable for various operational scenarios such as long-range surveillance and covert detection \cite{Mark2010RadarBook}. However, accurate synchronization between the transmitter and receiver is critical, as the receiver does not have direct access to the transmitted signal, which increases the system complexity \cite{Mark2010RadarBook}. External timing sources/techniques (e.g., direct signal reception, Global Positioning System (GPS) timing, or sophisticated signal tracking algorithms) are required to maintain accurate synchronization \cite{Mark2010RadarBook}. As the receiver is separated from the transmitter, it is harder for adversaries to detect and locate the radar system, making bi-static radar more suitable for military surveillance and stealth applications. The transmitter-receiver separation can reduce clutter caused by direct-path reflections (e.g., from the ground or other environmental features) \cite{Mark2010RadarBook}. On the other hand, the RCS of targets can change dramatically with different bi-static angles, making them sensitive to target orientation and geometry \cite{Mark2010RadarBook}. Several works \cite{Brunner2025, Park2024, Kai2024} utilize bistatic sensing in ISAC networks.  

\subsection{Multi-static radar (MSR)} 
Multiple transmitters and receivers are distributed across different locations in a multi-static radar (MSR) configuration (Figure~\ref{fig_RadarConfigurations}c) \cite{Mark2010RadarBook}. MSR leverages spatial diversity to enhance target identification, tracking accuracy, and resistance to jamming and interference \cite{Mark2010RadarBook}. By using multiple receivers, MSR can triangulate target positions more precisely than mono-static or bi-static systems, which is particularly beneficial in cluttered or complex environments. It can also generate high-resolution target images using techniques such as SAR and inverse SAR through multiple observational perspectives \cite{Mark2010RadarBook}. Jamming is more difficult, requiring simultaneous disruption of several radar links. Additionally, the spatial separation of receivers mitigates interference from environmental clutter and multi-path reflections \cite{Mark2010RadarBook}. However, this configuration demands tight coordination among radar nodes, increasing hardware, signal processing, and communication complexity. Precise synchronization between distributed transmitters and receivers remains a significant challenge, especially in dynamic or mobile scenarios \cite{Mark2010RadarBook}. CF-ISAC networks utilize mono-static sensing with distributed APs \cite{Mao2023, Demirhan2023, Huang2022Coordinated, Cao2023Design, Wang2023, Sakhnini2022Uplink, Silva2023, Behdad2022, Behdad2024Interplay}. More details on multi-static sensing can be found in Chapter~\ref{chp_CF_isac}.

\section{Target Parameter Estimation Techniques}
Radar systems detect the presence of targets and estimate key parameters such as range, velocity, and angle of arrival (AoA), which are essential for localization and tracking \cite{Mark2010RadarBook}. These estimations rely on signal processing techniques that extract information from radar echoes in noisy and cluttered environments \cite{richards2005fundamentals}.

\subsection{Range Estimation}
This is typically achieved using matched filtering, which maximizes the signal-to-noise ratio (SNR) by correlating the received signal with a transmitted signal template. In pulsed radar systems, this correlation process identifies a peak at a specific time delay corresponding to the round-trip travel time of the pulse. The received signal, expressed in its baseband (low-pass equivalent) form, can be modeled as  
\begin{eqnarray}
    y(t) = \alpha s(t-\tau) + n(t),
\end{eqnarray}\par \vspace{-0mm}
\noindent where $\alpha$ is the reflection coefficient, $\tau$ is the time delay, $s(t)$ is the transmitted signal,  and $n(t)$ is additive noise. The matched filter's peak at $t=\tau$ determines the target's range \cite{Mark2010RadarBook, richards2005fundamentals}.

The matched filter is defined as the time-reversed complex conjugate of $s(t)$, and its output is given by
\begin{eqnarray}
    z(t) = \int y(u) s^*(u-t) du.
\end{eqnarray}\par \vspace{-0mm}
\noindent The peak of $z(t)$ occurs at $t=\tau$, indicating the estimated time delay. The target range is then computed as
\begin{eqnarray}
    R = \frac{c\hat{\tau}}{2},
\end{eqnarray}\par \vspace{-0mm}
\noindent where $\hat{\tau}$ is the estimated round-trip time delay. In discrete-time implementations (e.g., digital radar), this operation is often performed via cross-correlation or FFT-based convolution \cite{Mark2010RadarBook, richards2005fundamentals}.

\subsection{Velocity Estimation}
This utilizes the Doppler effect, i.e., the motion of a target relative to the radar induces a frequency shift in the returned signal.  The  shift $\Delta f_{D}$ is proportional to the radial velocity of the target ($v_t$) and the carrier frequency of the transmitted signal ($f$), and is given by \cite{Mark2010RadarBook}
\begin{eqnarray}
    \Delta f_{D} = \frac{2v_t f}{c}.
\end{eqnarray}\par \vspace{-0mm}
\noindent To estimate velocity, radar systems often use coherent pulse-Doppler processing. It is a radar signal processing technique that exploits the phase coherence of transmitted pulses to measure the Doppler shift caused by moving targets. By transmitting a series of equally spaced pulses and coherently processing the received echoes (i.e., preserving their phase information), the radar can distinguish moving targets from stationary ones and estimate their radial velocity. This technique enables high-resolution velocity discrimination, clutter suppression (by filtering out stationary objects), and effective tracking of targets in dynamic environments. It is widely used in surveillance, tracking, and airborne radar systems.  Let $y_n(t)$ denote the received signal from the $n$-th pulse. A Doppler FFT is applied across a sequence of pulses at fixed range cells as
\begin{eqnarray}
    Y(k) = \sum_{n=0}^{N-1} y_n(t) e^{-j2\pi kn/N}, \quad \text{for} \quad k=\{1, \ldots, N-1\},
\end{eqnarray} \par \vspace{-0mm}
\noindent where $N$ is the total number of pulses. The index $k$ with the maximum magnitude $\vert Y(k) \vert$ corresponds to a Doppler frequency bin, which is mapped to a radial velocity, allowing for efficient estimation of the radial speed of moving targets \cite{Mark2010RadarBook, skolnik2008radar}. In high-resolution applications, phase differences between pulses can also be used to directly estimate Doppler shifts as follows:
\begin{eqnarray}
    \Delta \phi  = 2\pi \Delta f_D T_p,
\end{eqnarray} \par \vspace{-0mm}
\noindent where $T_p$ is the pulse repetition interval.

\subsection{AoA Estimation}
This involves determining the direction from which the reflected signal arrives at the radar receiver. Several techniques have been developed for AoA estimation, depending on the radar architecture and resolution requirements. These methods include: 
 \begin{enumerate}

    \item \textbf{Monopulse radar} estimates the AoA by forming two types of antenna patterns, i.e., one being the sum of signals from multiple antenna elements, and the other being their difference. These two beams are processed simultaneously to achieve accurate angular resolution, even in a single radar pulse \cite{skolnik2008radar}. The received signals are combined to form the sum and difference channels as
    \begin{eqnarray}
        A_{\mathrm{sum}} &=& \sum_{m=1}^{M} x_m, \\
        A_{\mathrm{diff}} &=& \sum_{m=1}^{M} w_m x_m,
    \end{eqnarray}\par \vspace{-0mm}
    \noindent where $x_m$ is the received signal at the $m$-th antenna element, and $w_m$ is the corresponding weighting factor that creates the difference beam (e.g., $+1$ for elements on one side of the array and $-1$ for elements on the other). 

    The AoA is then estimated as
    \begin{eqnarray}
        \theta \approx \frac{A_{\mathrm{sum}}- A_{\mathrm{diff}}}{A_{\mathrm{sum}}+ A_{\mathrm{diff}}},
    \end{eqnarray}\par \vspace{-0mm}
    \noindent This approach is widely used in practical radar systems due to its robustness and accuracy in a single look \cite{skolnik2008radar, richards2005fundamentals}.

    \item \textbf{Beamforming}, in phased-array radars, electronically steers the antenna beams and analyzes the received signals (received signal power or correlation over a scanning angle) to localize the target direction.  
    
    \item \textbf{Subspace-based methods}, more advanced techniques, such as multiple signal classification (MUSIC) and estimation of signal parameters via rotational invariance (ESPRIT), exploit the eigenstructure of the signal covariance matrix ($\q{R} = \E{\q{x} \q{x}^{\mathrm{H}}}$) to achieve super-resolution AoA estimation. For instance, MUSIC estimates AoA by finding peaks in the pseudospectrum:
    \begin{eqnarray}
        P_{\mathrm{MUSIC}}(\theta) = \frac{1}{\q{a}^{\mathrm{H}}(\theta) \q{E}_n \q{E}_n^{\mathrm{H}} \q{a}(\theta)},
    \end{eqnarray} \par \vspace{-0mm}
    \noindent where $\q{E}_n$ is the noise subspace and $\q{a}(\theta)$ is the steering vector.
\end{enumerate}

These methods can resolve multiple closely spaced targets even under noise and interference, making them ideal for high-resolution applications such as radar imaging and tracking in dense environments \cite{Mark2010RadarBook, richards2005fundamentals}. As the computational backbone of modern radar systems, they enable precise estimation of target location, motion, and orientation. While ISAC research has made significant progress in target detection and waveform design, relatively less attention has been given to accurate and efficient target parameter estimation, i.e., range, velocity, and AoA estimation. As ISAC systems advance, developing robust estimation techniques, especially those that jointly optimize sensing and communication objectives, emerges as a critical area for future research.


	\chapter{Integrated Sensing and Communication}\label{chp_isac}

Before delving into CF-ISAC, it is essential to establish a clear understanding of ISAC, the foundational concept briefly introduced in Chapter 1. ISAC refers to a unified system design that enables sensing and communication to operate concurrently within the same infrastructure \cite{liu2023integratedbook}. Rather than treating these functions as isolated, ISAC integrates them at the signal processing and system levels, facilitating a tighter coordination between information acquisition and transmission.

This chapter introduces the fundamentals of ISAC, covering key concepts, signal design principles, and system architectures. It discusses sensing performance metrics such as sensing SE, CRB, and beampattern gain, which are used to evaluate system effectiveness. The chapter also examines the emerging field of NF ISAC, driven by the adoption of large antenna arrays. Building on these foundations, it presents diverse ISAC use cases across various sectors, including autonomous vehicles, healthcare, smart manufacturing, and precision agriculture. Industry developments and standardization efforts by IEEE, 3GPP, and ITU are reviewed, highlighting ISAC's transition from theory to practice. Finally, the chapter contrasts system- and network-level ISAC architectures, outlining trade-offs in coverage, scalability, coordination, and energy efficiency. Together, these topics offer a comprehensive overview of ISAC technologies and their potential to transform future wireless systems.

\section{Fundamentals of ISAC}
With the shared hardware, spectrum, and signal processing resources for sensing and communication functionalities, ISAC leverages the dual nature of EM waves, i.e., capable of both conveying data and probing the environment via reflected echoes, to enable simultaneous perception and connectivity \cite{liu2023integratedbook}. However, a key design challenge arises from the conflicting waveform requirements for the two tasks: communication typically benefits from stochastic (random) signals such as Gaussian-distributed modulations to maximize channel capacity per Shannon's theorem \cite{Xiong2022}, while sensing requires deterministic, coherent signals (e.g., chirps, pulse trains) for optimal target resolution and estimation accuracy \cite{Mark2010RadarBook}. This random-deterministic trade-off defines a central theoretical and practical problem in ISAC system design.

Integration can be complete or partial, depending on the application. For instance, wireless sensor networks require hardware integration, surveillance radar systems need signaling integration, and cognitive radars rely on spectrum integration \cite{liu2023integratedbook}. Accordingly, ISAC architectures are commonly classified into the following four levels based on their degree of integration \cite{liu2023integratedbook}:
\begin{itemize}
    \item \textit{Level 1 - Spectral coexistence:} Communication and sensing systems share the same spectral band while operating independently. Coordination mechanisms are used to mitigate mutual interference (e.g., via scheduling or beam management), but signal design remains separate.

    \item \textit{Level 2 - Co-located hardware:} Communication and sensing functionalities are implemented on the same physical/hardware platform (e.g., same antennas or RF front-end), improving hardware efficiency. However, waveforms and processing chains are still disjoint.

    \item \textit{Level 3 - Joint signaling and processing:} A common waveform is utilized to carry both information and sensing functions. Let $s(t)$ denote the transmitted signal. Then, $s(t)$ is optimized to satisfy both:
    \begin{eqnarray}
        \text{Communication:} \quad  && \max_{s(t)} \; I(s(t); y_c(t)), \\
        \text{Sensing:}\quad   && \min_{s(t)} \; \text{MSE}(\hat{\theta} - \theta),
    \end{eqnarray} \par \vspace{-0mm}
    \noindent where $I(\cdot)$ denotes mutual information (communication metric), $y_c(t)$ is the received signal at the communication receiver, and MSE refers to the mean squared error of estimated sensing parameters $\hat{\theta}$. The waveform $s(t)$ must therefore be jointly designed to balance performance trade-offs.

    \item \textit{Level 4 - Perceptive networks:} The highest level of integration, where sensing is natively embedded into the entire network infrastructure (e.g., 6G networks). Communication nodes (e.g., base stations, user equipment) perform distributed environmental sensing alongside data transfer, forming a cognitive, perception-aware wireless network.
\end{itemize}

As a cornerstone of 6G and future wireless systems, ISAC offers significant potential for applications such as vehicular networks, unmanned aerial systems, smart factories, and remote health monitoring, where both high data throughput and situational awareness are critical \cite{liu2023integratedbook}.

\section{ISAC Design Philosophy}
It is about jointly designing communication and sensing functionalities to maximize SE, hardware reuse, and mutual information gain \cite{liu2023integratedbook, Azar2024, Ma2020}. This allows for different  trade-offs and synergies between these traditionally separate tasks.

To formalize the ISAC signal model, consider a BS equipped with an $M$-element uniform linear array (ULA), simultaneously serving $K$ single-antenna users and sensing surrounding targets. The transmit signal at the $l$-th time slot, denoted as $\q{x}(l) \in \mathbb{C}^{M\times 1}$, can be expressed as
\begin{eqnarray}\label{eqn_tx_signal}
    \q{x}(l) = \sum_{k =1}^{K} \sqrt{\rho} \q{w}_k q_k(l) + \sqrt{(1-\rho)} \q{s}(l),
\end{eqnarray}\par \vspace{-0mm}
\noindent where $\q{w}_k \in \mathbb{C}^{M \times 1}$  and $q_k(l)$ denote the communication beamforming vector and the $l$-th time slot data for the $k$-th user. The sensing signal, $\q{s}(l) \sim \mathcal{CN}(\boldsymbol{0}, \q{R}_s)$, follows a complex Gaussian distribution with covariance matrix $\q{R}_s = \E{\q{s} \q{s}^{\mathrm{H}}} \succeq 0$. This covariance matrix is designed to extend the DoF of the BS transmit signal, enhancing the system's sensing performance \cite{liu2023integratedbook}. Finally, $\rho \in [0,1]$ represents the power allocation or priority factor between communication and sensing, determining the balance and priority between these two functionalities \cite{liu2023integratedbook}.

Depending on system requirements, ISAC signal design can be categorized into three philosophies \cite{Ma2020}: (i) Communication-centric design with high $\rho$ ($\rho \rightarrow 1$), (ii) Sensing-centric design with low $\rho$ ($\rho \rightarrow 0$), and (iii) Joint design with moderate $\rho$.
\begin{enumerate}
    \item \textit{Communication-centric design ($\rho \rightarrow 1$):} Communication is the primary function, and existing communication waveforms (e.g., OFDM) are repurposed for sensing. Since these waveforms are typically non-ideal for radar processing, signal processing enhancements (e.g., super-resolution algorithms) are applied to extract useful sensing information from echoes \cite{Ma2021}.
    
    \item \textit{Sensing-centric design ($\rho \rightarrow 0$):} Radar sensing is prioritized, and communication functionality is embedded into radar signals. For instance, phase or amplitude modulation can encode communication bits onto radar pulses without significantly degrading sensing performance \cite{Chen2022}. The challenge lies in preserving radar accuracy while meeting communication constraints.

    \item \textit{Joint design ($0 <\rho <1$):} Both sensing and communication are jointly considered during waveform, beamforming, and resource allocation design. This strategy enables flexible trade-offs and provides increased degrees of freedom (DoF) for signal design, leading to more efficient and balanced ISAC operations \cite{Liu2020}.
\end{enumerate}

ISAC design concepts transform traditional wireless networks into perceptive systems that can sense their environment while serving users \cite{liu2023integratedbook}. Sensing is expected to become a standard feature in future networks, supporting a wide range of applications. In turn, sensing data can enhance communication performance, enabling innovations such as sensing-aided beamforming in vehicular networks, environment-aware resource allocation, and monitoring of traffic, weather, and human activity. These advances position sensing as both a byproduct and enabler of connectivity, making it integral to 6G and beyond \cite{liu2023integratedbook}.

\section{Sensing Metrics in ISAC} 
Sensing rate/SE, the  Cram\'{e}r-Rao Bound (CRB), and transmit/receive beampattern gain are key sensing metrics for ISAC \cite{Bell1993, Tang2010, Zhang2021Radar, Mark2010RadarBook, Kay1998, He2022, Stoica2007, Cui2014, Hua2023}. They quantify the system's ability to detect and estimate  target range, velocity, and direction of arrival (DoA), while also highlighting trade-offs with communication performance \cite{Tang2010, Zhang2021Radar}.

\subsection{Sensing SE}\label{sec_sensing_SE}
The sensing SE, measured in \qty{}{bps/\Hz}, quantifies the information acquired per unit bandwidth and time about the environment, e.g., target parameters such as range, velocity, or angle \cite{Tang2010, Zhang2021Radar}. It is formally defined as the mutual information (MI) between the received echoes and the target parameters, conditioned on the transmitted waveform \cite{Bell1993}. For example, consider an ISAC BS with $M$ transmit and $M$ receiver antennas. The received echo signal at the receiver over $L$ symbols can be given as 
\begin{equation} \label{eqn_obs_sig}
    \q{Y} = \q{G} \q{X} + \q{Z} \in \mathbb{C}^{M \times L},
\end{equation}\par \vspace{-0mm}
\noindent where $\q{X} \in \mathbb{C}^{M \times L}$ is the transmitted signal, $\q{G} \in \mathbb{C}^{M\times M}$ is sensing channel or the target response matrix \cite{Mark2010RadarBook}, and $\q{Z}$ is the additive white Gaussian noise (AWGN) matrix at the ISAC BS with independently distributed elements, i.e., $\sim \mathcal{CN}(0, \sigma^2)$. 

The MI between the received signal $\q{Y}$ and the sensing channel $\q{G}$, conditioned on $\q{X}$, is given by \cite{Zhang2021Radar}
\begin{eqnarray}
    I(\q{Y}; \q{G}\vert \q{X}) = M \log_2 \left(\det \left(\frac{1}{\sigma^2} \q{X}^{\mathrm{H}} \q{R}_G \q{X} + \q{I}_L\right)  \right),
\end{eqnarray}\par \vspace{-0mm}
\noindent where $I(X;Y\vert Z)$ is the MI between $X$ and $Y$ conditioned on $Z$, $\q{I}_L$ is an identity matrix of size $L\times L$, and $\q{R}_G = \E{\q{G} \q{G}}/M$. Thus, the achievable sensing SE in \qty{}{bps/\Hz} is defined as \cite{Tang2019, Zhenyao2023, Cui2014, Ouyang2022}
\begin{eqnarray}
    \mathcal{S}^{\mathrm{Sen}} = \max_{\tr(\q{X} \q{X}^{\mathrm{H}})\leq p_{\max}}  \frac{M}{L} \log_2 \left(\det \left(\frac{1}{\sigma^2} \q{X}^{\mathrm{H}} \q{R}_G \q{X} + \q{I}_L\right)  \right),
\end{eqnarray}\par \vspace{-0mm}
\noindent where $p_{\max}$ is the maximum allowable transmit power at the ISAC BS. 

In practice, this metric can be approximated using the sensing signal-to-interference-plus-noise ratio (SINR) as
\begin{eqnarray}
\mathcal{S}^{\mathrm{Sen}} \approx \log_2 \left( 1 + \text{SINR}^{\mathrm{Sen}} \right),
\end{eqnarray}\par \vspace{-0mm}
\noindent where $\text{SINR}^{\mathrm{Sen}}$ depends on waveform design, echo processing, and interference levels \cite{Tang2019, Zhenyao2023, Cui2014, Ouyang2022}. 

\begin{figure}[!t]\vspace{-0mm}
    \centering
    \includegraphics[width=0.7\textwidth]{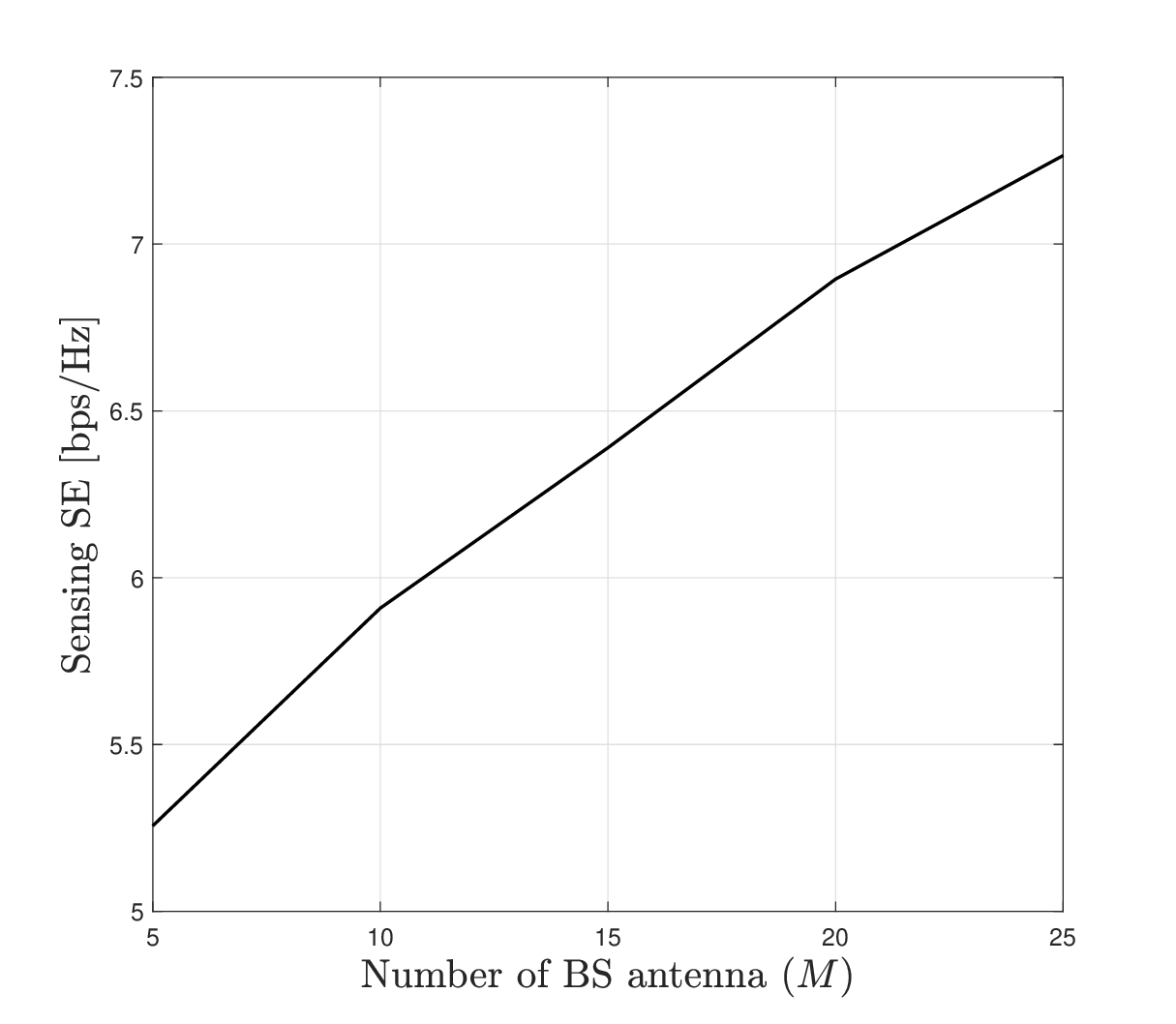}
    \vspace{-0mm}
    \caption{The sensing SE as a function of the number of BS antennas ($M$).}
    \label{fig_SensingSE_example} \vspace{-0mm}
\end{figure}

Figure~\ref{fig_SensingSE_example} shows sensing SE as a function of the number of BS antennas ($M$), revealing that increasing $M$ improves sensing SE by leveraging the spatial multiplexing gains of larger antenna arrays.

A higher sensing SE indicates greater information extraction per unit time, which is critical for accurate and rapid estimation of target parameters \cite{Tang2019, Zhenyao2023, Cui2014, Ouyang2022}. On the other hand, the optimal sensing waveform based on maximizing the sensing rate has the same estimation performance as the optimal sensing waveform based on minimizing the mean-square error (MSE) \cite{Tang2019}. It offers a broader perspective than accuracy metrics like CRB \cite{Tang2019}. Consequently, a high sensing SE is essential for many real-time applications, including autonomous vehicles, surveillance radar systems, healthcare monitoring, and environmental monitoring. In these applications, the volume of sensed information is often more crucial for decision-making than individual measurement precision. For example, autonomous vehicles require frequent updates to avoid collisions and navigate dynamic environments. A high sensing rate in surveillance radar systems allows for quick identification and tracking of fast-moving objects. In healthcare monitoring, sensing efficiency is crucial for accurate and continuous measurements of vital signs. 

In addition, the quality of target parameter estimation is also proportional to the sensing SE or corresponding sensing SINR \cite{Tang2019, Zhenyao2023, Cui2014}. Improved sensing SE enhances target parameter estimation through proper echo signal processing \cite{Tang2019, Zhenyao2023, Cui2014, Diluka2023}. Moreover, sensing SE enables parameter estimation using both transmit and receive beampatterns, helping to reduce interference between targets. However, achieving a high sensing rate necessitates careful allocation of system resources such as power and bandwidth, which are frequently shared with communication in ISAC systems \cite{Tang2019, Zhenyao2023, Cui2014, Ouyang2022}. 

\subsection{CRB}
The CRB establishes a lower bound on the variance of any unbiased estimator for a given parameter, serving as a fundamental benchmark in statistical signal processing \cite{Kay1998}. In ISAC systems, the CRB is widely used to assess the accuracy limits for estimating critical target parameters such as range, velocity, and DoA \cite{Mark2010RadarBook, Kay1998}. A lower CRB implies higher estimation accuracy, making it a key performance indicator for sensing designs.

The CRB is derived from the Fisher Information Matrix (FIM), which measures the sensitivity of the likelihood function to small changes in the parameter vector $\boldsymbol{\theta} = [\theta_1,\ldots,\theta_N]^{\mathrm{T}}$. Mathematically, the CRB for estimating the parameter $\theta_n$ is given by \cite{Kay1998}
\begin{eqnarray}
    \mt{CRB}(\theta_n) = \left[\q{F}^{-1}(\boldsymbol{\theta})\right]_{nn},
\end{eqnarray}\par \vspace{-0mm}
\noindent where $\q{F}(\boldsymbol{\theta})$ is the $N \times N$ FIM. This implies that the variance of any unbiased estimator $\hat{\theta}_n$ satisfies:
\begin{eqnarray}
    \text{Var}(\hat{\theta}_n) \geq \mt{CRB}(\theta_n).
\end{eqnarray}\par \vspace{-0mm}
\noindent where ${\mathrm{Var}}(\hat{\theta}_n)$ is the estimation variance and $\hat{\theta}_n$ is the estimated value of $\theta_n$ \cite{Mark2010RadarBook, Kay1998}. The elements of the FIM are computed as the expected value of the second derivatives of the log-likelihood function $p(\q{Y}; \boldsymbol{\theta})$ with respect to $\boldsymbol{\theta}$ and is given as 
\begin{eqnarray}
    [\q{F}(\boldsymbol{\theta})]_{n,m} =  - \E{\frac{\partial^2 \ln(p(\q{Y}; \boldsymbol{\theta}))}{\partial \theta_n  \partial \theta_m } },
\end{eqnarray}\par \vspace{-0mm}
\noindent  where $\q{Y}$ is the received echo signal \eqref{eqn_obs_sig}. From \eqref{eqn_obs_sig}, $\q{Y} \sim \mathcal{CN}(\q{G} \q{X}, \sigma^2 \q{I}_M)$, and hence the $\{n, m\}$-th element of $\q{F}(\boldsymbol{\theta})$ can be expressed as \cite{Bekkerman:TSP:2006, dan2024}
\begin{eqnarray}\label{eqn_F_nm}
    [\q{F}(\boldsymbol{\theta})]_{n,m} = \frac{2M}{\sigma^2} \Re \left\{\tr \left(\frac{\partial (\q{G} \q{X})}{\partial \theta_n} \frac{\partial (\q{G} \q{X})^{\mathrm{H}}}{\partial \theta_m} \right) \right\}.
\end{eqnarray}\par \vspace{-0mm}
\noindent This indicates how strongly the parameters $\boldsymbol{\theta}$ influence the observations $\q{Y}$ \cite{Mark2010RadarBook, Kay1998}. A higher Fisher information corresponds to more precise parameter estimation and a lower CRB \cite{Mark2010RadarBook, Kay1998}. 

\begin{figure}[!t]\vspace{-0mm}
    \centering
    \includegraphics[width=0.7\textwidth]{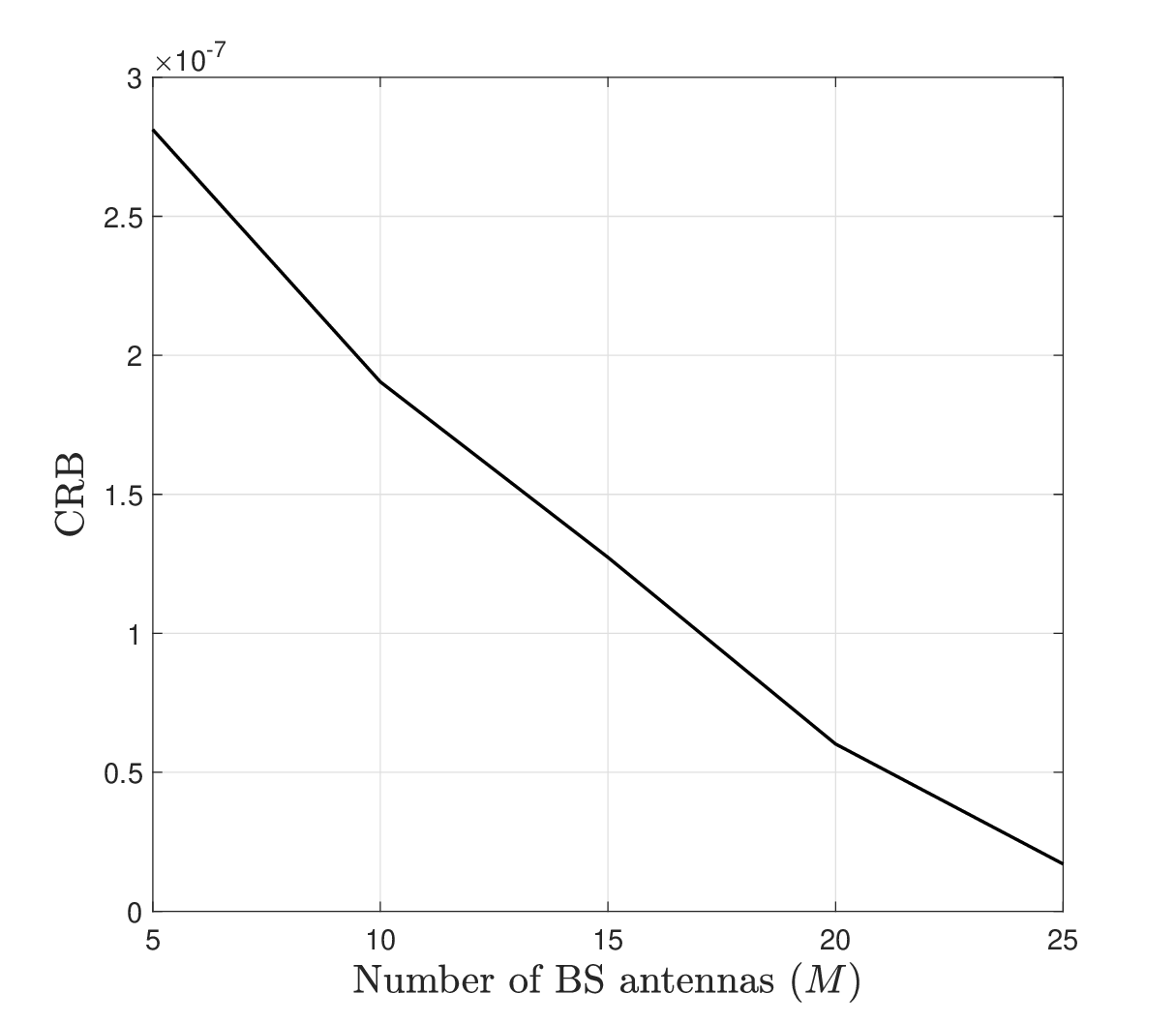}
    \vspace{-0mm}
    \caption{The CRB as a function of the number of BS antennas ($M$).}
    \label{fig_CRB_M} \vspace{-0mm}
\end{figure}

In Figure~\ref{fig_CRB_M}, the CRB is plotted as a function of the number of BS antennas $M$. Figure~\ref{fig_CRB_M} reveals that increasing $M$ reduces CRB, resulting in better sensing parameter estimation.

In sum,  the CRB is a key metric for evaluating sensor performance, designing signal processing schemes, and optimizing ISAC systems \cite{Li2007, Liu2020Radar, Rivetti:WCNC:2024}. It sets theoretical benchmarks for estimation algorithms, resource allocation, and waveform design. For instance, minimizing the CRB for sensing parameters while preserving communication quality helps balance communication-sensing trade-offs \cite{Li2007, Liu2020Radar, Rivetti:WCNC:2024}.

\subsection{Beampattern Gain}
Sensing beampattern gains are a key ISAC metric, determining how energy is directed (transmit) and signals are analyzed (receive) \cite{He2022, Stoica2007, Cui2014, Hua2023}. The transmit beampattern shapes energy radiation for efficient target illumination, while the receive beampattern, optimized via sensing combiners, enhances echo reception \cite{He2022, Stoica2007, Cui2014, Hua2023}. These patterns are crucial for optimizing sensing, target detection, and communication efficiency in ISAC systems \cite{He2022, Stoica2007, Cui2014, Hua2023, Zhenyao2023, galappaththige2024RSMA, zargari2024ISABC}. Thus, the three pivotal beampattern gains are  
\begin{subequations}\label{eqn_beamgain}
\begin{eqnarray} 
p_t(\theta) &&= \left| \q{a}^{\mathrm{H}}(\theta) \q{x} \right|^2, \label{eqn_tx_beamgain} \\ 
p_r(\theta) &&= \left| \q{u}^{\mathrm{H}} \q{b}(\theta) \right|^2, \label{eqn_rx_beamgain} \\
p_c(\theta) &&= \left| \q{u}^{\mathrm{H}} \q{b}(\theta) \q{a}^{\mathrm{H}}(\theta) \q{x} \right|^2, \label{eqn_com_beamgain} 
\end{eqnarray}
\end{subequations}\par
\begin{figure}[!t]\vspace{-0mm}
    \centering
    \includegraphics[width=0.7\textwidth]{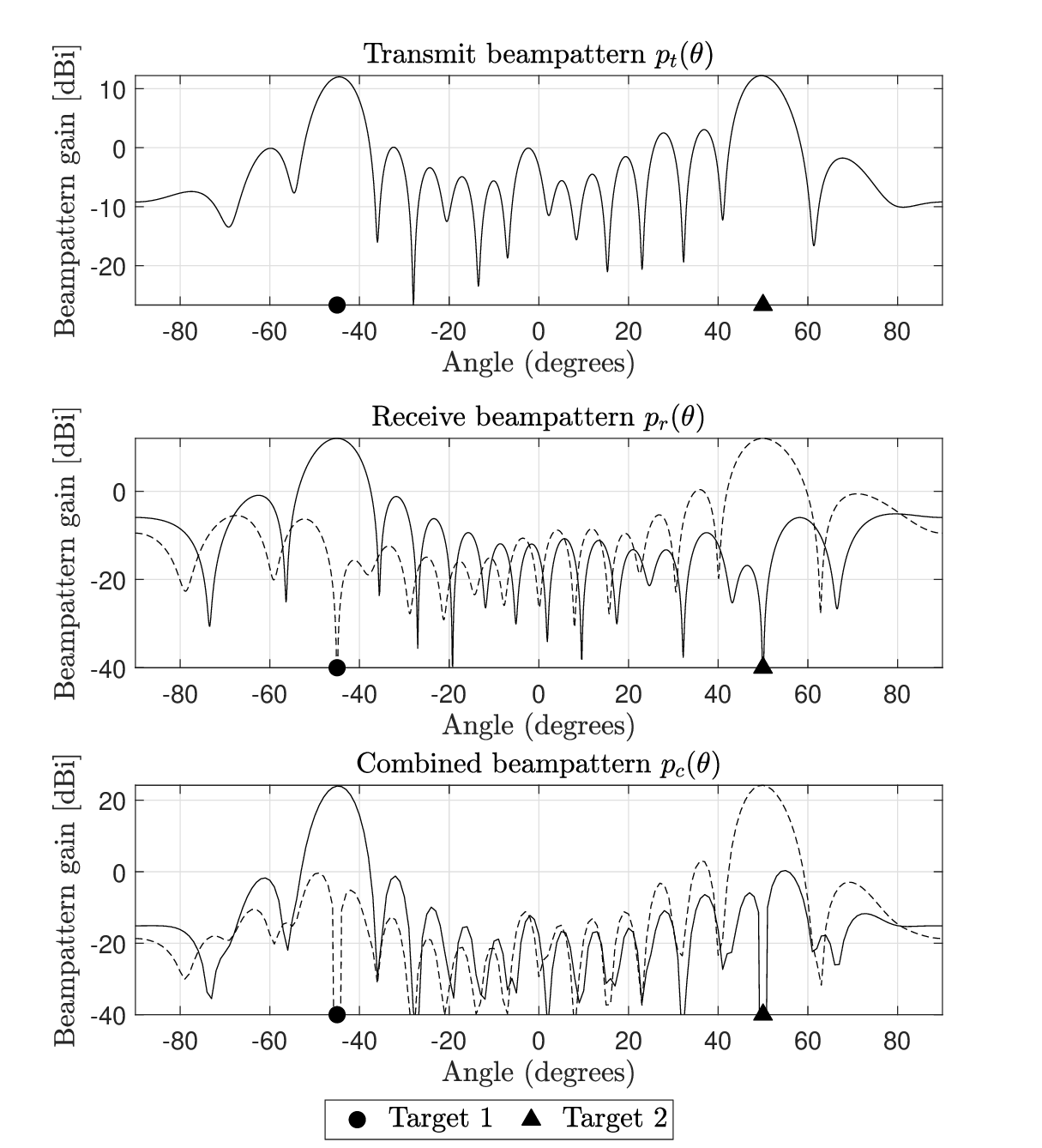}
    \vspace{-0mm}
    \caption{The directional beampattern gain profiles over a \qty{\pm 90}{\degree} angular spread.}
    \label{fig_BeamGain_Example} \vspace{-0mm}
\end{figure}

 \vspace{-0mm}
\noindent where $\q{x} \in \mathbb{C}^{M \times 1}$ is the transmitted signal, $\q{a}(\theta) \in \mathbb{C}^{M \times 1}$ and $\q{b}(\theta) \in \mathbb{C}^{M \times 1}$ are the transmit and receiver steering vectors that capture the array responses from direction $\theta$, and $\q{u} \in \mathbb{C}^{M \times 1}$ is the combiner at the receiver. First, \eqref{eqn_tx_beamgain} is the transmit beampattern gain and illustrates how the transmitted energy disperses as a function of angle $\theta$. Second, \eqref{eqn_rx_beamgain} is the receiver beampattern gain and encapsulates the sensitivity of the ISAC system across different angles during the reception of reflected energy. Finally, \eqref{eqn_com_beamgain} is the combined beampattern gain and offers a combined representation, integrating the effects of transmission and subsequent reflection processing. Figure~\ref{fig_BeamGain_Example} plots these three beampattern gains, demonstrating their energy concentration towards the target directions. 

Note that in CF-ISAC,  the beampattern gains, eq. \eqref{eqn_beamgain},  must be computed at multiple APs. These distributed gains from the APs provide a significant advantage over traditional co-located ISAC \cite{He2022, zargari2024CFISAC}. Unlike co-located ISAC, which relies on a single viewpoint (e.g., a BS) and primarily estimates target angles, CF-ISAC leverages spatial diversity to refine both angular and distance localization. By coherently combining beampattern gains -- whether from transmission, reception, or both --  CF-ISAC mitigates directional ambiguities and improves localization accuracy \cite{zargari2024CFISAC}. In a multi-static scenario, optimizing transmit beampatterns at transmitting APs and enhancing received beampatterns at sensing APs provide reliable data for CPU-level processing. This spatial diversity enhances sensing accuracy, robustness, and detection reliability in dynamic and multi-target environments \cite{zargari2024CFISAC}.

Since each AP must evaluate local beampattern gains and exchange information with the CPU for precise target localization, this advantage comes at the cost of increased computational complexity and coordination overhead \cite{He2022, zargari2024CFISAC}. However, computing beampattern gains at all APs may be unnecessary as a well-placed subset can achieve near-optimal localization while reducing system overhead \cite{He2022, zargari2024CFISAC}. The required number of APs depends on factors such as target count, visibility, desired localization precision, and AP distribution. Generally, three or more APs with sufficient angular separation are needed for 2D localization, while additional APs enhance robustness in multi-target scenarios and mitigate uncertainties from noise and obstructions.

\section{Near-Filed ISAC}
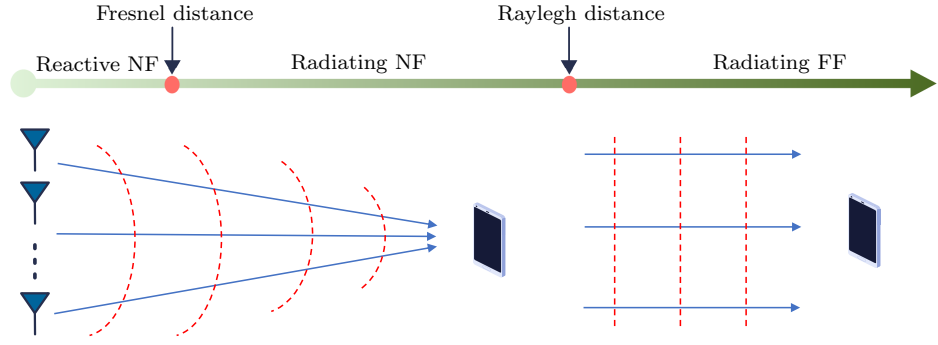
\begin{figure}[!h]\centering \vspace{0mm}
    \def\svgwidth{350pt} 
    \fontsize{8}{8}\selectfont 
    \graphicspath{{Figures/}}
\begingroup%
  \makeatletter%
  \providecommand\color[2][]{%
    \errmessage{(Inkscape) Color is used for the text in Inkscape, but the package 'color.sty' is not loaded}%
    \renewcommand\color[2][]{}%
  }%
  \providecommand\transparent[1]{%
    \errmessage{(Inkscape) Transparency is used (non-zero) for the text in Inkscape, but the package 'transparent.sty' is not loaded}%
    \renewcommand\transparent[1]{}%
  }%
  \providecommand\rotatebox[2]{#2}%
  \newcommand*\fsize{\dimexpr\f@size pt\relax}%
  \newcommand*\lineheight[1]{\fontsize{\fsize}{#1\fsize}\selectfont}%
  \ifx\svgwidth\undefined%
    \setlength{\unitlength}{604.59951782bp}%
    \ifx\svgscale\undefined%
      \relax%
    \else%
      \setlength{\unitlength}{\unitlength * \real{\svgscale}}%
    \fi%
  \else%
    \setlength{\unitlength}{\svgwidth}%
  \fi%
  \global\let\svgwidth\undefined%
  \global\let\svgscale\undefined%
  \makeatother%
  \begin{picture}(1,0.38318526)%
    \lineheight{1}%
    \setlength\tabcolsep{0pt}%
    \put(0,0){\includegraphics[width=\unitlength]{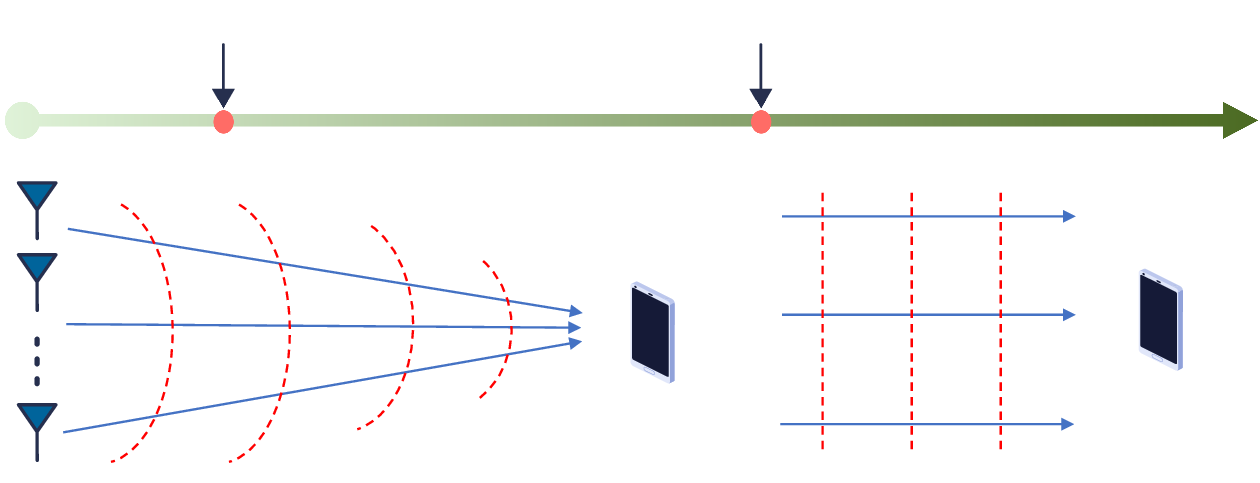}}%
    \put(0.09513553,0.3563834){\color[rgb]{0,0,0}\makebox(0,0)[lt]{\lineheight{1.25}\smash{\begin{tabular}[t]{l}Fresnel distance\end{tabular}}}}%
    \put(0.03034155,0.30152067){\color[rgb]{0,0,0}\makebox(0,0)[lt]{\lineheight{1.25}\smash{\begin{tabular}[t]{l}Reactive NF\end{tabular}}}}%
    \put(0.30437963,0.30284684){\color[rgb]{0,0,0}\makebox(0,0)[lt]{\lineheight{1.25}\smash{\begin{tabular}[t]{l}Radiating NF\end{tabular}}}}%
    \put(0.75839916,0.30284684){\color[rgb]{0,0,0}\makebox(0,0)[lt]{\lineheight{1.25}\smash{\begin{tabular}[t]{l}Radiating FF\end{tabular}}}}%
    \put(0.52682624,0.3563834){\color[rgb]{0,0,0}\makebox(0,0)[lt]{\lineheight{1.25}\smash{\begin{tabular}[t]{l}Raylegh distance\end{tabular}}}}%
  \end{picture}%
\endgroup%
 \vspace{0mm}
    \caption{FF planar wavefront versus NF spherical wavefront.}\vspace{0mm} \label{fig_NF_FF}
\end{figure}

Recent advances in wireless technologies, such as the use of extremely large-scale antenna arrays (ELAAs) with hundreds to thousands of elements (typically more than 100 antennas) and the adoption of higher-frequency bands (e.g., mmWave at \qtyrange{10}{100}{\giga\hertz} and THz at \qtyrange{100}{10000}{\giga\hertz}), have led to a fundamental shift in EM propagation models (Figure~\ref{fig_NF_FF}) \cite{Azar2024}. These developments move system modeling from the traditional far-field (FF) assumptions to the more accurate near-field (NF) regime \cite{Azar2024}.

The transition from FF to NF alters the fundamental wave propagation characteristics. In the FF regime, signals propagate as planar waves, and path loss decays proportionally to the inverse square of the distance. In contrast, NF propagation is characterized by spherical wavefronts and a stronger distance dependency, where signal strength typically decays with the fourth to sixth power of the distance (Figure~\ref{fig_NF_FF}) \cite{Azar2024}.

Despite the more complex fading behavior, NF propagation introduces several advantages for ISAC:
\begin{itemize}
    \item Spherical wave modeling enables joint estimation of range and angle from a single observation, enhancing sensing capabilities.

    \item Beam focusing in both angular and radial domains increases echo SNR, improving estimation precision.

    \item Higher spatial resolution from wider array apertures supports finer discrimination of closely spaced targets.

    \item NF allows distinguishing multiple targets at the same angle but different distances (i.e, target separability), which is infeasible under FF assumptions.

    \item Finally, thanks to the range-resolution benefits of NF sensing, fewer distributed nodes are needed for full-scene coverage, reducing the need for distributed synchronization.
\end{itemize}
Overall, NF ISAC not only enhances communication reliability via spatial focusing but also significantly improves sensing granularity and precision, making it a foundational technology for next-generation networks and applications such as holographic communications, micro-scale radar sensing, and ultra-dense environments \cite{Azar2024, Zhang2022NF, Diluka2024NF, Wang2023NF, Diluka2025NF}.

\subsection{NF for Communication} 
The distinct properties of NF spherical wavefronts can significantly enhance communication performance, particularly in terms of system capacity and available channel DoFs \cite{Liu2023Survey, qu2023nearfieldSurvey}.

In conventional FF communication, the incident wavefront is approximated as planar, leading to a linear phase progression across antenna elements. The phase shift observed between antenna elements is primarily governed by the angle of arrival (AoA) or angle of departure (AoD) and the inter-element spacing \cite{Liu2023Survey, qu2023nearfieldSurvey}. Due to the large distance between transmitter and receiver, the propagation distance increment between successive antenna elements is nearly constant, causing the resulting LoS channel vector to lie in the steering vector subspace. This geometric uniformity results in limited spatial DoFs, typically yielding only a single dominant mode in LoS communication \cite{Liu2023Survey, qu2023nearfieldSurvey}.

In contrast, NF communication deviates from these assumptions due to spherical wave propagation. In this regime, the propagation distance from each transmit antenna to a receive antenna is non-uniform, resulting in nonlinear and element-specific phase shifts \cite{Liu2023Survey}. The channel matrix in NF is no longer a simple outer product of steering vectors but encodes unique geometric information for each transceiver pair. This yields:
\begin{itemize}
    \item Increased spatial DoFs, as the system can exploit both angular and radial variations.
    
    \item Higher channel rank, supporting more independent spatial streams in LoS conditions.

    \item User separability in angle-range space, allowing discrimination of users located at the same angle but different distances.
\end{itemize}
This last property addresses a key limitation of FF systems, i.e., their inability to resolve users positioned along the same angular direction \cite{Liu2023Survey}. NF systems leverage the range sensitivity of spherical waves to separate such users, substantially reducing inter-user interference and improving spatial multiplexing capability \cite{Liu2023Survey}. In summary, NF propagation introduces richer spatial structures in communication channels, which can be harnessed for enhanced capacity, better user separation, and improved interference management \cite{Liu2023Survey}.

\subsection{NF for Sensing} 
NF propagation fundamentally enhances sensing capabilities by enabling beam-focusing, which produces tightly confined spatial beams that concentrate signal energy at specific locations. This spatial concentration increases the received signal power at the target, thereby improving the SNR of echo signals--an essential factor for precise sensing and parameter estimation \cite{qu2023nearfieldSurvey, Wang2023}.

Unlike FF systems that rely solely on angular information, NF systems leverage the inherent characteristics of spherical wavefronts, which encode both range (distance) and angle information. This dual dependency enables range-aware reception, facilitating selective signal capture from specific sources while suppressing interference from others. As a result, NF reception not only boosts sensing accuracy but also enhances interference mitigation through spatial filtering \cite{qu2023nearfieldSurvey}.

By combining NF reception with advanced parameter estimation algorithms and spectrum search techniques, significantly higher sensing precision can be achieved compared to FF methods. For example, antenna arrays with known spatial layouts can utilize signal attributes such as received power, time of arrival (ToA), or phase differences to infer target distance. NF propagation further enhances this process by embedding richer spatial information in the observed signals. These capabilities extend to various emerging applications, such as fine-grained localization, gesture and human activity recognition, and object classification, which require high spatial resolution \cite{qu2023nearfieldSurvey}.

However, realizing the full benefits of NF sensing (and communication) hinges on appropriately designed transmission architectures. Conventional FF-based designs fail to capture the unique characteristics of NF channels, leading to suboptimal performance. For instance, FF beams tend to diverge at close ranges, increasing inter-user interference and reducing angular resolution \cite{qu2023nearfieldSurvey}. Applying standard FF beamforming in NF settings often results in significant degradation in sensing accuracy and communication throughput. Therefore, to unlock the advantages of NF, it is crucial to develop dedicated beamforming strategies and transmission designs that account for the nonlinear, distance-dependent phase variations and spatial diversity of NF channels \cite{qu2023nearfieldSurvey}. These include methods tailored for NF channel modeling, angle-range coupling, and beam codebook construction \cite{qu2023nearfieldSurvey}.

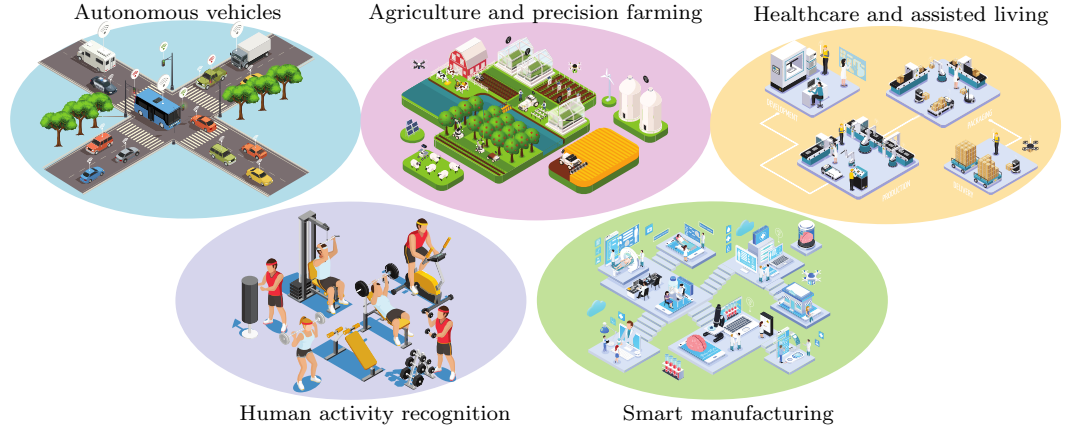
\begin{figure*}[!t]\centering \vspace{0mm}
\centering
    \def\svgwidth{400pt} 
    \fontsize{8}{8}\selectfont 
    \graphicspath{{Figures/}}
\begingroup%
  \makeatletter%
  \providecommand\color[2][]{%
    \errmessage{(Inkscape) Color is used for the text in Inkscape, but the package 'color.sty' is not loaded}%
    \renewcommand\color[2][]{}%
  }%
  \providecommand\transparent[1]{%
    \errmessage{(Inkscape) Transparency is used (non-zero) for the text in Inkscape, but the package 'transparent.sty' is not loaded}%
    \renewcommand\transparent[1]{}%
  }%
  \providecommand\rotatebox[2]{#2}%
  \newcommand*\fsize{\dimexpr\f@size pt\relax}%
  \newcommand*\lineheight[1]{\fontsize{\fsize}{#1\fsize}\selectfont}%
  \ifx\svgwidth\undefined%
    \setlength{\unitlength}{311.12908936bp}%
    \ifx\svgscale\undefined%
      \relax%
    \else%
      \setlength{\unitlength}{\unitlength * \real{\svgscale}}%
    \fi%
  \else%
    \setlength{\unitlength}{\svgwidth}%
  \fi%
  \global\let\svgwidth\undefined%
  \global\let\svgscale\undefined%
  \makeatother%
  \begin{picture}(1,0.41736934)%
    \lineheight{1}%
    \setlength\tabcolsep{0pt}%
    \put(0,0){\includegraphics[width=\unitlength]{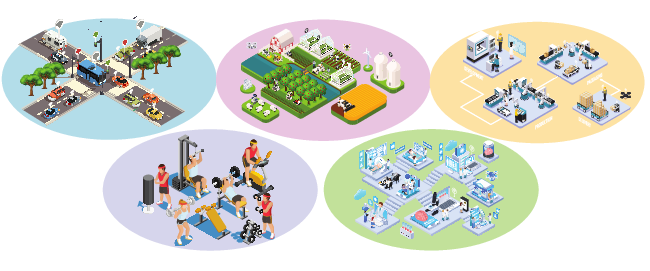}}%
    \put(0.21876735,0.01298682){\makebox(0,0)[lt]{\lineheight{1.25}\smash{\begin{tabular}[t]{l}Human activity recognition\end{tabular}}}}%
    \put(0.70203083,0.38756271){\makebox(0,0)[lt]{\lineheight{1.25}\smash{\begin{tabular}[t]{l}Healthcare and assisted living\end{tabular}}}}%
    \put(0.33996983,0.38970845){\makebox(0,0)[lt]{\lineheight{1.25}\smash{\begin{tabular}[t]{l}Agriculture and precision farming\end{tabular}}}}%
    \put(0.57927796,0.01285408){\makebox(0,0)[lt]{\lineheight{1.25}\smash{\begin{tabular}[t]{l}Smart manufacturing\end{tabular}}}}%
    \put(0.06265176,0.38994947){\makebox(0,0)[lt]{\lineheight{1.25}\smash{\begin{tabular}[t]{l}Autonomous vehicles\end{tabular}}}}%
  \end{picture}%
\endgroup%
 \vspace{0mm}
    \caption{ISAC applications.}\vspace{0mm} \label{fig_ISACApplications}
\end{figure*}

\section{ISAC Applications}
ISAC has a wide range of applications, including automotive, healthcare, smart cities, and industrial automation (Figure~\ref{fig_ISACApplications}) \cite{liu2023integratedbook, Wang2022ISAC}. 

\subsection{ISAC for Autonomous Vehicles} 
Self-driving vehicles need both high-resolution radar sensing for obstacle detection and reliable communication for V2X connectivity. This fusion can transform the transportation industry by enhancing navigation safety, highway capacity, and traffic flow while reducing fuel consumption, pollution, and accident rates \cite{Zeng2019, Milanes2012}. In particular,  autonomous vehicles acquire ambient information while exchanging data with roadside units (RSUs), other vehicles, and pedestrians \cite{Zeng2019, Milanes2012}. Conversely, ISAC-aided V2X can offer environmental information for quick vehicle platooning, secure and seamless access, and simultaneous localization and mapping,  addressing EM compatibility and spectrum congestion challenges \cite{Zeng2019, Milanes2012}. 

\subsection{Human Activity Recognition}
Computing systems can monitor, evaluate, and assist people daily by recording their behaviors, making activity recognition essential \cite{Yongsen2019}. Wireless signal fluctuations, influenced by static and moving objects and human actions, can detect activities such as presence, proximity, falls, sleep, breathing, and more. These capabilities have significant applications in healthcare and transportation \cite{Yongsen2019}. For instance, detecting a driver's blink rate with high-resolution sensors can help identify drowsy driving, enhancing road safety. Additionally, integrating sensing capabilities into commercial wireless devices like Wi-Fi can identify occupant behaviors, creating innovative, human-centric environments \cite{Yongsen2019}.

\subsection{Smart Manufacturing (Industry 4.0)}Communication and sensing enable the automation of production lines in smart manufacturing \cite{Popovski2019}. Modern factories feature interconnected machinery, robotic arms, and autonomous systems collaborating in real-time to ensure efficiency and precision. Machines communicate wirelessly and sense their environment, detecting factors like vibrations, temperature changes, and product quality. This dual functionality supports predictive maintenance, allowing machines to identify potential failures proactively, reducing downtime, and minimizing repair costs \cite{Popovski2019}.

\subsection{Healthcare and Assisted Living}Integrated solutions for continuous monitoring and real-time communication are vital for improving patient outcomes and elderly care \cite{Philip2021}. These systems enable remote monitoring by collecting health data such as heart rate, blood pressure, and oxygen levels and transferring it to healthcare providers for real-time evaluation. This allows personalized care and timely interventions without requiring hospital visits \cite{Philip2021}. In assisted living, sensors track daily activities, detect falls, and identify abnormal behavior, such as prolonged inactivity, alerting caregivers or medical professionals during emergencies. Hospitals can also benefit by optimizing patient flow, medical equipment usage, and staff coordination, enhancing healthcare delivery \cite{Philip2021}.

\subsection{Agriculture and Precision Farming}Critical environmental conditions, such as soil moisture, nutrient levels, weather, and crop health, play a key role in agriculture \cite{Chander2023}. For instance, soil sensors can monitor moisture content in real-time, enabling irrigation systems to adjust water usage automatically. Drones equipped with ISAC technology can survey large fields, detect plant diseases, pest infestations, and growth patterns, and use this data to apply pesticides or fertilizers precisely. This integration of communication and sensing reduces resource waste, such as water and chemicals, while boosting crop yields and sustainability \cite{Chander2023}.

\section{Industry Progress and Standardization}
As \num{6}G research gains momentum, ISAC has emerged as a focal point, drawing significant attention from key industries as well as standardization organizations, such as the Institute of Electrical and Electronics Engineers (IEEE) standardization association, the International Telecommunication Union (ITU), and the 3rd Generation Partnership Project (3GPP). 

The IEEE and 3GPP standardization bodies have significantly contributed to developing ISAC-related specifications. In \num{2019}, IEEE \num{802.11} formed the Wireless Local-Area Network (WLAN) Sensing Topic Interest Group and Study Group, followed by the establishment of the formal Task Group IEEE \num{802.11}bf in \num{2020}. These projects aim to define the necessary modifications to existing Wi-Fi standards, aligning them with \num{802.11}-compliant waveforms to improve sensing capabilities \cite{Meneghello2023}. Moreover, the ITU-R M.2516 technical report identifies ISAC as one of six key next-generation wireless technology developments \cite{IMT2023}.

The 3GPP has introduced new ISAC study items, highlighting potential applications across various domains, including indoor environments (e.g., home, office, factory), highways (e.g., automotive, traffic monitoring, intrusion detection), high-speed railways (e.g., autopilot, intrusion detection), weather forecasting (e.g., rainfall, flooding), UAVs (e.g., flight trajectory tracking, collision, and intrusion detection), traffic management (e.g., tourist/sports hotspot detection, car parking), health monitoring (e.g., heart rate, breathing, sleep tracking), and extended reality (e.g., gaming, metaverse) \cite{3GPPISAC2024}. In addition, six main ISAC configurations/topologies are considered in the 3GPP Release \num{19} RAN-1 study to facilitate sensing in communication systems \cite{3GPPISACTopology23}: (i) mono-static sensing via user equipment (UE), (ii) mono-static sensing via next-generation Node B (gNB), (iii) bi-static sensing from gNB to UE, (iv) bi-static sensing from UE to gNB, (v) bi-static sensing from gNB to gNB, and (vi) bi-static sensing from UE to UE. 

Conversely, leading wireless industries, such as Ericsson, NTT DOCOMO, ZTE, China Mobile, China Unicom, Intel, and Huawei, have emphasized the critical role of sensing in their \num{6}G white papers and Wi-Fi 7 visions, highlighting its importance in shaping the future of wireless technology \cite{Ericssonwhitepaper2022, CarlosIntel2020, Pin2021}. In particular, the technical report \cite{Waring_2020a} identifies harmonized sensing and communication as one of the three emerging opportunities for beyond \num{5}G networks, with the primary objective of leveraging the sensing capabilities of existing mMIMO BSs and enabling future UAVs and automotive vehicles. In \cite{Pin2021},  Huawei expects the \num{6}G air interface to handle simultaneous wireless communication and sensing signals,  enabling ISAC-enabled cellular networks to ``see" the physical world. In \cite{Alloulah2019}, Nokia proposes a unified mmWave system as a blueprint for future indoor ISAC technologies.

\section{Evolution of the ISAC Networks}\label{subsec_CF_isac}
Conventional ISAC systems, also referred to as \textit{link-level} or \textit{system-level} ISAC, primarily focus on single- or dual-BS cellular architectures~\cite{liu2023integratedbook, Liu2022ISAC, zargari2024riemannian, Rahman2020, Liu2020, Diluka2023, Diluka2024NF, Diluka2025NF, Zhenyao2023, Ouyang2022Uplink, Ouyang2022}. However, these systems face several limitations, including inter-cell interference, restricted coverage due to physical obstructions and high-frequency attenuation, and mismatches between sensing and communication ranges caused by two-hop path loss~\cite{Babu:TWC:2024, Wang:CLET:2024}.

To overcome these challenges, \textit{network-level ISAC} systems have emerged. By leveraging multiple coordinated ISAC transceivers across cells, these architectures improve both communication reliability and sensing accuracy~\cite{Li:TWC:2024, Meng:TWC:2024}. This collaborative framework enables high-throughput, ultra-reliable communications and high-resolution, wide-area sensing.

For sensing tasks, network-level ISAC extends the coverage area through \textit{multi-static sensing}, in which each BS processes both locally reflected echoes and target-reflected signals originating from other BSs or users. On the communication side, collaboration among transceivers enables techniques such as CoMP transmission to mitigate inter-cell interference and ensure robust connectivity. Additionally, integrating sensing and communication at the signal, data, and task levels improves overall system performance in terms of resource utilization, signal quality, interference management, and coverage enhancement~\cite{Meng:WL:2024}.

Despite these advantages, network-level ISAC introduces challenges such as scalability constraints, high fronthaul overhead, increased computational complexity, and sensitivity to CSI imperfections ~\cite{Li:TWC:2024, Meng:TWC:2024}. 
\begin{itemize}
    \item  \textit{Coverage limitations at cell edges:} Users near the boundaries of adjacent cells may experience degraded performance due to imperfect alignment of coordinated transmissions.
    
    \item \textit{Limited scalability:} As the number of users and BSs increases, system scalability becomes increasingly complex and resource-intensive, especially in real-time CoMP coordination.
    
    \item \textit{High fronthaul overhead:} CoMP requires real-time exchange of CSI and user data across BSs, placing substantial demand on the fronthaul infrastructure.
    
    \item \textit{Sensitivity to imperfect CSI:} The effectiveness of coordination relies heavily on accurate and timely CSI, which is difficult to maintain in dynamic environments due to estimation errors, feedback delays, and quantization.
    
    \item \textit{High computational complexity:} Coordinated beamforming, joint power allocation, and user scheduling significantly increase computational demands as the system scales.
    
    \item \textit{EE:} Centralized coordination and high-power transmissions can increase energy consumption, making CoMP-based ISAC architectures less energy efficient in practice.
\end{itemize}  
These limitations stem from the need for real-time data sharing, centralized coordination, and precise synchronization across distributed BSs. A detailed comparison of system-level and network-level ISAC is summarized in Table~\ref{tab_isac_compare}, highlighting the trade-offs in performance, complexity, and practical deployment considerations.

\begin{table}[t!]
\centering
\caption{Comparison of system-level and network-level ISAC.}\label{tab_isac_compare}
\begin{tabular}{|p{3cm}|p{5cm}|p{5cm}|}
\hline
\textbf{Aspect} & \textbf{System-Level ISAC} & \textbf{Network-Level ISAC} \\ \hline \hline
Architecture     &  Single or dual BSs  &  Multiple coordinated BSs/APs \\ \hline
Coverage   & Limited by individual BS range    &  Extended via multi-static sensing and BS collaboration  \\ \hline
Interference handling   & Inter-cell interference is a challenge    & Mitigates interference using CoMP and coordinated transmission   \\ \hline
Sensing range   &  Limited, often mismatched with communication range   & Enhanced through cooperative sensing across BSs   \\ \hline
DoF/Spatial diversity   &  Limited due to independent operation   & Increased via joint processing of spatially distributed signals   \\ \hline
Resource allocation   &  Local optimization only   & Joint resource allocation across the network   \\ \hline
Fronthaul dependency   &  Low or none   &  High: Requires real-time CSI and data sharing  \\ \hline
Computational complexity   &  Moderate   &  High: Requires joint beamforming, power control, and scheduling  \\ \hline
Scalability   &  More scalable in simple deployments   &  Challenging to scale with user and BS density  \\ \hline
EE   &  Typically better due to simpler coordination   &  Lower EE due to increased coordination overhead and transmission power  \\ \hline
Robustness to CSI errors   &  Less sensitive   &  Highly sensitive; requires accurate, low-latency CSI  \\ \hline
\end{tabular}
\end{table}

This comparative analysis underscores that while network-level ISAC offers substantial performance gains, its benefits come at the cost of system complexity and infrastructure demands. Accordingly, future ISAC system designs must balance these trade-offs, considering both architectural capabilities and operational constraints.
	\chapter{Cell-Free ISAC}\label{chp_CF_isac}
This chapter provides a comprehensive overview of CF-ISAC, covering its system architecture, sensing principles, radar configurations, and advantages over conventional ISAC designs. It emphasizes the benefits of multi-static sensing enabled by distributed APs, including broader coverage, higher accuracy via spatial diversity, improved interference management, and robustness to node failures. Key design aspects such as synchronization, beamforming coordination, and scalability are also addressed. A comparative evaluation with link-level ISAC demonstrates the superiority of  CF-ISAC in complex or large-scale environments, paving the way for high-precision, scalable ISAC implementations.

\section{Key Features of Cell-Free ISAC}

ISAC architectures, often relying on centralized or CoMP multi-BS coordination, are constrained by limited coverage, high inter-cell interference, and dependency on fronthaul capacity. CF-ISAC, by contrast, builds on the distributed CFMM framework introduced in Chapter 2 for foundational concepts. In CF-ISAC, distributed APs collaborate through centralized processing, naturally supporting multi-static sensing configurations without requiring full-duplex hardware at each AP. This spatial separation allows APs to act as independent transmitters and receivers, yielding uncorrelated and diverse observations that enhance target detection and environmental awareness~\cite{Mao2023, Demirhan2023, Huang2022Coordinated, Cao2023Design, Wang2023, Sakhnini2022Uplink, Silva2023, Behdad2022, Behdad2024Interplay}.

CF-ISAC's decentralized architecture enables refined spatial resolution and enhanced spectrum reuse, while its cooperative APs provide the spatial diversity gains necessary for robust sensing and high-throughput communication \cite{Mao2023, Yuanyuan2024, Demirhan2023, demirhan2024cellfree, Elfiatoure2023, Mao2024, Liu2024, Huang2022Coordinated, Cao2023Design, Behdad2024Interplay, elfiatoure2024multiple}. Compared to traditional co-located ISAC systems, CF-ISAC utilizes wireless resources more effectively by decoupling coverage from AP location constraints and exploiting network-wide coordination. These capabilities make CF-ISAC a compelling enabler for advanced applications in 6G networks, such as autonomous vehicles, smart infrastructure, industrial automation, IoT, and healthcare systems.

\subsection{Distributed Antenna Systems}
A cornerstone of CF-ISAC is its DAS architecture, in which geographically distributed APs collaboratively serve users and sense the environment~\cite{Moerman2022, You2010, Haimovich2008}. Unlike traditional cellular systems with co-located BSs, CF-ISAC distributes antennas over a wide area, which brings several key advantages for both communication and sensing. These include leveraging the spatial diversity of distributed access points (APs) across the entire coverage area to mitigate spatially correlated fading, overcome shadowing from obstacles, and shorten the average end-to-end transmission distances \cite{Demir2021book, Ngo2017, Zhang2019cellfree}.

\textit{Spatial Diversity and Robustness:} DAS offers robust protection against spatially correlated fading and shadowing effects. Since each AP experiences an independent channel realization, the effective channel gain to a user or a target is averaged across multiple links, i.e., 
\begin{eqnarray}
    h_k = \frac{1}{M_k} \sum_{m\in \mathcal{A}_k} h_{mk},
\end{eqnarray} \par \vspace{-0mm}
\noindent where $h_{mk}$ denotes the channel between the $k$-th user/target and the $m$-th AP, $\mathcal{A}_k$ is the set of APs serving the $k$-th user/target, and $M_k = \vert \mathcal{A}_k \vert$ is the number of serving APs. This spatial diversity reduces deep fades and enhances communication reliability and sensing detectability \cite{Mao2023, Yuanyuan2024, Demirhan2023, demirhan2024cellfree, Elfiatoure2023, Mao2024, Liu2024, Huang2022Coordinated, Cao2023Design}.

\textit{Multi-Static Sensing and Spatial Resolution:} For sensing, DAS enables multi-static radar configurations where each AP acts as a spatially distinct transmitter or receiver. Consider the received echo at the $m$-th AP from a target due to the transmitted signal from the $n$-th AP ($s(t)$). It can be modeled as
\begin{eqnarray}
    y_{mn}(t) = \alpha_{mn} s(t-\tau_{mn}) + w_{mn}(t),
\end{eqnarray} \par \vspace{-0mm}
\noindent where $\alpha_{mn}$ and $\tau_{mn}$ are the complex gain and round-trip delay, respectively, and $w_{mn}(t)$ is the AWGN noise. These observations from multiple such paths, i.e., $n$-th AP $\rightarrow$ target $\rightarrow$ $m$-th AP, allow for coherent combining, improving target localization accuracy and Doppler estimation \cite{Mao2023, Yuanyuan2024, Demirhan2023, demirhan2024cellfree, Elfiatoure2023, Mao2024, Liu2024, Huang2022Coordinated, Cao2023Design}.

\textit{Uniform Quality of Service and Low Latency:} In DAS, there are no fixed cell boundaries~\cite{Demir2021book, Ngo2017, Zhang2019cellfree}. Each user or sensing target is jointly served by multiple nearby APs. This proximity reduces the average path loss. Consequently, latency and energy consumption are minimized, key requirements in real-time ISAC applications~\cite{Mao2023, Yuanyuan2024, Demirhan2023, demirhan2024cellfree, Elfiatoure2023, Mao2024, Liu2024, Huang2022Coordinated, Cao2023Design, Behdad2024Interplay}.

\textit{Improved Coverage and Sensing Field:} Since APs are scattered throughout the environment, CF-ISAC ensures both dense communication coverage and rich sensing observation geometries~\cite{Mao2023, Yuanyuan2024, Demirhan2023, demirhan2024cellfree, Elfiatoure2023, Mao2024, Liu2024, Huang2022Coordinated, Cao2023Design, Behdad2024Interplay}. This enhances robustness in cluttered or obstructed environments where single-site ISAC systems may suffer blind spots.

The distributed architecture of CF-ISAC transforms traditional wireless systems by merging high-throughput communication and high-resolution sensing into a unified framework. DAS underpins this transformation, providing spatial diversity, reducing latency, eliminating cell boundaries, and enabling robust multi-static sensing capabilities~\cite{liu2023integratedbook}.

The benefits of DAS in CF-ISAC compared to the conventional co-located ISAC systems will be investigated in Chapter~\ref{chp_CF_isac_resource}.

\subsection{Seamless Handover and User/Target-Centric Operation}
In conventional cellular networks, as users move across cell boundaries, handover procedures must be initiated to transfer service responsibility between BSs~\cite{Demir2021book, Ngo2017, Zhang2019cellfree}. These handovers can introduce signaling overhead, latency, and service disruptions, particularly problematic in high-mobility scenarios such as autonomous driving or UAV control.

In contrast, CF-ISAC allows for handover-free operation, where service responsibility naturally transitions to nearby APs as the user or target moves through the environment~\cite{Demir2021book, Ngo2017, Zhang2019cellfree}. As a result, CF-ISAC provides reduced control signaling overhead, lower end-to-end latency, higher link reliability and robustness, and persistent sensing coverage and communication quality. This capability is vital for mobility-intensive ISAC applications, including vehicular networks, mobile robotics, and drone swarms, where both uninterrupted data transmission and continuous environmental sensing are essential~\cite{liu2023integratedbook}.

\textit{User/Target-Centric Operation:} Beyond mobility, CF-ISAC enables a user-/target-centric network design, which dynamically tailors network resources to individual users (for communication) or targets (for sensing) based on spatial proximity, channel state, and task requirements~\cite{Demir2021book, Ammar2022}. Rather than assigning fixed BSs or cells, a subset of APs is dynamically clustered around each user or target. This cluster of cooperating APs is responsible for delivering data and extracting sensing information. A two-stage process can be employed for efficient AP selection \cite{Demir2021book, Ammar2022}. 
\begin{itemize} 
\item \textit{Stage 1 - Long-term association:} A set of candidate APs, i.e., $\mathcal{A}_k^{\mathrm{long}}$, is chosen for the $k$-th user/target based on long-term channel statistics, e.g., path loss $\beta_{mk}$:
\begin{eqnarray}
    \mathcal{A}_k^{\mathrm{long}} = \{m \in \mathcal{M} \vert \beta_{mk} \geq \gamma\},
\end{eqnarray} \par \vspace{-0mm}
\noindent where $\gamma$ is the path-loss threshold and $\mathcal{M}$ is the set of APs.

\item \textit{Stage 2 - Short-term refinement:} At each time slot, the cluster is refined using small-scale fading $h_{mk}$, instantaneous traffic loads, or beam alignment metrics to form $\mathcal{A}_k^{\mathrm{short}}  \subseteq \mathcal{A}_k^{\mathrm{long}}$. This refined cluster is then used for beamforming and power control, scheduling of communication and sensing tasks, and load balancing and interference coordination.
\end{itemize}
This adaptive, fine-grained association enables the system to maximize communication SE, sensing resolution, and EE \cite{Demir2021book, Ammar2022}. CF-ISAC's handover-free architecture and user/target-centric design enable uninterrupted service and environment tracking in dynamic settings. By forming dynamic AP clusters optimized for each user or target, the system adapts in real-time to changes in location, channel quality, and sensing needs, making it particularly well-suited for 6G applications that require high mobility and situational awareness \cite{Demir2021book, Ammar2022}.

Despite the conceptual benefits of handover-free operation in CF-ISAC systems, the practical implications under varying mobility patterns, network densities, and dynamic user/target behavior remain insufficiently explored. Although user/target-centric operations have been examined in some early studies \cite{Cao2023, Buzzi2024}, a unified framework capturing their full potential across diverse ISAC scenarios is still lacking. Key challenges -- such as optimal AP clustering, latency-performance trade-offs, and the joint optimization of communication and sensing in dynamic environments -- remain open and present promising avenues for future research to fully exploit CF-ISAC capabilities in next-generation wireless systems.

\subsection{Interference Management and Resource Allocation} 
Interference is a critical challenge in co-located ISAC systems, where sensing and communication functions share spectral and hardware resources~\cite{liu2023integratedbook}. These dual functions often impose contradictory requirements on waveform design, power distribution, and beamforming strategies. For example, waveforms optimized for high SE in communication may exhibit poor range resolution or ambiguity functions in sensing, and vice versa. Moreover, the presence of multiple users and sensing targets introduces inter-user and cross-task interference, necessitating sophisticated joint optimization techniques to ensure reliable performance~\cite{liu2023integratedbook}.

In contrast, CF-ISAC mitigates many of these limitations through its spatially distributed and cooperative architecture~\cite{Demir2021book, Ngo2017, Zhang2019cellfree}. Geographically dispersed APs collaboratively manage interference by sharing CSI, sensing data, and coordination signals in real time. This distributed nature facilitates fine-grained spatial control and opens the door to advanced interference mitigation strategies across both communication and sensing domains \cite{Demir2021book,  Ngo2017, Zhang2019cellfree}.

\textit{Advanced Interference Management Techniques:} Several powerful strategies can be applied in CF-ISAC to handle both intra-task (user-user or target-target) and inter-task (sensing-communication) interference \cite{Demir2021book,  Ngo2017, Zhang2019cellfree}. These include:
\begin{itemize} 
    \item \textit{CoMP processing:} Joint transmission/reception across APs mitigates inter-user interference and enables coherent gain for both communication and radar tasks. Mathematically, for the $k$-th user, the received signal can be given as
    \begin{eqnarray}
        y_k = \sum_{m \in \mathcal{A}_k} \q{h}_{mk}^{\mathrm{H}} \q{w}_m q_k + \sum_{i\neq k} \sum_{m \in \mathcal{A}_i} \q{h}_{mk}^{\mathrm{H}} \q{w}_m q_j + n_k,
    \end{eqnarray} \par \vspace{-0mm}
    \noindent where $\q{w}_m$ is the beamforming vector at the $m$-th AP, $\q{h}_{mk}$ is the channel from the $m$-th AP to the $k$-th user, and $q_k$ is the symbol intended for the $m$-th user. CoMP seeks to design $\{\q{w}_m \}$ to maximize the SINR of the $k$-th user while suppressing cross-user interference.

    \item \textit{Interference alignment:} By aligning interference into orthogonal subspaces, residual DoFs are preserved for the desired signals. Particularly beneficial when APs and users have multiple antennas, this method leverages null-space projections.

    \item \textit{Joint beamforming design:} The transmit beamforming vector $\q{w}_m$ can be jointly optimized for communication QoS (e.g., SINR constraints) and sensing metrics such as beampattern sharpness or CRLB minimization. This leads to Pareto-optimal trade-offs.

    \textit{Dynamic resource allocation:} Resources, such as power $P_m$, time slots $T$, and bandwidth segments $B$, are allocated using utility-driven optimization. For example, one may solve
    \begin{eqnarray}
        \max_{\q{w}, \q{P}} \quad \sum_{k} \alpha_k \mathcal{S}_k^{\mathrm{Com}} + \sum_{t} \beta_t \mt{SINR}_t^{\mathrm{Sen}},
    \end{eqnarray} \par \vspace{-0mm}
    \noindent  
    where $\mathcal{S}_k^{\mathrm{Com}}$ is the $k$-th user communication rate and $\mt{SINR}_t^{\mathrm{Sen}}$ is the sensing SINR for the $t$-th target with $\alpha_k$ and $\beta_t $ capturing task priorities. 
\end{itemize}

In addition to these centralized strategies, the spatial granularity of CF-ISAC enables task-aware power shaping at the AP level. For instance, APs closer to a communication user may boost power in that direction, while others form low-power beams for side-lobe radar observations. This fine-grained control is particularly effective in environments that are bandwidth-limited, interference-prone, or rapidly changing, such as factories with moving obstacles or urban vehicular scenarios~\cite{Demir2021book, Ngo2017, Zhang2019cellfree}.

Chapter~\ref{chp_CF_isac_resource}, along with Chapters \ref{chp_CF_isac_security} and \ref{chp_NA_CF_isac}, explores CF-ISAC interference management and resource allocation.

\subsection{AP Cooperation and Synchronization}
In a CF-ISAC network, a large number of geographically distributed APs operate collaboratively to support both high-throughput communication and high-resolution sensing~\cite{Demir2021book, Ngo2017, Zhang2019cellfree}. This cooperative framework enables infrastructure sharing across tasks, leading to enhanced spatial coverage, effective interference suppression, and increased SE and EE~\cite{Demir2021book, Ngo2017, Zhang2019cellfree}.

A key enabler of this cooperation is cooperative beamforming, where APs jointly design transmit (and possibly receive) beam patterns. When synchronization is achieved at the level of carrier phase, timing, and frequency, the signals from multiple APs coherently combine at the target receiver or sensing point. For a communication user $k$, the received signal power is enhanced through constructive combining, i.e., 
\begin{eqnarray}
    y_k = \sum_{m=1}^{M} \q{h}_{mk}^{\mathrm{H}} \q{w}_{mk} q_k + \underbrace{\text{interference}}_{\text{Com + Sen}}  + n_k, \quad
\end{eqnarray} \par \vspace{-0mm}
\noindent where $\q{h}_{mk}$ is the channel from AP $m$ to user $k$, $\q{w}_{mk}$ is the beamforming vector, and $q_k$ is the data symbol. Coherent summation of these terms increases SINR, improving throughput and reliability.

In sensing, cooperation between APs enables multi-static radar configurations, where widely separated APs act as non-co-located transmitters and receivers observing the same target from different angles. This spatial diversity improves angular resolution, target localization, and resilience against multipath and occlusions. For example, the sensing signal reflected from a target located at $\q{x}_t$ and received at the $m$-th AP has a propagation delay $\tau_m = \frac{\Vert \q{x}_t - \q{x}_m \Vert}{c}$, where $\q{x}_m$ is the location vector of the $m$-th AP and $c$ is the speed of light. The joint processing across multiple APs allows precise estimation of $\q{x}_t$ by exploiting the delay and Doppler diversity~\cite{Mao2023, Yuanyuan2024, Demirhan2023, demirhan2024cellfree}. Such diversity is especially valuable in complex or cluttered environments where a single AP's LoS to the target may be obstructed.

\textit{Centralized Coordination via CPU:} The CPU plays a critical role in enabling coherent AP cooperation~\cite{Demir2021book, Ngo2017, Zhang2019cellfree}. It aggregates global CSI and sensing observations, enabling centralized (or semi-centralized) signal processing such as joint beamforming and scheduling, power and bandwidth allocation, and adaptive clustering and AP assignment. 

Let $\q{W} = \{\q{w}_{mk}\}$ be the aggregate beamforming matrix. The CPU can solve a multi-objective optimization problem:
\begin{eqnarray}
    \max_{\q{W}} \quad \sum_{k} \alpha_k \mathcal{S}_k^{\mathrm{Com}}(\q{W}) + \sum_{t} \beta_t \eta_t(\q{W}),
\end{eqnarray} \par \vspace{-0mm}
\noindent where $\mathcal{S}_k^{\mathrm{Com}}(\q{W})$ is the $k$-th user communication rate and $\eta_t(\q{W})$ the sensing metric (e.g., Cram\'{e}r-Rao bound for delay/angle estimation), with  $\alpha_k$ and $\beta_t $ as task-specific weights. This ensures a task-aware design that minimizes cross-task interference while maximizing joint performance.

\textit{Synchronization Challenges} To realize coherent combining, APs must be tightly synchronized in time (to align symbol boundaries and radar pulses), frequency (to avoid carrier offsets that degrade coherence), and phase (to enable constructive signal summation) \cite{Demir2021book,  Ngo2017, Zhang2019cellfree}. This may require dedicated synchronization protocols, such as network time protocol extensions, over-the-air synchronization beacons, GPS-disciplined local oscillators, or centralized reference signals from the CPU \cite{Demir2021book,  Ngo2017, Zhang2019cellfree}. Failure to maintain synchronization degrades beamforming gains and sensing precision, especially in mmWave and sub-THz systems where phase coherence is critical \cite{Demir2021book,  Ngo2017, Zhang2019cellfree}.

The cooperative capabilities of CF-ISAC networks, particularly joint beamforming and multi-static sensing, are further explored in Chapter~\ref{chp_CF_isac_perfor}, Chapter~\ref{chp_CF_isac_resource}, Chapter~\ref{chp_CF_isac_security}, and Chapter~\ref{chp_NA_CF_isac}. These chapters examine various resource management and security considerations in cooperative AP configurations. However, despite these advancements, the practical challenges and performance impacts of tight synchronization among distributed APs remain insufficiently addressed. Issues such as synchronization overhead, robustness under mobility and hardware impairments, and scalability in large-scale deployments remain open problems. Future research is needed to develop robust, low-complexity synchronization mechanisms that can support the stringent coherence requirements of joint communication and sensing in CF-ISAC systems.

\section{Cell-Free ISAC Versus Link-Level ISAC Design}\label{sec_sensing_CFvsCL}
The CF-ISAC designs with multi-static sensing provide significant advantages over link-level mono-static or bi-static ISAC designs. These advantages stem primarily from their distributed architecture and the ability of multiple APs to participate in both communication and sensing~\cite{Mao2023, Demirhan2023, Huang2022Coordinated, Cao2023Design, Wang2023, Sakhnini2022Uplink, Silva2023, Behdad2022, Behdad2024Interplay}. 

\begin{table}[!t]
\centering
\caption{A comparison of sensing in CF-ISAC and conventional ISAC.}\label{tab_features}
\begin{tabular}{|p{3cm}|p{5cm}|p{5cm}|}
\hline
\textbf{Feature} & \textbf{CF-ISAC sensing (Multi-static)} & \textbf{Conventional ISAC sensing (Mono/bi-static)} \\ \hline \hline
Sensing coverage   &   Broader with fewer blind spots  & Limited to single/double viewpoint  \\ \hline
Sensing accuracy  &  High due to triangulation from multiple points &  Lower and limited by fewer viewpoints  \\ \hline
Robustness  &  High resilience to node failure & Vulnerable to single-point failure \\ \hline
SNR and sensitivity  & Improved through signal diversity &  Limited SNR with sensitive to noise \\ \hline
Scalability  & Easily scalable across APs  &  Difficult to scale \\ \hline
Multi-path exploitation   & Can utilize multi-path signals for extra information  &  Often treats multi-path as interference \\ \hline
Multi-target tracking   & 	More effective and can track multiple targets & Limited in tracking multiple targets simultaneously \\ \hline
Interference management   &  Better due to spatial diversity &  More susceptible to interference  \\ \hline
Resource optimization    & Highly flexible for joint communication-sensing  & Limited flexibility     \\ \hline
NLoS handling    &  More effective due to multiple angles &  Often requires LoS or favorable geometry \\ \hline
\end{tabular}
\end{table}

\subsection{Improved Sensing Coverage} CF-ISAC with multi-static sensing involves multiple distributed sensing transmitter-receivers (APs) covering a wide area. This provides a larger coverage than link-level ISAC design, which normally relies on a single or two locations for sensing. Furthermore, link-level ISAC design is frequently restricted by LoS conditions. CF-ISAC avoids blind spots and provides reliable coverage across broader regions. 

\subsection{Improved Sensing Accuracy} CF-ISAC utilizes measurements from multiple distributed nodes, resulting in improved triangulation and more accurate localization or target detection. Furthermore, cross-correlating observations from multiple sources can also greatly improve resolution, precision, and detection. On the other hand, because link-level ISAC design relies on one or two points of view, their precision is restricted, and they are more likely to overlook fine data details of the target, e.g., position, movement, or characteristics. 

\subsection{Enhanced Robustness and Resilience} 
Multi-static sensing in CF-ISAC improves system robustness by enabling continued operation even in the event of node failures or interference, ensuring the system remains functional under challenging conditions. The distributed structure provides robustness and reliability when mono-static/bi-static systems fail due to obstruction, occlusion, or physical limitations. In link-level ISAC design, mono-static/bi-static sensing is more susceptible to single-point failures, i.e., if the BS fails to perform sensing, the system loses all sensing capabilities. 

\subsection{Improved SNR and Detection Sensitivity} CF-ISAC systems employ coherent combining and diversity gain to improve sensing SNR, leveraging multiple observations through received signals. This can assist in detecting weak or distant targets that may have been overlooked in mono-static or bi-static systems. Mono-static/bi-static systems, on the other hand, are limited by the signal SNR or dual channels, i.e., transmitting and receiving, reducing detection sensitivity, particularly in challenging environments (e.g., dense urban areas and cluttered environments).

\subsection{Distributed and Scalable System Architecture} Multiple APs in CF-ISAC perform joint communication and sensing in a distributed and scalable manner, i.e., each AP contributes to both tasks, optimizing network resources and providing seamless scalability. In co-located ISAC, communication and sensing tasks are centralized and less flexible, i.e., they lack inherent scalability compared to CF networks when expanding coverage or adding new sensing nodes. 

\subsection{Multi-Path (Macro-Diversity) Exploitation} CF-ISAC can leverage multi-path propagation or macro-diversity, which involves signals reflected from multiple points and surfaces, providing more information about the environment and targets. This is especially beneficial in indoor or cluttered environments with plenty of reflections.  However, link-level ISAC design may fail to effectively utilize multi-path signals, often treating them as interference rather than meaningful information. 

\subsection{Simultaneous Multi-Target Tracking} By leveraging the user-centric (UC) nature of the CFMM architecture, CF-ISAC allows different APs to focus on distinct targets and combine the gathered data, enabling simultaneous monitoring of multiple targets. This approach enhances the efficiency and accuracy of multi-target tracking, particularly in dynamic environments with moving objects. In contrast, the link-level ISAC design with a single BS has limited capacity to manage multiple targets simultaneously, leading to performance degradation due to signal overlap or occlusion.

\subsection{Interference Management} CF-ISAC can use the spatial diversity of the distributed architecture to reduce interference. In particular, observations/measurements from several angles and distances aid in differentiating targets from noise. Because of the limited number of observations in the link-level ISAC design, interference from other communication channels or ambient clutter is more difficult to eliminate. 

\subsection{Joint Communication-Sensing Optimization} In CF-ISAC systems, the network may optimize resources (e.g., power, bandwidth) among scattered APs to balance communication and sensing demands. The distributed design provides greater flexibility in distributing resources depending on dynamic demands. For example, sensing in a specific region while maintaining communication quality elsewhere. Nonetheless, co-located ISAC systems have fewer DoF for joint resource allocation, making it more difficult to accomplish effective communication-sensing trade-offs, particularly in dynamic environments.

\subsection{Non-LoS Capability} In CF-ISAC, multi-static sensing can successfully manage non-LoS (NLoS) conditions as APs provide alternate viewpoints of the target, potentially circumventing obstructions or blockages.  In contrast, mono-static/bi-static sensing in the link-level ISAC design relies on LoS or favorable positioning. Thus, obstacles or structures that conceal the target are significant limitations.  

In conclusion, multi-static sensing in CF-ISAC systems offers significant advantages in sensing coverage, accuracy, robustness, and flexibility compared to the link-level mono-static or bi-static ISAC design. These benefits make CF-ISAC particularly well-suited for complex, dense, or large-scale future-generation wireless networks. A summary of these aspects is provided in Table~\ref{tab_features}.

The benefits of CF-ISAC outlined above, such as improved sensing coverage, accuracy, resilience, and scalability, are not merely theoretical. Chapters~\ref{chp_CF_isac_security} through~\ref{chp_NA_CF_isac} delve into the practical realization and evaluation of these features across a range of scenarios. These chapters showcase how CF-ISAC's architectural strengths translate into performance gains under diverse operating conditions.  
	\chapter{State of the Art and Open Challenges in in Cell-Free ISAC}\label{chp_state}
This chapter surveys the state of the art in CF-ISAC, highlighting advances and open challenges in performance analysis, resource allocation, security, and scalability. It reviews analytical frameworks that capture the interplay between communication and sensing, covering detection performance, signal models, and optimization for robust operation. Resource allocation approaches are then examined, including beamforming, power control, and AP selection, to jointly improve SE, fairness, and sensing accuracy in multi-user and multi-target scenarios. Security is addressed next, with emphasis on vulnerabilities from distributed architectures and integrated sensing, along with recent physical-layer defenses. Finally, user- and target-centric clustering strategies are presented as scalable solutions to reduce coordination overhead while enabling adaptive, high-resolution sensing. Collectively, these efforts provide a comprehensive view of current research directions and guide the design of next-generation CF-ISAC networks.

\section{Performance Analysis in Cell-Free ISAC}
Analytical performance evaluation facilitates an understanding of the performance and limitations of wireless networks under various operating conditions \cite{1199275,1556835,545899,4694096}. It predicts performance without requiring extensive simulations or real-world testing. For CF-ISAC systems, such analysis is crucial to optimize the interplay between communication and sensing, guide efficient resource allocation, assess trade-offs between conflicting objectives, ensure scalability, and determine the feasibility of practical deployments. Despite its importance, analytical performance evaluations for CF-ISAC remain limited. To date, only a few works, including \cite{Sakhnini2022, Behdad2024, Kulathunga2025}, have made notable contributions.

Reference \cite{Sakhnini2022} proposes a communication and sensing protocol for a CF-ISAC system with a single target, where a designated AP acts as a virtual user, transmitting radar signals to facilitate UL sensing. Thus, the cost imposed on the communication system is the loss of an AP, as it is treated as a virtual user in the communication system. The approach models the system as a distributed bi-static radar and introduces two radar sensing modes, i.e., sensing during either the UL training period or the data payload segment of the communication frame. A subspace-based signal model is employed, and a generalized likelihood ratio test (GLRT) is developed to derive expressions for detection and false alarm probabilities, offering analytical insight into radar detection performance in CF environments.

In \cite{Behdad2024}, a centralized CF-ISAC mMIMO system is investigated for single-target detection, where transmit APs simultaneously serve DL users and direct beams toward a sensing target in a multi-static sensing setup. It develops a maximum a-posteriori (MAP) detector for target identification under clutter, using sensing SE as a key performance metric. Two ISAC signal designs are compared, i.e., one that reuses communication beams for sensing and another that includes dedicated sensing beams, along with a power allocation algorithm to maximize the sensing SINR while meeting minimum communication QoS constraints.

Reference \cite{Kulathunga2025} analyzes a multi-user, single-target CF-ISAC system using MRT precoding at transmit APs, each employing locally estimated CSI to construct superimposed ISAC waveforms. A subset of APs is assigned to sensing, processing reflected echoes for target detection. The study proposes a max-min power optimization strategy to ensure fairness across users while maintaining a minimum signal-to-clutter-plus-noise ratio (SCNR) at the sensing APs. Analytical expressions for achievable user rates and the two-dimensional (2D) MUSIC spectrum are derived, accounting for CSI imperfections, spatially correlated Rician fading, and clutter. Results confirm the feasibility of robust sensing and communication performance using lightweight precoding techniques.

\section{Resource Allocation in Cell-Free ISAC}
Efficient resource allocation is central to optimizing bandwidth, power, time, and antenna usage across communication systems. In ISAC networks, these challenges are further intensified by the need to balance dual-functionality (sensing and communication) while supporting multiple users with diverse service requirements. Different networks, such as Wi-Fi, LTE, sensor networks, and energy harvesting systems, demand specialized resource allocation approaches. Similarly, CF-ISAC systems require tailored strategies beyond conventional methods due to their distributed architecture and joint objectives.

In CF-ISAC, resource allocation encompasses power distribution, spectrum management, AP selection, and particularly beamforming design, all of which impact both communication quality and sensing precision. Recent works have focused on these aspects through algorithmic and optimization-based approaches to achieve joint performance goals \cite{Mao2023, Yuanyuan2024, Demirhan2023, demirhan2024cellfree, Elfiatoure2023, Mao2024, Liu2024, Huang2022Coordinated, Cao2023Design, Behdad2024Interplay, behdad2024Joint, Silva2023, Behdad2022, Wang2023, Sakhnini2022Uplink, zargari2024CFISAC}.

Many studies focus on AP beamforming design to facilitate efficient resource allocation in CF-ISAC networks \cite{Mao2023, Demirhan2023, demirhan2024cellfree, Cao2023Design, Wang2023, Silva2023, Mao2024, Liu2024, zargari2024CFISAC}. Although traditionally considered a signal processing technique, beamforming in CF-ISAC inherently involves resource management as the AP transmit power is embedded within the beamforming vectors. This integration directly affects how system power is distributed across APs to optimize communication and sensing performance. Thus, beamforming plays a critical role in CF-ISAC resource allocation frameworks.

In \cite{Mao2023}, a robust beamforming design is proposed for a multi-user, single-target system under imperfect CSI. The beamforming problem is formulated as a sensing beampattern matching MSE minimization, subject to per-AP power budget and user rate constraints. With lower bound user rates over the imperfect CSI, a successive convex approximation (SCA)-based algorithm is presented. For a single-target CF-ISAC system, max-min fairness-based joint beamforming designs are further explored in \cite{Demirhan2023, demirhan2024cellfree}. The approach utilizes two benchmarks: communication-prioritized sensing beamforming and sensing-prioritized communication beamforming, to balance performance between communication and sensing tasks. Reference \cite{Huang2022Coordinated} investigates a coordinated power control scheme for CF-ISAC transmitters. It minimizes the total AP transmit power while meeting the minimum communication SINR for users and the maximum Cram\'{e}r-Rao lower bound (CRLB) for target location estimation. Two efficient algorithms based on semidefinite relaxation (SDR) and CRLB approximation are proposed.

For multi-target scenarios, reference \cite{Cao2023Design} proposes vector orthogonal frequency division multiplexing (OFDM) signals to improve SE, detection resolution, and latency/Doppler estimation accuracy. A low-complexity grid-searching approach is presented to estimate multi-target AoAs with AP beamforming and power allocation optimized via an alternative Lagrange multiplier algorithm. In \cite{Wang2023}, a hybrid beamforming strategy is used to maximize the communication sum rate while controlling beampattern MSE for sensing via a semi-distributed AO algorithm in a multi-user, multi-target cooperative CF dual-function radar-communication network. Reference \cite{Sakhnini2022Uplink} examines virtual UL-based CF-ISAC where a subset of APs are dedicated to radar transmission alongside user communications. The study uses interference-aware power control under sum-SE and sum-log-SNR policies and proposes a low-complexity heuristic based on large-scale fading.

Reference~\cite{Silva2023} studies a single-target, multi-user CF-ISAC system where a user acts as an adversary attempting to infer the target's position. To overcome this, a transmit beamforming design is proposed using an expectation-maximization algorithm to estimate the transmitted signal, improving sensing security. Reference \cite{Behdad2022} introduces a power allocation algorithm to maximize sensing SNR while ensuring minimum user SINR, complemented by a maximum a posteriori ratio test detector for target detection using signals received at distributed APs. The works \cite{Behdad2024Interplay, behdad2024Joint} focus on DL CF-ISAC with ultra-reliable low-latency communication (URLLC). Using SCA-based methods, they propose power control schemes to maximize EE while satisfying the sensing SINR and communication decoding error probability requirements.

In~\cite{Yuanyuan2024}, a multi-user, single-target CF-ISAC mmWave mMIMO system with capacity-limited fronthaul links is studied. A power allocation scheme with a fronthaul compression design is presented with a hybrid precoding scheme. Digital precoders are computed at the CPU, while analog precoders are locally designed at APs. Two block coordinate descent (BCD)-based algorithms are developed for power and compression optimization. For a multi-user, single-target CF-ISAC system, reference \cite{Elfiatoure2023} presents an AP operating mode selection approach, i.e., some APs are dedicated to DL communication, while the remaining APs are utilized for sensing. A max-min fairness problem is formulated to maximize the worst-case user SE based on closed-form SE and beampattern metrics. In \cite{Mao2024}, different beamforming strategies, i.e., joint, sensing-only, and communication-only, are investigated for a multi-user, multi-target CF-ISAC system, using Lagrangian dual transform,  the quadratic fractional transform technique, the BCD method, and the SCA method. 

Reference \cite{Liu2024} explores the joint AP mode selection, transmit beamforming, and receive filter designs for cooperative CF-ISAC networks, where the APs cooperatively serve multiple communication users and detect targets. Three heuristic AP mode selection techniques and an efficient joint beamforming design method are presented. Reference \cite{zargari2024CFISAC} proposes a novel, efficient, and low-complexity beamforming design for generic CF-ISAC networks. The proposed algorithm uses the augmented Lagrangian model-based Riemannian manifold optimization technique to maximize the communication sum rate while satisfying sensing beampattern gains of multiple targets and per-AP transmit power constraints.

\section{Security Challenges of  Cell-Free ISAC}
Compared to traditional wireless systems, CF-ISAC networks face significant security threats due to their integrated communication and sensing functionalities. In particular, the use of information beams to facilitate high-resolution sensing can inadvertently expose sensitive information, especially when the targets themselves are adversarial, such as eavesdropping UAVs~\cite{Qu2024}. They not only intercept communication signals but may also exploit sensing returns to infer system operations or actively disrupt them.

Conventional physical-layer security techniques, such as beamforming, sensing covariance matrix design, and artificial noise (AN) injection, offer viable means of enhancing secrecy~\cite{Zhu:TuT:2024}. However, these solutions must be re-engineered to accommodate the decentralized and cooperative nature of CF networks. While increasing the number of APs and distributed receivers enhances spatial diversity and robustness, it also introduces more potential points of vulnerability. Attackers can exploit this structure via active eavesdropping, pilot spoofing, or signal manipulation, thereby undermining legitimate sensing and communication processes.

Recent studies have addressed some of these emerging challenges in CF-ISAC security \cite{Rivetti:WCNC:2024, Nasir2024, Ren2024}. In particular, the study \cite{Rivetti:WCNC:2024} investigates a multi-user, single-target secure CF-ISAC system under the assumption that the sensing target may act as an eavesdropper. It aims to maintain reliable communications and detect the target while simultaneously limiting its ability to intercept information. To achieve this, the authors embed artificial noise within the ISAC waveform to degrade the eavesdropper's reception. A joint optimization problem is formulated to minimize the CRB for direction estimation, subject to user SINR constraints and an upper bound on the eavesdropper's SNR. A SDR-based solution is developed, and the results show that the CRB is inversely proportional to the distance between the user and the eavesdropper, demonstrating how spatial geometry influences the system's secrecy performance.

In \cite{Nasir2024}, a multi-static sensing architecture is investigated, wherein multiple distributed APs collaboratively serve communication users while protecting against multiple eavesdroppers attempting to intercept transmissions. The authors propose a joint beamforming design that maximizes the sensing SNR while ensuring the secrecy rate for all users remains above a given threshold. This work highlights the trade-offs between spatial resolution in sensing and secrecy capacity in communication, providing a beamforming-based defense mechanism against passive interception.

The study in \cite{Ren2024} addresses the security of both communication and sensing in the presence of information and sensing eavesdroppers, who seek to intercept confidential communications and extract target information, respectively. To mitigate these risks, a transmit beamforming optimization problem is formulated to maximize the probability of correct target detection. Constraints are imposed to ensure minimum SINR for users, maximum tolerable SNR for information eavesdroppers, bounded detection probability for sensing eavesdroppers, and per-AP power limits. An SDR-based algorithm is employed to find the globally optimal solution. This study is notable for jointly considering both communication and sensing eavesdropping within a unified security framework.

\section{User/Target-Centric Cell-Free ISAC}
Scalability remains a major hurdle in deploying CF networks due to the increasing demands on computational resources and fronthaul capacity as the number of users grows. Specifically, having every AP serve all users leads to significant overhead in processing and coordination. To address this, the UC-CF architecture has been proposed \cite{Emil:TCOM:2020}, where each user is served by a selected subset of APs with the strongest channels. This strategy forms overlapping AP clusters tailored to each user, enabling more efficient resource utilization and reducing coordination complexity.

In CF-ISAC, the UC paradigm can be extended beyond communications to include sensing-centric clustering. Two clustering paradigms are currently explored in the literature:
\begin{itemize}
    \item \textit{UC CF-ISAC:} Each communication user is served by a personalized subset of APs offering favorable links, while all APs jointly participate in sensing operations \cite{Cao2023}. This hybrid strategy prioritizes communication performance and simplifies sensing coordination.
    \item \textit{Target-centric CF-ISAC:} The surveillance region is divided into non-overlapping spatial sectors, each monitored by a subset of APs \cite{Buzzi2024}. This localized approach enhances multi-target detection and resolution by allocating sensing responsibilities based on spatial proximity or sensing accuracy requirements.
\end{itemize}
While these approaches improve scalability and efficiency, they also introduce new coordination and design challenges. In particular, forming and managing dynamic AP clusters that consider user/target mobility, channel variations, and inter-cluster interference requires adaptive algorithms and real-time decision-making strategies. Additionally, collaborative sensing across overlapping clusters remains an open problem, especially in dynamic environments with multiple users and targets.

	\chapter{Performance Analysis in Cell-Free ISAC}\label{chp_CF_isac_perfor}

This chapter investigates the performance of a generalized CF-ISAC system supporting multiple users and targets, thereby extending prior analyses that focused solely on single-target configurations (Figure~\ref{fig_SystemModelPerfomance}). The evaluation explores the dual functionalities of sensing and communication, emphasizing the inherent trade-offs between them.

\section{System and Channel Models} 
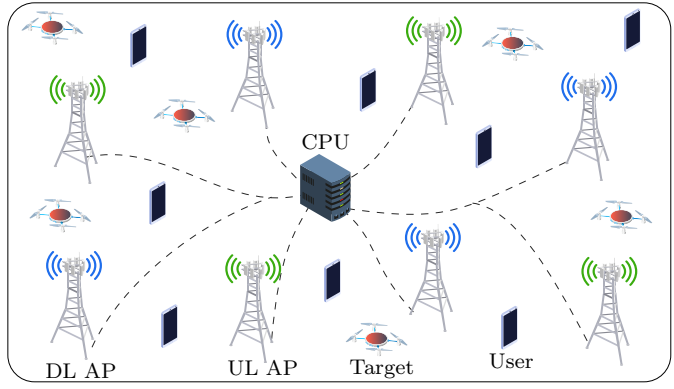
\begin{figure}[!t]\vspace{0mm}
    \centering 
    \def\svgwidth{250pt} 
    \fontsize{8}{8}\selectfont 
    \graphicspath{{Figures/}}
\begingroup%
  \makeatletter%
  \providecommand\color[2][]{%
    \errmessage{(Inkscape) Color is used for the text in Inkscape, but the package 'color.sty' is not loaded}%
    \renewcommand\color[2][]{}%
  }%
  \providecommand\transparent[1]{%
    \errmessage{(Inkscape) Transparency is used (non-zero) for the text in Inkscape, but the package 'transparent.sty' is not loaded}%
    \renewcommand\transparent[1]{}%
  }%
  \providecommand\rotatebox[2]{#2}%
  \newcommand*\fsize{\dimexpr\f@size pt\relax}%
  \newcommand*\lineheight[1]{\fontsize{\fsize}{#1\fsize}\selectfont}%
  \ifx\svgwidth\undefined%
    \setlength{\unitlength}{647.41088867bp}%
    \ifx\svgscale\undefined%
      \relax%
    \else%
      \setlength{\unitlength}{\unitlength * \real{\svgscale}}%
    \fi%
  \else%
    \setlength{\unitlength}{\svgwidth}%
  \fi%
  \global\let\svgwidth\undefined%
  \global\let\svgscale\undefined%
  \makeatother%
  \begin{picture}(1,0.5722109)%
    \lineheight{1}%
    \setlength\tabcolsep{0pt}%
    \put(0,0){\includegraphics[width=\unitlength]{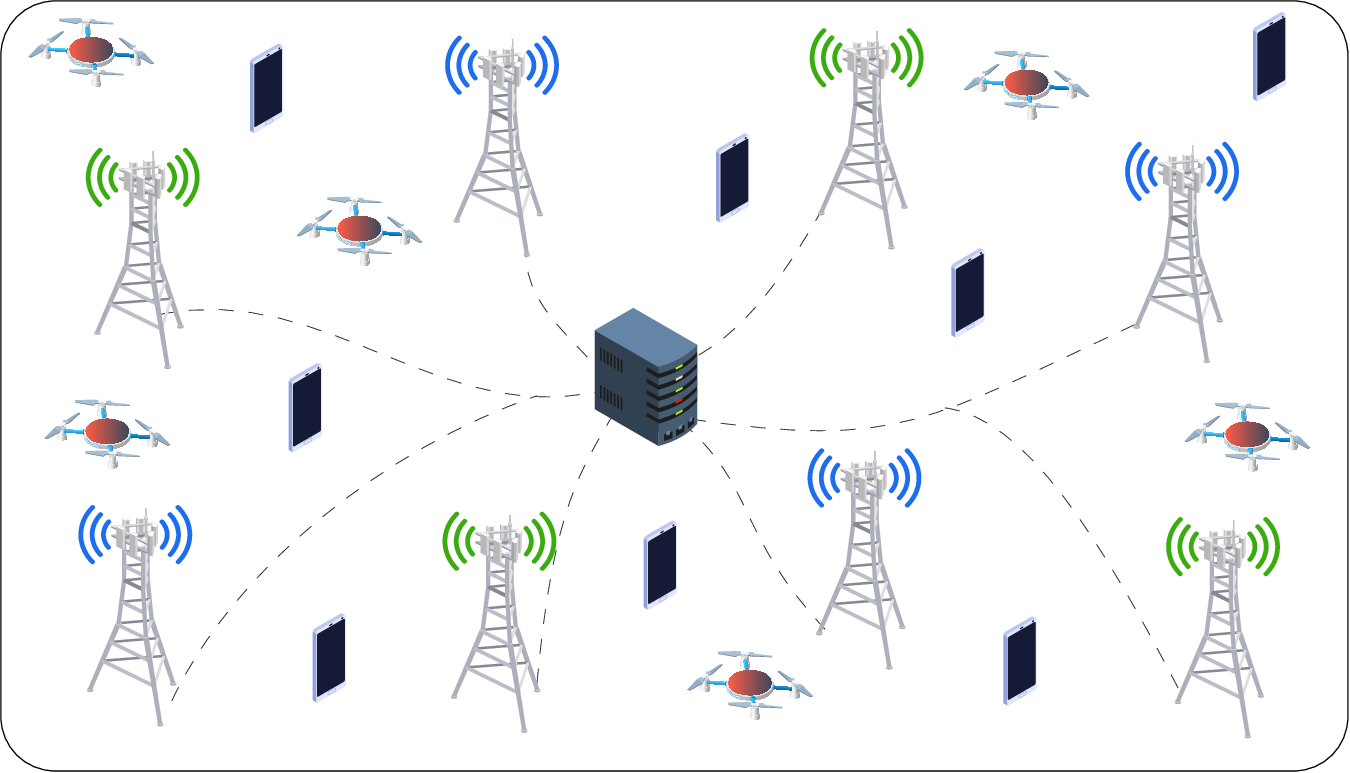}}%
    \put(0.05716431,0.00697471){\color[rgb]{0,0,0}\makebox(0,0)[lt]{\lineheight{1.25}\smash{\begin{tabular}[t]{l}DL AP\end{tabular}}}}%
    \put(0.44270859,0.35176811){\color[rgb]{0,0,0}\makebox(0,0)[lt]{\lineheight{1.25}\smash{\begin{tabular}[t]{l}CPU\end{tabular}}}}%
    \put(0.33112427,0.01156734){\color[rgb]{0,0,0}\makebox(0,0)[lt]{\lineheight{1.25}\smash{\begin{tabular}[t]{l}UL AP\end{tabular}}}}%
    \put(0.72467572,0.02254578){\color[rgb]{0,0,0}\makebox(0,0)[lt]{\lineheight{1.25}\smash{\begin{tabular}[t]{l}User\end{tabular}}}}%
    \put(0.51499612,0.01346496){\color[rgb]{0,0,0}\makebox(0,0)[lt]{\lineheight{1.25}\smash{\begin{tabular}[t]{l}Target\end{tabular}}}}%
  \end{picture}%
\endgroup%
 \vspace{-0mm} 
    \caption{A CF-ISAC system setup with UL and DL APs. }  \label{fig_SystemModelPerfomance}\vspace{-0mm}
\end{figure}

We consider a CF-ISAC massive MIMO system composed of $M$ transmit (DL) APs and $N$ receiver (UL) APs, each equipped with $L$ antennas. The network serves $K$ single-antenna DL users and simultaneously performs sensing on $T$ targets, as illustrated in Figure~\ref{fig_SystemModelPerfomance}. All DL APs jointly transmit to all users while concurrently directing sensing beams toward the targets, operating over the same time-frequency resources.

\textit{Backhaul/Fronthaul Requirement:}
In the CFMM architecture, each AP is connected to a CPU through high-capacity backhaul or fronthaul links~\cite{Interdonato2019}. UL channel estimation is performed locally at each AP using pilot signals, enabling them to execute local precoding for both data decoding and sensing tasks without requiring inter-AP CSI exchange. This significantly lowers coordination overhead compared to centralized systems like network MIMO or CoMP. However, DL data transmission necessitates coordination with the CPU, introducing a trade-off between system performance and the bandwidth demands of the fronthaul/backhaul links~\cite{Interdonato2019}.

Block flat-fading channel models are adopted, where the channel remains constant over each coherence interval. During each fading block, $\q{h}_{mk} \in \mathbb{C}^{L \times 1}$, $\q{g}_{mt}^{\mt{d}} \in \mathbb{C}^{L \times 1}$, and $\q{g}_{nt}^{\mt{u}} \in \mathbb{C}^{L \times 1}$ are the channel vectors from the $m$-th DL AP to the $k$-th user, the $m$-th DL AP to the $t$-th target, and the $t$-th target to the $n$-th UL AP, respectively. In addition, $\q{F}_{mn} \in \mathbb{C}^{L\times L}$ characterizes the direct channel between the $m$-th DL AP and the $n$-th UL AP. 

All channel links are assumed to follow independent quasi-static Rayleigh fading. Within each coherence interval, the channel coefficients are constant and independently change across intervals. A unified representation of all channels is given as
\begin{eqnarray}\label{eqn_chnl_model}
    \q{a} = \zeta_{\q{a}}^{1/2} \tilde{\q{a}}, 
\end{eqnarray} \par \vspace{-0mm}
\noindent where $\q{a} \in \{ \q{h}_{mk}, \q{g}_{mt}^{\mt{d}}, \q{g}_{mt}^{\mt{u}}\}$, $\tilde{\q{a}} \sim \mathcal{CN}\left(\q{0},\q{I}_L \right)$ captures the small-scale Rayleigh fading, which is static during one coherence interval, and ${\zeta}_{\q{a}}$ accounts for the large-scale path-loss and shadowing. Since large-scale fading coefficients remain constant over many coherence intervals, they only need to be estimated once every few tens or even hundreds of intervals~\cite{Marzettabook2016}. In a time-division duplexing (TDD) system, used for both channel estimation and data transmission, accurate CSI can be obtained through orthogonal pilot sequences~\cite{Demir2021book}. Given the high accuracy of this method, we assume perfect CSI availability throughout the analysis.

\section{Transmission Model} 
The DL APs transmit signals that are jointly optimized to serve two purposes: delivering information to users and simultaneously illuminating targets for sensing. Accordingly, the transmitted signal from the $m$-th DL AP, denoted as $\q{x}_{m} \in \mathbb{C}^{L \times 1}$, is  given as
\begin{eqnarray}
    \q{x}_m = \sum_{k = 1}^{K} \q{w}_{mk} q_k + \sum_{t=1}^{T} \q{s}_{mt},
\end{eqnarray} \par \vspace{-0mm}
\noindent where $q_k \in \mathbb{C}$ denotes the data symbol intended for the $k$-th user, assumed to have unit power, i.e., $\mathbb{E}{\vert q_k\vert^2}=1$. The vector $\q{w}_{mk} \in \mathbb{C}^{L\times 1}$ is the transmit beamforming vector at the $m$-th DL AP for the $k$-th user, while $\q{s}_{mt} \in \mathbb{C}^{L\times 1}$ represents the sensing signal from the $m$-th DL AP directed at the $t$-th target~\cite{Zhenyao2023}. It is assumed that $q_k$ and $\q{s}_{mt}$ are statistically independent~\cite{Zhenyao2023}.

Each user receives the combined transmissions from all DL APs. Assuming negligible propagation delay differences between signals from different DL APs~\cite{Demir2021book}, the received signal at the $k$-th user can be expressed as
\begin{eqnarray} \label{eqn_rx_DL_k}
    y_k &=& \sum_{m=1}^{M} \q{h}_{mk}^{\mathrm{H}} \q{x}_m + z_k \nonumber \\
    &=& \underbrace{\sum_{m=1}^{M}  \q{h}_{mk}^{\mathrm{H}} \q{w}_{mk} q_k}_{\text{Desired signal }} + \underbrace{\sum_{i\neq k}^{K} \sum_{m=1}^{M} \q{h}_{mk}^{\mathrm{H}} \q{w}_{mi} q_i }_{\text{Multi-user interference}}  + \underbrace{\sum_{t=1}^{T} \sum_{m=1}^{M} \q{h}_{mk}^{\mathrm{H}} \q{s}_{mt} }_{\text{Sensing signal interference }}  + \underbrace{z_k}_{\text{AWGN}}, \quad
\end{eqnarray} \par \vspace{-0mm}
\noindent where $z_k \sim \mathcal{CN}(0,\sigma^2)$ is the AWGN at the $k$-th user with \num{0} mean and $\sigma^2$ variance.

The UL APs capture the echoes reflected from the targets to infer their state information for sensing purposes~\cite{Zhenyao2023}. Accordingly, the received signal at the $n$-th UL AP, denoted by $\q{y}_n \in \mathbb{C}^{L\times 1}$, is expressed as
\begin{eqnarray}\label{eqn_rx_AP_n}
    \q{y}_n &&= \underbrace{\sum_{m=1}^{M} \q{F}_{mn} \q{x}_m}_{\text{Direct-link interference}} + \underbrace{\sum_{t=1}^{T} \q{g}_{nt}^{\mt{u}} \alpha_t \sum_{m=1}^{M} (\q{g}_{mt}^{\mt{d}})^{\mathrm{H}} \q{x}_m}_{\text{Target echoes}}  + \underbrace{\q{z}_n}_{\text{AWGN}},
\end{eqnarray} \par \vspace{-0mm}
\noindent where $\q{z}_n \sim \mathcal{CN}(\q{0},\sigma^2 \q{I}_L)$ denotes the AWGN at the $n$-th UL AP, and $\alpha_t \in \mathbb{C}$ represents the complex reflection coefficient of the $t$-th target, incorporating both round-trip path-loss and the target's RCS~\cite{Liu2022}. The path-loss captures signal attenuation over distance, while the RCS quantifies how much energy is reflected back toward the receiver, influenced by the target's physical characteristics such as size, geometry, and material. It is also assumed that UL APs implement clutter suppression methods to reduce interference caused by unwanted environmental reflections~\cite{Mark2010RadarBook}.

Since the APs are connected to the CPU via backhaul links, it is assumed that the UL APs are aware of direct-link interference (DLI), which is the interference between APs. As a result, the UL APs are able to remove the DLI before applying the sensing combiner, $\q{u}_{nt} \in \mathbb{C}^{L\times 1}$, to process the target's sensing data. The processed signal used to extract the $t$-th target's sensing information at the $n$-th UL AP is expressed as
\begin{eqnarray}\label{eqn_rx_Sens_t}
    {y}_{nt} &=& \q{u}_{nt}^{\mathrm{H}} \left(\q{y}_{n}-\sum_{m=1}^{M} \q{F}_{mn} \q{x}_m \right)\nonumber\\
    &=& \underbrace{\alpha_t \q{u}_{nt}^{\mathrm{H}} \q{g}_{nt}^{\mt{u}} \sum_{m=1}^{M} (\q{g}_{mt}^{\mt{d}})^{\mathrm{H}} \left( \sum_{i=1}^{K} \q{w}_{mi} q_i + \sum_{l=1}^{T} \q{s}_{ml}\right)}_{\text{the $t$-th target's desired reflection}} \nonumber \\
    &&+ \underbrace{\sum_{j\neq t}^{T} \alpha_j \q{u}_{nt}^{\mathrm{H}} \q{g}_{nj}^{\mt{u}} \sum_{m=1}^{M} (\q{g}_{mj}^{\mt{d}})^{\mathrm{H}} \left( \sum_{i=1}^{K} \q{w}_{mi} q_i + \sum_{j=1}^{T} \q{s}_{mj}\right)}_{\text{Multi-target interference reflections}} + \q{u}_{nt}^{\mathrm{H}} \q{z}_n. \quad
\end{eqnarray}

\section{Communication SE} 
The users utilize the received signal from the DL APs to decode their intended data. To this end, from \eqref{eqn_rx_DL_k}, the received communication SINR at the $k$-th user can be obtained as
\begin{eqnarray}\label{eqn_user_sinr_k}
    \mt{SINR}_k^{\mathrm{Com}} = \frac{{\mathrm{DS}}_k}{\sum_{i\neq k}^{K} {\mathrm{MUI}}_{ki} +  \sum_{t=1}^{T} {\mathrm{SSI}}_{kt} + \sigma^2},
\end{eqnarray} \par \vspace{-0mm}
\noindent where 
\begin{subequations}
\begin{eqnarray}
    {\mathrm{DS}}_k &&= \E{\left\vert \sum_{m=1}^{M}  \q{h}_{mk}^{\mathrm{H}} \q{w}_{mk} \right\vert^2}, \\
    {\mathrm{MUI}}_{ki} &&= \E{\left\vert \sum_{m=1}^{M}  \q{h}_{mk}^{\mathrm{H}} \q{w}_{mi} \right\vert^2}, \\
    {\mathrm{SSI}}_{kt} &&= \E{\left\vert \sum_{m=1}^{M}  \q{h}_{mk}^{\mathrm{H}} \q{s}_{mt} \right\vert^2}.
\end{eqnarray}
\end{subequations} \par \vspace{-0mm}
\noindent To derive the closed-form solution, we assume that the DL APs adopt MRT beamforming for both communication and sensing, i.e., $\q{w}_{mk} = \q{h}_{mk}$ and $\q{s}_{mt} = \q{g}_{mt}^{\mt{d}}$. To this end, by evaluating the expectation terms in \eqref{eqn_user_sinr_k}, the closed-form solution of the SINR at the $k$-th user is given by
\begin{eqnarray}\label{eqn_SINR_userk_close}
    \mt{SINR}_k^{\mathrm{Com}} = \frac{L(L+1)\sum\limits_{m=1}^{M} \zeta_{\q{h}_{mk}}^2 + L^2 \sum\limits_{m=1}^{M} \sum\limits_{m'\neq m}^{M} \zeta_{\q{h}_{mk}} \zeta_{\q{h}_{m'k}} }{ L\sum\limits_{i\neq k}^{K} \sum\limits_{m=1}^{M} \zeta_{\q{h}_{mk}} \zeta_{\q{h}_{mi}} + L \sum\limits_{t=1}^{T} \sum\limits_{m=1}^{M} \zeta_{\q{h}_{mk}} \zeta_{\q{g}_{mt}^{\mt{d}}} + \sigma^2}.
\end{eqnarray} \par \vspace{-0mm}
\noindent Thus, the SE of the $k$-th user is given as
\begin{eqnarray}\label{eqn_DL_rate}
    \mathcal{S}_k^{\mathrm{Com}} = \log_2 \left(1+ \mt{SINR}_k^{\mathrm{Com}} \right).
\end{eqnarray}

\section{Sensing SE} 
To assess sensing performance, we utilize sensing spectral efficiency (SE) as the primary metric (Section~\ref{sec_sensing_SE}). While traditional metrics, such as transmit beampattern gain or the MSE of the transmit beampattern, are widely used due to their simplicity, they exhibit significant limitations~\cite{He2022, Stoica2007}. Notably, these metrics neglect the impact of the receive beam pattern and fail to account for interference among multiple targets. As a result, such simplifications can impair the system's ability to accurately detect and distinguish multiple targets, as interference from overlapping signal reflections can degrade performance~\cite{Zhenyao2023, Cui2014}. Alternatively, the CRB has been employed to evaluate estimation accuracy~\cite{Tang2019, Cui2014}. While CRB offers insight into the minimum achievable error for parameter estimation, such as range or velocity, it does not reflect the amount of environmental information collected over time~\cite{Zhenyao2023, Cui2014}.

To overcome these shortcomings, sensing SINR and SE have emerged as more comprehensive and informative performance indicators. The detection probability of a target is directly related to its sensing SINR, which considers both transmit and receive beamforming and helps mitigate mutual interference among targets~\cite{Zhenyao2023, Cui2014}. Hence, sensing SE delivers a more robust and interference-aware evaluation framework for sensing tasks.

The UL APs utilize the targets' echoes to perform the sensing. From  \eqref{eqn_rx_Sens_t}, the sensing SE of the $t$-th target at the $n$-th UL AP is obtained as 
\begin{eqnarray}
    \mathcal{S}_{nt}^{\mathrm{Sen}} \approx \log_2 \left(1+ \mt{SINR}_{nt}^{\mathrm{Sen}} \right),
\end{eqnarray} \par \vspace{-0mm}
\noindent where the $t$-th target's sensing SINR an the $n$-th UL AP is given as
\begin{eqnarray}\label{eqn_target_sinr_t}
    \mt{SINR}_{nt}^{\mathrm{Sen}} = \frac{\vert \alpha_t \vert^2 {\mathrm{TDS}}_{nt}}{\sum_{j\neq t}^{T} \vert \alpha_j\vert^2 {\mathrm{MTI}}_{n,tj} + \sigma^2 \E{\Vert \q{u}_{nt} \Vert^2}},
\end{eqnarray} \par \vspace{-0mm}
\noindent where
\begin{subequations}
\begin{eqnarray}
    {\mathrm{TDS}}_{nt} &&=  \E{ \Bigg\vert \q{u}_{nt}^{\mathrm{H}} \q{g}_{nt}^{\mt{u}}  \sum_{m=1}^{M} (\q{g}_{mt}^{\mt{d}})^{\mathrm{H}}  \left(  \sum_{i=1}^{K} \q{w}_{mi}  +  \sum_{l=1}^{T} \q{s}_{ml} \right) \Bigg\vert^2 }, \qquad \\
    {\mathrm{MTI}}_{n,tj} &&= \E{ \left\vert \q{u}_{nt}^{\mathrm{H}} \q{g}_{nj}^{\mt{u}}  \sum_{m=1}^{M} (\q{g}_{mj}^{\mt{d}})^{\mathrm{H}} \left( \sum_{i=1}^{K} \q{w}_{mi}  + \sum_{j=1}^{T} \q{s}_{mj} \right)   \right\vert^2 }. \qquad
\end{eqnarray}
\end{subequations} \par \vspace{-0mm}
\noindent By assuming the $n$-th UL AP employ the MRC to extract the $t$-th target's state information, i.e., $\q{u}_{nt} = \q{g}_{nt}^{\mt{u}}$, the closed-form expression of the sensing SINR, $\mt{SINR}_{nt}^{\mathrm{Sen}}$, is given as
\begin{align}\label{eqn_SINR_targett_close}
    \mt{SINR}_{nt}^{\mathrm{Sen}} = \frac{\vert \alpha_t \vert^2 L(L+1) \zeta_{\q{g}_{nt}^{\mt{u}}}^2  \sum\limits_{m=1}^{M}  \left( \vartheta_{mtt} + L \sum\limits_{l\neq t}^{T} \zeta_{\q{g}_{mt}^{\mt{d}}} \zeta_{\q{g}_{ml}^{\mt{d}}}  \right) }{\sum\limits_{j\neq t}^T \vert \alpha_j\vert^2 L \zeta_{\q{g}_{nt}^{\mt{u}}} \zeta_{\q{g}_{nj}^{\mt{u}}}  \sum\limits_{m=1}^{M} \left( \vartheta_{mtj} + L \sum\limits_{l\neq t}^{T} \zeta_{\q{g}_{mj}^{\mt{d}}} \zeta_{\q{g}_{ml}^{\mt{d}}}  \right) + L \sigma^2 \zeta_{\q{g}_{nt}^{\mt{u}}}},
\end{align}\par \vspace{-0mm}
\noindent where 
\begin{eqnarray}
    \vartheta_{mtj} = L \sum_{i=1}^{K}  \zeta_{\q{g}_{mt}^{\mt{d}}} \zeta_{\q{h}_{mi}} + L(L+1) \zeta_{\q{g}_{mj}^{\mt{d}}}^2 + L^2  \sum_{m'\neq m}^{M}  \zeta_{\q{g}_{mj}^{\mt{d}}} \zeta_{\q{g}_{m'j}^{\mt{d}}}.
\end{eqnarray}

\section{Simulation Example}
The large-scale fading $\zeta_{\q{a}}$, where $\q{a} \in \{ \q{h}_{mk}, \q{g}_{mt}^{\mt{d}}, \q{g}_{mt}^{\mt{u}} \}$, is modeled using the 3GPP Urban Micro (UMi) path loss model, assuming a carrier frequency of $f_c=\qty{3}{\GHz}$ \cite[Table B.1.2.1]{3GPP2010}. The AWGN power is modeled as $\sigma^2 = 10 \log_{10}{(N_0 B N_f)}$,\qty{}{\dB m}, where $N_0 = \qty{-174}{\dB m/\Hz}$ is the noise power spectral density, $B = \qty{10}{\MHz}$ is the system bandwidth, and $N_f = \qty{10}{\dB}$ represents the receiver noise figure. The uplink and downlink APs are placed uniformly across the area, while users and targets are randomly distributed over a $\num{200} \times \qty{200}{\m^2}$ region.

\begin{figure}[!t]\vspace{-0mm}
    \centering
    \includegraphics[width=0.7\textwidth]{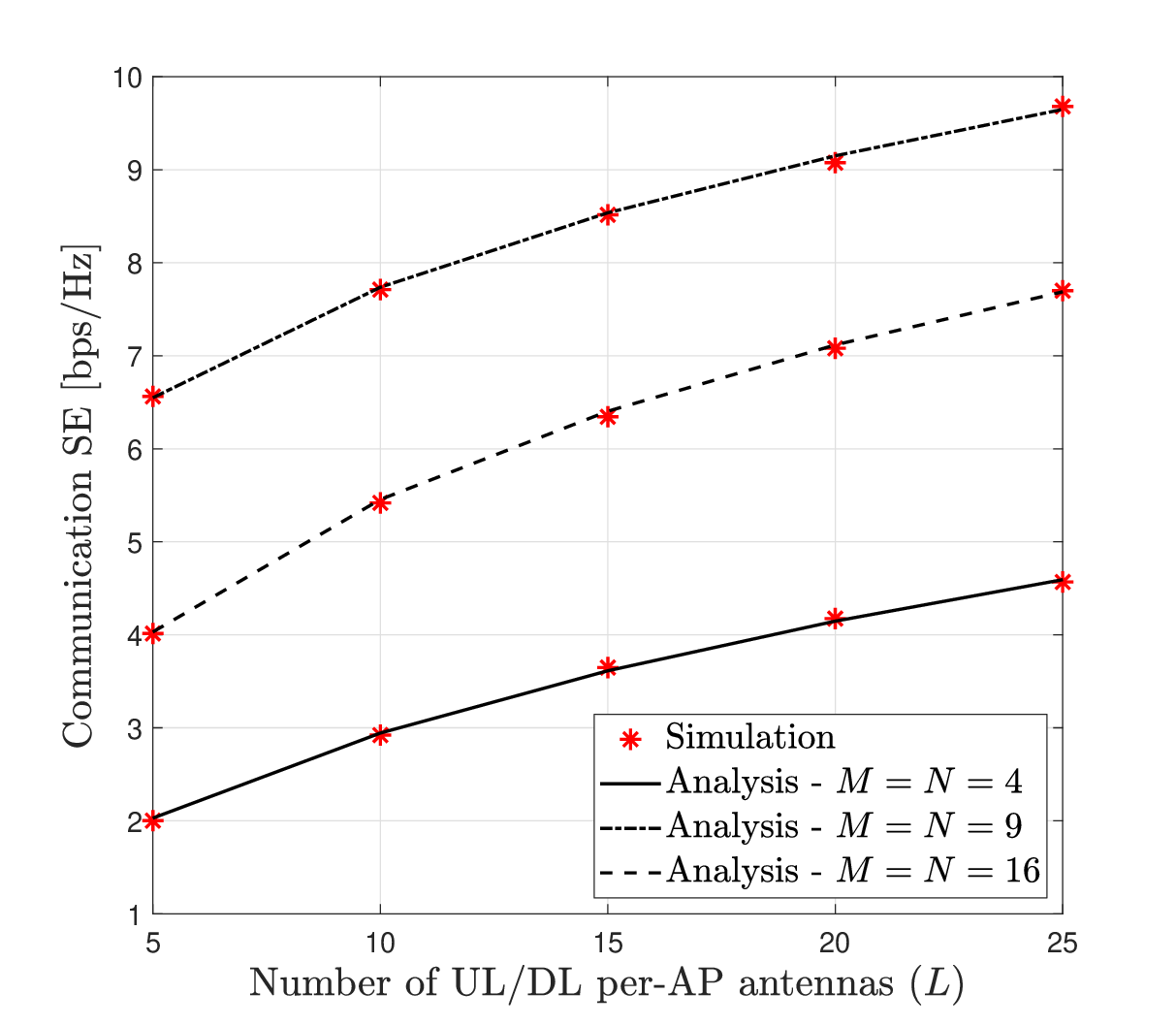}
    \vspace{-0mm}
    \caption{Communication SE versus the number of AP antennas.}
    \label{fig_ComSumRate_perf} \vspace{-0mm}
\end{figure}

\begin{figure}[!t]\vspace{-0mm}
    \centering
    \includegraphics[width=0.7\textwidth]{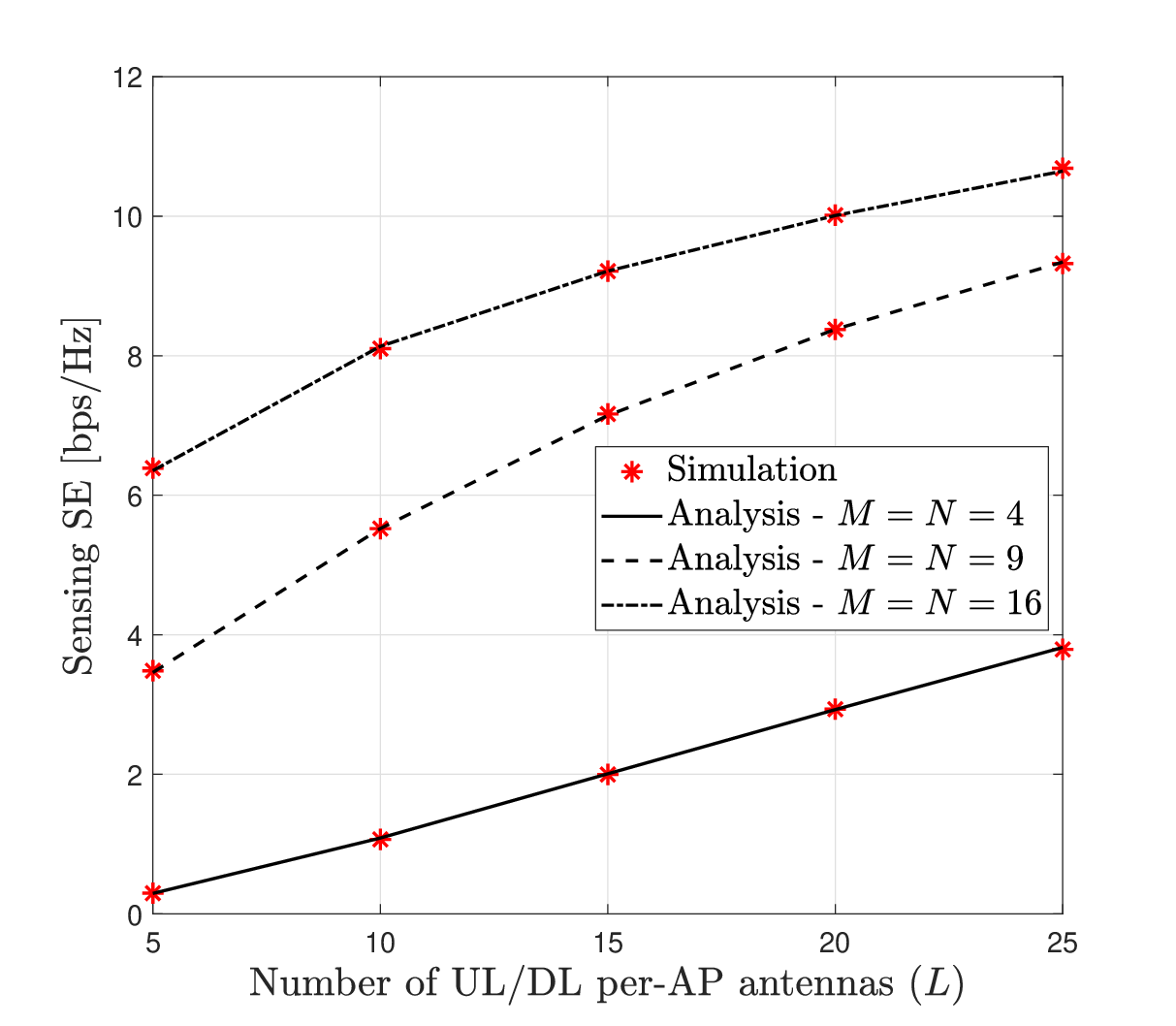}
    \vspace{-0mm}
    \caption{Sensing SE versus the number of AP antennas.}
    \label{fig_SensSumRate_perf} \vspace{-0mm}
\end{figure}

Figure~\ref{fig_ComSumRate_perf} and Figure~\ref{fig_SensSumRate_perf} illustrate the communication and sensing SEs, respectively, as functions of the number of antennas per UL/DL AP ($L$), for various AP configurations with $M=N=\{4, 9, 16\}$, $K=2$, and $T=3$. The accuracy of the analytical SEs is validated using Monte-Carlo simulations. In particular, the analytical communication and sensing SE curves and respective Monte-Carlo simulation curves coincide regardless of the simulation setup, validating the accuracy of the derived analytical rate expressions. The figures show that increasing the number of antennas at the APs significantly enhances both communication and sensing SEs, attributed to improved spatial resolution, stronger beamforming, and lower interference. For instance, with $M=N=9$, doubling the antenna count from \num{10} to \num{20} yields approximately \qty{31.3}{\percent} and \qty{51.7}{\percent} improvements in communication and sensing SE, respectively. Furthermore, deploying more APs amplifies spatial diversity and extends coverage, leading to additional performance gains. These results underscore the importance of high antenna densities and AP deployment in maximizing SE in CF-ISAC systems by effectively utilizing distributed resources.

	\chapter{Beamforming Designs in Cell-Free ISAC}\label{chp_CF_isac_resource}

This chapter presents and evaluates two beamforming design strategies for a generalized CF-ISAC system (Figure~\ref{fig_SystemModelResourceAllocation}): one based on traditional convex optimization techniques and the other using manifold optimization (MO), supported by numerical simulations.

\section{System and Channel Models}
\begin{figure}[!t]\vspace{4mm}
    \centering 
    \def\svgwidth{250pt} 
    \fontsize{8}{8}\selectfont 
    \graphicspath{{Figures/}}
\begingroup%
  \makeatletter%
  \providecommand\color[2][]{%
    \errmessage{(Inkscape) Color is used for the text in Inkscape, but the package 'color.sty' is not loaded}%
    \renewcommand\color[2][]{}%
  }%
  \providecommand\transparent[1]{%
    \errmessage{(Inkscape) Transparency is used (non-zero) for the text in Inkscape, but the package 'transparent.sty' is not loaded}%
    \renewcommand\transparent[1]{}%
  }%
  \providecommand\rotatebox[2]{#2}%
  \newcommand*\fsize{\dimexpr\f@size pt\relax}%
  \newcommand*\lineheight[1]{\fontsize{\fsize}{#1\fsize}\selectfont}%
  \ifx\svgwidth\undefined%
    \setlength{\unitlength}{647.06634521bp}%
    \ifx\svgscale\undefined%
      \relax%
    \else%
      \setlength{\unitlength}{\unitlength * \real{\svgscale}}%
    \fi%
  \else%
    \setlength{\unitlength}{\svgwidth}%
  \fi%
  \global\let\svgwidth\undefined%
  \global\let\svgscale\undefined%
  \makeatother%
  \begin{picture}(1,0.57198316)%
    \lineheight{1}%
    \setlength\tabcolsep{0pt}%
    \put(0,0){\includegraphics[width=\unitlength]{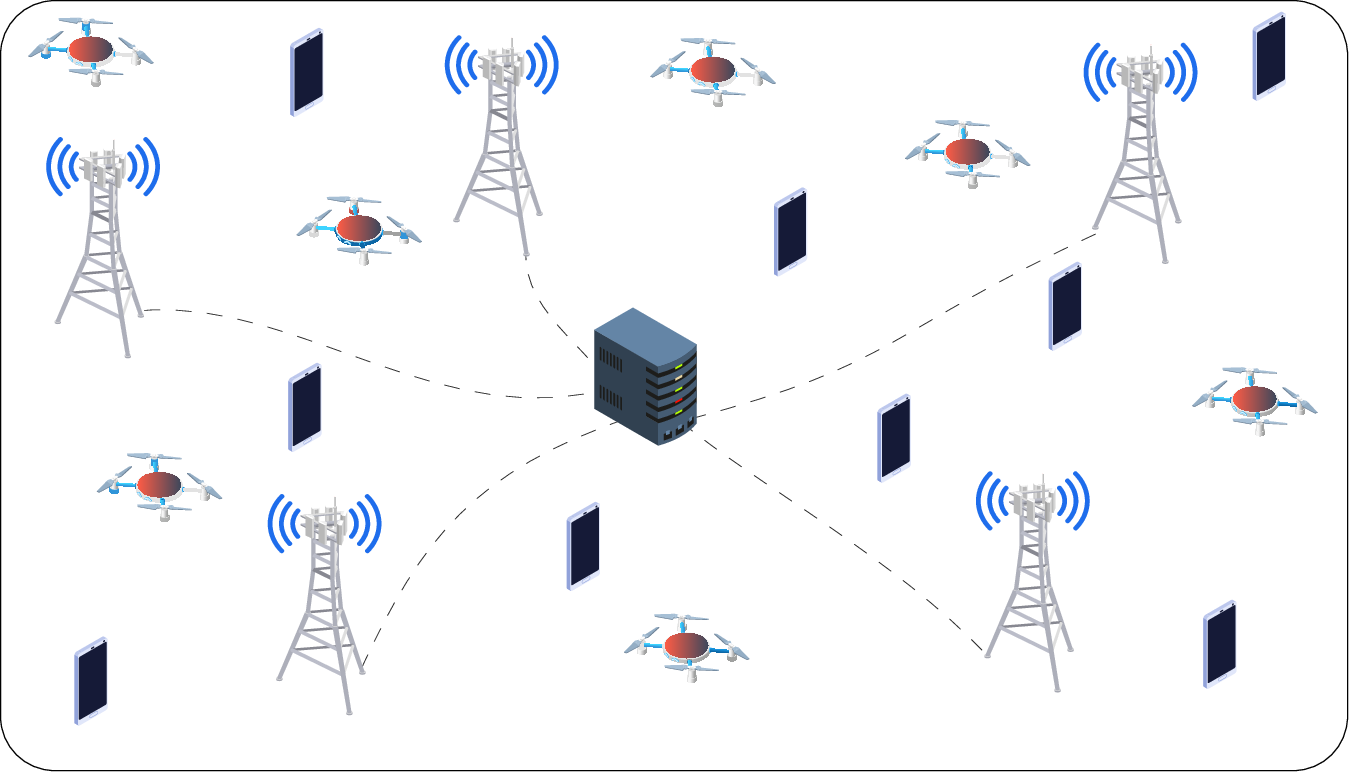}}%
    \put(0.21845862,0.01689264){\color[rgb]{0,0,0}\makebox(0,0)[lt]{\lineheight{1.25}\smash{\begin{tabular}[t]{l}AP\end{tabular}}}}%
    \put(0.44267809,0.35168914){\color[rgb]{0,0,0}\makebox(0,0)[lt]{\lineheight{1.25}\smash{\begin{tabular}[t]{l}CPU\end{tabular}}}}%
    \put(0.87334232,0.03450071){\color[rgb]{0,0,0}\makebox(0,0)[lt]{\lineheight{1.25}\smash{\begin{tabular}[t]{l}User\end{tabular}}}}%
    \put(0.46824042,0.04054456){\color[rgb]{0,0,0}\makebox(0,0)[lt]{\lineheight{1.25}\smash{\begin{tabular}[t]{l}Target\end{tabular}}}}%
  \end{picture}%
\endgroup%
 \vspace{-0mm} 
    \caption{A CF-ISAC system setup.}  \label{fig_SystemModelResourceAllocation}\vspace{-0mm}
\end{figure}
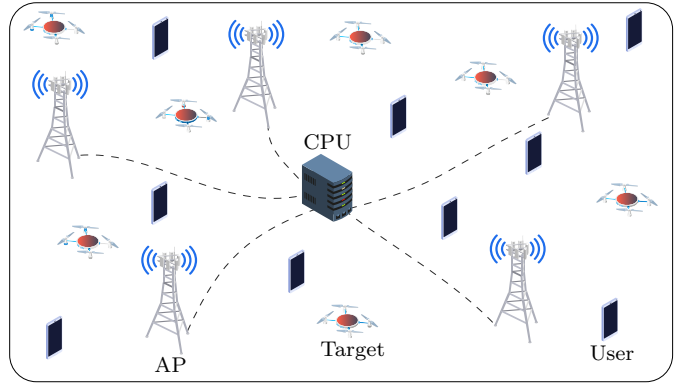

Figure~\ref{fig_SystemModelResourceAllocation} considers a CF-ISAC setup that consists of $M\geq 1 $ APs, each equipped with $L \geq 1 $ antennas, serving $K$ single-antenna users and tracking $T\geq 1 $ potential targets. Each AP employs a ULA configuration with half-wavelength spacing between elements, supporting accurate joint communication and sensing capabilities \cite{Zhenyao2023}. All APs are connected to a central CPU through fronthaul/backhaul links. The CPU centrally manages coordination and beamforming across the APs, ensuring full synchronization and streamlined operation \cite{Ngo2017}. With fully digital beamforming, where each antenna is paired with its own RF chain, the APs can simultaneously transmit integrated communication and sensing signals, enabling efficient user service and target detection.

Under a block-flat fading channel model, each coherence block assumes static channels. Within such a block, $\q{h}_{mk} \in \mathbb{C}^{L\times 1}$ and $\q{a}(\theta_{mt})$ denote the channel vectors from the $m$-th AP to the $k$-th user and to the $t$-th target, respectively. The communication channels, specifically $\q{h}_{mk}$ for $m \in \{1, \dots, M\}$ and $k \in \{1, \dots, K\}$, are modeled as
\begin{eqnarray}
    \q{h}_{mk} = \zeta_{\q{h}_{mk}}^{1/2} , \tilde{\q{h}}_{mk}
\end{eqnarray} \par \vspace{-0mm}
\noindent where $\zeta_{\q{h}_{mk}}$ denotes the large-scale fading coefficient, accounting for path-loss and shadowing, which stays constant across multiple coherence intervals. Moreover, $\tilde{\q{h}}_{mk} \sim \mathcal{CN}(\q{0}, \q{I}_{L})$ captures the small-scale Rayleigh fading, representing rapid channel fluctuations within a single coherence block \cite{Ngo2017}.

In contrast, the sensing channels are described differently. By following the echo signal modeling used in MIMO radar systems, the sensing channels are assumed to follow a LoS model \cite{Zhenyao2023}. Accordingly, the transmit array steering vector pointing toward the direction $\theta_{mt}$ is expressed as
\begin{eqnarray}
\q{a}(\theta_{mt}) =  \frac{1}{\sqrt{L}} \left[1, e^{j\pi \sin(\theta_{mt})}, \ldots, e^{j\pi (L-1) \sin(\theta_{mt})} \right]^{\mathrm{T}},
\end{eqnarray} \par \vspace{-0mm}
\noindent where $\theta_{mt}$ (for $m\in\{1,\dots, M\}$ and $t\in\{1,\dots, T\}$) denotes the direction of the $t$-th target relative to the $x$-axis from the perspective of the $m$-th AP.

The analysis proceeds under the following assumptions: 
\begin{itemize} 
    \item[(i)] $\theta_{mt}$ is assumed to be known at the CPU, having been estimated in advance via prior scanning procedures \cite{Tsinos2021Joint, Zhenyao2023, Wu2018}. 
    \item[(ii)] A dedicated channel estimation phase is executed before commencing communication and sensing, ensuring CSI is available for beamforming. 
    \item[(iii)] The APs and users are connected via a controlled link, facilitating the exchange of necessary commands \cite{positioningLTE}.  
\end{itemize} 
CF systems rely on TDD mode for channel estimation, where CSI can be obtained using techniques like least squares or minimum mean-squared error (MMSE) estimators \cite{Marzettabook2016, Nayebi2018, Ngo2024}. Therefore, it is assumed that CSI is accessible at both the APs and the users, consistent with standard practice in recent works \cite{demirhan2024cellfree, Demirhan2023, Liu2024}. Meanwhile, control links function separately from the main communication link, handling critical information exchange, such as CSI, beamforming vectors, and synchronization signals, between users and APs \cite{positioningLTE}. These links carry lightweight, low-rate messages and contribute minimal overhead while enabling smooth and coordinated system operation.

\section{Transmission Model} 
The $m$-th AP transmit signal $\q{x}_m \in \mathbb{C}^{L\times 1}$ to perform  communication and sensing. It is given as  \cite{Zhao2022, zargari2024riemannian}
\begin{eqnarray}\label{eqn_tx_signal}
\q{x}_m = \sum_{k=1}^{K} \q{w}_{mk} q_{k} + \q{s}_m,
\end{eqnarray}
where $q_{k} \in \mathbb{C}$ is the data symbol intended for the $k$-th user with $\mathbb{E}\{\vert q_{k}\vert^2\} = 1$, $\q{w}_{mk} \in \mathbb{C}^{L\times 1}$ denotes the $m$-th AP beamforming vector for the $k$-th user, and $\q{s}_m  \in \mathbb{C}^{L\times 1}$ is the dedicated sensing signal at the $m$-th AP. The received signal at the $k$-th user is expressed as
\begin{eqnarray} \label{eqn_rx_user_k}
    y_k &&= \sum_{m=1}^{M} \q{h}_{mk}^{\mathrm{H}} \q{x}_m + z_k \nonumber \\
    &&= \underbrace{\sum_{m=1}^{M}  \q{h}_{mk}^{\mathrm{H}} \q{w}_{mk} q_k}_{\text{Desired signal }} + \underbrace{\sum_{i\neq k}^{K} \sum_{m=1}^{M} \q{h}_{mk}^{\mathrm{H}} \q{w}_{mi} q_i }_{\text{Multi-user interference}}  + \underbrace{ \sum_{m=1}^{M} \q{h}_{mk}^{\mathrm{H}} \q{s}_{m} }_{\text{Sensing signal interference }}  + \underbrace{z_k}_{\text{AWGN}}, \quad
\end{eqnarray} \par \vspace{-0mm}
\noindent where $z_k \sim \mathcal{CN}(0, \sigma^2)$ is the AWGN at the $k$-th user. Here, it is also assumed that clutter rejection techniques are applied to mitigate reflected interference from the targets and the surrounding environment \cite{Mark2010RadarBook}.

\section{Communication SE} 
Using \eqref{eqn_rx_user_k}, the $k$-th user SINR is given as
\begin{eqnarray}\label{eqn_gamma}
\mt{SINR}_k^{\mathrm{Com}}&&=\frac{\big\vert \sum_{m =1}^{M} \q{h}_{mk}^{\mathrm{H}}  \q{w}_{mk} \big\vert^2}{\sum_{i\neq k}^{K} \left\vert \sum_{m=1}^{M} \q{h}_{mk}^{\mathrm{H}} \q{w}_{mi} \right\vert^2 + \left\vert \sum_{m=1}^{M} \q{h}_{mk}^{\mathrm{H}} \q{s}_{m} \right\vert^2 + \sigma^2}.
\end{eqnarray} \par \vspace{-0mm}
\noindent The SE of the $k$-th user can be approximated as
\begin{eqnarray}
\mathcal{S}_k^{\mathrm{Com}} \approx \log_2(1+ \mt{SINR}_k^{\mathrm{Com}}).   
\end{eqnarray}\par \vspace{-0mm}
\noindent The SE quantifies the effectiveness of data decoding by users.

\section{Sensing Beampattern Gain}
The transmit beampattern gain is a commonly used metric for assessing the design of sensing signals \cite{Stoica2007}. These gains, which describe how the transmit signal's power is distributed across different sensing angles $\theta$, can greatly enhance detection, recognition, and accuracy in sensing tasks \cite{Stoica2007}. The transmit sensing beampattern gain at the $m$-th AP towards the $t$-th target can be expressed as \cite{He2022}.
\begin{eqnarray}\label{eqn_beamgain_mt}
p(\theta_{mt}) &&= \mathbb{E}\left\{| \q{a}^{\mathrm{H}}(\theta_{mt}) \q{x} |^2 \right\} \nonumber\\
&&= \sum_{i=1}^{K} \q{a}^{\mathrm{H}}(\theta_{mt}) \q{w}_{mi} \q{w}_{mi}^{\mathrm{H}} \q{a}(\theta_{mt}) + \q{a}^{\mathrm{H}}(\theta_{mt}) \q{s}_{m} \q{s}_{m}^{\mathrm{H}} \q{a}(\theta_{mt}).
\end{eqnarray}\par \vspace{-0mm}
\noindent  This metric is tailored to address the specific needs of target sensing. When the directions of targets are uncertain, a uniformly distributed beampattern is the most effective. However, in scenarios where the approximate directions of the targets are known, the beampattern gain should be concentrated on these directions to improve the effectiveness of target detection and tracking \cite{Stoica2007}.

Our ISAC design focuses on two key metrics: (1) communication SINR and (2) sensing beampattern gain. The first metric ensures accurate symbol detection and reduces communication errors. Meanwhile, sensing efficiency is primarily determined by the beampattern gain, which plays a crucial role in enhancing target detection probability. A carefully crafted sensing beampattern gain is essential for achieving effective target identification and overall success in sensing \cite{Stoica2007}.

\section{Problem Formulation}
The objective is to maximize the communication sum SE for the users while satisfying sensing beampattern gain requirements for each target and per-AP transmit power constraints. The problem is thus formulated as follows: 
\begin{subequations}\label{prob_P1}
\begin{eqnarray}
(\mathcal{P}1):~&& \max_{\{\q{w}_{mk}, \q{s}_m\}} \sum_{k=1}^{K} \log_2\left(1 + \mt{SINR}_k^{\mathrm{Com}} \right), \label{prob_P1_obj}  \\
\text{s.t.} \quad &&   p(\theta_{mt}) \geq \Gamma_{t}^{\mathrm{th}}, ~\forall m, t, \label{prob_P1_beamgain}\\
&& \sum_{i=1}^{K} \Vert \q{w}_{mi} \Vert^2 + \Vert \q{s}_{m} \Vert^2 \leq p_{\mathrm{max}}, ~\forall m, \label{prob_P1_tx_pow}
\end{eqnarray}
\end{subequations} \par \vspace{-0mm}
\noindent where \eqref{prob_P1_beamgain} ensures the sensing beampattern each target, in which $\Gamma_{t}^{\mathrm{th}}$ is the $t$-th target's required seeing threshold, and \eqref{prob_P1_tx_pow} sets the the  maximum allowable transmit power at $m$-th AP as $p_{\mathrm{max}}$.

\section{Proposed Solution 1: CCPA-Based Beamforming}
This section develops the beamforming design to $(\mathcal{P}1)$, utilizing convex-concave procedure algorithm (CCPA) based on SDR and SCA to determine optimal $\q{w}_{mk}$ and $\q{s}_m$.

We first define $\q{W}_k = \q{w}_k \q{w}_k^{\mathrm{H}}$ and $\q{S} = \q{s} \q{s}^{\mathrm{H}}$, where $\q{w}_k = [\q{w}_{1k}, \ldots, \q{w}_{Mk}] \in \mathbb{C}^{LM \times 1}$ and $\q{s} = [\q{s}_{1}, \ldots, \q{s}_{M}] \in \mathbb{C}^{LM \times 1}$ are the communication beamforming for the $k$-the user and sensing signal from all APs. Here, $\q{W}_{k}$ and $\q{S}$ are semi-definite matrices, i.e., $\q{W}_{k} \succeq 0$ and $\q{S} \succeq 0$, and $\q{W}_{k}$ must satisfy rank one constraint, i.e., ${\mathrm{Rank}}(\q{W}_{k}) = 1$. Then, utilizing the SDR techniques to relax the highly non-convex rank one constraint, $(\mathcal{P}1)$ can be reformulated into a standard semi-definite problem (SDP) as follows:
\begin{subequations}\label{prob_Qr1}
\begin{eqnarray}
    (\mathcal{P}_c2):~&& \max_{\q{W}_{k}, \q{S}} \quad \sum_{k=1}^{K} f(\q{W}_k, \q{S}), \label{prob_Qr1_obj}  \\
    \text{s.t.} \quad && \tr\left(\sum_{i=1}^{K} \q{Q}_m \q{g}_t \q{g}_t^{\mathrm{H}} \q{Q}_{m} \q{W}_{i} \right) + \tr\left( \q{Q}_m \q{g}_t \q{g}_t^{\mathrm{H}} \q{Q}_{m} \q{S} \right) \nonumber\\
    && - \Gamma_{t}^{\mathrm{th}} \geq 0, ~\forall m, t, \\
    && \sum_{i=1}^{K} \tr\left( \q{W}_{i} \q{Q}_m \right) + \tr\left( \q{S} \q{Q}_m \right) \leq p_{\mathrm{max}}, ~\forall m, \label{prob_Qr1_tx_pow} \\
    && \q{W}_k,  \q{S} \succeq 0,~\forall k,
\end{eqnarray}
\end{subequations} \par \vspace{-0mm}
\noindent where $\q{Q}_m \in \mathbb{R}^{LM \times LM}$ is a block diagonal selection matrix, consisting of $M$ blocks, each corresponding to an AP, i.e.,  $\q{Q}_m = \text{diag}(\q{0}, \dots, \q{I}_{L},  \dots, \q{0})$ with $\q{I}_{L}$ in the $m$-th block on the diagonal. Moreover,  $\q{f}_k = [\q{h}_{1k}^{\mathrm{T}}, \dots, \q{h}_{Mk}^{\mathrm{T}}]^{\mathrm{T}} \in \mathbb{C}^{LM \times 1}$ and $\q{g}_t = [\q{a}^{\mathrm{T}}(\theta_{1t}), \dots, \q{a}^{\mathrm{T}}(\theta_{Mt})]^{\mathrm{T}} \in \mathbb{C}^{LM \times 1}$, are the cascaded channel vectors. In \eqref{prob_Qr1}, as the objective is not a convex function, we use the SCA method to linearize it, which is given by 
\begin{eqnarray}\label{eqn_Qr1_obj_def_}
    f(\q{W}_k, \q{S}) &=& \log_2 \left( \sum_{i=1}^{K} \tr\left( \q{f}_k \q{f}_k^{\mathrm{H}} \q{W}_i \right) + \tr\left( \q{f}_k \q{f}_k^{\mathrm{H}} \q{S} \right) + \sigma^2 \right)  \nonumber\\
    &&- \log_2 \left( \sum_{i\neq k}^{K} \tr\left( \q{f}_k \q{f}_k^{\mathrm{H}} \q{W}_i^{(l)} \right) + \tr\left( \q{f}_k \q{f}_k^{\mathrm{H}} \q{S}^{(l)} \right) + \sigma^2 \right) \nonumber \\
    && - \frac{\sum_{i\neq k}^{K} \tr\left( \q{f}_k \q{f}_k^{\mathrm{H}} (\q{W}_i - \q{W}_i^{(l)}) \right)  }{\ln(2) \left( \sum_{i\neq k}^{K} \tr\left( \q{f}_k \q{f}_k^{\mathrm{H}} \q{W}_i^{(l)} \right) + \tr\left( \q{f}_k \q{f}_k^{\mathrm{H}} \q{S}^{(l)} \right) + \sigma^2 \right)} \nonumber\\
    &&- \frac{\tr\left( \q{f}_k \q{f}_k^{\mathrm{H}} (\q{S} - \q{S}^{(l)}) \right)  }{\ln(2) \left( \sum_{i\neq k}^{K} \tr\left( \q{f}_k \q{f}_k^{\mathrm{H}} \q{W}_i^{(l)} \right) + \tr\left( \q{f}_k \q{f}_k^{\mathrm{H}} \q{S}^{(l)} \right) + \sigma^2 \right)}, \qquad
\end{eqnarray}\par \vspace{-0mm}
\noindent where $(\cdot)^{(l)}$ denotes the previous iteration values of respective variables. The relaxed problem $(\mathcal{P}_c2)$ can be solved via the interior-point method using the SDP \cite{boyd2004convex}. Finally, to impose the relaxed rank one constraint, Gaussian randomization (GR) is utilized \cite{Qingqing2019}. 

In particular, let the solution to the relaxed problem $(\mathcal{P}_c2)$ to be $\{\q{W}_k^*, \q{S}^*\}$. The optimal sensing beamforming vector $\q{s}^*$ is obtained by eigenvalue decomposition (EVD) \cite{Luo2010}. Let the EVD of $\q{S}^*$ to be $\q{S}^* = \q{V}_s \boldsymbol{\Sigma}_s \q{V}_s^{\mathrm{H}}$, where $\q{V}_s = [\q{v}_{s,1}, \dots, \q{v}_{s, ML}]$ is a unitary matrix and $\boldsymbol{\Sigma}_s  = \text{diag}(\lambda_{s,1}, \dots, \lambda_{s,ML})$ is a diagonal matrix. Then,  $\q{s}^*$, is the eigenvector for the maximum eigenvalue, i.e., $\q{s}^*= \q{v}_{s,1}$. Conversely, let the EVD of $\q{W}_k^*$ to be $\q{W}^* = \q{V}_k \boldsymbol{\Sigma}_k \q{V}_k^{\mathrm{H}}$, where $\q{V}_k = [\q{v}_{k,1}, \dots, \q{v}_{k, ML}]$ and $\boldsymbol{\Sigma}_k  = \text{diag}(\lambda_{k,1}, \dots, \lambda_{k,ML})$. If ${\mathrm{Rank}}(\q{W}_{k}) = 1$, then the optimal $\q{w}_k^* = \q{v}_{k,1}$. Otherwise, the GR process is utilized to enforce the relaxed rank-one constraint \cite{Luo2010}. In particular, a  solution to $(\mathcal{P}_c2)$ is generated as $\q{W}_k' = \q{V}_k \boldsymbol{\Sigma}_k^{1/2} \q{e}_k$, where $\q{e}_k \in \mathcal{CN}(\q{0}, \q{I}_{ML})$. Such solutions are generated for $10^5$ times, and the best one is selected. These numerous random realizations of $\q{r}$ ensure an $\frac{\pi}{4}$-approximation to the optimal value of $(\mathcal{P}_m2)$ \cite{Luo2010}. Algorithm~\ref{alg_ccpa_beamforming} outlines the steps to solve $(\mathcal{P}_c2)$ \eqref{prob_Qr1}.

\begin{algorithm}[!t]
\caption{: CCPA-Based CF-ISAC Beamforming Algorithm} 
\begin{algorithmic}[1] \label{alg_ccpa_beamforming}
\STATE \textbf{Initialization}: Begin - CVX.
\STATE Solve the convex problem $(\mathcal{P}_c2)$ in \eqref{prob_Qr1}.
\STATE End - CVX.
\STATE EVD  $\q{S}^*$ as $\q{S}^*=\q{V}_s \boldsymbol{\Sigma}_{s} \q{V}_s^{\mathrm{H}}$, where $\q{P}=[\q{v}_{s,1}, \dots, \q{v}_{s,LM}]$.
\STATE \textbf{return}: $\q{s}^*= \q{v}_{s,1}$.
\FOR{$k=\{1, \dots, K\}$}
\STATE EVD  $\q{W}_k^*$ as $\q{W}_k^*=\q{V}_k \boldsymbol{\Sigma}_k \q{V}_k^{\mathrm{H}}$, where $\q{V}_k=[\mathbf{v}_{k,1}, \dots, \q{v}_{k,LM}]$.
\IF{$\mathrm{Rank}(\mathbf{W}_k^*)=1$,} 
    \STATE \textbf{return}: $\q{w}_k^*= \q{v}_{k,1}$.
\ELSE
    \FOR{$d=1,\ldots,D$}
        \STATE Generate random  $\q{r}_{k,d} = \q{V}_k \boldsymbol{\Sigma}_k^{1/2} \q{e}_{k,d}$, where $\q{e}_{k,d}  \sim \mathcal{CN}(\q{0}, \q{I}_{LM})$.
        \STATE Check if $(\mathcal{P}_c2)$ is feasible with $\q{r}_{k,d}$. 
    \ENDFOR
    \STATE \textbf{return}: $\q{w}_k^*= \q{r}_{k}$, where  $\q{r}_{k}= \underset {d=\{1,\ldots,D\}}{\mathrm{arg \,min}} \,\, \q{r}_{k,d}$.
\ENDIF 
\ENDFOR
\STATE \textbf{Output}: Optimal communication and sensing beamforming, i.e.,  $\{\q{w}_k^*\}_{k=1}^{K}, \q{s}^*$.
\end{algorithmic}
\end{algorithm}

\subsubsection{Computational Complexity}
According to \cite[Th. 3.12]{polik2010interior}, the order of complexity for a SDP problem with $m$ SDP constraints, which includes a $n \times n$ positive semi-definite matrix is given by $\mathcal{O}\left( \sqrt{n} \log\left(\frac{1}{\epsilon}\right) (mn^3 + m^2n^2 + m^3) \right)$ with $\epsilon > 0$. For problem $(\mathcal{P}_c2)$, with $n = LM$ and $m = N + K + 1$, the approximate computational complexity is given by
\begin{eqnarray}
    \mathcal{O}\left( \sqrt{LM} \log\left(\frac{1}{\epsilon}\right) \left((N+K+1)(LM)^{3}\right) \right).
\end{eqnarray}

\section{Proposed Solution 2: MO-Based Beamforming}
Before presenting the solution, we introduce some mathematical notations for clarity. First, the beamforming vectors for the $m$-th AP are consolidated into a single vector $\q{w}_m = [\q{w}_{m1}^{\mathrm{T}}, \ldots, \q{w}_{mK}^{\mathrm{T}}, \q{s}_{m}^{\mathrm{T}}]^{\mathrm{T}} \in \mathbb{C}^{L(K+1) \times 1}$. Then, organizing all AP beamforming vectors, the matrix $\q{W} = [\q{w}_1, \ldots, \q{w}_M] \in \mathbb{C}^{L(K+1) \times M}$ is constructed. Next, selection matrices, $\q{E}_k = [\q{0}, \ldots, \q{I}_L, \ldots, \q{0}] \in \mathbb{R}^{L \times L(K+1)}$ for the $k$-th user, where the $k$-th block containing $L\times L$ identity matrix, i.e., $\q{I}_L$, and $\q{D} = \q{I}_M \in \mathbb{R}^{M \times M}$ for APs are introduced. Combining $\q{W}$, $\q{E}_k$, and $\q{D}$ allows for representation of any $\q{w}_{mk}$, i.e., $\q{w}_{mk} = \q{W} \q{E}_k \q{D}_m$, where $\q{D}_m$ is the $m$-th column of matrix $\q{D}$.

To deal with the non-convex sum-log terms in the objective, we use fractional programming (FP) \cite[\textit{Theorem 3}]{Shen2018FPpart2}. To this end, each SINR term in \eqref{prob_P1} is replaced with auxiliary variables $\mu_k$ such that $\mu_k \leq \mt{SINR}_k^{\mathrm{Com}}$, and $(\mathcal{P}_m1)$ is reformulated as 

\begin{subequations}\label{prob_P2}
\begin{eqnarray}
(\mathcal{P}_m2):~&& \max_{\q{W}, \boldsymbol{\mu}} ~f(\q{W}, \boldsymbol{\mu}) = \frac{1}{\ln(2)} \sum_{k=1}^{K} \ln(1 + \mu_k)  \nonumber \label{prob_P2_obj}\\
&&\qquad + \frac{1}{\ln(2)} \sum_{k=1}^{K} \left( - \mu_k + \frac{(1 + \mu_k)\mt{SINR}_k^{\mathrm{Com}}}{1 + \mt{SINR}_k^{\mathrm{Com}}} \right),  \\
\text{s.t.} \quad && \eqref{prob_P1_beamgain}-\eqref{prob_P1_tx_pow},
\end{eqnarray}
\end{subequations} \par \vspace{-0mm}
\noindent where $\boldsymbol{\mu} = [\mu_1, \dots, \mu_K]$ is the auxiliary variable vector. The problem $(\mathcal{P}_m2)$ is a two-part optimization problem: (i) an outer optimization over $\q{W}$ with a fixed $\boldsymbol{\mu}$ and (ii) an inner optimization over $\boldsymbol{\mu}$ with a fixed $\q{W}$. To solve $(\mathcal{P}_m2)$, $\q{W}$ and $\boldsymbol{\mu}$ are iteratively optimized until the objective function stabilizes.

Note that the original problem is reformulated into an equivalent form, where $\q{W}$ solves $(\mathcal{P}1)$ if and only if it also solves $(\mathcal{P}_m2)$ \cite[\textit{Theorem 3}]{Shen2018FPpart2}. Furthermore, it has been established that the optimal objective values for $(\mathcal{P}1)$ and $(\mathcal{P}_m2)$ are the same \cite{Shen2018FPpart2}.

\subsection{Optimization of $\boldsymbol{\mu}$ with Fixed $\q{W}$}
For a fixed $\q{W}$, the objective $f(\q{W}, \boldsymbol{\mu})$ \eqref{prob_P2_obj} is a concave and differentiable function over $\boldsymbol{\mu}$. Thus, the optimal $\boldsymbol{\mu}$ can be obtained by setting $\frac{\partial f(\q{V}, \boldsymbol{\mu})}{\partial \mu_k}=0$ for each $\mu_k$ as \cite{Shen2018}
\begin{eqnarray}\label{eqn_opt_mu}
    \mu_k^* =\mt{SINR}_k^{\mathrm{Com}}.
\end{eqnarray}

\subsection{Optimization of $\q{W}$  with Fixed  $\boldsymbol{\mu}$}
Next, for a given $\boldsymbol{\mu}$, by eliminating the the constant terms with respect to $\q{W}$ in $(\mathcal{P}_m2)$, it can be reformulated as
\begin{subequations}
\begin{eqnarray}\label{prob_P3}
(\mathcal{P}_m3):~&& \max_{\q{W}}  \sum_{k=1}^{K} \frac{\tilde{\mu}_k \vert  \sum_{m=1}^{M} \q{h}_{mk}^{\mathrm{H}} \q{E}_k \q{W} \q{D}_m \vert^2}{\sum_{i=1}^{K+1} \vert \sum_{m=1}^{M} \q{h}_{mk}^{\mathrm{H}} \q{E}_i \q{W} \q{D}_m \vert^2 + \sigma^2},  \\
\text{s.t.} \quad &&   \eqref{prob_P1_beamgain}-\eqref{prob_P1_tx_pow}, 
\end{eqnarray}
\end{subequations} \par \vspace{-0mm}
\noindent where $\tilde{\mu}_k = 1+\mu_k$ for $k \in \{1,\dots, K\}$. To solve $(\mathcal{P}_m3)$, the MO technique is employed.  The per-AP power constraint is normalized such that $\sum_{i=1}^{K+1} \Vert \q{E}_i \q{W} \q{D}_m \Vert^2 \leq 1$. A modified vector $\q{v}_m = [\q{v}^{\mathrm{T}}_{m1}, \ldots, \q{v}^{\mathrm{T}}_{m(K+1)}]^{\mathrm{T}} \in \mathbb{C}^{(L+1)(K+1) \times 1}$ is introduced, where $\q{v}_{mk} = [\q{w}^{\mathrm{T}}_{mk}, z_k]^{\mathrm{T}} \in \mathbb{C}^{(L+1) \times 1}$. Here, $\mathbf{z} = [z_1, \ldots, z_{K+1}]$ is an auxiliary vector, resulting in the expanded matrix $\q{V}=[\q{v}_1, \ldots, \q{v}_M] \in \mathbb{C}^{(L+1)(K+1) \times M}$. These modifications lead to the power normalization condition $\sum_{k=1}^{K+1} \Vert \tilde{\q{E}}_k \q{V} \q{D}_m \Vert^2 = 1$, where $\tilde{\q{E}}_k=[\q{0},\ldots, \q{I}_{L+1},\ldots, \q{0}] \in \mathbb{R}^{(L+1) \times (L+1)(K+1)}$. A complex oblique manifold is defined  as follows:
\begin{eqnarray}\label{eqn_M}
\mathcal{M} = \left\{\q{V} \in \mathbb{C}^{(L+1) (K+1) \times M} \:|\:  \Vert\q{V}_{:1} \Vert^2 = \cdots= \Vert\q{V}_{:M}\Vert^2  = 1 \right\}.
\end{eqnarray}

Consequently, $(\mathcal{P}_m3)$ cab be reformulated as a constrained optimization problem on $\mathcal{M}$, yielding  $(\mathcal{P}_m4)$ \eqref{prob_P4},
\begin{subequations}\label{prob_P4}
\begin{eqnarray}
(\mathcal{P}_m4):~&& \min_{\q{V} \in \mathcal{M}} ~ f(\q{V}) = -\sum_{k=1}^{K} \frac{\tilde{\mu}_k \vert \sum_{m=1}^{M} \hat{\q{h}}_{mk}^{\mathrm{H}} \tilde{\q{E}}_k \q{V} \q{D}_m  \vert^2}{\sum_{i=1}^{K+1} \vert \sum_{m=1}^{M} \hat{\q{h}}_{mk}^{\mathrm{H}} \tilde{\q{E}}_i \q{V} \q{D}_m \vert^2 + \sigma^2}, \label{prob_P4_obj} \qquad \\ 
\text{s.t.} \quad &&   u_{mt}(\q{V}) = \Gamma_{t}^{\mathrm{th}} \nonumber\\
&&- \sum_{i=1}^{K+1} \hat{\q{a}}^{\mathrm{H}}(\theta_{mt}) \tilde{\q{E}}_i \q{V}  \q{D}_m (\tilde{\q{E}}_i \q{V} \q{D}_m)^{\mathrm{H}} \hat{\q{a}}(\theta_{mt}) \leq 0 ,~\forall m, t,\label{prob_P4_sens} \qquad
\end{eqnarray}
\end{subequations}\par \vspace{-0mm}
\noindent where $\hat{\q{h}}_{mk} = \sqrt{p_{\mathrm{max}}}[\q{h}_{mk}, 0]$ and $\hat{\q{a}}(\theta_{mn}) = \sqrt{p_{\mathrm{max}}}[\q{a}(\theta_{mn}), 0]$ are adjusted to fit the dimensionality and scaling of the problem. 

The constraint \eqref{prob_P4_sens} in problem $(\mathcal{P}_m4)$ is beyond the MO. To address it, we use the ALM \cite{zargari2024riemannian, Birgin2014book}. Thus, the constraint \eqref{prob_P4_sens} is added to the objective function via a penalty. Hence, the new cost function is expressed as \cite{zargari2024riemannian, Birgin2014book} 
\begin{eqnarray}\label{eqn_Lag_penalty}
\mathcal{L}_\rho(\q{V} , \boldsymbol{\lambda}) = f(\q{V}) + \frac{\rho}{2} \sum_{t=1}^{T} \sum_{m=1}^{M} \max\left\{0, \frac{\lambda_{mt}}{\rho} + {u_{mt}(\q{V})}\right\}^2,
\end{eqnarray} \par \vspace{-0mm}
\noindent where $\rho > 0$ is the penalty parameter and $\boldsymbol{\lambda}  \in \mathbb{R}^{M\times T}$ is the Lagrange multiplier matrix. The problem $(\mathcal{P}_m4)$ is thus recast as
\begin{eqnarray}\label{prob_P5}
(\mathcal{P}_m5):~~ \min_{\q{V} \in \mathcal{M},\boldsymbol{\lambda}} \quad  \mathcal{L}_\rho(\q{V}, \boldsymbol{\lambda}).
\end{eqnarray}

\begin{algorithm}[!t]
\caption{: ALMCI Bemforming Algorithm} 
\begin{algorithmic}[1] \label{alg_beamforming}
\STATE \textbf{Initialization}: $\q{V}_0 \in \mathcal{M}$,  $\boldsymbol{\lambda}^0 \in \mathbb{R}^{M\times T}$, accuracy tolerance $\epsilon_{\min}$, convergence tolerance $\upsilon_1>0$, $\epsilon_0 > 0$,  $\rho_0$, reduction factors $\theta_\epsilon \in (0, 1)$ and $\theta_\rho > 1$,  $\lambda^{\min}, \lambda^{\max} \in \mathbb{R}$ with $\lambda^{\min} \leq \lambda^{\max}$, $\tau \in (0, 1)$,  minimum acceptable distance $d_{\min}$, and final convergence tolerance $\upsilon_2>0$.
\STATE Set $l = 0$.
\WHILE{$\text{dist}(f(\q{V}_l), f(\q{V}_{l+1})) \geq \upsilon_2$}
\STATE Set $j = 0$.
    \WHILE{$\text{dist}(\q{V}_j, \q{V}_{j+1}) \geq d_{\min}$ or $\epsilon_j > \epsilon_{\min}$}
        \STATE Set $i = 0$.
        \STATE Calculate $\boldsymbol{\eta}_0 = -{\mathrm{grad}}_{\q{V}_0} \mathcal{L}_\rho(\q{V}, \boldsymbol{\lambda})$.
        \WHILE{$\|{\mathrm{grad}}_{\q{V}_i} \mathcal{L}_\rho(\q{V}, \boldsymbol{\lambda})\|_2 > \upsilon_1$}
            \STATE  Calculate the search step $\alpha_i$ according to \cite{Shewchuk1994}.
            \STATE Update $\q{V}_{i+1}$ using retraction $R_{\q{V}_i}(\alpha_i\boldsymbol{\eta}_i)$.
            \STATE Compute the gradient at the new point ${\mathrm{grad}}_{\q{V}_{i+1}} \mathcal{L}_\rho(\q{V}, \boldsymbol{\lambda})$.
            \STATE Compute the vector transport $\mathcal{T}_{\q{V}_i \rightarrow \q{V}_{i+1}}(\boldsymbol{\eta}_i)$.
            \STATE Compute $\beta_i$ according to \cite{Shewchuk1994}.
            \STATE Update the gradient direction.
            \STATE $i \leftarrow i + 1$;
        \ENDWHILE
        \STATE Update the Lagrange multiplier.
        \STATE Set $\sigma_{t}^{j+1} = \max \left\{ {u_{mt}(\q{V}_{j+1})}, -\frac{\lambda_{mt}^{j+1}}{\rho_l} \right\}$.
        \STATE Adjust the accuracy tolerance $\epsilon_{j+1} = \max \{\epsilon_{\min}, \theta_\epsilon \epsilon_j\}$.
        \IF{$j = 0$ or $\underset{t}{\max} \{|\sigma_{t}^{j+1}| \} \leq \tau \underset{t}{\max} \{ |\sigma_{t}^{j}| \}$}
            \STATE $\rho_{j+1} = \rho_j$.
        \ELSE
            \STATE $\rho_{j+1} = \theta_\rho \rho_j$.
        \ENDIF
        \STATE $j \leftarrow j + 1$
    \ENDWHILE
    \STATE Calculate $\mu_k$ using \eqref{eqn_opt_mu}.
    \STATE Obtain $\q{V}_{l+1}$.
    \STATE $l \leftarrow l + 1$
\ENDWHILE
\STATE \textbf{Output}: $\q{W}^* = \q{V}^{*}(1:L(K+1), M)$.
\end{algorithmic}
\end{algorithm}

Algorithm~\ref{alg_beamforming}, which represents the augmented Lagrangian model-based iterative MO optimization (ALMCI) for CF-ISAC, outlines the essential steps to optimize \eqref{prob_P5} on $\mathcal{M}$. The process includes the following key steps \cite{liu2020simple, zargari2024riemannian}:

\noindent (a) \textbf{Riemannian gradient:}
This computes the Riemannian gradient of $f(\q{V})$ on $\mathcal{M}$ by projecting the Euclidean gradient onto the tangent space $T_{\q{V}_l}\mathcal{M}$ at the current point $\q{V}_l$, and is given as 
\begin{eqnarray}
    {\mathrm{grad}}_{\q{V}_l} \mathcal{L}_\rho(\q{V}, \boldsymbol{\lambda}) = \nabla_{\q{V}_l} \mathcal{L}_\rho(\q{V}, \boldsymbol{\lambda}) - \Re\{\nabla_{\q{V}_l} \mathcal{L}_\rho(\q{V} , \boldsymbol{\lambda}) \circ \q{V}^*_l\}\circ \q{V}_l, \qquad
\end{eqnarray} \par \vspace{-0mm}
\noindent where  $T_{\q{V}_l}\mathcal{M} = \left\{ \mathbf{c} \in \mathbb{C}^{(L+1) (K+1)} \mid \Re\{\mathbf{c} \circ \q{V}^*_l\} = \mathbf{0}_{(L+1) (K+1)} \right\}$ and $\mathbf{c} \in \mathbb{C}^{(L+1) (K+1)}$. The Euclidean gradient of $\mathcal{L}_\rho(\q{V} , \boldsymbol{\lambda})$, i.e., $\nabla_{\q{V}_l} \mathcal{L}_\rho(\q{V} , \boldsymbol{\lambda})$, is given by 
\begin{eqnarray}  \label{derivtive_eq}
&&\nabla_{\q{V}_l} \mathcal{L}_\rho(\q{V} , \boldsymbol{\lambda}) =  \sum_{k=1}^{K} -\tilde{\mu}_k 
\Bigg( \frac{2 \sum_{m=1}^{M}\hat{\q{h}}_{mk}^{\mathrm{H}}  \tilde{\q{E}}_{k} \q{V}  \q{D}_m  \sum_{m=1}^{M}  \tilde{\q{E}}_{k}^{\mathrm{H}}  \hat{\q{h}}_{mk}  \q{D}_m^{\mathrm{H}}   }{\sum_{j=1}^{K+1} \vert \sum_{m=1}^{M} \hat{\q{h}}_{mk}^{\mathrm{H}} \tilde{\q{E}}_j \q{V} \q{D}_m \vert^2 + \sigma^2}\nonumber\\
&&\quad - \sum_{i=1}^{K+1} \frac{2\vert \sum_{m=1}^{M} \hat{\q{h}}_{mk}^{\mathrm{H}} \tilde{\q{E}}_k \q{V} \q{D}_m  \vert^2 \sum_{m=1}^{M}\hat{\q{h}}_{mk}^{\mathrm{H}}  \tilde{\q{E}}_{i} \q{V}  \q{D}_m  \sum_{m=1}^{M}  \tilde{\q{E}}_{i}^{\mathrm{H}}  \hat{\q{h}}_{mk}  \q{D}_m^{\mathrm{H}} }{\left(\sum_{j=1}^{K+1} \vert \sum_{m=1}^{M} \hat{\q{h}}_{mk}^{\mathrm{H}} \tilde{\q{E}}_j \q{V} \q{D}_m \vert^2 + \sigma^2 \right)^2} \Bigg) \nonumber \\
&&\quad - 2 \rho \sum_{t=1}^{T} \sum_{m=1}^{M}  \mathbf{1}_{\left\{\lambda_{mt} + \frac{u_{mt}(\q{V})}{\rho}\right\}} \left(\frac{\lambda_{mt}}{\rho} + {u_{mt}(\q{V} )}\right) \nonumber \\
&&\quad \times \sum_{i=1}^{K+1}\tilde{\q{E}}_i^{\mathrm{H}} \hat{\q{a}}(\theta_{mt}) \left(\hat{\q{a}}^{\mathrm{H}}(\theta_{mt})\tilde{\q{E}}_i  \q{V}\q{D}_m \q{D}_m^{\mathrm{H}} \right)
\end{eqnarray}	

\noindent (b) \textbf{Search direction:} This obtains the search direction by choosing a descent direction in $T_{\q{V}_l}\mathcal{M}$, and is given by $\boldsymbol{\eta}_{l+1} = -{\mathrm{grad}}_{\q{V}_{l+1}} \mathcal{L}_\rho(\q{V}, \boldsymbol{\lambda}) + \beta_l \mathcal{T}_{\q{V}_l \rightarrow \q{V}_{l+1}}(\boldsymbol{\eta}_{l})$, where $ \boldsymbol{\eta}_{l} $ is the current search direction and $ \beta_l $ is computed using the Hestenes-Stiefel approach \cite{Shewchuk1994}, and $\mathcal{T}_{\q{V}_l \rightarrow \q{V}_{l+1}}(\boldsymbol{\eta}_{l})$ is mapping a vector from the tangent space at $\q{V}_l$ to the tangent space at $\q{V}_{l+1}$ \cite{zargari2024riemannian}. 

\noindent (c) \textbf{Retraction (Mapping):} This maps the updated point, initially in the tangent space, back onto the manifold $\mathcal{M}$, ensuring that the next iterate stays on the manifold after the update. The retraction operation is given as $R_{\q{V}_l} (\alpha_l\eta_l) = \text{unt} (\alpha_l\eta_l)$, where $\alpha_l$ is the step size. 

\noindent (d) \textbf{Updating the Lagrange multipliers:} This updates each Lagrange multiplier at iteration $l$ to meet the constraints. The updating rule is given as $
\lambda_{mt}^{l+1} = \text{clip}_{[\lambda^{\min}, \lambda^{\max}]}\left(\lambda_{mt}^{l} + \rho_l u_{mt}(\q{V}_{l+1})\right)$ where $\rho_l > 0$ denotes the penalty parameter and the clipping limits each $\lambda_{mt}^{l+1}$ to a predetermined range \cite{liu2020simple, zargari2024riemannian}. 

\subsubsection{Computational Complexity}
The computational complexity is dominated by the iterations of Algorithm \ref{alg_beamforming}, yielding an approximate complexity of $\mathcal{O}(I_t(LM(K+1) + LM(K+1)^3))$, where $I_t$ denotes the total iterations \cite{liu2020simple, zargari2024riemannian}.

\section{Simulation Example}
The large-scale fading path-loss, denoted as $\zeta_{\q{h}{mk}}$, is modeled using the 3GPP UMi model, with an operating frequency of $f_c = \qty{3}{\GHz}$ \cite[Table B.1.2.1]{3GPP2010}. The AWGN variance is given by $\sigma^2 = 10 \log{10}{(N_0 B N_f)}$,\qty{}{\dB m}, where $N_0=\qty{-174}{\dB m/\Hz}$, $B=\qty{10}{\MHz}$ represents the bandwidth, and $N_f=\qty{10}{\dB}$ is the noise figure. Unless otherwise stated, the simulation parameters are listed in Table \ref{table_simulation_para}. In the simulation, the APs are uniformly distributed, while users and targets are randomly positioned within a $\num{200} \times \qty{200}{\m^2}$ area.

\begin{table}[t]
\centering
\caption{Simulation and algorithm parameters.}\vspace{-0mm}
\label{table_simulation_para}
\begin{tabular}{|c |c |c |c |}    
\hline
\textbf{Parameter}& \textbf{Value} & \textbf{Parameter}& \textbf{Value}\\  \hline \hline
$L$ & \num{8}  & $\Gamma_{t}^{\mathrm{th}}$  & \qty{10}{\dB m}  \\ \hline
$K$ & \num{2} & $p_{\mathrm{max}}$  & \qty{30}{\dB m} \\ \hline
$T$ & \num{3} & $\upsilon_1,\upsilon_2$  & \num{e-6}  \\ \hline
$d_{\min}$ & \num{e-10} & $\epsilon_0$  & \num{e-3} \\\hline
$\epsilon_{\min}$ & \num{e-6} & $\tau$  & \num{0.5}  \\\hline
$\theta_\epsilon$ & \num{0.5} & $\theta_\rho$ & \num{0.25} \\\hline
$\rho_0$ & \num{1} & $\{\lambda^{\min},\lambda^{\max}\}$ & \{0,100\} \\ \hline
\end{tabular}\vspace{-0mm}
\end{table}

\begin{figure}[!t]\vspace{-0mm}
    \centering
    \includegraphics[width=0.7\textwidth]{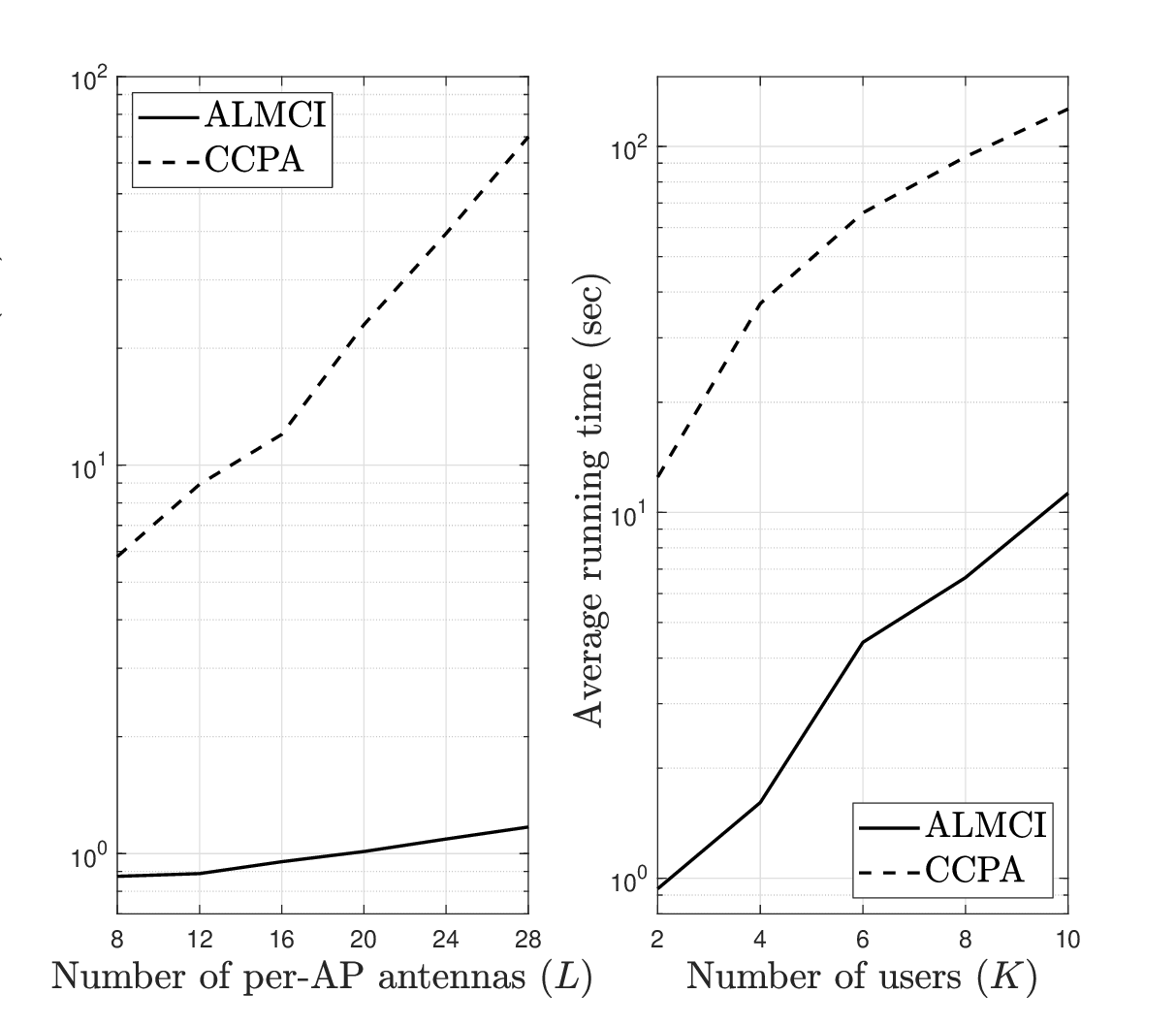}
    \vspace{-0mm}
    \caption{Average running time versus number of users and per-AP antennas for CCPA and ALMCI algorithms.}
    \label{fig_time_combine_resource} \vspace{-0mm}
\end{figure}

Figure~\ref{fig_time_combine_resource} shows the average running times for the ALMCI and CCPA algorithms, evaluating their performance with respect to varying numbers of per-AP antennas ($L$) and different numbers of users ($K$). The first graph demonstrates that the average running time for CCPA increases as the number of antennas grows. This trend reflects the algorithm's iterative structure and the computational complexity involved in the SCA/SDR process, which becomes more demanding as the antenna array size increases. In contrast, ALMCI shows a more gradual and moderate rise in running time, highlighting its superior computational efficiency and capability to handle larger antenna arrays. For instance, with \num{4} APs, each equipped with \num{16} antennas and a transmit power of \qty{30}{\dB m}, ALMCI completes the algorithm execution \num{12} times faster than CCPA.

The second graph illustrates the average running time as a function of the number of users ($K$). While all algorithms show an increasing trend, CCPA exhibits a sharp rise, indicating poor scalability as the number of users grows. This is due to the CCPA's need to solve larger optimization problems and perform more complex matrix operations. In contrast, ALMCI shows a slower rate of increase in running time, demonstrating that its MO approach scales more efficiently and manages additional users with better performance.

The reduced execution time and computational complexity of ALMCI arise from its search within a manifold that has $(L+1)KM$ dimensions, in contrast to the larger Euclidean space of $MLKT$ dimensions employed by conventional methods like CCPA. This dimensionality reduction streamlines gradient computations, step size adjustments, memory usage, and ensures numerical stability, all while eliminating the need for explicit constraint handling and iterative feasibility checks. By leveraging Riemannian gradients and manifold curvature, ALMCI achieves precise updates with large steps, facilitating rapid convergence \cite{liu2020simple, hu2020brief, boumal2023introduction, lee2006riemannian}. These attributes make ALMCI particularly suitable for real-time ISAC systems, especially in large and dynamic networks.

\begin{figure}[!t]\vspace{-0mm}
    \centering
    \includegraphics[width=0.7\textwidth]{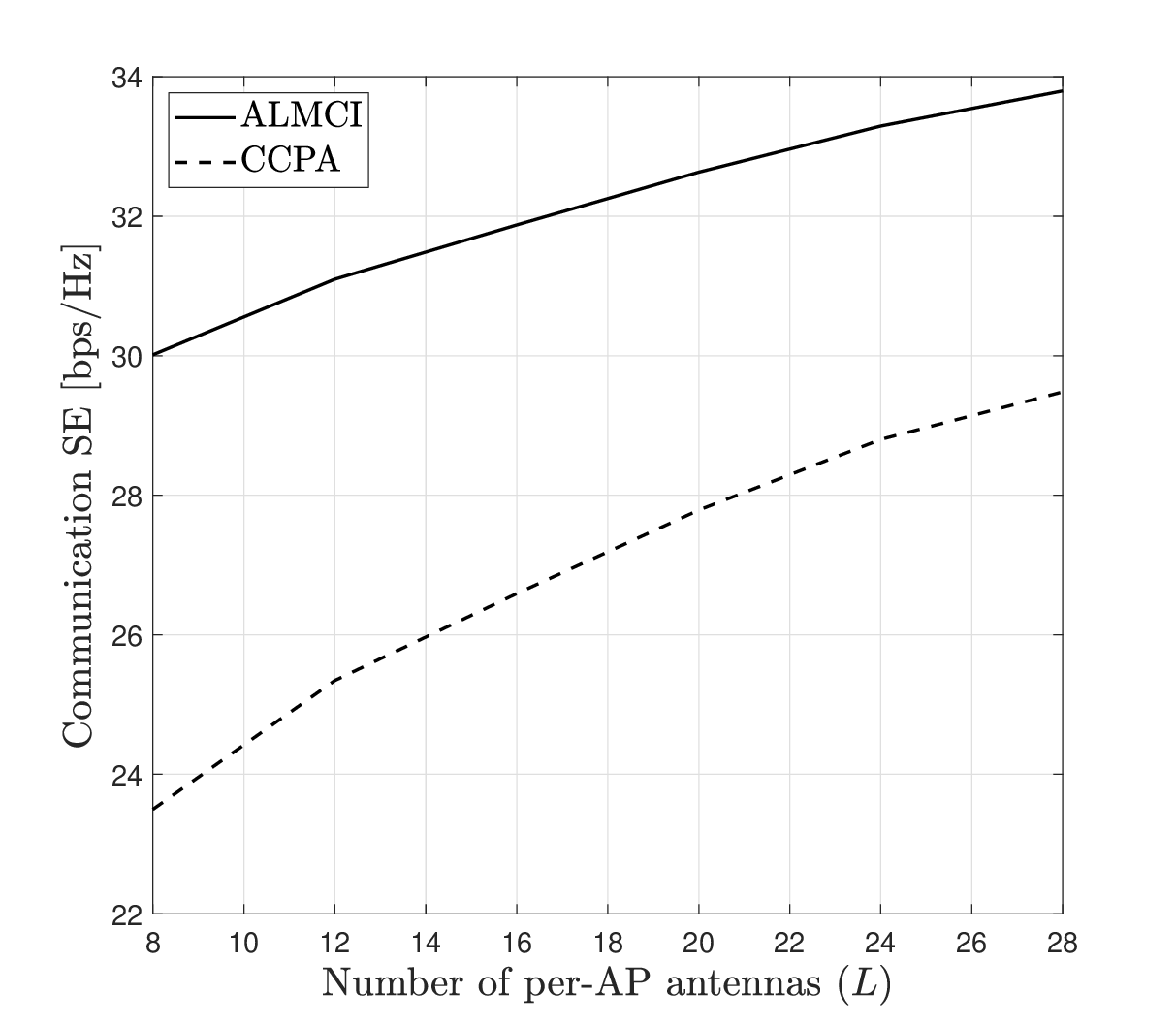}
    \vspace{-0mm}
    \caption{Communication SE compression between CCPA and ALMCI algorithms as a function of the number of per-AP antennas.}
    \label{fig_Sumrate_antennas_resource_comp} \vspace{-0mm}
\end{figure}

Figure \ref{fig_Sumrate_antennas_resource_comp} compares the communication SE as a function of the number of per-AP antennas ($L$) with $p_{\mathrm{max}} = \qty{30}{\dB m}$. While all algorithms benefit from spatial multiplexing gains, ALMCI consistently outperforms the others, exhibiting significant improvements as $L$ increases. Specifically, at $L = \num{12}$, ALMCI achieves a \qty{22.7}{\percent} higher communication SE than CCPA. This performance gap highlights the inferior performance of CCPA, which is attributed to its reliance on approximations and rank relaxations.

\begin{figure}[!t]\vspace{-0mm}
    \centering
    \includegraphics[width=0.7\textwidth]{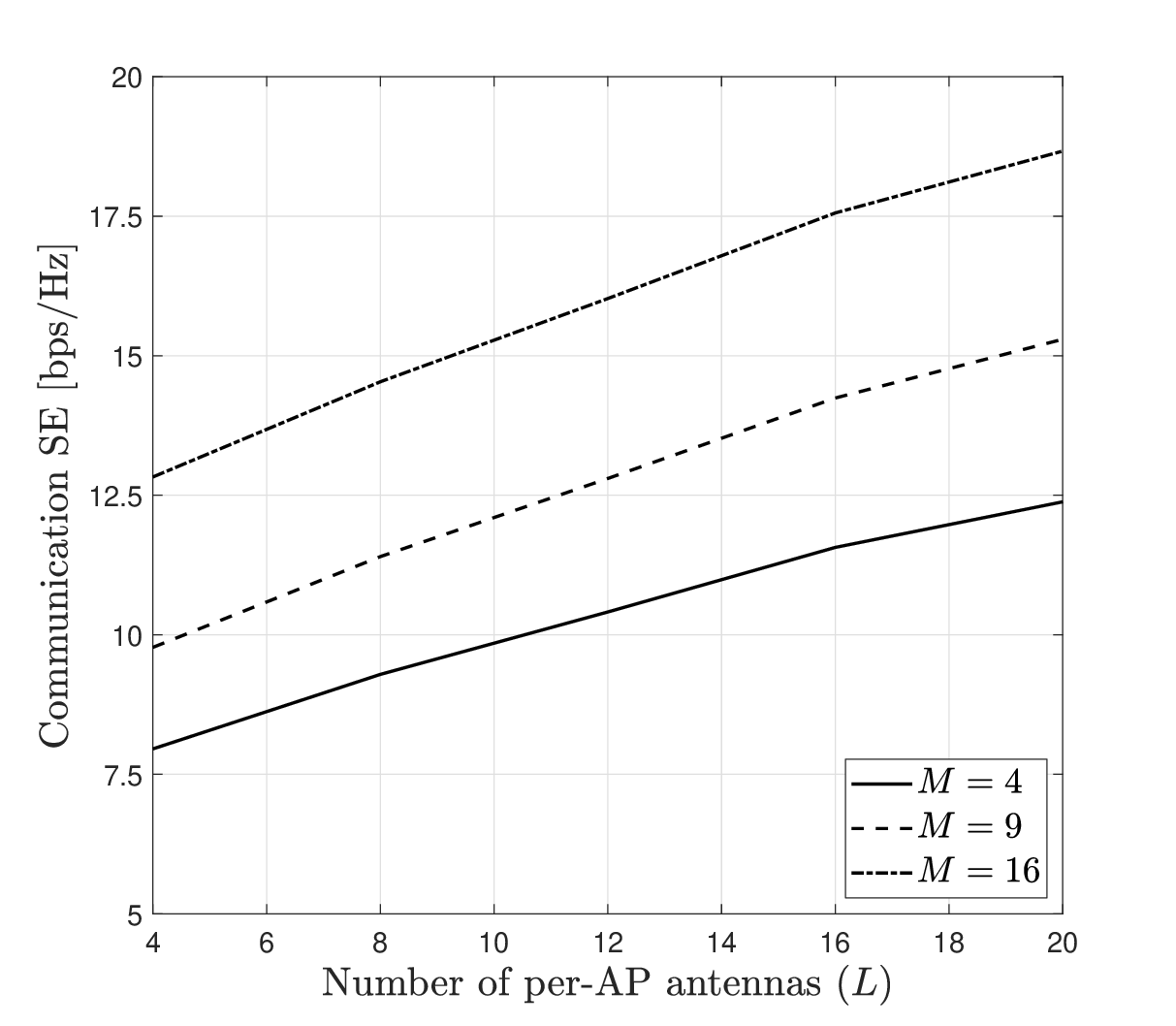}
    \vspace{-0mm}
    \caption{Communication SE versus the number of per-AP antennas for $K=2$ and $T=3$.}
    \label{fig_Sumrate_antennas_resource} \vspace{-0mm}
\end{figure}

Figure~\ref{fig_Sumrate_antennas_resource} shows the communication SE as a function of the number of AP antennas ($L$) for different numbers of APs, $M = \{4, 9, 16\}$, with $K = 2$ and $T = 3$. The plot demonstrates that, for all values of $M$, the communication SE increases as the number of AP antennas grows. Moreover, for a fixed $L$, a higher value of $M$ leads to a greater communication SE. For instance, when $M = \num{16}$ and $L = \num{12}$, the communication SE is \qty{53.9}{\percent} and \qty{25.2}{\percent} higher compared to the $M = \num{4}$ and $M = \num{9}$ cases, respectively.

This performance improvement emphasizes the scalability and versatility of CF-ISAC systems. Increasing the number of APs enhances spatial diversity, which in turn enables more efficient interference management and better resource allocation across multiple users. Moreover, adding more antennas at each AP boosts the beamforming gain, facilitating finer spatial multiplexing and improved signal quality. The distributed architecture of CF-ISAC ensures that users benefit from the collective cooperation of multiple APs, helping to mitigate the effects of path loss and shadowing.

\begin{figure}[!t]\vspace{-0mm}
    \centering
    \makebox[\textwidth][c]{
    \includegraphics[width=1.1\textwidth]{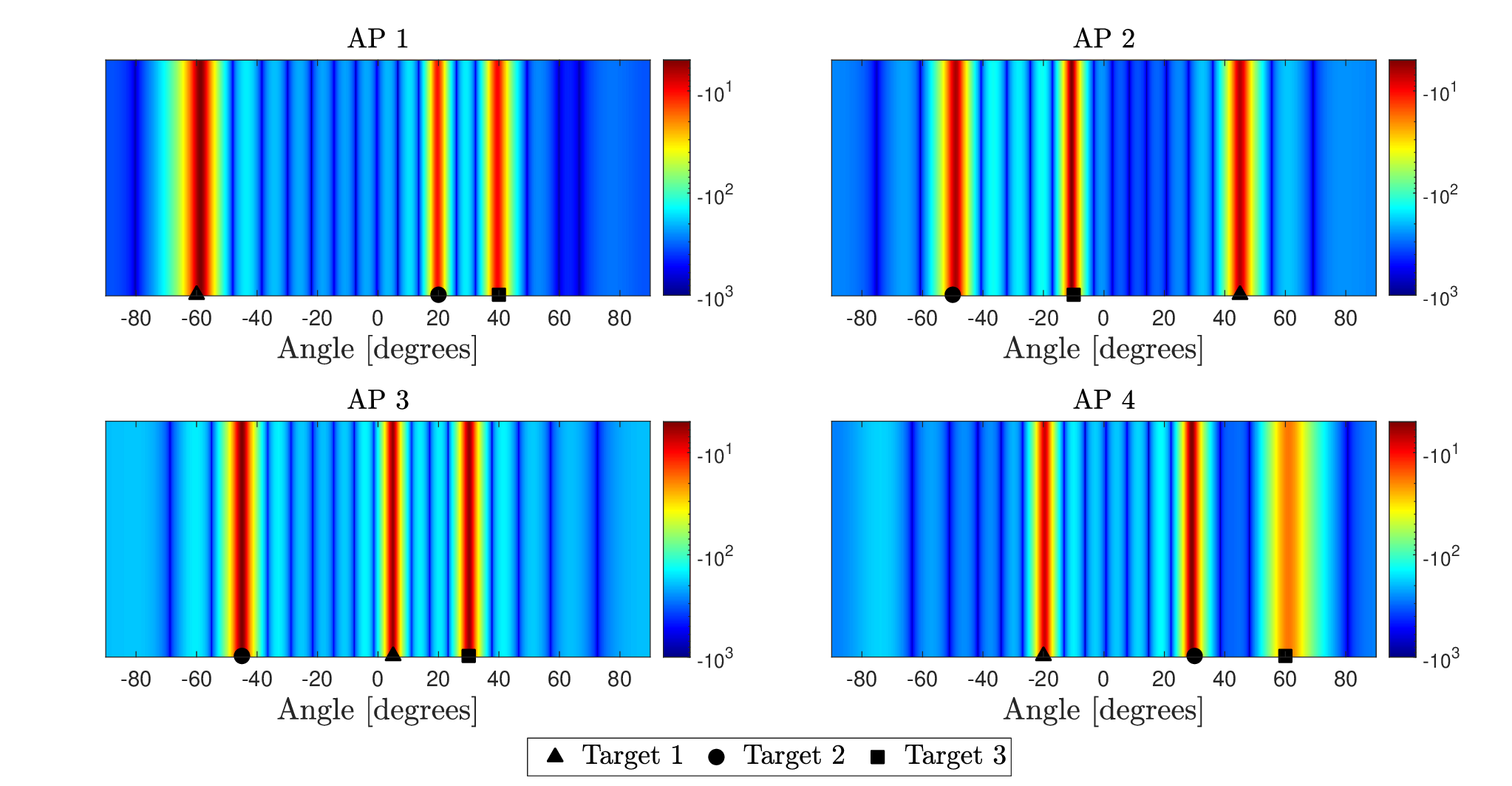}}
    \vspace{-0mm}
    \caption{Beampattern gain profiles over a \qty{\pm 90}{\degree} angular spread at different APs, illustrating the gain variations and directivity in a color-coded scale for $L=8$, $M=4$, $K=2$, and $T=3$.}
    \label{fig_HeatMap_resource} \vspace{-0mm}
\end{figure}

Figure~\ref{fig_HeatMap_resource} presents the effects of beamforming gains utilizing $L = \num{8}$ AP antennas with $M=\num{4}$ APs for $K=2$ and $T=3$. In particular, Figure~{\ref{fig_HeatMap_resource}} plots directional gain profiles in a color-coded scale to evaluate the performance of beamforming gains at each APs. The direction angles for sensing targets from AP 1, AP 2, AP 3, and AP 4, are set to $\{-60, 20, 40\}\qty{}{\degree}$, $\{25, -70, -10\}\qty{}{\degree}$, $\{25, -45, 75\}\qty{}{\degree}$, and $\{-20, 30, 60\}\qty{}{\degree}$, respectively, ensuring coverage of a broad angular range.

\begin{figure}[!t]\vspace{-0mm}
    \centering
    \includegraphics[width=0.7\textwidth]{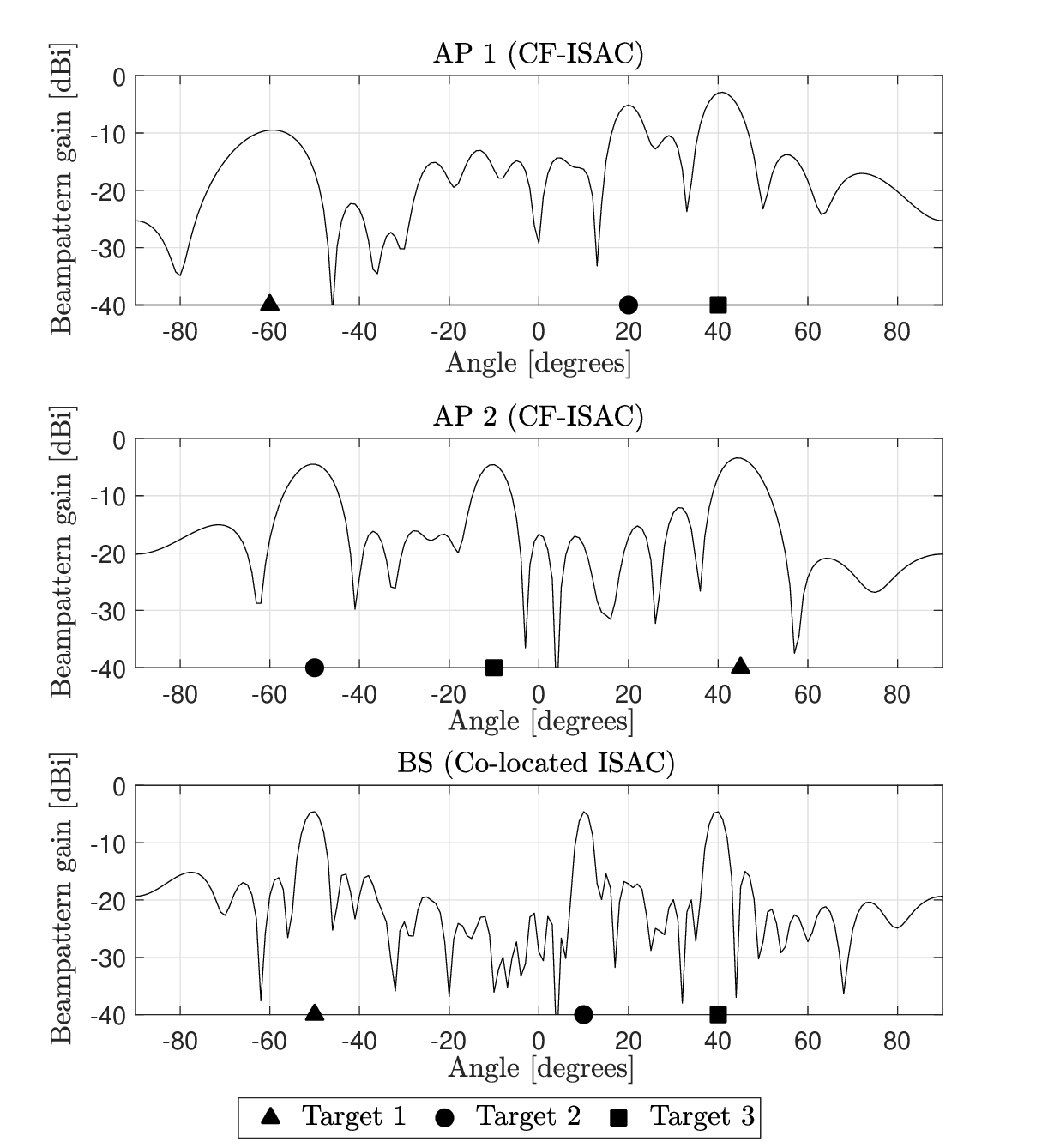}
    \vspace{-0mm}
    \caption{A comparison of directional beampattern gain profiles over a \qty{\pm 90}{\degree} angular spread between CF-ISAC and co-located ISAC systems.}
    \label{fig_BeamGani_Comp} \vspace{-0mm}
\end{figure}

The color gradient in the figure represents the magnitude of the beamforming gain, where red indicates strong gains due to focused energy, and blue represents areas with minimal radiated power. The intersection of beampattern gain directions across multiple APs enables precise target localization, a significant advantage of CF-ISAC's multi-static sensing, as compared to mono-static or bi-static methods used in co-located ISAC systems (Figure~\ref{fig_BeamGani_Comp}). The distributed nature of the APs offers various viewpoints, improving spatial resolution and reducing multi-path fading. The overlap of beampattern gains from different APs enhances triangulation accuracy using AoA information from multiple perspectives, resulting in better localization precision and robustness against blockages and interference. Furthermore, if one AP experiences weak signal quality, the others can compensate, ensuring consistent sensing performance across the entire coverage area.

\begin{figure}[!t]\vspace{-0mm}
    \centering
    \includegraphics[width=0.7\textwidth]{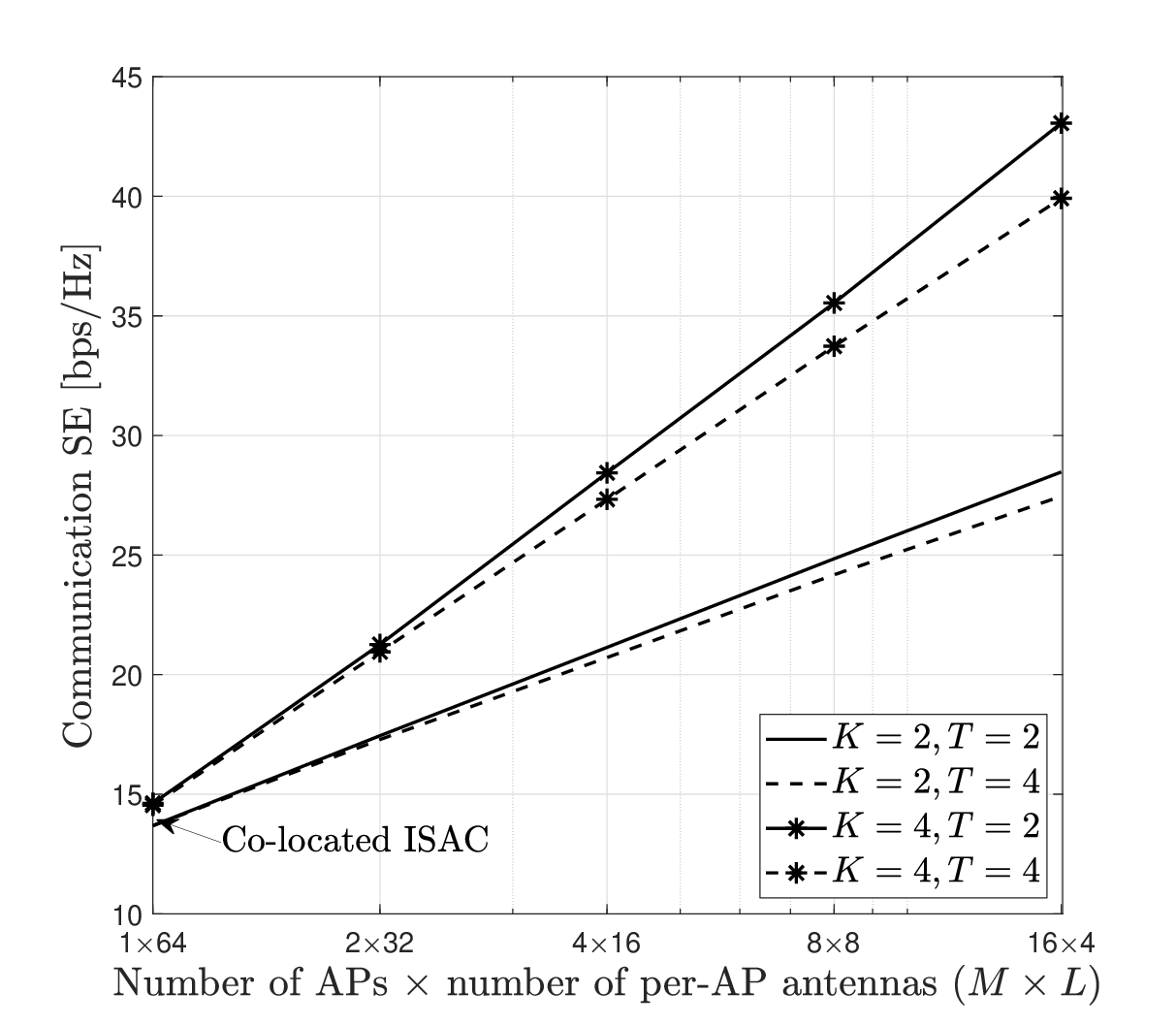}
    \vspace{-0mm}
    \caption{Communication SE comparison between CF-ISAC and co-located ISAC systems for $M\times L =\num{64}$.}
    \label{fig_Comm_SE_Comp} \vspace{-0mm}
\end{figure}

Figure~\ref{fig_Comm_SE_Comp} compares the communication SE between CF-ISAC and co-located ISAC systems. Specifically, the product of the number of APs and per-AP antennas, $M \times L$, is kept constant at \num{64}. In the co-located ISAC system, the configuration is $M=\num{1}$ and $L=\num{64}$, whereas CF-ISAC configurations have $M > \num{1}$. The figure demonstrates that CF-ISAC systems consistently outperform their co-located counterparts. This performance improvement is primarily due to the micro-diversity gains realized by spatially distributed APs, which reduce path loss and shadowing effects. The distributed nature of CF-ISAC ensures more uniform coverage and better signal quality, providing a significant advantage over co-located ISAC systems.

Additionally, Figure~\ref{fig_Comm_SE_Comp} highlights the trade-off between communication and sensing performance. As the number of targets increases for a fixed number of users, the communication SE decreases. This occurs because more system resources, such as transmit power for target detection and tracking, are allocated to sensing tasks, leaving fewer resources available for communication. This emphasizes the need to balance communication and sensing tasks in ISAC systems to optimize overall performance. The CF-ISAC architecture, with its distributed APs, provides flexibility in dynamically managing this trade-off by leveraging spatial diversity and adaptive resource allocation \cite{zargari2024CFISAC}.

Figure~\ref{fig_BeamGani_Comp} compares beampattern gains between CF-ISAC and co-located ISAC systems for $T=3$. The former uses two 16-antenna  APs (AP 1 and AP 2) to perform communication and sensing. In contrast, a 32-antenna, single BS at the center of the coverage area serves the latter. The direction angles for sensing targets from AP1 and AP2 are set to $\{-60, 20, 40\}\qty{}{\degree}$ and $\{45, -50, -10\}\qty{}{\degree}$.  In contrast, the co-located ISAC system targets at directions $\{-50, 10, 40\}\qty{}{\degree}$ from the BS.

As shown in Figure~\ref{fig_BeamGani_Comp}, both CF and co-located configurations direct their main lobes toward targets, indicating their locations. However, CF-ISAC offers a significant advantage over co-located ISAC by utilizing distributed APs. By combining the beampattern gains from multiple APs, CF-ISAC enables precise target localization in both angular and distance dimensions. In contrast, co-located ISAC, relying on a single viewpoint (i.e., BS), can only estimate the angular directions. The spatial diversity provided by CF-ISAC helps resolve directional ambiguities and improves position estimation, offering superior accuracy.

Additionally, CF-ISAC enables more flexible and adaptive beamforming through AP cooperation, which enhances localization accuracy and provides greater robustness against blockages and interference. In contrast, the single BS in a co-located ISAC system may face challenges in accurately estimating distances. The distributed architecture of CF-ISAC is especially advantageous in dynamic environments, as it ensures more precise and resilient target tracking, even in the presence of obstacles or interference. 
	\chapter{Security Challenges of  Cell-Free ISAC}\label{chp_CF_isac_security}

This chapter builds upon the CF-ISAC beamforming framework presented in Figure~\ref{fig_SystemModelResourceAllocation} of Chapter~\ref{chp_CF_isac_resource}, with a focus on enhancing its security capabilities. Specifically, it explores how the distributed nature of CF-ISAC systems can be harnessed to design secure beamforming strategies that safeguard both communication and sensing functions against potential threats.

\section{Eavesdropper Model}
In Figure~\ref{fig_SystemModelResourceAllocation}, the targets are modeled as untrusted entities or devices. Specifically, one or more targets serve a dual role: they are objects of interest for sensing and potential eavesdroppers aiming to intercept confidential communication intended for legitimate users \cite{Bazzi2024, He2024, Sun2024}. It is assumed that a malicious target attempts to decode any user's data from its received signal. If successful, it may further exploit the successive interference cancellation (SIC) technique to extract the information of all remaining users.

The received signal at the $t$-th target is given as
\begin{eqnarray} \label{eqn_Tt_rx_signal}
    y_t &&= \sum_{m=1}^{M} \q{a}^{\mathrm{H}}(\theta_{mt}) \q{x}_m + z_t \nonumber \\
    &&=\sum_{i=1}^{K} \sum_{m=1}^{M} \q{a}^{\mathrm{H}}(\theta_{mt}) \q{w}_{mi} q_i  + \sum_{m=1}^{M} \q{h}_{mk}^{\mathrm{H}} \q{s}_{m}   + z_t,
\end{eqnarray} \par \vspace{-0mm}
\noindent where $z_t \sim \mathcal{CN}(0, \sigma^2)$ denotes the AWGN at the $t$-th target. Consequently, the leakage SE at the $t$-th target for decoding the $k$-th user's data can be expressed as
\begin{eqnarray}
    \mathcal{S}_{t,k}^{\mathrm{Leak}} \approx \log_2(1+ \mt{SINR}_{t,k}^{\mathrm{Leak}}),  
\end{eqnarray} \par \vspace{-0mm}
\noindent where $\mt{SINR}_{t,k}^{\mathrm{Leak}}$ is the SINR at the $t$-th target for decoding the $k$-th user's data and computed using \eqref{eqn_Tt_rx_signal} as
\begin{eqnarray}\label{eqn_leakage_SINR}
\mt{SINR}_{t,k}^{\mathrm{Leak}} = \frac{\big\vert \sum_{m =1}^{M} \q{a}^{\mathrm{H}}(\theta_{mt})  \q{w}_{mk} \big\vert^2}{\sum\limits_{i\neq k}^{K} \left\vert \sum\limits_{m=1}^{M} \q{a}^{\mathrm{H}}(\theta_{mt}) \q{w}_{mi} \right\vert^2 + \left\vert \sum\limits_{m=1}^{M} \q{a}^{\mathrm{H}}(\theta_{mt}) \q{s}_{m} \right\vert^2 + \sigma^2}.
\end{eqnarray}

\section{Problem Formulation}
Our objective is to maximize the sum communication SE for legitimate users, while simultaneously satisfying sensing beampattern gain requirements and minimizing information leakage to targets acting as eavesdroppers. This ensures robust physical-layer security by preventing unauthorized targets from successfully decoding any user's data. The resulting optimization problem is formulated to strike a balance between communication performance, sensing accuracy, and secrecy constraints, and is expressed as follows: 
\begin{subequations}\label{prob_Q1}
\begin{eqnarray}
(\mathcal{Q}1):~&& \max_{\q{w}_{mk}, \q{s}_m} \sum_{k=1}^{K} \log_2\left(1 + \mt{SINR}_k^{\mathrm{Com}} \right), \label{prob_Q1_obj}  \\
\text{s.t.} \quad &&   p(\theta_{mt}) \geq \Gamma_{t}^{\mathrm{th}}, ~\forall m, t, \label{prob_Q1_beamgain}\\
&& \mathcal{S}_{t,k}^{\mathrm{Leak}} \leq \delta_{\mathrm{max}}, ~\forall t, k, \label{prob_Q1_leak_rate}\\
&& \sum_{i=1}^{K} \Vert \q{w}_{mi}\Vert^2 + \Vert \q{s}_{m}\Vert^2 \leq p_{\mathrm{max}}, ~\forall m, \label{prob_Q1_tx_pow}
\end{eqnarray}
\end{subequations} \par \vspace{-0mm}
\noindent where $\mt{SINR}_k^{\mathrm{Com}}$ is the communication SINR at the $k$-th user, given in \eqref{eqn_gamma},  \eqref{prob_Q1_beamgain} is the sensing beampattern gain requirement for each target, with a required sensing threshold of $\Gamma_{t}^{\mathrm{th}}$, and $p(\theta_{mt})$ is the transmit beampattern gain at the $m$-th AP toward target the $t$-th target and is given in \eqref{eqn_beamgain_mt}. Moreover, \eqref{prob_Q1_leak_rate} defines the maximum allowable leakage SE at all targets/eavesdroppers, with a maximum leakage SE of $\delta_{\mathrm{max}}$, while \eqref{prob_Q1_tx_pow} sets the maximum allowable transmit power at the $m$-th AP.

\section{Proposed Solution}
Problem $(\mathcal{Q}1)$ is non-convex due to both the non-convex objective function and constraints. To solve this, we employ the SDR technique. First, we define the matrices $\q{W}_k \triangleq \q{w}_k \q{w}_k^{\mathrm{H}}$ and $\q{S} \triangleq \q{s} \q{s}^{\mathrm{H}}$, where $\q{w}_k = [\q{w}_{1k}^{\mathrm{T}}, \dots, \q{w}_{Mk}^{\mathrm{T}}]^{\mathrm{T}} \in \mathbb{C}^{LM \times 1}$ and $\q{s} = [\q{s}_{1}^{\mathrm{T}}, \dots, \q{s}_{M}^{\mathrm{T}}]^{\mathrm{T}} \in \mathbb{C}^{LM \times 1}$. Here, $\q{W}_k$ is a semi-definite matrix with a rank one constraint, i.e., ${\mathrm{Rank}}(\q{W}_k) = 1$, and $\q{S}$ is also a semi-definite matrix. Next, we define the cascaded channel vectors $\q{f}_k = [\q{h}_{1k}^{\mathrm{T}}, \dots, \q{h}_{Mk}^{\mathrm{T}}]^{\mathrm{T}} \in \mathbb{C}^{LM \times 1}$ and $\q{g}_t = [\q{a}^{\mathrm{T}}(\theta_{1t}), \dots, \q{a}^{\mathrm{T}}(\theta_{Mt})]^{\mathrm{T}} \in \mathbb{C}^{LM \times 1}$, and the selection matrix $\q{Q}_m \in \mathbb{R}^{LM \times LM}$, which is a block diagonal matrix consisting of $M$ blocks, each corresponding to an AP, i.e.,  $\q{Q}_m = \text{diag}(\q{0}, \dots, \q{I}_{L},  \dots, \q{0})$ with $\q{I}_{L}$ in the $m$-th block on the diagonal. This matrix selects the transmit antennas of the $m$-th AP.  

By utilizing SDR techniques to relax the highly non-convex rank one constraints, the resultant problem can be formulated as follows:
\begin{subequations}\label{prob_Q2}
\begin{eqnarray}
(\mathcal{Q}2):~&& \max_{\q{W}_{k}, \q{S}} \sum_{k=1}^{K} f(\q{W}_k, \q{S}), \label{prob_Q2_obj}  \\
\text{s.t.} \quad && \tr\left(\sum_{i=1}^{K} \q{Q}_m \q{g}_t \q{g}_t^{\mathrm{H}} \q{Q}_{m} \q{W}_{i} \right) + \tr\left( \q{Q}_m \q{g}_t \q{g}_t^{\mathrm{H}} \q{Q}_{m} \q{S} \right) \nonumber\\
&& - \Gamma_{t}^{\mathrm{th}} \geq 0, ~\forall m, t, \\
&& \delta_{\mathrm{max}}' \left( \sum_{i\neq k}^{K} \tr\left( \q{g}_t \q{g}_t^{\mathrm{H}} \q{W}_i \right) + \tr\left( \q{g}_t \q{g}_t^{\mathrm{H}} \q{S} \right) + \sigma^2 \right) \nonumber \\
&& - \tr\left( \q{g}_t \q{g}_t^{\mathrm{H}} \q{W}_k \right) \geq 0, ~\forall t, k, \label{prob_Q2_leak_rate}\\
&& \sum_{i=1}^{K} \tr\left( \q{W}_{i} \q{Q}_m \right) + \tr\left( \q{S} \q{Q}_m \right) \leq p_{\mathrm{max}}, ~\forall m, \label{prob_Q2_tx_pow} \\
&& \q{W}_k,  \q{S} \succeq 0,~\forall k,
\end{eqnarray}
\end{subequations} \par \vspace{-0mm}
\noindent where $\delta_{\mathrm{max}}' = 2^{\delta_{\mathrm{max}}} -1$. In \eqref{prob_Q2}, as the objective is not a convex function, we use the SCA method to linearize it which is given by 
\begin{eqnarray}\label{eqn_Q2_obj_def}
    f(\q{W}_k, \q{S}) &=& \log_2 \left( \sum_{i=1}^{K} \tr\left( \q{f}_k \q{f}_k^{\mathrm{H}} \q{W}_i \right) + \tr\left( \q{f}_k \q{f}_k^{\mathrm{H}} \q{S} \right) + \sigma^2 \right)  \nonumber\\
    &&- \log_2 \left( \sum_{i\neq k}^{K} \tr\left( \q{f}_k \q{f}_k^{\mathrm{H}} \q{W}_i^{(l)} \right) + \tr\left( \q{f}_k \q{f}_k^{\mathrm{H}} \q{S}^{(l)} \right) + \sigma^2 \right) \nonumber \\
    && - \frac{\sum_{i\neq k}^{K} \tr\left( \q{f}_k \q{f}_k^{\mathrm{H}} (\q{W}_i - \q{W}_i^{(l)}) \right)  }{\ln(2) \left( \sum_{i\neq k}^{K} \tr\left( \q{f}_k \q{f}_k^{\mathrm{H}} \q{W}_i^{(l)} \right) + \tr\left( \q{f}_k \q{f}_k^{\mathrm{H}} \q{S}^{(l)} \right) + \sigma^2 \right)} \nonumber\\
    &&- \frac{\tr\left( \q{f}_k \q{f}_k^{\mathrm{H}} (\q{S} - \q{S}^{(l)}) \right)  }{\ln(2) \left( \sum_{i\neq k}^{K} \tr\left( \q{f}_k \q{f}_k^{\mathrm{H}} \q{W}_i^{(l)} \right) + \tr\left( \q{f}_k \q{f}_k^{\mathrm{H}} \q{S}^{(l)} \right) + \sigma^2 \right)}, \qquad
\end{eqnarray}\par \vspace{-0mm}
\noindent where $(\cdot)^{(l)}$ denotes the previous iteration values of respective variables.

Moreover, as SE is a monotonically increasing function of its argument SINR, we replace $\mathcal{S}_{t,k}^{\mathrm{Leak}}$ with $\mt{SINR}_{t,k}^{\mathrm{Leak}}$ to obtain \eqref{prob_Q2_leak_rate}. The relaxed problem $(\mathcal{Q}2)$ is a standard semidefinite programming (SDP) problem, which can be solved using the CVX MATLAB tool \cite{boyd2004convex}. 

If the SDR solutions meet the conditions ${\mathrm{Rank}}(\q{W}_k) = 1$ the optimal transmit communication beamforming can be derived through eigenvalue decomposition (EVD) \cite{Qingqing2019}. Let the eigenvalue decomposition of $\q{W}_k$ be represented as $\q{W}_k = \q{U}_k \boldsymbol{\Sigma}_k \q{U}_k^H$, where $\q{U}$ is a unitary matrix and $\boldsymbol{\Sigma}_k = \text{diag}(\lambda_{k,1}, \dots, \lambda_{k, LM})$ is a diagonal matrix, both of size $LM \times LM$. If $\q{W}_k$ is rank one, the optimal transmit beamforming $\q{w}_k^*$ corresponds to the eigenvector associated with the maximum eigenvalue. Otherwise, we employ GR to obtain a near-optimal solution for $(\mathcal{Q}2)$ \cite{Qingqing2019}. However, applying GR to satisfy the rank-one constraint may result in a slight degradation in the SE of $(\mathcal{Q}2)$. To solve $(\mathcal{Q}2)$ \eqref{prob_Q2}, an algorithm similar to Algorithm~\ref{alg_ccpa_beamforming} can be employed.


\section{Simulation Example}

\begin{figure}[!t]\vspace{-0mm}
    \centering
    \includegraphics[width=0.7\textwidth]{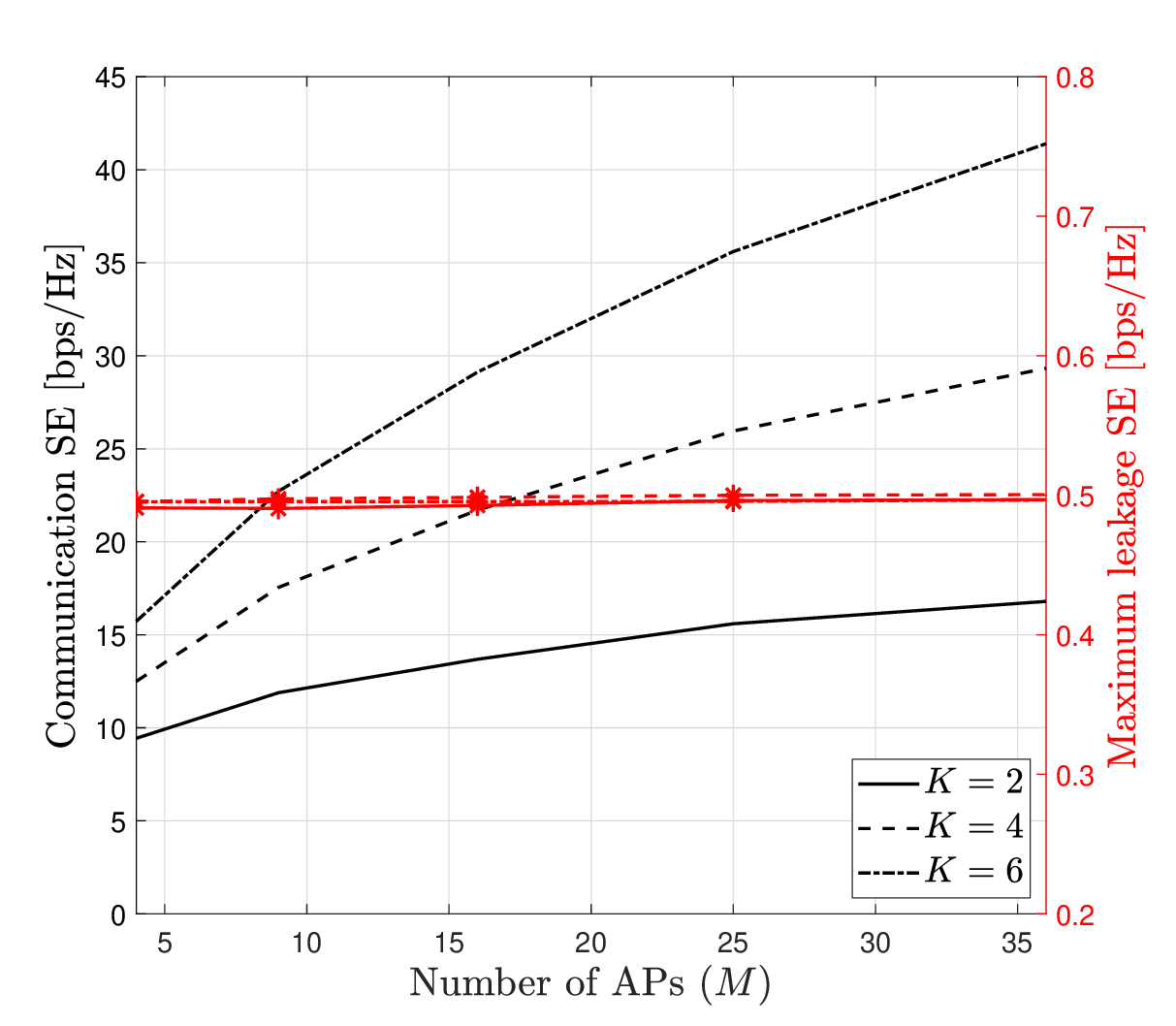}
    \vspace{-0mm}
    \caption{Communication SE and maximum leakage SE as functions of the number of APs.}
    \label{fig_ComRate_LeakRate_M_security} \vspace{-0mm}
\end{figure}

Figure~\ref{fig_ComRate_LeakRate_M_security} illustrates how the number of APs ($M$) influences both the communication SE (plotted on the left $y$-axis) and the maximum leakage SE at the targets or eavesdroppers (plotted on the right $y$-axis), for various user counts $K = \{2, 4, 6\}$. As $M$ increases, the communication SE improves markedly, driven by greater spatial diversity, enhanced beamforming gains, reduced path loss and shadowing, and more effective interference mitigation. The CF-ISAC system demonstrates strong scalability with increasing user numbers, capitalizing on spatial multiplexing and the distributed nature of the APs. For example, with $M = \num{16}$, the system configured for $K = \num{6}$ achieves communication SE improvements of \qty{112.8}{\percent} and \qty{34.2}{\percent} compared to configurations with $K = \num{2}$ and $K = \num{4}$, respectively. These results highlight CF-ISAC's ability to efficiently support higher user densities without compromising performance.

The secrecy performance, quantified by the maximum leakage SE at potential eavesdroppers, remains consistent across varying AP densities, demonstrating the CF-ISAC system's ability to suppress information leakage effectively. This stability underscores the role of secure beamforming in preserving data confidentiality, even as the network becomes denser. Although adding more APs significantly improves communication SE through spatial diversity and enhanced beamforming, maintaining strong secrecy performance necessitates careful beamforming optimization to mitigate leakage. These results affirm CF-ISAC's potential for secure and scalable wireless deployments, particularly in mission-critical applications such as the IoT, defense systems, and surveillance networks.

\begin{figure*}[!t]\vspace{-0mm}
    \centering
    \makebox[\textwidth][c]{
    \includegraphics[width=1.1\textwidth]{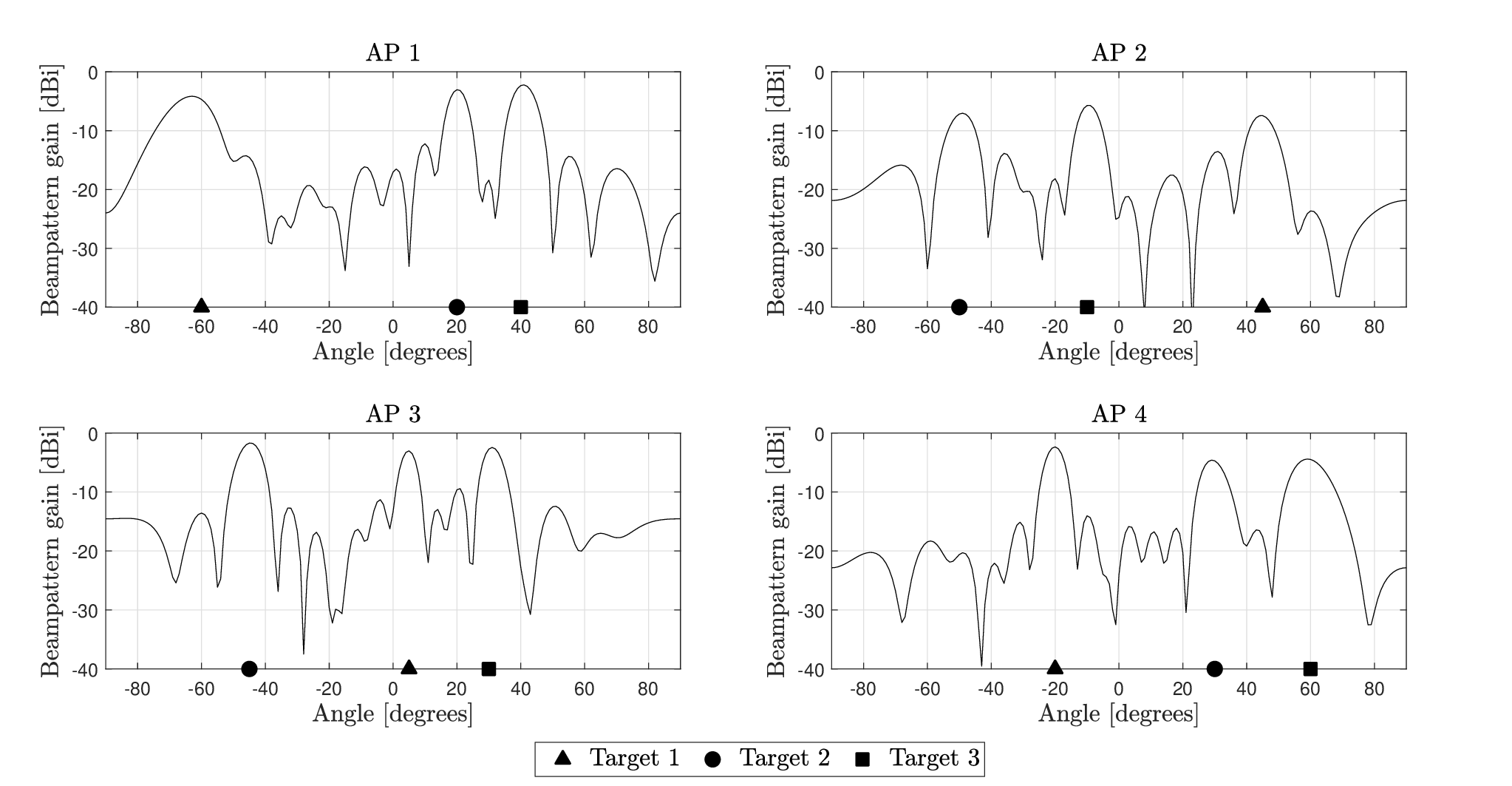}}
    \vspace{-0mm}
    \caption{Directional beampattern gain profiles over a \qty{\pm 90}{\degree} angular spread at different APs.}
    \label{fig_BeamGain_Secure} \vspace{-0mm}
\end{figure*}

Figure~\ref{fig_BeamGain_Secure} illustrates the directional beampattern gain profiles of the secure CF-ISAC system, configured with $L=\num{8}$ antennas per AP, $M=\num{4}$ APs, $K=\num{2}$ users, and $T=\num{3}$ targets. Each subfigure corresponds to the beampattern gain at a specific AP (AP 1 through AP 4), with target directions consistent with those defined in Figure~\ref{fig_HeatMap_resource}. The figure demonstrates how secure beamforming effectively steers power toward designated target directions at each AP. This is achieved by jointly optimizing communication and sensing beamforming strategies to maximize directional gain, mitigate mutual interference between users and targets, and suppress information leakage to eavesdroppers. The result is a highly focused and secure transmission pattern that enhances both sensing accuracy and communication confidentiality.

The sharp and focused main lobes demonstrate the system's ability to achieve precise beam steering, which is essential for accurate sensing and dependable communication. At the same time, the deep nulls and strong sidelobe suppression reflect the system's effectiveness in minimizing unintended signal leakage, thereby lowering the risk of eavesdropping. The variation in beampatterns among APs highlights the distributed and collaborative architecture of CF-ISAC, where each AP adaptively tunes its beamforming to balance performance and security. These findings validate the system's capability to simultaneously ensure high sensing precision and robust communication secrecy, making it highly suitable for mission-critical applications that demand both reliable target detection and stringent data protection. 
	\chapter{Network-Assisted Cell-Free ISAC}\label{chp_NA_CF_isac}

This chapter investigates a network-assisted full-duplex (NAFD) CF ISAC by presenting an AP mode selection and beamforming framework.

\section{System and Channel Models}
\begin{figure}[!t]\vspace{0mm}
    \centering 
    \def\svgwidth{250pt} 
    \fontsize{8}{8}\selectfont 
    \graphicspath{{Figures/}}
\begingroup%
  \makeatletter%
  \providecommand\color[2][]{%
    \errmessage{(Inkscape) Color is used for the text in Inkscape, but the package 'color.sty' is not loaded}%
    \renewcommand\color[2][]{}%
  }%
  \providecommand\transparent[1]{%
    \errmessage{(Inkscape) Transparency is used (non-zero) for the text in Inkscape, but the package 'transparent.sty' is not loaded}%
    \renewcommand\transparent[1]{}%
  }%
  \providecommand\rotatebox[2]{#2}%
  \newcommand*\fsize{\dimexpr\f@size pt\relax}%
  \newcommand*\lineheight[1]{\fontsize{\fsize}{#1\fsize}\selectfont}%
  \ifx\svgwidth\undefined%
    \setlength{\unitlength}{647.06634521bp}%
    \ifx\svgscale\undefined%
      \relax%
    \else%
      \setlength{\unitlength}{\unitlength * \real{\svgscale}}%
    \fi%
  \else%
    \setlength{\unitlength}{\svgwidth}%
  \fi%
  \global\let\svgwidth\undefined%
  \global\let\svgscale\undefined%
  \makeatother%
  \begin{picture}(1,0.57198316)%
    \lineheight{1}%
    \setlength\tabcolsep{0pt}%
    \put(0,0){\includegraphics[width=\unitlength]{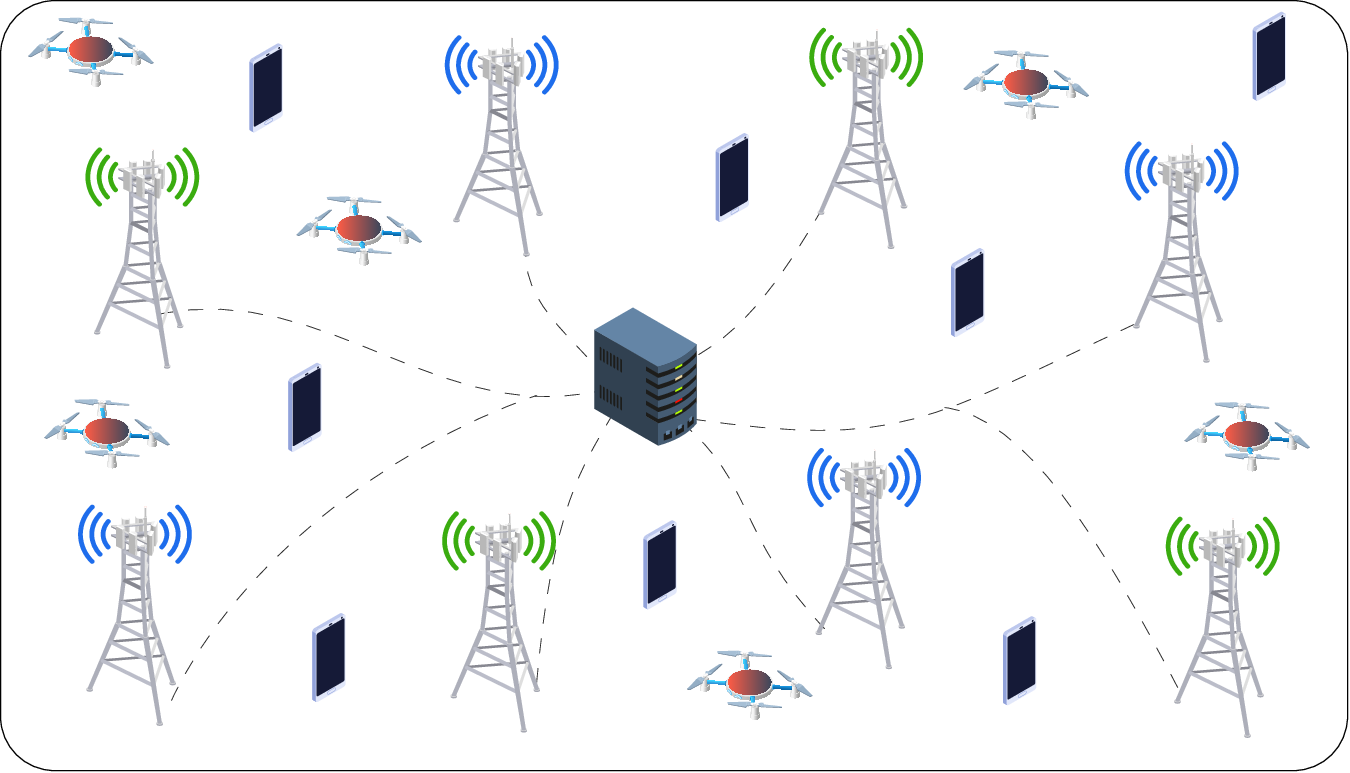}}%
    \put(0.05229232,0.00671221){\color[rgb]{0,0,0}\makebox(0,0)[lt]{\lineheight{1.25}\smash{\begin{tabular}[t]{l}AP (DL)\end{tabular}}}}%
    \put(0.44267811,0.35168919){\color[rgb]{0,0,0}\makebox(0,0)[lt]{\lineheight{1.25}\smash{\begin{tabular}[t]{l}CPU\end{tabular}}}}%
    \put(0.32639815,0.01130732){\color[rgb]{0,0,0}\makebox(0,0)[lt]{\lineheight{1.25}\smash{\begin{tabular}[t]{l}AP (UL)\end{tabular}}}}%
    \put(0.72479537,0.02229163){\color[rgb]{0,0,0}\makebox(0,0)[lt]{\lineheight{1.25}\smash{\begin{tabular}[t]{l}User\end{tabular}}}}%
    \put(0.51500412,0.01320592){\color[rgb]{0,0,0}\makebox(0,0)[lt]{\lineheight{1.25}\smash{\begin{tabular}[t]{l}Target\end{tabular}}}}%
  \end{picture}%
\endgroup%
 \vspace{-0mm} 
    \caption{A NAFD CF ISAC system setup with multiple users and targets.}  \label{fig_SystemModelNAFD}\vspace{-0mm}
\end{figure}
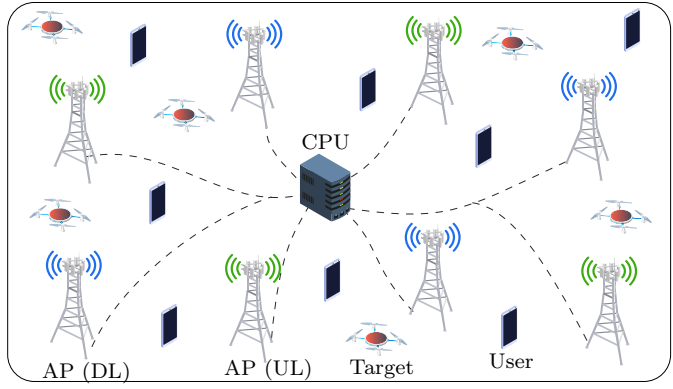

We consider an NAFD CFMM system consisting of $M$ APs, $K$ single-antenna users, and $T$ potential targets (Figure~\ref{fig_SystemModelNAFD}). Each AP is equipped with $L$ ULA antennas spaced at half-wavelengths \cite{Zhenyao2023}. All APs are connected to a CPU via fronthaul/backhaul links with infinite capacity, facilitating seamless necessary information exchange and synchronization among APs. All APs and users are half-duplex (HD) devices. Based on the assigned operation mode, i.e., either UL or DL, the APs perform simultaneous DL communication or UL sensing over the same frequency band. In particular, the DL APs jointly communicate with the users while the UL APs perform radar sensing towards $T$ potential target directions.

We consider the block-fading channel model. Within each fading block, the channel between the $m$-th AP and the $k$-th user is denoted as $\q{h}_{mk} \in \mathbb{C}^{L \times 1}$, and modeled as 
\begin{eqnarray}
    \q{h}_{mk} = \zeta_{mk}^{1/2} \tilde{\q{h}}_{mk},
\end{eqnarray} \par \vspace{-0mm}
\noindent where $\zeta_{mk}$ represents the large-scale path-loss and remains constant over several coherence intervals, and  $\tilde{\q{h}}_{mk} \sim \mathcal{CN}(\mathbf{0}, \mathbf{I}_{L})$ is the small-scale Rayleigh fading and follows a complex normal distribution \cite{Goldsmith_2005, Tse_Viswanath_2005}. Moreover, $\q{G}_{mn} \in \mathbb{C}^{L \times L}$ represents the channel between the $m$-th DL AP and the $n$-th UL AP, and is modeled as $\q{G}_{mn} = \q{D}_{mn}^{1/2} \tilde{\q{G}}_{mn}$, where $\q{D}_{mn} = \zeta_{mn} \q{I}_L$, $\zeta_{mn}$ is the the large-scale path-loss, and $\tilde{\q{G}}_{mn} \sim \mathcal{CN}(\q{0}_{L\times L}, \q{I}_{L} \otimes \q{I}_{L})$ is the small-scale Rayleigh fading \cite{Goldsmith_2005, Tse_Viswanath_2005}.

Conversely, following the MIMO radar echo signal representation \cite{Zhenyao2023}, the transmit channel between the $m$-th DL AP and the $t$-th target and the receiver channel between the $n$-th AP and the $t$-th target, i.e., $\q{a}(\theta_{mt}) \in \mathbb{C}^{L\times 1}$ and $\q{b}(\theta_{nt}) \in \mathbb{C}^{L\times 1}$, respectively, are modeled as LoS channels. Consequently, the transmit/receiver array steering vector to the direction $\theta_{dt}$ is thus given by 
\begin{eqnarray}
    \q{c}(\theta_{dt}) = \frac{1}{\sqrt{L}} \left[1, e^{j\pi \sin(\theta_{dt})}, \ldots, e^{j\pi (L-1) \sin(\theta_{dt})} \right]^{\mathrm{T}},
\end{eqnarray} \par \vspace{-0mm}
\noindent where $\q{c}\in \{\q{a}, \q{b}\}$, $d\in \{m,n\}$ and, $\theta_{dt}$ is the $t$-th target's direction from the $d$-th AP (i.e., the $m$-th DL AP or the $n$-th UL AP) with respect to the $x$-axis of the coordinate system.

\section{Transmission Model}
It is assumed that the APs are able to switch between the UL and DL modes. The operation mode of an AP is decided based on the system's SE as discussed in Section~\ref{sec_prob_formulation_na}. The binary variables to indicate the mode assigned to the $m$-th AP are defined as
\begin{eqnarray}
    a_m = \begin{cases}
        1, & \text{if AP $m$ operates in the DL mode},\\
        0, & \text{otherwise},        
    \end{cases} \quad \forall m, \label{eqn_am}\\
    b_m = \begin{cases}
        1, & \text{if AP $m$ operates in the UL mode},\\
        0, & \text{otherwise},
    \end{cases} \quad \forall m. \label{eqn_bm}
\end{eqnarray}\par \vspace{-0mm}
\noindent From \eqref{eqn_am} and \eqref{eqn_bm}, we have 
\begin{eqnarray}\label{eqn_sum_ambm}
    a_m + b_m = 1, \quad \forall m. 
\end{eqnarray} \par \vspace{-0mm}
\noindent Therefore, \eqref{eqn_sum_ambm} ensures that the $m$-th AP only operates in either the DL or UL mode.

Let the $m$-th AP operate in the DL mode. The $m$-th AP transmit signal $\q{x}_m \in \mathbb{C}^{L\times 1}$ to communicate with the users as well as to enable sensing at the UL APs and is given by 
\begin{eqnarray}\label{eqn_tx_signal_NA}
    \q{x}_m = \sum_{k=1}^{K} a_m \q{w}_{mk} q_{k} + a_m \q{s}_m,
\end{eqnarray} \par \vspace{-0mm}
\noindent where $q_{k} \in \mathbb{C}$ is the data symbol intended for the $k$-th user with $\mathbb{E}\{\vert q_{k}\vert^2\} = 1$, $\q{w}_{mk} \in \mathbb{C}^{L\times 1}$ is the $m$-th DL AP communication beamforming vector associated with the $k$-th user, and $\q{s}_m\in \mathbb{C}^{L\times 1}$ is the dedicated sensing signal at the $m$-th DL AP \cite{Zhenyao2023}. Each DL AP must meet the average transmit power constraint, i.e., $\mathbb{E}\{\Vert \q{x}_m \Vert^2 \} \leq p_{\mathrm{max}}$, which leads to the following per-AP power constraint:
\begin{eqnarray}
    \sum_{k=1}^{K} a_{m}^2 \Vert \q{w}_{mk} \Vert^2 + a_{m}^2 \Vert \q{s}_{m} \Vert^2 \leq p_{\mathrm{max}}, \quad  \forall m.
\end{eqnarray}

The received signal at the $k$-th user can be expressed as
\begin{eqnarray}\label{eqn_k_user_rx_signal_na}
    y_k &=& \sum_{m=1}^{M} \q{h}_{mk}^{\mathrm{H}} \q{x}_m + z_k \nonumber \\
    &=& \sum_{m=1}^{M} a_m \q{h}_{mk}^{\mathrm{H}} \q{w}_{mk} q_k  + \sum_{i\neq k}^{K} \sum_{m=1}^{M} a_m \q{h}_{mk}^{\mathrm{H}} \q{w}_{mi} q_i + \sum_{m=1}^{M} a_m \q{h}_{mk}^{\mathrm{H}} \q{s}_{m}  + z_k, \quad
\end{eqnarray}\par \vspace{-0mm}
\noindent where $z_k \sim \mathcal{CN}(0, \sigma^2)$ is the additive white Gaussian noise (AWGN) at the $k$-th user. In \eqref{eqn_k_user_rx_signal_na}, it is assumed that the users apply clutter rejection techniques to mitigate the reflected interference from the targets and surrounding environment \cite{Xiao2021, Mark2010RadarBook}.

The UL APs use the target echo signal (i.e., the reflected DL signals by the targets) to perform target sensing. To this end, the received multi-target echo signal at the $n$-th UL AP is given by
\begin{eqnarray}\label{eqn_echo_m_AP_na}
    \q{y}_n = \sum_{t=1}^{T} \beta_t \q{b}(\theta_{nt}) \sum_{m=1}^{M} a_m \q{a}^{\mathrm{H}}(\theta_{mt}) \q{x}_m + \sum_{m=1}^{M} a_m \q{G}_{mn} \q{x}_m + \q{z}_n,
\end{eqnarray} \par \vspace{-0mm}
\noindent where $\q{z}_n \sim \mathcal{CN}(\q{0}, \sigma^2\q{I}_L)$ is the AWGN vector at the $n$-th UL AP. In \eqref{eqn_echo_m_AP_na}, $\beta_t$ represents the complex scattering coefficient of the $t$-th target reflection and accounts for the round-trip path loss and the RCS of the target \cite{Mark2010RadarBook}. It is assumed that UL APs employ clutter rejection techniques to minimize the reflected clutter interference from the surrounding environment \cite{Mark2010RadarBook}. The UL APs apply SIC to mitigate the DL AP to UL AP interference before extracting target information. Without loss of generality, we assume imperfect SIC at the UL APs. Then, the $n$-th UL AP utilizes the sensing combiner, $\q{u}_{nt} \in \mathbb{C}^{L\times 1}$ to the received echo signal to capture the desired reflected signal of the $t$-th target. The post-processed signal for obtaining $t$-th target's sensing information at the $n$-th UL AP is thus given by
\begin{eqnarray}\label{eqn_rx_Sens_na}
    y_{nt} &=& b_n \beta_t  \q{u}_{nt}^{\mathrm{H}} \q{b} (\theta_{nt}) \sum_{m=1}^{M} a_m \q{a}^{\mathrm{H}}(\theta_{mt}) \q{x}_m  \nonumber \\
    &&+ b_n \sum_{j\neq t}^{T} \beta_j \q{u}_{nt}^{\mathrm{H}} \q{b} (\theta_{nj}) \sum_{m=1}^{M} a_m \q{a}^{\mathrm{H}}(\theta_{mj}) \q{x}_m \nonumber\\
    &&+ b_n \sqrt{\delta_{\mathrm{AP}}} \q{u}_{nt}^{\mathrm{H}} \sum_{m=1}^{M} a_m \q{G}_{mn} \q{x}_m + b_n \q{u}_{nt}^{\mathrm{H}} \q{z}_{n}, 
\end{eqnarray} \par \vspace{-0mm}
\noindent where $\delta_{\mathrm{AP}} \in [0,1]$ accounts for the SIC quality for the DL AP to UL AP signal. 

\section{Communication SE}
The $k$-th user utilizes the received signal \eqref{eqn_k_user_rx_signal_na} to decode its intended data from the DL APs. Thus, the received SINR at the $k$-th user is derived as
\begin{eqnarray}
    \mt{SINR}_k^{\mathrm{Com}} &=& \frac{ \vert \sum_{m=1}^{M} a_m \q{h}_{mk} \q{w}_{mk} \vert^2}{ \sum_{i\neq k}^{K}  \vert \sum_{m=1}^{M} a_m \q{h}_{mk} \q{w}_{mi} \vert^2 + \vert \sum_{m=1}^{M} a_m \q{h}_{mk} \q{s}_{m} \vert^2 + \sigma^2} \nonumber \\
    &=& \frac{ \vert \q{h}_{k} \q{A} \q{w}_{k} \vert^2}{ \sum_{i\neq k}^{K}  \vert \q{h}_{k} \q{A} \q{w}_{i} \vert^2 + \vert \q{h}_{k} \q{A} \q{s} \vert^2 + \sigma^2}, 
\end{eqnarray} \par \vspace{-0mm}
\noindent where $\q{h}_{k} = [\q{h}_{1k}^{\mathrm{T}}, \dots, \q{h}_{Mk}^{\mathrm{T}}]^{\mathrm{T}} \in \mathbb{C}^{LM \times 1}$ is the cascaded channel vector of the $k$-th user,  $\q{w}_{i} = [\q{w}_{1i}^{\mathrm{T}}, \dots, \q{w}_{Mi}^{\mathrm{T}}]^{\mathrm{T}} \in \mathbb{C}^{LM \times 1}$ is the cascaded beamforming vector for the $i$-th user, and $\q{A} = \diag(a_1, \ldots, a_M) \otimes \q{I}_L = \bdiag(a_1 \q{I}_L, \ldots, a_M \q{I}_L) \in \mathbb{R}^{LM \times LM}$. Thereby, the $k$-th user rate can be approximated as
\begin{eqnarray}
    \mathcal{S}_k^{\mathrm{Com}} \approx \log_2(1 + \mt{SINR}_k^{\mathrm{Com}}). 
\end{eqnarray}

\section{Sensing SE}
Here, we employ sensing SE (or SINR) to evaluate the sensing performance (Section~\ref{sec_sensing_SE}). From \eqref{eqn_rx_Sens_na}, the sensing SE of the $t$-th target at the $n$-th UL AP can be approximated as
\begin{eqnarray}
    \mathcal{S}_{nt}^{\mathrm{Sen}} \approx \log_2(1 + \mt{SINR}_{nt}^{\mathrm{Sen}}), 
\end{eqnarray}\par \vspace{-0mm}
\noindent where $\mt{SINR}_{nt}^{\mathrm{Sen}}$ is the sensing SINR of the $t$-th target at the $n$-th UL AP and given by
\begin{eqnarray}\label{eqn_sens_SINR_targ_na}
    \mt{SINR}_{nt}^{\mathrm{Sen}}   &=& \frac{b_n^2 \vert\beta_t \vert^2 \mathbb{E} \left\{ \vert \q{u}_{nt}^{\mathrm{H}} 
    \q{b}(\theta_{nt}) \q{a}_t^{\mathrm{H}} \q{A} \q{x}  \vert^2 \right\} }
    {\splitfrac{ b_n^2  \sum_{j\neq t}^{T} \vert\beta_j\vert^2 \mathbb{E} 
    \left\{ \vert \q{u}_{nt}^{\mathrm{H}} \q{b}(\theta_{nj}) \q{a}_j^{\mathrm{H}} \q{A} \q{x} \vert^2 \right\} }{
    \quad  + b_n^2 \delta_{\mathrm{AP}} \mathbb{E} \left\{ \vert \q{u}_{nt}^{\mathrm{H}} \q{G}_{n} \q{A} \q{x} \vert^2 \right\}  
    + b_n^2\mathbb{E} \left\{ \vert \q{u}_{nt}^{\mathrm{H}} \mathbf{z}_n \vert^2 \right\} }} \nonumber \\
    &=& \frac{ b_n^2 \vert\beta_t \vert^2  \q{u}_{nt}^{\mathrm{H}} \q{b}(\theta_{nt}) \q{a}_t^{\mathrm{H}} 
    \q{A} \q{R}_x \q{A}^{\mathrm{H}} \q{a}_t \q{b}^{\mathrm{H}}(\theta_{nt}) \q{u}_{nt}  }
    { \splitfrac{\q{u}_{nt}^{\mathrm{H}} b_n^2  \left( \sum_{j\neq t}^{T} \vert\beta_j\vert^2  \q{b}(\theta_{nj}) 
    \q{a}_j^{\mathrm{H}} \q{A} \q{R}_x \q{A}^{\mathrm{H}} \q{a}_j \q{b}^{\mathrm{H}}(\theta_{nj}) \right) \q{u}_{nt}}{ + \q{u}_{nt}^{\mathrm{H}} b_n^2  \left( \delta_{\mathrm{AP}}  \q{G}_{n} \q{A} \q{R}_x  \q{A}^{\mathrm{H}} \q{G}_{n}^{\mathrm{H}} 
    +  \sigma^2 \mathbf{I}_L  \right) \q{u}_{nt}} },  \qquad
\end{eqnarray}\par \vspace{-0mm}
\noindent where $\q{a}_{t} = [\q{a}^{\mathrm{T}}(\theta_{1t}), \ldots, \q{a}^{\mathrm{T}}(\theta_{Mt})]^{\mathrm{T}} \in \mathbb{C}^{LM \times 1}$, $\q{G}_n = [\q{G}_{1n}, \ldots, \q{G}_{Mn}] \in \mathbb{C}^{L \times LM}$, and $\q{x} = [\q{x}_1^{\mathrm{T}}, \ldots, \q{x}_M^{\mathrm{T}}]^{\mathrm{T}} \in \mathbb{C}^{LM \times 1}$. Moreover, $\q{R}_x \triangleq \mathbb{E} \{\q{x} \q{x}^{\mathrm{H}} \} =  \sum_{i =1}^{K} \q{w}_i \q{w}_i^{\mathrm{H}} + \q{S}$ is the covariance matrix of the cascaded transmitted signal, where $\q{S} \triangleq \mathbb{E} \{\q{s} \q{s}^{\mathrm{H}} \}$ \cite{Zhenyao2023}.

\section{Problem Formulation}\label{sec_prob_formulation_na}
Here, we aim to maximize the communication sum SE of the NAFD CFMM system. This objective is achieved by jointly optimizing the transmit communication and sensing beamforming at the DL APs, i.e., $\{\q{w}_k\}$ and $\q{s}$, sensing combining at the UL APs, i.e., $\{\q{u}_{nt}\}$, and AP mode selection variables, i.e., $\{a_m, b_m\}$. The set of all the optimization variables is denoted as $\mathcal{A} = \{\{\q{w}_k\}, \q{s}, \{\q{u}_{nt}\}, \{a_m, b_m\}\}$. Moreover, the communication sum SE is minimized while adhering to constraints on the minimum sensing SE and maximum per AP transmit power. The optimization problem is thus formulated as follows:

\begin{subequations}\label{prob_R1}
\begin{eqnarray}
    (\mathcal{R}1):~&&  \max_{\mathcal{A}} \quad \sum_{k=1}^{K} \log_2\left(1 + \mt{SINR}_k^{\mathrm{Com}}\right), \label{prob_R1_obj}  \\
    \text{s.t.} \quad &&   \log_2\left(1 + \mt{SINR}_{nt}^{\mathrm{Sen}}\right) \geq \Upsilon_{\mathrm{th}}^{\mathrm{Sen}}, ~\forall n, t, \label{prob_R1_sens_rate}\\
    && \eqref{eqn_am}-\eqref{eqn_sum_ambm},  \label{prob_R1_mode_sel}\\
    && \sum_{i=1}^{K} \Vert \q{w}_{mi}\Vert^2 + \Vert \q{s}_{m}\Vert^2 \leq p_{\mathrm{max}}, ~\forall m, \label{prob_R1_tx_pow} \\
    && \Vert \q{u}_{nt} \Vert^2 = 1, ~\forall n, t, \label{prob_R1_comb} 
\end{eqnarray}
\end{subequations} \par \vspace{-0mm}
\noindent where \eqref{prob_R1_sens_rate} is the sensing rate requirement for each target, with a required sensing rate threshold of $\Upsilon_{\mathrm{th}}^{\mathrm{Sen}}$, \eqref{prob_R1_mode_sel} defines the AP mode assignment, \eqref{prob_R1_tx_pow} sets the maximum allowable transmit power at the $m$-th DL AP, and \eqref{prob_R1_comb} is the normalization constraint for the $t$-th target sensing combiner at the $n$-th UL AP. 

\section{Proposed Solution}
This section presents a framework for solving the formulated optimization problem $(\mathcal{R}1)$. Problem $(\mathcal{R}1)$ is non-convex due to the involved products of optimization variables, i.e., $\mathcal{A}$, in the objective and constraints. To address this challenge, we utilize the AO technique \cite{bezdek2003convergence}. In particular, $(\mathcal{R}1)$ is divided into three sub-problems, i.e., optimizing $\{\q{w}_k, \q{s}\}$, $\{\q{u}_{nt}\}$ and $\{a_m, b_m\}$, and then solved one at a time while keeping the other variables fixed. The process repeats until a stopping criterion is met. This approach works when directly optimizing all variables is challenging or computationally expensive \cite{bezdek2003convergence}. 

Mathematically, to solve $\min_{x} f(x)$, where $x \in \mathbb{R}^{s}$ can be divided into $l > 1$ blocks, i.e., $x = (x_1, x_2, \ldots, x_l)^{\mathrm{T}}$ with $x_l \in \mathbb{R}^{s_k}$ and $\sum_{k=1}^l s_k = s$, $f(x)$ is minimized over $x_k (k=1,\ldots, l)$, while holding $x_m|m\neq k$ at the prior values. The cycle repeats until convergence \cite{bezdek2003convergence}.

\subsection{Sub-Problem 1: Optimization Over $\{\q{w}_k, \q{s}\}$}
For given $\{\{\q{u}_{nt}\}, \{a_m, b_m\}\}$, problem $(\mathcal{R}1)$ can be reformulated as the following beamforming problem:
\begin{subequations}\label{prob_R2}
\begin{eqnarray}
    (\mathcal{R}2):~&&  \max_{\q{w}_k, \q{s}} \quad \sum_{k=1}^{K} \log_2\left(1 + \mt{SINR}_k^{\mathrm{Com}}\right), \label{prob_R2_obj}  \\
    \text{s.t.} \quad &&   \eqref{prob_R1_sens_rate},  \eqref{prob_R1_tx_pow}. 
\end{eqnarray}
\end{subequations} \par \vspace{-0mm}
\noindent Problem $(\mathcal{R}2)$ is still non-convex due to the interference terms within the SINRs, i.e., communication and sensing. To handle this, we employ the SDR technique. To this end, we define the matrices $\q{W}_{k} \triangleq \q{w}_{k} \q{w}_{k}^{\mathrm{H}}$ and $\q{S} \triangleq \q{s} \q{s}^{\mathrm{H}}$, where $\q{W}_{k}$ and $\q{S}$ are semi-definite matrices, i.e., $\q{W}_{k} \succeq 0$ and $\q{S} \succeq 0$. Moreover, $\q{W}_{k}$ must satisfy rank one constraint, i.e., ${\mathrm{Rank}}(\q{W}_{k}) = 1$. Then, we utilize the SDR techniques to relax the highly non-convex rank one constraint and reformulate the problem as follows:
\begin{subequations}\label{prob_R3}
\begin{eqnarray}
    (\mathcal{R}3):~&& \max_{\q{W}_{k}, \q{S}} \quad \sum_{k=1}^{K} f(\q{W}_k, \q{S}), \label{prob_R3_obj}  \\
    \text{s.t.} \quad && \digamma_{nt} \leq 0, ~\forall n,t, \\
    && \sum_{i=1}^{K} \tr\left( \q{W}_{i} \q{Q}_m \right) + \tr\left( \q{S} \q{Q}_m \right) \leq p_{\mathrm{max}}, ~\forall m, \label{prob_R3_tx_pow} \\
    && \q{W}_k,  \q{S} \succeq 0,~\forall k,
\end{eqnarray}
\end{subequations} \par \vspace{-0mm}
\noindent where $\hat{\q{h}}_k = \q{A}^{\mathrm{H}} \q{h}_k$, $\hat{\q{a}}_t = \q{A}^{\mathrm{H}} \q{a}_t$, $\q{F}_{nj} = \q{A}^{\mathrm{H}} \q{a}_j \q{b}^{\mathrm{H}}(\theta_{nj})$, $\hat{\q{G}}_n = \q{A}^{\mathrm{H}} \q{G}_n$, $\hat{\Upsilon}_{\mathrm{th}}^{\mathrm{Sen}} = 2^{\Upsilon_{\mathrm{th}}^{\mathrm{Sen}}} -1$, and  $\hat{\delta}_{\mathrm{max}} = 2^{\delta_{\mathrm{max}}} -1$. Moreover, $\q{Q}_m \in \mathbb{R}^{LM \times LM}$ is a block diagonal selection matrix, consisting of $M$ blocks, each corresponding to an AP, i.e.,  $\q{Q}_m = \text{diag}(\q{0}, \dots, \q{I}_{L},  \dots, \q{0})$ with $\q{I}_{L}$ in the $m$-th block on the diagonal. This matrix selects the transmit antennas of the $m$-th AP. In $(\mathcal{R}3)$, as the objective and constraint \eqref{prob_R1_sens_rate} are not convex functions, we employ the SCA method to linearize them which are given by 
\begin{eqnarray}
    f(\q{W}_k, \q{S}) &=& \log_2 \left( \sum_{i=1}^{K} \tr\left( \hat{\q{h}}_k \hat{\q{h}}_k^{\mathrm{H}} \q{W}_i \right) + \tr\left( \hat{\q{h}}_k \hat{\q{h}}_k^{\mathrm{H}} \q{S} \right) + \sigma^2 \right) \nonumber \\
    &&- \log_2 \left( \sum_{i\neq k}^{K} \tr\left( \hat{\q{h}}_k \hat{\q{h}}_k^{\mathrm{H}} \q{W}_i^{(p)} \right) + \tr\left( \hat{\q{h}}_k \hat{\q{h}}_k^{\mathrm{H}} \q{S}^{(p)} \right) + \sigma^2 \right) \nonumber \\
    && - \frac{\sum_{i\neq k}^{K} \tr\left( \hat{\q{h}}_k \hat{\q{h}}_k^{\mathrm{H}} (\q{W}_i - \q{W}_i^{(p)}) \right)  }{\ln(2) \left( \sum_{i\neq k}^{K} \tr\left( \hat{\q{h}}_k \hat{\q{h}}_k^{\mathrm{H}} \q{W}_i^{(p)} \right) + \tr\left( \hat{\q{h}}_k \hat{\q{h}}_k^{\mathrm{H}} \q{S}^{(p)} \right) + \sigma^2 \right)} \nonumber\\
    &&- \frac{\tr\left( \hat{\q{h}}_k \hat{\q{h}}_k^{\mathrm{H}} (\q{S} - \q{S}^{(p)}) \right)  }{\ln(2) \left( \sum_{i\neq k}^{K} \tr\left( \hat{\q{h}}_k \hat{\q{h}}_k^{\mathrm{H}} \q{W}_i^{(p)} \right) + \tr\left( \hat{\q{h}}_k \hat{\q{h}}_k^{\mathrm{H}} \q{S}^{(p)} \right) + \sigma^2 \right)}, \qquad\\
    \digamma_{nt} &=& \hat{\Upsilon}_{\mathrm{th}}^{\mathrm{Sen}} \q{u}_{nt}^{\mathrm{H}} \left( b_n^2 \sum_{j\neq t}^{T} \vert \beta_j\vert^2 \left( \sum_{i=1}^{K} \tr(\q{F}_{nj}  \q{F}_{nj}^{\mathrm{H}} \q{W}_{i}) + \tr( \q{F}_{nj} \q{F}_{nj}^{\mathrm{H}} \q{S} ) \right)  \right. \nonumber \\
    &&\left. + b_n^2 \delta_{\mathrm{AP}} \left( \sum_{i=1}^{K} \tr(\hat{\q{G}}_{n}  \hat{\q{G}}_{n}^{\mathrm{H}} \q{W}_{i}) + \tr( \hat{\q{G}}_{n} \hat{\q{G}}_{n}^{\mathrm{H}} \q{S} ) \right) + \sigma^2 \q{I}_L  \right) \q{u}_{nt} \nonumber \\
    &&- b_n^2 \vert \beta_t \vert^2 \q{u}_{nt}^{\mathrm{H}} \left( \sum_{i=1}^{K} \tr(\q{F}_{nt}  \q{F}_{nt}^{\mathrm{H}} \q{W}_{i}) + \tr( \q{F}_{nt} \q{F}_{nt}^{\mathrm{H}} \q{S} ) \right)\q{u}_{nt}, \quad
\end{eqnarray} \par \vspace{-0mm}
\noindent where $(\cdot)^{(p)}$ denotes the previous iteration values of respective variables. Finally, the relaxed problem $(\mathcal{R}3)$ is a SDP problem and can be addressed using the CVX Matlab tool \cite{boyd2004convex, grant2014cvx}.

\subsection{Sub-Problem 2: Optimization Over $\{\q{u}_{nt}\}$}
For given $\{\{\q{w}_k\}, \q{s}, \{a_m, b_m\}\}$, problem $(\mathcal{R}1)$ is reduced to a sensing combining optimization problem at the UL APs. The resultant problem is a feasibility problem as the user communication rate is independent of $\q{u}_{nt}$. Hence, any feasible values of $\{\q{u}_{nt}\}$ that satisfy \eqref{prob_R1_sens_rate} and \eqref{prob_R1_comb} can be a potential solution.

Although $\{\q{u}_{nt}\}$ does not have a direct impact on objective function \eqref{prob_R1_obj}, it is crucial for optimizing the sensing SINR at the UL APs for each target. Thus, using the unique structure of the sensing SINR for each target at the UL APs in \eqref{eqn_sens_SINR_targ_na}, this sub-problem can be formulated as a generalized Rayleigh quotient optimization problem with a closed-form solution \cite{Stanczak2008book, Wan2016}. To this end, the corresponding problem can be given by

\begin{subequations}\label{prob_R4}
\begin{eqnarray}
    (\mathcal{R}4):~&&  \max_{\q{u}_{nt}} \quad \frac{\q{u}_{nt}^{\mathrm{H}} \q{f}_{nt} \q{f}_{nt}^{\mathrm{H}} \q{u}_{nt}}{\q{u}_{nt}^{\mathrm{H}} \q{Q}_{nt} \q{u}_{nt}}, \label{prob_R4_obj}  \\
    \text{s.t.} \quad &&  \eqref{prob_R1_comb},
\end{eqnarray}
\end{subequations} \par \vspace{-0mm}
\noindent where 
\begin{eqnarray}
\q{f}_{nt} &=& b_n \vert\beta_t \vert \q{b}(\theta_{nt}) \q{a}_t^{\mathrm{H}} \q{A} \left(\sum_{i=1}^{K} \q{w}_i + \q{s}\right), \\
\q{Q}_{nt} &=&  b_n^2 \sum_{j\neq t}^{T} \vert\beta_j\vert^2  \q{b}(\theta_{nj}) \q{a}_j^{\mathrm{H}} \q{A} \q{R}_x \q{A}^{\mathrm{H}} \q{a}_j \q{b}^{\mathrm{H}}(\theta_{nj}) \nonumber \\
&&+ b_n^2 \delta_{\mathrm{AP}}  \q{G}_{n} \q{A} \q{R}_x  \q{A}^{\mathrm{H}} \q{G}_{n}^{\mathrm{H}} +  b_n^2  \sigma^2 \mathbf{I}_L.
\end{eqnarray} \par \vspace{-0mm}
\noindent Problem $(\mathcal{R}4)$ thus becomes a generalized Rayleigh ratio quotient problem \cite{Stanczak2008book, Wan2016}, and the optimal sensing combiner for the $t$-th target at the $n$-th UL AP is given as \cite{Stanczak2008book, Wan2016} 
\begin{eqnarray}\label{opt_unt_na}
    \q{u}_{nt}^* = \frac{\q{Q}_{nt}^{-1} \q{f}_{nt} }{\Vert \q{Q}_{nt}^{-1} \q{f}_{nt} \Vert},~~ \forall n,t,
\end{eqnarray} \par \vspace{-0mm}
\noindent which is an MMSE filter \cite{Stanczak2008book, Wan2016}. 

\subsection{Sub-Problem 3: Optimization Over $\{a_m, b_m\}$}
For given $\{\{\q{w}_k\}, \q{s}, \{\q{u}_{nt}\}\}$, problem $(\mathcal{P}1)$ becomes an AP mode selection problem and can be given as
\begin{subequations}\label{prob_R5}
\begin{eqnarray}
    (\mathcal{R}5):~&&  \max_{a_m, b_m} \quad \sum_{k=1}^{K} \log_2\left(1 + \mt{SINR}_k^{\mathrm{Com}}\right), \label{prob_R5_obj}  \\
    \text{s.t.} \quad &&   \eqref{prob_R1_sens_rate},  \eqref{eqn_am}-\eqref{eqn_sum_ambm}.  
\end{eqnarray}
\end{subequations} \par \vspace{-0mm}
\noindent Problem $(\mathcal{R}5)$ is a non-convex problem due to the involved product of optimization variables and binary constraints \eqref{eqn_am} and \eqref{eqn_bm}. To handle these binary constraints, we observe that for any real number $x$, we have $x\in \{0,1\} \Leftrightarrow x-x^2 = 0 \Leftrightarrow (x\in [0,1] \Leftrightarrow x-x^2 \leq 0)$ \cite{Vu2018}. Thus, constraints \eqref{eqn_am} and \eqref{eqn_bm} can be transformed into the following equivalent constraints
\begin{eqnarray}
    C(\q{a}, \q{b}) &&\triangleq \sum_{m=1}^{M} a_m - a_m^2 + \sum_{m=1}^{M} b_m - b_m^2 \leq 0, \label{eqn_c_def} \\
    && \begin{cases}
    0 \leq a_m \leq 1,\\
    0 \leq b_m \leq 1,
    \end{cases} \quad \forall m, \label{eqn_am_bm_ineq}
\end{eqnarray}  \par \vspace{-0mm}
\noindent where $\q{a} = [a_1, \ldots, a_M]^{\mathrm{T}}$ and $\q{b} = [b_1, \ldots, b_M]^{\mathrm{T}}$. Next, to handle the resultant constraint \eqref{eqn_c_def}, we invoke the Lagrangian duality method \cite{liu2020simple}. To this end, the corresponding Lagrangian problem can be given as follows:
\begin{subequations}\label{prob_R6}
\begin{eqnarray}
    (\mathcal{R}6):~&&  \max_{\q{a}, \q{b}} \quad \sum_{k=1}^{K} f_k(\q{a}) + \lambda C(\q{a}, \q{b}), \label{prob_R6_obj}  \\
    \text{s.t.} \quad &&   \eqref{prob_R1_sens_rate}, \eqref{eqn_sum_ambm},  \eqref{eqn_am_bm_ineq}, 
\end{eqnarray}
\end{subequations}  \par \vspace{-0mm}
\noindent where $f_k(\q{a}) \triangleq \log_2\left(1 + \mt{SINR}_k^{\mathrm{Com}}\right)$ and $\lambda$ is the Lagrangian multiplier corresponding to constraint \eqref{eqn_c_def}. 

Problem $(\mathcal{R}6)$ is still non-convex due to the non-convex objective function and constraint \eqref{prob_R1_sens_rate}. To handle the communication rate in the objective function \eqref{prob_R6_obj}, i.e., $f_k(\q{a})$, we invoke FP techniques \cite{Shen2018}. Thus, by applying the quadratic transformation technique in \cite{Shen2018}, the resultant communication rate function can be given as
\begin{eqnarray}\label{eqn_FP_rate}
    f_k(\q{a}, \phi_k) = \log_2(1 + 2 \phi_k A_k(\q{a}) - \phi^2 B_k(\q{a})),
\end{eqnarray}  \par \vspace{-0mm}
\noindent where 
\begin{eqnarray}
    A_k(\q{a}) &\triangleq& \Re\left\{\sum_{m=1}^{M} a_m \q{h}_{mk} \q{w}_{mk}\right\} \\
    B_k(\q{a}) &\triangleq& \sum_{i\neq k}^{K}  \left\vert \sum_{m=1}^{M} a_m \q{h}_{mk} \q{w}_{mi} \right\vert^2 + \left\vert \sum_{m=1}^{M} a_m \q{h}_{mk} \q{s}_{m} \right\vert^2 + \sigma^2.
\end{eqnarray}  \par \vspace{-0mm}
\noindent Moreover, $\boldsymbol{\phi}=[\phi_1, \ldots, \phi_K]^{\mathrm{T}}$ is an auxiliary variable vector introduced by the quadratic transformation. For a given $\q{a}$, the optimal value of $\phi_k^*$ can be given in closed-form as \cite{Shen2018}
\begin{eqnarray}\label{eqn:opt_lambda}
    \phi_k^* = \frac{A_k(\q{a})}{B_k(\q{a})}.
\end{eqnarray}

For the second term in \eqref{prob_R6_obj}, i.e., $C(\q{a}, \q{b})$, using \eqref{eqn_bm}, we note that $b_m = 1- a_m$. Thus, we replace the variable $b_m$ by $1- a_m$ as 
\begin{eqnarray}
    C(\q{a}, \q{b}) = 2 \sum_{m=1}^{M} (a_m - a_m^2) \triangleq f_{\lambda}(\q{a}). 
\end{eqnarray} \par \vspace{-0mm}
\noindent However, $f(\q{a})$ is still non-convex due to the squared term. To address this, we employ the Taylor expansion-based SCA technique, leading to the following equivalent convex function
\begin{eqnarray}
    f_{\lambda}(\q{a}) = 2 \sum_{m=1}^{M} (a_m - 2 a_m^{(p)} a_m + (a_m^{(p)})^2), 
\end{eqnarray} \par \vspace{-0mm}
\noindent where $(\cdot)^{(p)}$ denotes the previous iteration values of the respective variable. The overall objective function is thus given as
\begin{eqnarray}
    f_{\mathrm{obj}}(\q{a}) \triangleq \sum_{k=1}^{K} f_k(\q{a}, \phi_k^*) + \lambda f_{\lambda}(\q{a}). 
\end{eqnarray}

Next, utilizing the monotonically increasing nature of $\log_2(\cdot)$ function, we can rewire the consraint \eqref{prob_R1_sens_rate} as follows: 
\begin{eqnarray}\label{eqn_ses_rate_const_na}
    &&\vert \beta_t \vert^2 \left\vert b_{ntt} \sum_{m=1}^{M} a_m q_{mt} \right\vert^2 \geq \nonumber \\
    && \qquad \hat{\Upsilon}_{\mathrm{th}}^{\mathrm{Sen}} \left(\sum_{j\neq t}^{T} \vert \beta_j \vert^2 A_{ntj}(\q{a}) + \delta_{\mathrm{AP}} G_{nt}(\q{a}) + \sigma^2 \Vert \q{u}_{nt} \Vert^2\right), 
\end{eqnarray} \par \vspace{-0mm}
\noindent where $A_{ntj}(\q{a}) \triangleq \vert b_{ntj} \sum_{m=1}^{M} a_m q_{mj} \vert^2$ and $G_{nt}(\q{a}) \triangleq \vert \sum_{m=1}^{M} a_m r_{mnt} \vert^2$. Moreover,  $b_ntj = \q{u}_{nt}^{\mathrm{H}} \q{b}(\theta_{nj})$, $q_{mj} = \q{a}^{\mathrm{H}}(\theta_{mj}) (\sum_{i=1}^{K} \q{w}_{mi} + \q{s}_m)$, and $r_{mnt} = \q{u}_{nt}^{\mathrm{H}} \q{G}_{mn} (\sum_{i=1}^{K} \q{w}_{mi} + \q{s}_m)$. However,  \eqref{eqn_ses_rate_const_na} is non-convex due to the quadratic terms $A_{ntj}(\q{a})$ and $G_{nt}(\q{a})$. To handle this, we use the second-order Taylor approximation to convexify them as follows:
\begin{subequations}
\begin{eqnarray}
    A_{ntj}(\q{a}) &\approx& \left\vert b_{ntj} \sum_{m=1}^{M} a_m^{(p)} q_{mj} \right\vert^2 \nonumber \\
    &&+ 2 \sum_{m=1}^{M} \Re \left\{b_{ntj} \left(\sum_{l=1}^{M} a_l^{(p)} q_{lj} \right) b_{ntj} q_{mj} \right\}(a_m - a_m^{(p)}), \qquad \\
    G_{nt}(\q{a}) &\approx& \left\vert \sum_{m=1}^{M} a_m^{(p)} r_{mnt} \right\vert^2 \nonumber \\
    &&+ 2 \sum_{m=1}^{M} \Re \left\{\left(\sum_{l=1}^{M} a_l^{(p)} r_{lnt} \right) r_{mnt} \right\}(a_m - a_m^{(p)}). 
\end{eqnarray}
\end{subequations} \par \vspace{-0mm}
\noindent Finally, the convex version of problem $(\mathcal{R}6)$ can be given as
\begin{subequations}\label{prob_R7}
\begin{eqnarray}
    (\mathcal{R}7):~&&  \max_{\q{a}} \quad f_{\mathrm{obj}}(\q{a}), \label{prob_R7_obj}  \\
    \text{s.t.} \quad &&   \eqref{eqn_ses_rate_const_na}, \eqref{eqn_am_bm_ineq}. 
\end{eqnarray}
\end{subequations} \par \vspace{-0mm}
\noindent Problem $(\mathcal{R}7)$ can thus be solved via CVX Matlab tool \cite{boyd2004convex, grant2014cvx}.

\section{Simulation Example}
We use the 3GPP UMi model with an operating frequency of $f_c = \qty{3}{\GHz}$ \cite[Table B.1.2.1]{3GPP2010} for large-scale fading path-loss, i.e., $\zeta_{mk}$. The AWGN variance is modeled as  $\sigma^2 = 10 \log_{10}{(N_0 B N_f)}$\,\qty{}{\dB m}, where $N_0=\qty{-174}{\dB m/\Hz}$, $B=\qty{10}{\MHz}$ is the bandwidth, and $N_f=\qty{10}{\dB}$ is the noise figure. Moreover, the APs are uniformly distributed, while users and targets are randomly placed within a $\num{200} \times \qty{200}{\m^2}$ area.

For comparative evaluation purposes, a benchmark with a random AP assignment is considered. In particular, this scheme assists in evaluating the impacts of optimally selecting/assigning AP modes in both communication and sensing performance.

\begin{figure}[!t]\vspace{-0mm}
    \centering
    \includegraphics[width=0.7\textwidth]{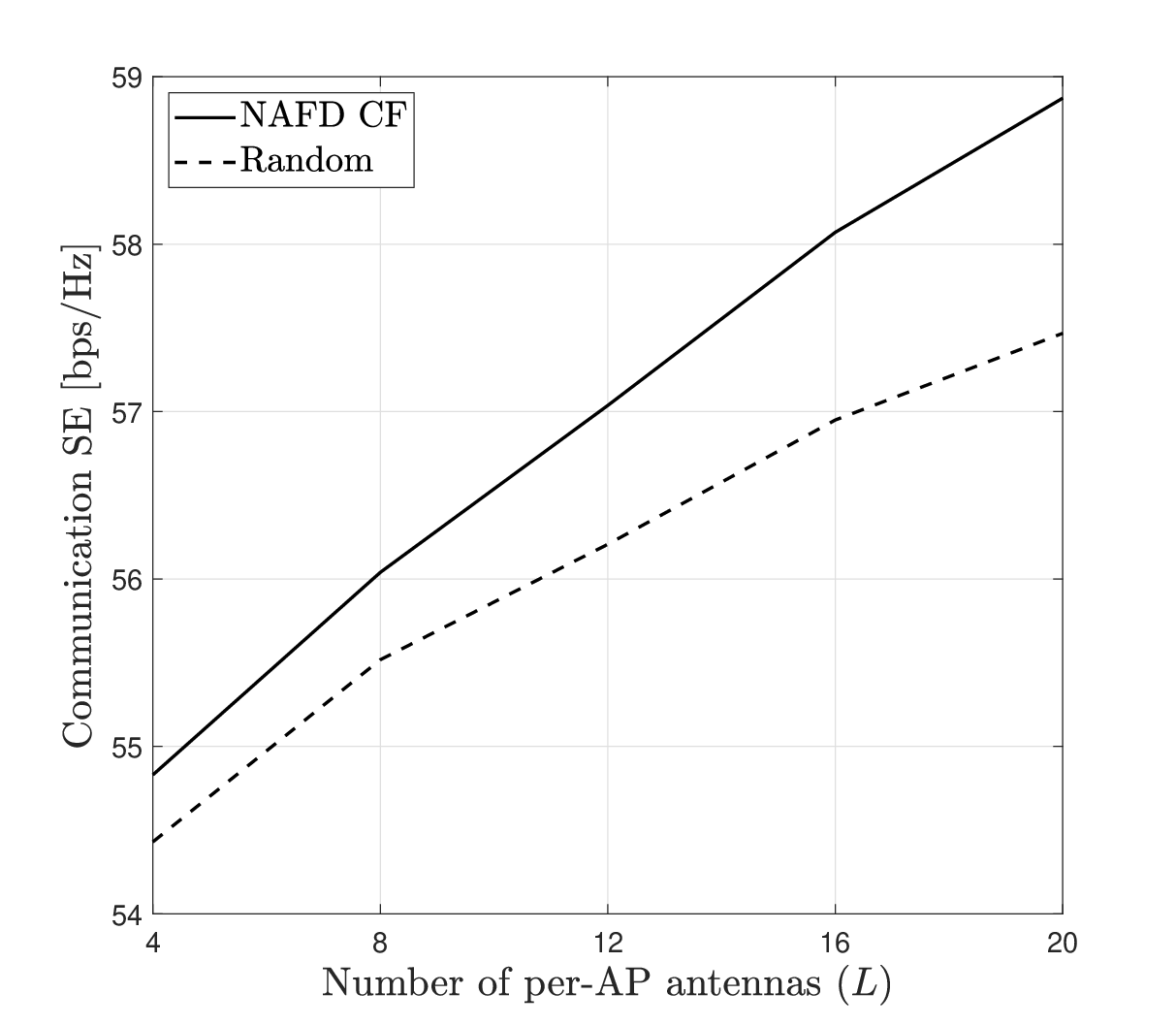}
    \vspace{-0mm}
    \caption{Communication SE versus the number of per-AP antennas for $M=9$, $K=2$, and $T=3$.}
    \label{fig_NAFD_CommRate} \vspace{-0mm}
\end{figure}

Figure~\ref{fig_NAFD_CommRate} investigates the communication SE as a function of the number of per-AP antennas ($L$) for $M=9$ APs, $K=2$ users, and $T=3$ sensing targets. As expected, both the NAFD CF and random schemes exhibit increasing SE trends with larger values of $L$, owing to the enhanced spatial degrees of freedom and array gain. However, the NAFD CF scheme consistently yields superior performance over the random AP allocation. This communication SE gain highlights the effectiveness of the proposed NAFD strategy in intelligently coordinating AP modes to jointly optimize communication and sensing trade-offs. The gain becomes more pronounced as $L$ increases, suggesting that the NAFD scheme better capitalizes on additional antenna resources by allocating them more judiciously. For example, with $L=12$, the NAFD CF scheme provides approximately a \qty{2}{\percent} communication SE gain compared to the random AP assignment scheme. This gain reflects not only better AP mode selection but also more efficient spatial resource utilization under the NAFD framework, i.e., it assigns APs based on the communication and sensing requirements. From a practical point of view, this behavior underscores the importance of intelligent AP management in CF ISAC systems where APs support both communication and sensing functionalities. Such strategies can be instrumental in environments with variable traffic demands or sensing priorities, allowing dynamic adaptation without hardware modifications.

\begin{figure}[!t]\vspace{-0mm}
    \centering
    \includegraphics[width=0.7\textwidth]{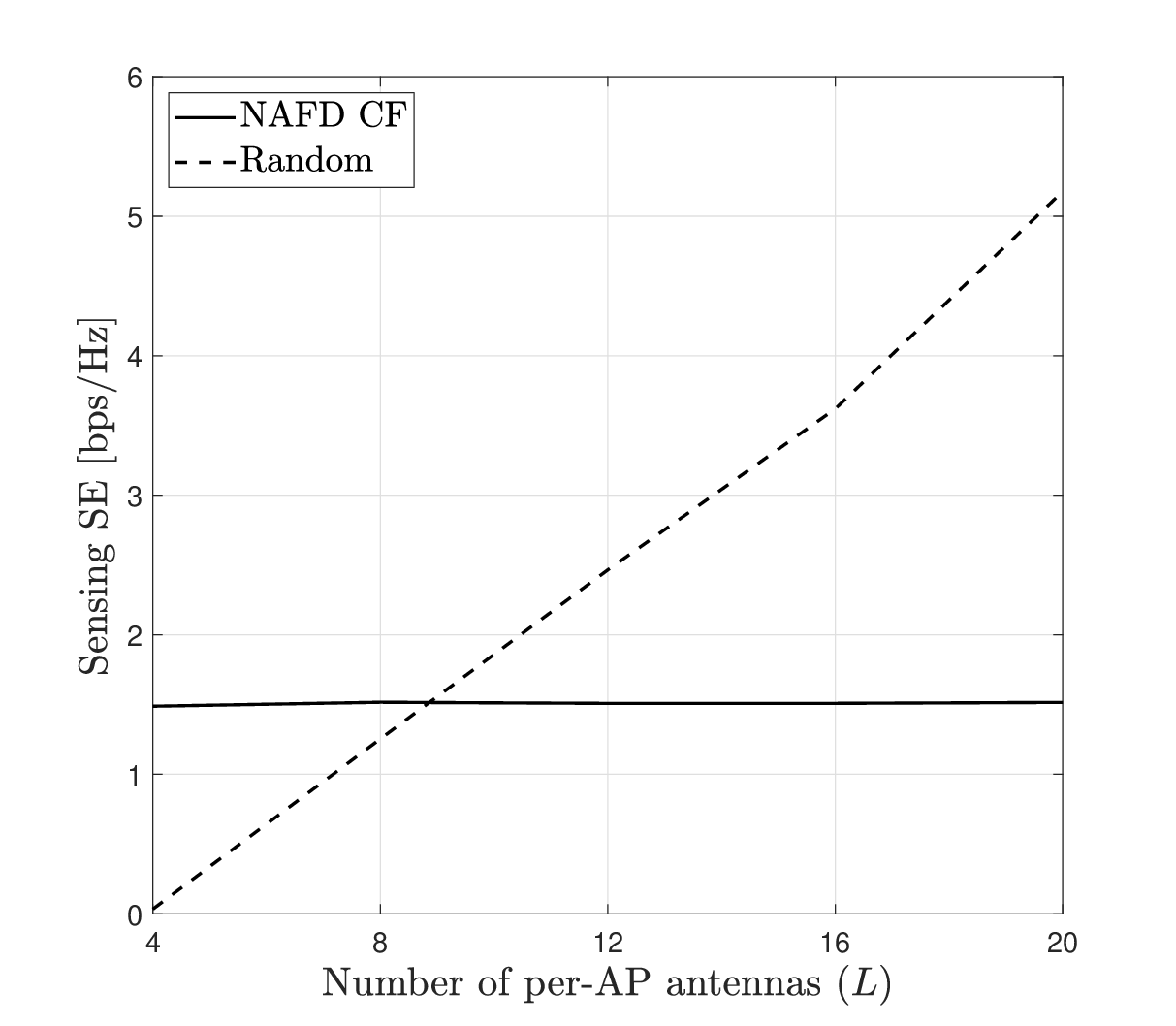}
    \vspace{-0mm}
    \caption{Sensing SE versus the number of per-AP antennas for $M=9$, $K=2$, and $T=3$.}
    \label{fig_NAFD_SensingRate} \vspace{-0mm}
\end{figure}

In Figure~\ref{fig_NAFD_SensingRate}, sensing SE is evaluated against the number of per-AP antennas ($L$) for both the NAFD CF and random AP selection strategies, under a target SE threshold of $\mt{SINR}_{nt}^{\mathrm{Sen}}=\qty{1.5}{bps/\Hz}$. It reveals that the NAFD CF scheme maintains a constant yet reliable sensing SE, consistently satisfying the threshold requirement across the entire range of $L$. This highlights its predictable sensing performance regardless of the number of per-AP antennas and target location. In contrast, the random mode selection exhibits a steep increase in sensing SE as $L$ increases. While this might initially seem advantageous, it actually indicates inefficiency, i.e., the system requires high antenna resources to meet the same sensing SE target achieved effortlessly by NAFD CF. For instance, the random strategy requires at least $L=9$ antennas per AP to exceed the minimum acceptable sensing SE, whereas NAFD CF meets it consistently even with the smallest $L$. This is because the modes of APs are not selected based on the sensing and communication requirements of the system.

\begin{figure*}[!t]\vspace{-0mm}
    \centering
    \makebox[\textwidth][c]{
    \includegraphics[width=1.1\textwidth]{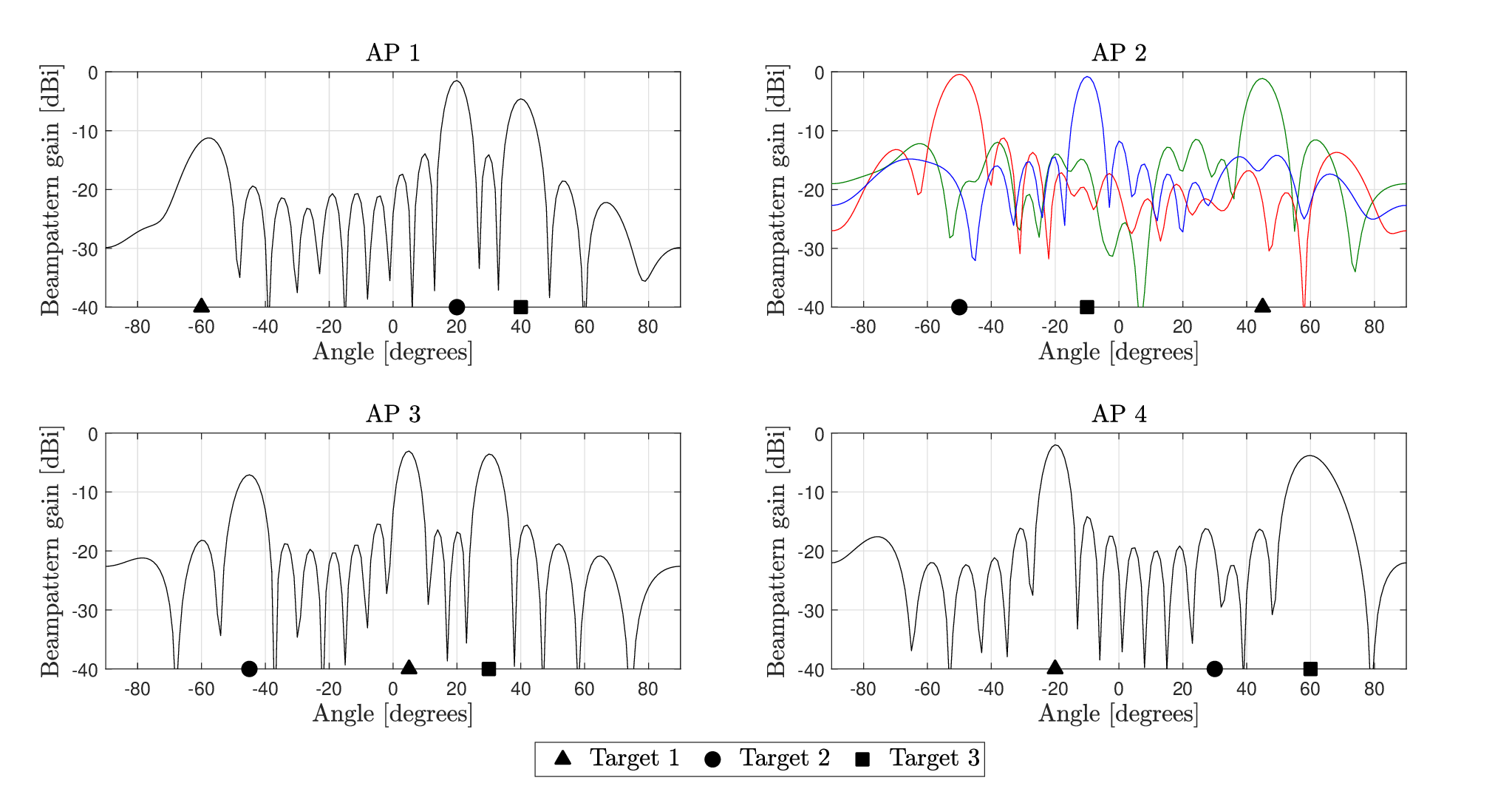}}
    \vspace{-0mm}
    \caption{Directional beampattern gain profiles over a \qty{\pm 90}{\degree} angular spread at different APs.}
    \label{fig_BeamGain_NA} \vspace{-0mm}
\end{figure*}

Figure~\ref{fig_BeamGain_NA} illustrates the directional beampattern gain profiles of the NAFD CF ISAC system for $M=\num{4}$ APs with $L=\num{8}$ antennas each, supporting $K=\num{2}$ users and $T=\num{3}$ sensing targets. The direction angles for sensing targets from AP 1, AP 2, AP 3, and AP 4, are set to $\{-60, 20, 40\}\qty{}{\degree}$, $\{45, -50, -10\}\qty{}{\degree}$, $\{5, -45, 30\}\qty{}{\degree}$, and $\{-20, 30, 60\}\qty{}{\degree}$, respectively. Each subplot corresponds to a different AP, with APs 1, 3, and 4 operating in DL mode and AP 2 in UL mode, as determined dynamically by the NAFD CF algorithm. The beam patterns are tailored to focus energy in DL mode or combined gain in the UL mode toward the angular positions of the sensing targets.

The DL APs (AP-1, -3, and -4) exhibit distinct lobes directed at the target angles, with sidelobes and nulls carefully controlled to suppress interference and leakage into undesired directions. This demonstrates the efficacy of the transmit beamforming design, which seeks to maximize gain at the intended sensing targets while maintaining spectral coexistence with communication users. In the case of AP 2 (UL mode), a composite beampattern is plotted that accounts for both the receive combining and the transmit beamforming. The presence of clearly defined lobes aligned with the target angles validates the NAFD CF's ability to generate effective receive-side combining filters. Importantly, the figure illustrates the spatial diversity and adaptability introduced by the NAFD CF framework. Each AP can be optimized independently to cater to network requirements, allowing flexible and scalable deployment. 
	\chapter{Key Challenges in Cell-Free ISAC}\label{chp_key_challenges}

\section{Multi-Target Sensing}
The accurate detection, localization, and tracking of multiple targets simultaneously is a fundamental yet challenging task in CF-ISAC systems. Unlike conventional mono-static or bi-static ISAC architectures, where sensing is performed at a single or dual vantage point, CF-ISAC networks leverage a dense deployment of spatially distributed APs that collaboratively transmit and receive sensing signals~\cite{Buzzi2024}. This distributed topology introduces both opportunities and complexities for multi-target sensing~\cite{Buzzi2024}. In particular, the spatial diversity inherent to CF architectures can significantly enhance the sensing resolution and robustness by offering uncorrelated multiple views of the targets. However, it also increases the complexity of target separation, especially in environments where echoes from multiple targets overlap temporally and spatially. Resolving such ambiguities requires not only fine-grained synchronization across APs but also joint processing of sensing data at the CPU~\cite{Buzzi2024}.

Another challenge in CF-ISAC multi-target sensing is echo incorporation, which involves matching reflections acquired at distinct APs to the adequate originating targets. This is particularly difficult when targets are closely spaced, have similar RCS, or are moving at comparable velocities. Interference between concurrent sensing operations further exacerbates this issue. Therefore, coordination among APs is essential~\cite{Buzzi2024}.

To manage interference and improve scalability, a potential solution is to partition the scan area into disjoint or partially overlapping radar subregions (or cells), with specific AP clusters responsible for each subregion. APs may participate in multiple overlapping cells to provide redundancy and improve spatial resolution. However, this requires spatial and temporal coordination protocols to prevent the concurrent illumination of neighboring subregions, which could lead to cross-subregion interference and degraded detection performance. Time-division or frequency-division scheduling, alongside spatial filtering techniques, may assist in preventing mutual interference.

Advanced signal processing techniques such as compressed sensing, multi-static radar signal fusion, and sparse recovery algorithms can also be employed to resolve closely spaced targets in cluttered environments~\cite{Buzzi2024}. These approaches capitalize on the sparsity of the target scene in the angle, range, and Doppler domains, allowing the CF-ISAC system to recover more targets than the number of individual AP observations~\cite{Buzzi2024}. Moreover, the centralized processing of CF-ISAC allows for global data fusion, where received signals from all APs are jointly processed at the CPU to estimate the number, position, and motion parameters of multiple targets. This centralized processing, however, imposes stringent requirements on fronthaul capacity and latency, especially when real-time tracking is required~\cite{Buzzi2024}.

\section{Synchronization}
Synchronization is essential for reliable operation in CF-ISAC systems~\cite{Demir2021book,  Ngo2017, Zhang2019cellfree}. With the distributed architecture of CF, synchronization ensures coherent transmission and reception among APs. It encompasses timing, frequency, and phase alignment, which are critical for achieving high data rates in communication and precise target detection in sensing applications~\cite{Demir2021book, Ngo2017, Zhang2019cellfree}. 

In communication, synchronization enables distributed beamforming, which relies on phase coherence across APs to focus energy constructively at the user locations \cite{Demir2021book,  Ngo2017, Zhang2019cellfree}. Any phase misalignment caused by oscillator drifts, clock offsets, or propagation delay disparities leads to destructive interference, reducing SE and compromising the quality-of-service (QoS). For sensing, particularly in multi-static configurations, synchronization is even more critical \cite{Cheng2024}. Precise timing alignment is required to accurately determine target range, velocity (via Doppler shift), and AoA based on the time delay and frequency shift of the received echoes \cite{Cheng2024}.

In CF-ISAC, the synchronization challenge is more severe than in traditional co-located systems due to the lack of a shared local oscillator and the spatial separation of APs. Practical issues such as hardware imperfections, frequency offsets, and oscillator phase noise can result in non-coherent signal transmission and reception, which significantly degrades both communication throughput and sensing accuracy \cite{Cheng2024}. These impairments are particularly harmful to joint beamforming and coherent signal fusion processes, key enablers of CF-ISAC performance.

To address these challenges, CF-ISAC systems may leverage Global Positioning Systems (GPS)-disciplined oscillators or use centralized synchronization protocols distributed via fronthaul links~\cite{Cheng2024}. GPS-based methods provide high timing accuracy but may be unsuitable for indoor or underground deployments. Alternatively, fronthaul-based approaches can distribute reference timing and phase information from the CPU to the APs. They require low-latency and high-bandwidth backhaul links to minimize synchronization errors. However, as the number of spatial distributed APs increases, achieving and maintaining precise synchronization becomes more complex. Non-ideal clock behaviors, such as phase noise and jitter, introduce additional errors that distort processed echo signals, degrading detection accuracy and overall system performance~\cite{Cheng2024}.

\section{Interference Management}
Interference management is a critical challenge in CF-ISAC systems due to the shared use of spectral, temporal, and spatial resources for simultaneous communication and sensing operation~\cite{Niu2024}. In contrast to conventional co-located systems with clearly defined cell borders, CF-ISAC networks depend on a dense deployment of distributed APs to jointly detect targets and serve all users. This dense and overlapping spatial coverage increases the risk of inter-AP and inter-functional interference, particularly when many APs simultaneously perform joint beamforming for DL communication while illuminating the environment for radar sensing~\cite{Meng2024}. 

A key source of interference is co-channel operation, in which communication signals and radar probing waveforms share the same frequency bands. In such cases, sensing signals may leak into communication receivers, interfering with DL communication signals, and and vice versa. The issue is exacerbated by the geographical proximity of APs as well as asynchronous operations caused by imperfect network coordination~\cite{Niu2024}.

To mitigate these issues, CF-ISAC systems can employ joint interference-aware adaptive beamforming, interference alignment, and spectrum sharing protocols~\cite{Niu2024, Meng2024}. Adaptive and coordinated beamforming is particularly effective in CF architectures, where APs can use their spatial DoF to steer nulls in interference-prone directions or shape beams to serve users while also illuminating regions of interest \cite{Demir2021book,  Ngo2017, Zhang2019cellfree}. The distributed nature of CF also enables spatial reuse and cooperative interference nulling across multiple APs. Moreover, interference alignment can be utilized in the spatial and temporal domains to constrain interference into a reduced-dimensional subspace, allowing both functionalities to coexist with minimal mutual interference \cite{Khodkar2022}. While interference alignment has been investigated in communication-centric CF systems \cite{Khodkar2022}, applying it to ISAC scenarios necessitates incorporating radar-specific requirements, such as waveform orthogonality and target resolvability, into the alignment framework.

Additionally, resource allocation strategies, including dynamic spectrum access and power control algorithms, are essential to optimize the coexistence of communication and sensing functionalities. For example, CF-ISAC systems can benefit from context-aware dynamic spectrum access, which schedules sensing and communication tasks based on target density, user distribution, or application priority. Consequently, the spectrum can be prioritized for high-resolution sensing in safety-critical zones or increased communication SE in high-demand areas. Also, power control algorithms can alter AP transmit power to balance interference mitigation, SE, and coverage \cite{Demir2021book,  Ngo2017, Zhang2019cellfree}. Finally, ML-based interference prediction and avoidance methods, such as reinforcement learning or federated learning, can be used to enable APs to adjust their transmission policies proactively based on previous interference patterns and local data \cite{Demirhan2024ML}.

\section{Fronthaul Capacity and Latency}
The performance of CF-ISAC systems is heavily dependent on the fronthaul infrastructure, which connects spatially distributed APs to the CPU or cloud controller. In CF-ISAC networks, this infrastructure allows not only traditional communication-related data sharing, such as UL/DL CSI, user data, and beamforming weights, but also time-sensitive radar sensing data, including reflected echoes and target parameter estimates \cite{Masoumi2020}. This dual-functional requirement imposes significant demands on the fronthaul in terms of bandwidth/capacity, latency, and reliability  \cite{Masoumi2020}.

Low fronthaul latency is particularly important for ensuring accurate synchronization across distributed APs, which is required for coherent transmission, joint beamforming, and multi-static radar operations. When latency is high, outdated CSI and delayed sensing feedback can cause ineffective beam steering and missed detections, especially in highly dynamic scenarios with mobile users and targets \cite{Masoumi2020}. On the other hand, limited bandwidth restricts the amount and frequency of CSI updates and sensing data exchange, reducing the system's resolution and adaptability.

These limitations are further exacerbated by the joint processing requirements of CF-ISAC. For example, multi-static radar sensing across several APs necessitates the rapid collection, fusion, and processing of echo signals from all relevant nodes \cite{Masoumi2020}. This is highly sensitive to both fronthaul delay and jitter, as a minor misalignment in time-of-arrival measurements can significantly reduce the accuracy of range, velocity, and angle estimations. Similarly, joint communication beamforming requires up-to-date CSI aggregated across the network, which becomes challenging under fronthaul bottlenecks.

To address these challenges, several mitigation strategies can be adopted. Specifically, edge computing and decentralized signal processing architectures can offload real-time tasks such as local beamforming, preliminary target detection, or local CSI estimation, reducing the dependency on high-capacity centralized fronthaul links. Moreover, fronthaul compression and quantization-aware signal design can reduce the payload while preserving key information. For sensing, this includes compressive radar processing where sparse or low-rank structures in the radar scene are exploited. For communication, schemes such as functional splits or CSI prediction can reduce the frequency and volume of CSI reporting \cite{Masoumi2020}. Additionally, latency-aware resource scheduling can dynamically prioritize data streams, e.g., by assigning higher priority to radar updates during sensing-critical windows or communication packets with stringent QoS constraints. Fronthaul-aware beamforming designs can also account for imperfect or delayed information, using robust optimization techniques to hedge against uncertainty.

Advanced high-capacity fronthaul technologies, such as millimeter-wave wireless links, free-space optics, and wavelength-division multiplexed optical fiber, can support the high bandwidth and low-latency requirements of CF-ISAC. These technologies, when combined with software-defined networking for dynamic fronthaul reconfiguration, pave the way for scalable and resilient CF-ISAC deployments~\cite{Masoumi2020}. 
	\chapter{Open Research Directions and Future Trends}\label{chp_future}

This chapter outlines emerging research directions to advance CF-ISAC systems. It explores network-assisted coordination for dynamic AP operation, novel antenna technologies like fluid antennas and holographic MIMO, and the potential of near-field propagation in upper mid-band frequencies. It also examines the integration of complementary technologies such as non-orthogonal multiple access (NOMA), reconfigurable intelligent surfaces (RIS), and UAVs, and highlights the role of machine learning in addressing CF-ISAC design and optimization challenges.

\section{Network-Assisted Cell-Free ISAC}
In network-assisted CFMM, the CPU plays a pivotal role in coordinating the operational modes of APs, determining whether they function in UL or DL mode. This coordination effectively enables FD operation within the CF architecture~\cite{Wang:TCOM.2020}. By dynamically assigning APs to transmit DL signals or receive UL signals and echoes based on real-time network demands, user distribution, and channel conditions, this approach can significantly enhance SE and reduce interference across the network~\cite{Wang:TCOM.2020, Mohammadi:TCOM:2024}.

Network-assisted CF-ISAC is particularly promising in meeting both UL and DL communication requirements, while simultaneously supporting sensing functions. This system enables flexible multi-static sensing, where APs can be adaptively assigned to transmit DL sensing signals or receive UL echoes, depending on the current network demands. Furthermore, the framework supports the simultaneous transmission and reception of both UL and DL signals~\cite{Zeng2023, zeng2024}, providing the system with greater flexibility and capacity.

However, interference between communication and sensing tasks remains a significant challenge. UL communication signals received by APs operating in UL mode contain not only echo-sensing signals but also cross-link interference from DL APs. Similarly, sensing signals at UL APs are subject to interference from both UL and DL communications, which can degrade the overall sensing performance. To mitigate these issues, the resource allocation in the network-assisted CF-ISAC system must be optimized to carefully design UL and DL power control coefficients, as well as configure AP mode operations to meet both communication and sensing demands.

One potential solution is to partition the UL APs into two distinct groups: one group dedicated solely to sensing tasks, while the other is reserved for UL communications. This division can help alleviate interference and ensure that the system operates efficiently while simultaneously supporting both communication and sensing functionalities.

\section{New Antenna Technologies and Cell-Free ISAC}
To fully exploit spatial resources and further enhance ISAC performance, novel antenna technologies, such as fluid antennas, also referred to as movable antennas, have recently been introduced in the literature~\cite{Zou:WCL:2024, Zhou:WCL:2024, Qin:WCL:2024}. The core concept behind fluid antennas is their ability to change position within a specified region, thus providing spatial diversity while requiring fewer RF chains in a relatively compact space~\cite{Wang:WL:2024}. This characteristic proves particularly beneficial at user terminals, where available space is often limited. By deploying fluid antennas at both APs and multi-antenna users in CFMM networks, new spatial DoF are introduced, enhancing the performance of ISAC applications.

However, several challenges must be addressed when integrating fluid antennas into such systems. Channel estimation and the optimization of position adjustments represent significant hurdles. Moreover, as the system involves high-dimensional optimization problems, the complexity of such tasks is expected to increase, especially when considering the unique requirements of ISAC scenarios. While fluid antenna systems can exploit spatially correlated signals, the use of artificial intelligence (AI) presents an opportunity to overcome these challenges. AI's ability to uncover hidden correlations could greatly assist in the design and optimization of fluid antenna systems within ISAC applications.

Another promising antenna technology that is gaining significant attention is holographic MIMO (HMIMO). This technology leverages affordable, transformative wireless planar structures composed of sub-wavelength metallic or dielectric scattering particles, which can manipulate EM waves in highly controlled ways to achieve specific objectives~\cite{Huang:WCL:2020}. HMIMO surfaces can function as transmitters, receivers, or reflectors, making them an extremely versatile solution for creating reconfigurable wireless environments. In~\cite{Adhikary:IoT:2024}, an AI-based framework has been developed to integrate HMIMO APs into CF networks, enabling efficient power allocation for beamforming in the desired direction, with a focus on ISAC. This framework optimally allocates power for beamforming by selectively activating the necessary grids within the serving HMIMO APs, based on user demands. The authors formulate an optimization problem aimed at maximizing the sensing utility function, which in turn improves the SINR of received signals, enhances the sensing SINR of reflected echo signals, and boosts EE. By ensuring effective power allocation, this approach enhances both communication and sensing performance in CF-ISAC systems.

\section{Cell-Free ISAC and Near-Field}
The growing academic and industrial interest in the upper mid-band spectrum ($7$-$24$ GHz), also referred to as FR3~\cite{Zhang:MCOM:2025, chaves2024coverage, Azar2024}, has brought renewed attention to extremely large-scale MIMO (XL-MIMO) antenna arrays. At these frequencies, either the significantly increased antenna aperture size or the shorter wavelengths cause the Fraunhofer distance, defined as $2D_{\text{array}}^2/\lambda$, where $D_{\text{array}}$ is the array's physical aperture and $\lambda$ is the wavelength, to become considerably large~\cite{Azar2024}. As a result, users operating at relatively short distances may fall into the near-field region, marking a paradigm shift from traditional far-field assumptions, especially as we transition from 5G to 6G networks.

In the near-field, beam propagation exhibits a spotlight-like focus due to the spherical wavefronts and their curvature. Unlike in the far field, where beams radiate conically, near-field beams maintain a tighter directional focus over limited distances. One of the key advantages of near-field beam focusing is that XL-MIMO systems can resolve users in both angular and radial (distance) dimensions. This enables richer spatial discrimination in multi-user scenarios, thereby enhancing channel capacity and user separation~\cite{Zhang:MCOM:2023, Cui:MCOM:2023}. Moreover, by leveraging the full geometric information contained in the spherical wavefronts, near-field systems can achieve high-precision 3D localization, which is particularly valuable for ISAC functionalities~\cite{Azar2024}. These benefits are especially relevant in the mmWave and sub-THz frequency bands, where shorter wavelengths make near-field conditions more prevalent. However, the practical Fraunhofer distance, which defines the boundary of the near-field region, remains limited to just a few tens of meters for typical 6G base station apertures operating in the upper mid-band~\cite{Cui:MCOM:2023, Lei:WC:2025}. This poses a critical question: How can CFMM systems, which consist of distributed APs with smaller antenna arrays, exploit near-field properties for ISAC?

A promising direction, discussed in~\cite{bjornson2024enabling}, involves deploying coordinated subarrays separated by only a few meters. This configuration allows for near-field beam focusing without relying on a single large continuous aperture. Each subarray captures locally planar wavefronts, but when signals are jointly processed across subarrays, the system reveals the global spherical wavefront structure. This concept aligns closely with the CFMM architecture, where spatially distributed APs can collectively emulate large-aperture near-field capabilities, thus enabling multi-static sensing and fine-grained localization.

However, effectively leveraging near-field gains in CF-ISAC systems comes with practical challenges. Accurate simulation of near-field behavior requires careful AP placement, precise synchronization, and high-capacity fronthaul to ensure coherent signal processing. These infrastructure requirements could lead to increased power consumption, potentially undermining the EE benefits anticipated in future 6G networks. As such, EE remains a key design consideration when incorporating near-field features into CF-ISAC architectures.

\section{Consolidation of Complementary Technologies into Cell-Free ISAC}
To further boost the performance and versatility of CF-ISAC systems, the integration of complementary technologies has become a prominent research direction. One such advancement is the incorporation of NOMA. In~\cite{Dong:IoT:2024}, a NOMA-assisted CF-ISAC framework was proposed to enhance both connectivity and spectral efficiency. Focusing on a single-radar sensing setup, the study jointly optimized user pairing and beamforming to maximize the minimum achievable communication rate while satisfying sensing constraints. This foundational work opens several future avenues, including the use of statistical CSI-based designs to improve robustness under imperfect channel knowledge, or the extension to multi-target sensing scenarios using hybrid instantaneous and statistical optimization techniques.

Another promising technology is the integration of RIS. In~\cite{Abdelaziz:TCOM:2024}, an FD CFMM system augmented with a single RIS was investigated. The RIS was utilized to enhance both radar and communication functionalities by jointly optimizing UL transmit power, DL sensing beamforming vectors, receive beamforming, and RIS reflection coefficients to maximize the weighted sum of SINRs. Future work may involve deploying multiple RISs to increase spatial coverage, improve signal quality, and enhance sensing resolution. Moreover, advanced RIS technologies, such as beyond-diagonal configurations and stacked metasurfaces, offer exciting opportunities to realize more dynamic and reconfigurable propagation environments.

The use of UAVs in CFMM-based ISAC architectures has also been explored. In~\cite{Flores:ICC:2024}, the authors investigated a UAV-assisted CF-ISAC framework with dedicated sensing signals under three deployment scenarios: mobile, tethered, and fixed UAVs. For each case, a joint transmit precoding strategy was proposed to simultaneously meet sensing and communication requirements while adhering to power constraints. This UAV-based architecture introduces aerial flexibility and rapid deployment capabilities, which are especially useful in dynamic or infrastructure-sparse environments.

Together, these complementary technologies, such as NOMA, RIS, and UAVs, can significantly extend the capabilities of CF-ISAC systems. Their integration promises not only performance gains in terms of throughput, sensing accuracy, and spatial coverage but also enhanced adaptability for emerging use cases such as smart cities, disaster response, and autonomous mobility.

\section{Cell-Free ISAC and BackCom}
Backscatter communication (BackCom) has emerged as a cornerstone technology for the ambient-powered IoT \cite{Rezaei2023Coding, Diluka2022, rezaei2023timespread, Galappaththige2023SR, Galappaththige2023RIS, Rezaei2024NOMA, Galappaththige2024passive}. Unlike conventional radios, BackCom tags do not actively generate RF signals. Instead, they modulate and reflect incident carrier waves emitted by surrounding RF sources such as cellular access points, Wi-Fi routers, or TV towers \cite{Rezaei2023Coding, Diluka2022, rezaei2023timespread, Galappaththige2023SR, Galappaththige2023RIS, Rezaei2024NOMA, Galappaththige2024passive}. This reflection-based mechanism allows BackCom devices to operate with ultra-low power consumption, often in the order of nanowatts to microwatts, enabling battery-free or semi-passive IoT operation \cite{Rezaei2023Coding, Diluka2022, rezaei2023timespread, Galappaththige2023SR, Galappaththige2023RIS, Rezaei2024NOMA, Galappaththige2024passive}. While BackCom supports massive device connectivity at minimal cost, it also faces limitations such as short communication range, low data rates, and increased vulnerability to interference in dense deployments \cite{Rezaei2023Coding, Diluka2022, rezaei2023timespread, Galappaththige2023SR, Galappaththige2023RIS, Rezaei2024NOMA, Galappaththige2024passive}.

The integration of BackCom into CF-ISAC systems gives rise to the emerging paradigm of CF integrated sensing and backscatter communication (ISABC) \cite{Diluka2025ISABC, Diluka2023, galappaththige2024RSMA, zargari2024ISABC}. Unlike conventional ISAC, where sensing relies primarily on echoes from random objects, ISABC exploits the structured backscattered signals of tags to jointly perform communication and sensing \cite{Diluka2025ISABC, Diluka2023, galappaththige2024RSMA, zargari2024ISABC}. This dual functionality with CF architecture can introduce new opportunities for low-power IoT networks, where distributed CF APs simultaneously facilitate data transfer, environmental sensing, and device state estimation \cite{Diluka2024CFBiBC}. The inherent advantages of the CF architecture are significant in this context, as the distributed APs can extend coverage, enable multi-static tag detection, and improve reliability in cluttered or obstructed environments \cite{Diluka2024CFBiBC}.

Several promising research directions naturally arise when combining CF-ISAC with BackCom. One key challenge is the joint optimization of communication and sensing, since the same weak backscattered signals must be decoded for data while also being leveraged for environment perception \cite{Diluka2025ISABC, Diluka2023, galappaththige2024RSMA, zargari2024ISABC}. Another avenue is scalability, as the cooperative nature of CF networks is well-suited to handle massive numbers of passive or semi-passive tags while reducing collisions and enhancing detection accuracy \cite{Diluka2024CFBiBC}. EE and SE also play a central role, as CF-ISABC eliminates the need for additional RF sources or sensing hardware by reusing existing communication signals for dual purposes. Finally, reliable deployment will depend heavily on advanced signal processing and learning-based methods for interference suppression, successive interference cancellation, and multi-tag detection in realistic environments.

\section{Machine Learning-Based Techniques for Cell-Free ISAC}
Integrating sensing and communication within CFMM systems enables resource-efficient designs, reducing spectrum congestion and hardware costs through shared infrastructure. However, realizing these benefits depends heavily on the optimal design of ISAC waveforms, precoders, detectors, and estimators. This involves solving complex multi-objective optimization problems, where inherent trade-offs between communication and sensing performance must be carefully balanced. Classical model-based approaches often struggle to provide satisfactory solutions in CF-ISAC contexts due to model inaccuracies, computational intractability, or a lack of generalizable analytical frameworks. In some cases, Pareto-optimal solutions for these multi-objective formulations may be unknown or infeasible for real-time deployment.

In this context, ML techniques, particularly deep learning, have emerged as powerful alternatives. ML-driven methods offer data-centric solutions that bypass the limitations of traditional model-based approaches by learning directly from observed data patterns \cite{Yu2022, Hu2021, Dai2020}. These methods are particularly effective in CF-ISAC settings, where multiple competing objectives, such as SE, sensing accuracy, and EE, must be jointly optimized \cite{Demirhan2024ML}. For instance, deep learning-based beamforming can effectively manage trade-offs by learning unified beamforming strategies that align with both communication and sensing goals \cite{Demirhan2024ML}. These methods capitalize on the underlying low-dimensional structures of high-dimensional, multimodal datasets generated in CFMM environments, extracting relevant features to improve decision-making and control \cite{Cheng2024a}. Moreover, by preserving the 3D structure of data, such as spatial, temporal, and frequency domains, learning-based techniques achieve higher generalization accuracy, even when training datasets are limited.

Beyond beamforming, deep learning facilitates joint estimation of communication and sensing channel parameters, including CSI, delay spreads, Doppler shifts, and AoA. These capabilities significantly enhance system performance in tasks such as data decoding, target localization, and environment perception. Furthermore, ML introduces a layer of intelligence and adaptability to CF-ISAC networks, making them more responsive to dynamic conditions. Key applications of ML-enhanced CF-ISAC include autonomous navigation, precise localization, human activity recognition, object tracking, smart surveillance, environmental monitoring, and wide-area imaging--all of which benefit from the system's ability to learn, predict, and adapt in real time.
%

\bibliographystyle{IEEEtran}
\bibliography{ref}

\end{document}